\documentclass[11pt]{book}

\usepackage[left=4.2cm,right=4.2cm,top=4.1cm, bottom=4.1cm]{geometry} 
\usepackage{graphicx}
\usepackage{amsmath}
\usepackage{amssymb}
\usepackage{epsfig}
\usepackage{conventions}
\usepackage{natbib}
\usepackage{fancyhdr}
\usepackage{sectsty}
\usepackage{txfonts} 
\usepackage{appendix}
\usepackage{axodraw}

\renewcommand{\chaptermark}[1]{\markboth{#1}{}} 
\renewcommand{\sectionmark}[1]{        
\markright{\thesection.\ #1}}         
\newcommand{\comment}[1]{}

\fancypagestyle{plain}{
                        \fancyfoot{}
                        \fancyhead{}
                      }

\makeatletter 
\def\cleardoublepage{\clearpage\if@twoside
  \ifodd\c@page\else    
    \hbox{} 
  \thispagestyle{empty}
  \newpage 
 \if@twocolumn\hbox{}\newpage\fi\fi\fi}
\makeatother

\newcommand{\unity}{1\hspace{-1.3mm}1}
\newcommand{\simordertwo}{\raisebox{-4pt} {$\, \stackrel{\textstyle
<}{\sim} \,$}} \def\sumint{\hbox{$\sum$}\!\!\!\!\!\!\int}
\def\isumdiff{\hbox{${\scriptstyle \Delta}$}\hspace{-0.275cm}\int}
\def\isumint{\hbox{${\scriptstyle \Sigma}$}\hspace{-0.275cm}\int}
\def\sumdiff{\hbox{$\Delta$}\!\!\!\!\!\!\int}

\subsectionfont{\mdseries\normalsize\bfseries}
\sectionfont{\mdseries\large\bfseries}
\chapterfont{\LARGE\bfseries}
\renewcommand{\theta}{\vartheta}
\newcommand{\donothing}[1]{ }

\title{{\bf Thermodynamics of QCD-inspired theories}}

\author{}
\date{}

\begin{document}

\maketitle
\thispagestyle{empty} 
\phantom{\citet{Andersen2004,Andersen2004b,Warringa2003,Warringa2004,Warringa2005, Warringa2005c, Andersen2006}}

%\phantom{\citet{Andersen2004,Andersen2004b,Warringa2003,Warringa2004,Warringa2005}}
\thispagestyle{empty}
\begin{center}
        {VRIJE UNIVERSITEIT}
        \\
        \vspace*{2.5cm} {\bf
        {\LARGE Thermodynamics of QCD-inspired theories\par}}
        \vspace*{2.5cm}
        {ACADEMISCH PROEFSCHRIFT}
        \\
        \vspace*{1.5cm}
%       \parbox{8.2cm}{
        ter verkrijging van de graad Doctor aan\\
        de Vrije Universiteit Amsterdam,\\
        op gezag van de rector magnificus\\
        prof.dr.\ T.~Sminia,\\
        in het openbaar te verdedigen \\
        ten overstaan van de promotiecommissie \\
        van de faculteit der Exacte Wetenschappen\\
        op dinsdag 28~februari~2006 om 13.45 uur\\
        in de aula van de universiteit, \\
        De Boelelaan 1105 %}
        \vspace*{1.5cm}

        door \\
        \vspace*{1.5cm}

        {\large \bf Harmen Jakob Warringa} \\
        \vspace*{1cm}
        geboren te Emmen
\end{center}

\pagebreak

\thispagestyle{empty}
\noindent
\begin{tabular}{ll}
promotor: & prof.dr.\ P.J.G.\ Mulders \\
copromotor: & dr.\ D.\ Boer 
\end{tabular}
\indent

\pagebreak
\thispagestyle{empty}
This thesis is based on the following publications

\begin{itemize}
\item[-]{Jens~O.~Andersen, Daniel Boer and Harmen~J. Warringa,
{\it Thermodynamics of the O(N) nonlinear sigma model in (1+1)-dimensions},
Phys.~Rev.~{\bf D69}, 076006 (2004), hep-ph/0309091.}
\item[-]{Jens~O.~Andersen, Daniel Boer and Harmen~J. Warringa, {\it The effect of
quantum instantons on the thermodynamics 
of the $\mathbb{C}P^{N-1}$ model}, hep-th/0602082.}
\item[-]{Jens O.~Andersen, Daniel Boer and Harmen J.~Warringa, 
{\it Thermodynamics of O(N) sigma models: 1/N corrections},
Phys.~Rev.~{\bf D70}, 116007 (2004), hep-ph/0408033.}
\item[-]{Harmen~J.~Warringa, Daniel~Boer and Jens~O.~Andersen, {\it
Color superconductivity vs. pseudoscalar condensation in a
three-flavor NJL model}, Phys. Rev. {\bf D72}, 014015 (2005), hep-ph/0504177.}
\item[-]{Harmen~J.~Warringa, {\it Heating the O(N) nonlinear sigma model}, 
in Proceedings of the 43rd Cracow school of theoretical physics,
Acta Phys. Polon. {\bf B34}, 5857
(2003), hep-ph/0309277.}
\item[-]{Harmen J. Warringa, {\it Thermodynamics of the 1+1-dimensional nonlinear
sigma model through next-to-leading order in 1/N}, in Proceedings of
the SEWM2004 meeting, World Scientific (2005),
hep-ph/0408257.} 
\item[-]{Harmen~J.~Warringa, {\it Phase diagrams of the NJL model
with color superconductivity and pseudoscalar condensation}, to appear in proceedings
of the XQCD'05 workshop, hep-ph/0512226.}
\end{itemize}
\pagebreak %intro

\fancyhf{}
\fancyhead[LE,RO]{\thepage} 
\fancyhead[RE]{{\it Contents}} 

\tableofcontents

\newpage
\fancyhf{}
\fancyhead[LE,RO]{\thepage} 
\fancyhead[RE]{\itshape\leftmark}
\fancyhead[LO]{\itshape\rightmark}

\chapter{Introduction}

What happens to matter when you squeeze it further and further? And
what if it will be heated more and more?  Liquid water for example
will turn at some point into a different phase called steam when it is
heated. If instead the density is increased at room temperature by
applying an external pressure, water will subsequently turn into
different types of ice, called ice VI, ice VII, etc.  Such phase transitions
are not specific for water, but they can take place in any interacting
substance, like for example hadronic matter.

Hadronic matter is matter built out of quarks and gluons. The neutron
for instance is a form of hadronic matter, it is a bound state of two
down-quarks, one up-quark and gluons which keep the quarks
together. Imagine a hypothetical situation in which one has lots of
these neutrons in a box. Then increase the temperature. What will
happen? At some point the kinetic energy of the quarks which build up
the neutron will become larger than the energy that is gained by confining
the quarks inside the neutron. At this point the neutrons will cease
to exist. The matter in the box is now in a new phase, which is called
the quark gluon plasma. An order of magnitude estimate of this
transition temperature can easily be made. Classically the kinetic
energy of a quark at temperature $T$ is about $3 T/2$\footnote{In this
thesis natural units are used, so $c=1$, $\hbar = 1$ and $k_B = 1$.}.
Since the masses of the quarks are much smaller than the neutron mass,
the mass of the neutron is almost completely due to confining energy.
So the confining energy per quark is about $m_N / 3
\approx 300\;\mathrm{MeV}$.
This implies that the transition
temperature $T_c$ to the quark gluon plasma should be around $T_c =
200\;\mathrm{MeV}$, which is very hot (actually it is about $10^5$
times the temperature of the solar core). In nature these temperatures
were achieved in the early universe and are possible in relativistic
heavy ion collisions for extremely short periods of time. 

Now start again from scratch at low temperatures and squeeze the box
further and further. At some point the neutrons will start to overlap
and at even higher densities they will cease to exist as separate
entities. Around this point it is expected that quarks will form
Cooper pairs. The matter will transform into a so-called
color-superconducting state. An order of magnitude estimate shows that
this will occur around densities of about $m_N / V_N
\approx 160\;
\mathrm{MeV}\cdot\mathrm{fm}^{-3} \approx 3 \times
10^{17}\;\mathrm{kg} \cdot \mathrm{m}^{-3}$. This density is quite
large, and takes only place in extremely dense objects like for
example neutron stars. 

The theory which describes the interactions between the quarks
mediated by gluons is called quantum chromodynamics (QCD) and is
treated in more detail in the following section. Using QCD one could
in principle predict the behavior of matter under these extreme
circumstances by calculating its {\it equation of state} (that is the relation
between its pressure and energy density) and its {\it phase
diagram}, which are discussed in Secs.~1.2 and 1.3 respectively. The
situations in which these extreme circumstances are realized in nature
are reviewed in Sec.~1.4. Since it turns out that QCD is very
complicated at the energy scales around the phase transition, this
thesis will deal with models inspired by QCD to describe matter at
high temperatures and densities, as will be discussed in more detail
in Secs.~1.5 and 1.6.  A more extensive review on QCD at high
temperatures and densities can be found in for example
\citet{Meyer-Ortmanns1996},
\citet{Rajagopal2000} and
\citet{Rischke2003}.

\section{Quantum chromodynamics}
Quantum chromodynamics (QCD) is a non-Abelian $\mathrm{SU}(3)$ gauge
field theory which describes the interactions between the quarks. It
is a generalization of Maxwells theory of electromagnetism. Like the
electrons, quarks carry a charge, called color. Unlike the
photons in electromagnetism, the gluons, which are the force carriers of QCD
carry a color charge as well. As a result the gluons interact
with themselves and with quarks. Due to the $\mathrm{SU}(3)$
gauge symmetry QCD has three different color charges, named red, blue and
green. Together with the electroweak theory, QCD is one of the
building blocks of the Standard Model of elementary particle physics.
QCD is defined by the following Lagrangian density
\begin{equation}
\begin{split}
 \mathcal{L} &= \bar \psi \left(i \gamma^\mu \mathcal{D}_\mu - m_{0} +
 \mu \gamma_0 \right) \psi - \frac{1}{4} F^{\mu \nu}_a
 F_{\mu \nu}^a \;,
\\
 \mathcal{D}_\mu &= \partial_\mu - i g
 A^a_\mu T_a \;, \;\;\;\;\;\;\;
F_{\mu \nu}^a = \partial_\mu A^a_\nu-
\partial_\nu A^a_\mu + g f^{a}_{bc} A^b_\mu A^c_\nu \;,
 \label{eq:QCDLagrangian}
\end{split}
\end{equation}
where $g$ is the QCD coupling constant, $T_a$ is a hermitian generator
of $\mathrm{SU}(3)$ and $f^a_{bc}$ denotes its corresponding structure
constant. The matrices $m_0$ and $\mu$ are diagonal and contain the
current quark masses and the quark chemical potentials respectively.
There are six different quark flavors. The up, down and
strange quark are relatively light, while the charm, bottom and top
quark are heavy. Since the masses of the heavy quarks are so much
larger than the estimated transition temperature of about 200 MeV,
these quarks will play a minor role at these energies and will
therefore be neglected in this thesis. The discussion in this thesis
will mainly deal with two ($N_f=2$) and three-flavor ($N_f=3$)
situations.  

The chemical potentials are necessary to describe a
system at finite density. The baryon chemical potential $\mu_B = 
\mu_u + \mu_d + \mu_s$ for example, is basically the energy it takes to add
one additional baryon to the system. Temperature is introduced by
considering a Euclidean space in which the $x_0$ direction is made
periodic (for bosons) or antiperiodic (for fermions) with periodicity
$1/T$.  Field theory at finite temperature and densities is discussed
in more detail in Chapter~2.

QCD has different symmetries which are reflected in the hadron
spectrum as a consequence. First of all it is invariant under local
$\mathrm{SU}(3)$ transformations. This implies for example that red up
quarks are as heavy as blue up ones. In addition, in absence of quark
masses and chemical potentials, QCD has a global chiral $\mathrm{SU}(N_f)_L
\times \mathrm{SU}(N_f)_R$ symmetry. Moreover it has a global $\mathrm{U}(1)_B$
symmetry related to baryon number conservation and a global
$\mathrm{U}(1)_A$ (axial) symmetry. At low temperatures and chemical
potentials it turns out that the chiral symmetry is spontaneously
broken down to $\mathrm{SU}(N_f)_V$ giving rise to $N_f^2 - 1$
massless pseudoscalar Goldstone modes.  For $N_f = 2$ these are the
three pions, for $N_f=3$ also the four kaons and the $\eta$ particle
are among the pseudoscalar Goldstone modes. If chiral symmetry is
broken the $\langle \bar u u \rangle$, $\langle \bar d d \rangle$, and
$\langle \bar s s \rangle$ condensates obtain a vacuum expectation
value.

However, in reality the quarks have a small mass. Therefore chiral
symmetry is only an approximate symmetry, as a result the pions, the
kaons and the $\eta$ particle become massive. This remaining
(approximate) $\mathrm{SU}(N_f)_V$ symmetry is the reason why the
constituent quark model of Gell-Mann works as well as it does.
Particles are eigenstates of the QCD Hamiltonian which due to the
symmetry commutes with $\mathrm{SU}(N_f)_V$. Hence the particles can
be classified by the representations of $\mathrm{SU}(N_f)_V$.  At high
temperatures and/or densities the chiral symmetry is approximately
restored, giving rise to a phase transition.  Although the
$\mathrm{U}(1)_A$ symmetry is broken due to nonzero quark masses as
well, it also has another reason of breakdown.  The non-trivial
topological vacuum structure of QCD due to instantons is causing
$\mathrm{U}(1)$ axial symmetry breaking too, which explains the
relatively high mass of the $\eta$ meson \citep{tHooft1976}.

One of the mysteries in QCD is confinement. It turns out
experimentally that hadrons, which are bound states of quarks, carry
no color. Colored objects, like freely moving quarks, do not occur in
nature at low energies. In numerical computations (lattice QCD) it is
confirmed that QCD has this confinement property. But a detailed
understanding of the confinement mechanism is still lacking. At high
temperatures and/or chemical potentials it is expected that matter
will be in a deconfined phase, which means that in that situation
quarks are liberated from the hadrons. Whether the deconfinement phase
transition for light quarks coincides with the chiral symmetry
restoration transition is an important issue which has not yet been
resolved.

QCD is asymptotically free, this implies that the effective coupling
of quarks to gluons becomes smaller at high energies. So at high
energy scales, larger than about $1\;\mathrm{GeV}$, QCD is a theory of
weakly interacting quarks and gluons. Due to the small coupling
constant it is possible to perform calculations in this regime using
perturbation theory. But, at lower energies QCD becomes strongly
coupled and perturbation theory breaks down. The order of magnitude estimate
of the transition temperature $T_c \approx 200\;\mathrm{MeV}$ from the
first paragraph tell us that QCD is likely to be strongly coupled
around the phase transition. Hence it is expected that perturbation
theory will fail to describe QCD near $T_c$. This will be illustrated
next by comparing perturbative and lattice
calculations of the QCD pressure.

\section{QCD equation of state}
Two important macroscopic thermodynamical quantities are the pressure
$\mathcal{P}$ and the energy density $\mathcal{E}$ of QCD.  The
relation between $\mathcal{E}$ and $\mathcal{P}$ is called the
equation of state and determines for example the behavior of matter
created in a relativistic heavy ion collision, the properties of a
(neutron) star and the evolution of the early stage of the universe
shortly after the big bang, see Sec.~1.4.  Due to asymptotic freedom,
QCD describes a gas of weakly interacting quarks and gluons in the
limit of very high temperature. In that case the coupling constant
is small, and hence perturbation theory should be
applicable. However as will be shown next perturbation theory does not
work for the temperatures of in the neighborhood of $T_c$.

\subsection*{Perturbative calculations}
\begin{figure}[t]
\begin{center}
\scalebox{0.5}{\includegraphics{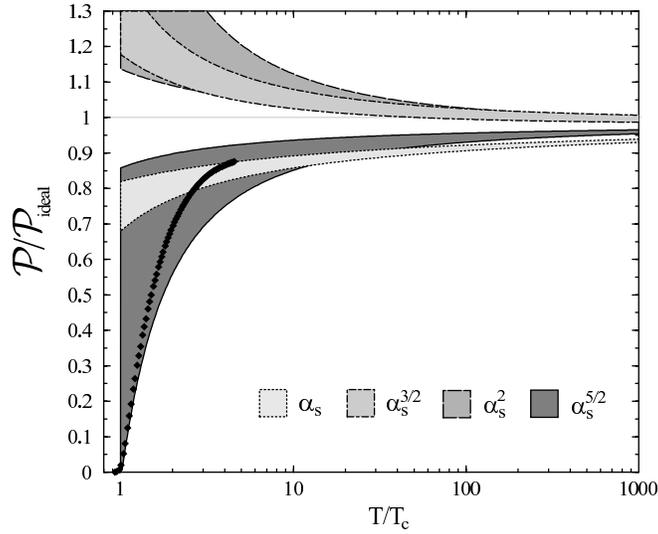}}
\end{center}
\caption{Perturbative results (indicated by their orders) and lattice
results of \citep{Boyd1996} (indicated with black diamonds) of the
pure glue QCD pressure normalized to the ideal gas value, as a
function of $T/T_c$. The shaded regions arise due to varying the
$\overline{MS}$ renormalization scale between $\pi T$ and $4 \pi T$
and are hence an indication of the error that is inherent to
truncating the perturbative series. The order $\alpha$ calculation was
performed by \citet{Shuryak1978}, $\alpha^{3/2}$ by
\citet{Kapusta1979}, $\alpha^{3/2} \log \alpha$ by
\citet{Toimela1983}, $\alpha^2$ by \cite{Arnold1995}, $\alpha^{5/2}$
by \citet{Zhai1995} and \citet{Braaten1996} and order $\alpha^3 \log
\alpha$ (not shown in figure) by \citet{Kajantie2002}. This figure is
adapted from \citet{Andersen2002b}.  }
\label{fig:pertqcdpres}
\end{figure}

The results of the perturbative calculation of the pure glue (which
means no quarks) QCD pressure compared to numerical calculations
(lattice QCD) are displayed in Fig.~\ref{fig:pertqcdpres} up to order
$\alpha^{5/2} = (g^2 / 4\pi)^{5/2}$. It can be seen in the figure that
for temperatures in the neighborhood of the transition temperature
$T_c$, the results are varying a lot upon inclusion of higher order
corrections. This implies that the perturbative expansion breaks down,
even for very high temperatures of around $1000 T_c$.  However, by
reorganizing the perturbative series using resummation methods
\citep{Andersen1999, Andersen2002b} and self-consistent
approaches \citep{Blaizot1999} based on hard thermal
loops \citep{Braaten1990} 
it is possible to obtain reliable results for $T
\gtrapprox 3T_c$.  In perturbation theory it is impossible to
calculate the order $g^6$ contribution because an infinite number of
diagrams are contributing in this order and cannot be resummed as
was argued by \citet{Linde1980}.

It is possible to perform perturbation theory for finite chemical
potentials as well, but it again fails for densities where the phase
transition occurs. Also this perturbation series can be improved by
applying the so-called hard dense loop resummation method
\citep{Andersen2002c}.

\subsection*{Lattice calculations}
The best known method to obtain the QCD thermodynamical quantities
from first principles near the phase transition temperature is by
lattice calculations. In these calculations, space-time is discretized
and replaced by a lattice of a finite size. The fermion fields live on
the vertices of this lattice, the gauge fields are replaced by links
connecting the different vertices. All thermodynamical quantities can
be obtained by numerically calculating the partition function
(discussed in Chapter~2). This requires integration over the fermion
fields and links. Since this results in a huge number of integrations,
typical lattice sizes are taken to be rather small, in the order of ten
points for each dimension. Due to this modest lattice sizes, the
particles with low mass (like the pion), which can propagate over
longer distances are not very well described. The integrations in
lattice calculations are performed statistically, using
importance sampling Monte-Carlo methods. This works fine for QCD at
zero chemical potential. But at finite baryon chemical potential, the
contribution from the fermions, the so-called fermionic determinant,
becomes imaginary (see also Chapter~7). As a result, the integrand of
the partition function becomes oscillatory, which hampers the importance
sampling methods. This complication is called the fermion sign
problem. Small baryon chemical potentials are, however, accessible by
making a Taylor expansion around $\mu_B = 0$ \citep{Allton2002, Fodor2001,
deForcrand2002, DElia2003}.

\begin{figure}[t]
\begin{center}
\scalebox{0.8}{\includegraphics{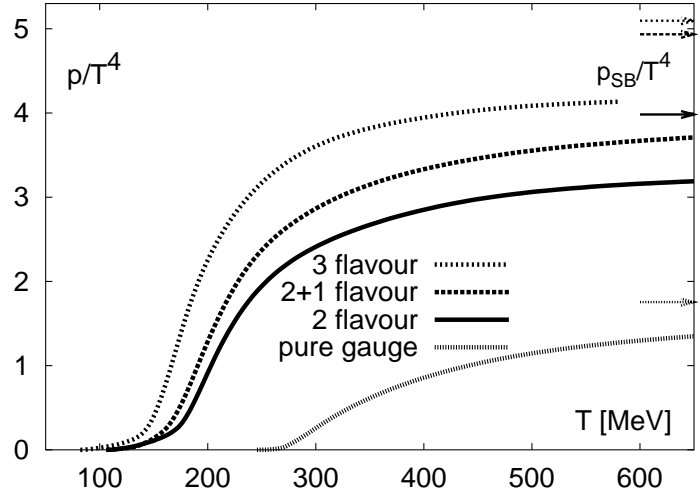}}
\end{center}
\caption{Lattice calculation by \citet{Karsch2000} 
of the pure glue and full QCD pressure (for 2 light quark flavors, 2
light \& 1 heavy flavor and 3 light flavors), normalized to $T^4$, as
a function of $T$.  The arrows indicate the limit of the pressures
when $T \rightarrow \infty$ and are calculated in Sec.~2.4. 
This figure is adapted from \citet{Karsch2002}.}
\label{fig:fullQCDpres}
\end{figure}

In Fig.~\ref{fig:fullQCDpres} the lattice results of the pressure of
QCD for different numbers of quark flavors are displayed as a function
of $T$ for $\mu=0$. Here $T_c \approx\;170\;\mathrm{MeV}$. From the figure it
can be seen that the QCD pressure rises quickly after passing the
transition point. This may be an indication that many new degrees of
freedom are formed, just what one would expect if the quarks become
deconfined from the hadrons.  The figure shows that even around $4
T_c$ the pressure is still far away from that of a freely interacting
gas of quarks and gluons $\mathcal{P}=\mathcal{P}_\mathrm{SB}$. While
the lattice gives reliable results for $T> T_c$, the data at $T<T_c$
can not be trusted. This is because the pion, the lightest particle of
QCD, which is expected to dominate the pressure of QCD at low
temperatures is still far too heavy on the lattice.

For $T > T_c$ the lattice data for the QCD pressure can be fitted to
quasiparticle models \citep{Peshier1995, Levai1998, Schneider2001}. The
results of these fits can be extended to finite chemical potential
\citep{Rebhan2003, Thaler2004}. In this way the equation of state of
QCD at finite chemical potential for $T>T_c$ can be predicted.

It is generally believed that the lattice calculations of the pressure
are reliable for temperatures around $T_c$ and higher.  The results at
high temperatures can be predicted by hard-thermal-loop resummation
methods. At low temperatures lattice QCD is still unreliable due to
the unrealistically large pion mass. In order to predict the pressure
of QCD at low temperatures one can use a low-energy effective theory
in order to describe the hadron gas phase as is done in this thesis in
Chapter~6 and will be discussed in more detail in Sec~1.5.

\section{QCD phase diagram}
By considering the behavior of order parameters (parameters which
vanish in one phase and are non-vanishing in another) one can determine
a phase diagram. One typically distinguishes between two different
types of phase transitions, a first order phase transition in which
the order parameter changes discontinuously and a second order phase
transition in which the derivative of the order parameter changes
discontinuously. Higher order phase transitions are also possible,
but are often called second order transitions as well.
In addition, cross-over transitions in which the
order parameter changes smoothly can occur. The different possibilities
are displayed in Fig.~\ref{fig:trans}.

In the limit of zero quark masses the chiral condensates, $\left<\bar u
u\right>$, $\langle \bar d d \rangle$ and $\left < \bar s s
\right>$ are order parameters for the breaking
of chiral symmetry. In Sec.~1.1 it was mentioned that the chiral
symmetry is already explicitly broken in the QCD Lagrangian density
due to the non-zero quark masses. In that case the chiral condensates
are only approximate order parameters.  The order parameter for the
confinement/deconfinement transition in the limit of infinitely heavy
quarks is the trace of the so-called Polyakov loop. For finite quark
masses no order parameter for this transition is known, see for
example \citet{Weiss1993}.

\begin{figure}[t]
\begin{center}
\input{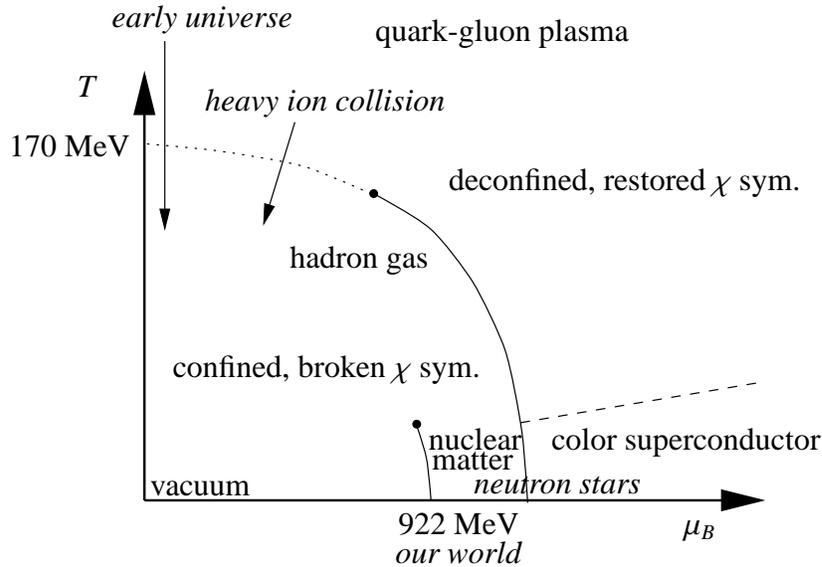}
\end{center}
\caption{Schematic structure of the current understanding of the phase diagram of QCD.
First order phase transitions are indicated with a solid line, second
order with a dashed line and a cross-over with a dotted line. In the
diagram the cooling trajectory of the early universe and of matter
produced in heavy ion collisions is sketched. It is also roughly
indicated in which phase the matter inside neutron stars can be.}
\label{fig:qcdphase}
\end{figure}

In Fig.~\ref{fig:qcdphase} the current understanding of the QCD phase
diagram is displayed schematically. As was discussed in the previous
section, only results for zero and small baryon chemical potential can
be obtained from lattice QCD. Lattice calculations find a cross-over
transition at $T_c =170\;\mathrm{MeV}$. The rest of the phase diagram
is not yet obtained from first principles QCD but can be estimated by
means of effective models like the NJL model studied in Chapter~7.
However, the phases with temperatures and densities much higher than
the densities and temperatures where the phase transition takes place
are accessible by hard-thermal loop and hard-dense loop resummation
techniques as was discussed in the previous section. The NJL phase
diagram as a function of $\mu_B$ and $T$ is displayed in
Fig.~\ref{fig:mubT}.

As mentioned before, finite chemical potentials are needed to describe
a system at finite density. However, it is important to keep in mind
that the relation between chemical potential and number density is not
linear. Especially at a first order phase transition, a single value
of a chemical potential can correspond to a whole range of densities,
as is illustrated for the NJL model in Fig~\ref{fig:numberdens}. In
this case one also speaks of a mixed phase, two phases can occur
together. The world we live in is an example of a mixed phase of
nuclear matter and vacuum.  One should be aware that the real phase
diagram of matter is not just the QCD phase diagram. To obtain the
complete phase diagram of matter, one should also take into account
the electromagnetic and weak interactions.

The best known point in the QCD phase diagram, is the transition from
the vacuum to the nuclear matter phase, there is a first order phase
transition at $\mu_B = 922\;\mathrm{MeV}$, see also
\citet{Halasz1998}.

It is illustrated in Fig.~\ref{fig:qcdphase} that matter at low
chemical potentials and temperatures matter is in a confined phase in
which chiral symmetry is broken as well. If the temperature is
increased, matter goes according to the current understanding via a
cross-over transition to the deconfined and chirally symmetric phase
at low chemical potentials, and via a first-order transition at higher
chemical potentials. This deconfined phase is called the quark-gluon
plasma. The point in which the first-order transition goes over to a
cross-over is called the critical endpoint. It still is uncertain
where this critical endpoint lies exactly in the phase diagram. At low
temperatures and high chemical potentials due to an attractive
interaction, quarks can form Cooper pairs just like electrons in
ordinary superconductivity. This phenomenon will be discussed in more
detail in Chapter~7.

The phase diagram in Fig.~\ref{fig:qcdphase} is displayed as a
function of baryon chemical potential and temperature. Of course it is
interesting to investigate other phase diagrams as well, for example
as a function of the quark masses, as is discussed in
\citet{Laermann2003}.  In Chapter~7 phase diagrams of the NJL model
will be investigated for unequal chemical potentials and
temperature. In that case a new possibility appears which is not
present in Fig.~\ref{fig:qcdphase}, namely quarks can form
pseudoscalar condensates, like the pion condensate $\langle
\bar u i \gamma_5 d \rangle$.

In Fig.~\ref{fig:qcdphase} it is also indicated which part of the
phase diagram can be investigated using relativistic heavy ion
collisions, which kind of matter neutrons stars are presumably made
of, and through which phases the early universe went shortly after the
big bang. In the following section these three situations will be
examined in somewhat more detail.

\section{Matter under extreme conditions}
Typical situations in which temperatures and densities could be high
enough for deconfinement to occur, are the universe just after the big
bang, during heavy ion collisions and inside very compact neutron
stars. In this section a short overview of these situations is
given. More extensive discussions can be found in for example
\citet{Ellis2005} (the big bang in relation to heavy ion collisions),
\citet{Gyulassy2005} (heavy ion collisions) and \citet{Weber2005}
(neutron stars).

\subsection*{The big bang}
In one of the earliest stages of the universe, about $10^{-6}$ seconds
after the big bang, the universe was still so hot that the matter
inside was in the deconfined phase, i.e.\ the quark gluon
plasma. Since at that time the particles and antiparticles had not
annihilated yet, the baryon chemical potential was very small. As is
indicated in Fig.~\ref{fig:qcdphase} when the universe cooled, it
probably went through a cross-over transition to the confined
phase. Since a cross-over transition is smooth, it is unlikely that
the expansion of the universe was modified substantially during this
transition.

To describe the evolution of the universe one has to use an equation
of state. Most often a simple equation of state is used,
$\mathcal{P} = \mathcal{E}/3$ for the radiation dominated era at early
times, and $\mathcal{P} = 0$ for the matter dominated era which
occured later. The description can be made more realistic by using the QCD
equation of state.

\subsection*{Relativistic heavy ion collisions}
The behavior of matter under extreme circumstances can be studied
experimentally using heavy ion collisions. Such experiments have been
performed at the Super-Proton-Synchrotron (SPS) at CERN and are being
performed at the Relativistic Heavy Ion Collider (RHIC) at
BNL. Currently new accelerators which, among other things, will be
used for relativistic heavy ion collision studies are being build at
CERN (the Large Hadron Collider (LHC)), and at GSI
(Schwerionen-Synchrotron (SIS 200)).

In a typical relativistic heavy ion collision two incoming heavy
nuclei (for example gold with 197 nucleons) collide at relativistic
energies. At RHIC these energies are up to 200 GeV per nucleon.
During the collision a large fraction of the kinetic energy is
converted into particles. Therefore statistical methods can be used to
describe the system. High temperatures and energy densities are
achieved during these collisions. The baryon chemical potential
remains low in a heavy ion collision. The reason for this is that due
to the large production of particle antiparticle pairs, the initial
dominance of particles over antiparticles is washed out. Since SIS 200
will operate at a lower energy than RHIC and LHC, a higher baryon
chemical potential can be achieved. However as a result the final
temperature will be lower, which could make it more difficult to probe
the phase transition.

At RHIC it seems that the produced matter quickly achieves thermal
equilibrium. After that moment, relativistic hydrodynamics can
describe the evolution using the QCD equation of state. This indicates
that the matter created during the collisions at RHIC has a very low
viscosity, possibly the most perfect fluid ever made.

\begin{figure}[t]
\begin{center}
 \scalebox{0.5}{\includegraphics{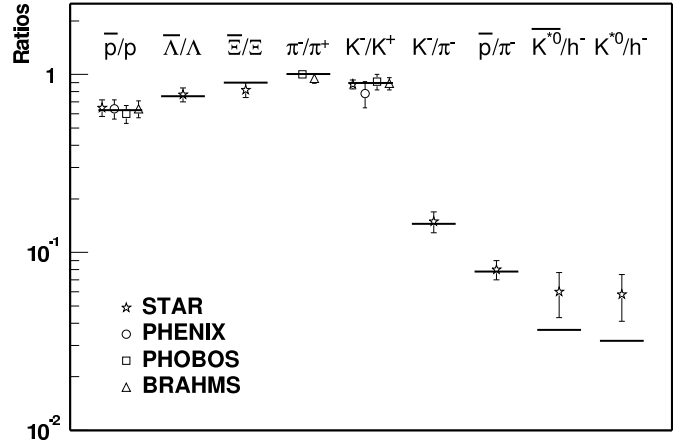}}
\end{center}
\label{fig:partyields}
\caption{Prediction of ratios of produced particles for 
gold-gold collisions at $\sqrt{s} = 130\;\mathrm{GeV}$ by a
statistical model compared to results of the four different experiments at
RHIC. This figure is adapted from
\citet{Braun-Munzinger2001}.}
\end{figure}

That a statistical description to model heavy ion collisions works, is
illustrated in Fig.~\ref{fig:partyields}. In that figure, a
prediction of the ratio of particle yields is compared to the
experimental data of the four different experiments at RHIC
\citep{Braun-Munzinger2001}. The statistical model only has two fit
parameters, $T \approx 174\;\mathrm{MeV}$ and $\mu_B \approx
46\;\mathrm{MeV}$.  It can also be seen from the figure that the
baryon chemical potential is not that big compared to the temperature,
because the ratio between particles and antiparticles of the same
hadronic species are close to unity.

The main objective of these heavy ion collision experiments is to
produce the quark gluon plasma and measure its properties. One of the
clearest signals for the creation of a very dense and hot state of
matter comes from the suppression of back-to-back correlations between two high
transverse-momentum jets. In a collision at low energy densities, such a jet
should be correlated with a jet produced in the opposite direction due
to momentum conservation. This is indeed observed in heavy ion
collision experiments. However, at energies of 200 GeV per nucleon in
central gold-gold collisions at RHIC, this correlation has suddenly
disappeared. This indicates that the momentum of the opposite jet is
absorbed in a very hot and dense medium, probably the quark-gluon
plasma.

Using heavy ion collisions it is difficult to reach high chemical
potentials, so in order to investigate that situation one has to look
to extremely compact objects like neutron stars.

\subsection*{Neutron stars}
If the mass of a star is larger than about ten times the solar mass,
the fusion process can continue until an iron-nickel core is
formed. Then fusion will stop since iron is the nucleus which has the
lowest binding energy per nucleon. As a result the temperature of the
core will drop, hence the pressure will go down. Then the
gravitational interactions cause the core to collapse until nuclear
densities are reached. At this point the collapse stops because it
takes a lot of energy to squeeze the core further. This creates a
shock wave which as a result emits all of the matter from the shells
surrounding the collapsed core. This event is called a supernova
explosion. The remaining core cools down and becomes a neutron star or
at even higher densities a black hole. Typical densities could be so
high that the core of a neutron star is a color superconductor.  Using
the QCD equation of state at high-densities, the so-called
Tolman-Oppenheimer-Volkov relation can be applied to calculate the
mass as a function of the radius of a neutron star, see for example
\citet{Fraga2001} and \citet{Andersen2002c}. These mass-radius
relationships can be compared to observations. However, until now
there has not yet been discovered a neutron star from which one is
certain that the inner core is so dense that it must be in a color
superconducting phase.

\section{QCD inspired theories}
As was argued in the previous sections, due to the nonperturbative
nature of QCD near the phase transition, it is not known how to obtain
for example the order of the phase transition, the equation of state
and the phase diagram using analytical methods from first principles
QCD for all temperatures and chemical potentials. Also the
confinement/deconfinement mechanism is not understood analytically.
However, toy models (models which share features with QCD) and
low-energy effective theories (theories which describe QCD in the
non-perturbative regime) can
be studied to learn about certain aspects to be addressed below.

\subsection*{Toy models}
Toy models for QCD are models which have features in common with
QCD. One can study these models in order to learn about
nonperturbative methods and phenomena like for example the behavior of 
thermodynamical quantities, the generation of a mass gap, confinement
and the importance of topological configurations like instantons.  In
this thesis two of these toy models are studied, the $\mathrm{O}(N)$
nonlinear sigma model in $1+1$ dimensions in Chapter~3 and the
$\mathbb{C}P^{N-1}$ model in $1+1$ dimensions in Chapter~4. The
nonlinear sigma model is a real scalar field theory which like QCD is
asymptotically free and has a dynamically generated mass gap. The
$\mathbb{C}P^{N-1}$ model contains complex scalar fields and
$\mathrm{U}(1)$ gauge fields. This model is also asymptotically free,
has a dynamically generated mass gap, contains instanton
configurations and has the confinement property. In these two models
it is possible to expand in the number of fields $N$. This expansion
is called the large-$N$ approximation and is a method which can give
insight in the nonperturbative behavior of these theories. This method
will be used throughout this thesis and is discussed in more detail in
Chapter~3. In this thesis the pressure for both models is calculated
to next-to-leading order in $1/N$. In this way it is possible to check
the validity of the $1/N$ expansion and to investigate the effects of
instantons on the thermodynamical quantities in the
$\mathbb{C}P^{N-1}$ model. 

\subsection*{Low-energy effective theories}
Low-energy effective theories describe QCD in the non-perturbative
regime where perturbation theory is no longer applicable.  The
simplest examples of such an effective theory are the $\mathrm{O}(4)$
linear and nonlinear sigma model in $3+1$ dimensions. The
$\mathrm{O}(4)$ symmetry (which is locally isomorphic to the chiral
$\mathrm{SU}(2)_L \times \mathrm{SU}(2)_R$ symmetry of the two-flavor
QCD Lagrangian) of these models is spontaneously broken to
$\mathrm{O}(3)$ (which is isomorphic to the remaining chiral
$\mathrm{SU}(2)_V$ symmetry of the QCD vacuum state). Since these
models have the same symmetry breaking pattern as two-flavor QCD, they
serve as a low-energy effective theory for 2-flavor QCD in which only
the pion and the sigma meson can occur. Moreover for the same reason,
the phase transition of the $\mathrm{O}(4)$ (non)linear sigma model
falls in the same universality class as the 2-flavor QCD chiral phase
transition. This allows one to study the order of the phase transition
and critical exponents of QCD using the $\mathrm{O}(4)$ (non)linear
sigma model. In this thesis the pressure of these models is calculated
to next-to-leading order in $1/N$ in Chapter~6. In this way a
prediction for the pressure of QCD at temperatures below $T_c$ is
made, where the lattice calculations are not reliable.

Another effective theory studied in Chapter 7 of this thesis, is the
Nambu--Jona-Lasinio (NJL) model. In this model the gluon exchange
between the quarks is replaced by a 4-point quark interaction. As a
result some features of QCD like confinement and asymptotic freedom
are lost. However, low-energy properties like the meson masses are
described very well using this model. Furthermore this model has the
same pattern of chiral symmetry breaking as QCD. Therefore, it is
expected that the NJL model gives a realistic qualitative description
of the QCD phase diagram for low temperatures and chemical
potentials. In Chapter~7 phase diagrams with different up, down and
strange quark chemical potentials are calculated, in order to study
the competition between phases in which color superconductivity is
possible and in phases where the pions or kaons condense.

\section{Overview of this thesis}
To summarize, the thermodynamics of QCD inspired theories is studied
in this thesis. In Chapter~2 a short introduction to finite density
and temperature field theory is given. Particular emphasis is put on
the analytic and numerical calculation of a combination of a sum and
an integral, which will be required frequently throughout this thesis.
In Chapter~3 the effective potential is derived from which by
minimization one can derive the thermodynamical quantities like the
pressure and determine the phase diagram. Almost all calculations in
this thesis are based on evaluating this effective
potential. Moreover, Chapter~3 discusses the $1/N$ approximation which
can give insight into non-perturbative physics. This is also a basic
ingredient for all the calculations performed in this thesis.  The
thermodynamics of the $\mathrm{O}(N)$ nonlinear sigma model in $1+1$
dimensions is studied in Chapter~4. In Chapter~5, the effect of
quantum instantons on the thermodynamical quantities is investigated
using the $\mathbb{C}P^{N-1}$ model in $1+1$ dimensions. Chapter 6 is
devoted to the study of the thermodynamics of the $\mathrm{O}(N)$
linear and nonlinear sigma model in $3+1$ dimensions, which can be
used to predict the pressure of QCD at low temperatures. The NJL model
and its phase diagrams are discussed in Chapter~7.

\pagebreak
\section{Notations}
Several notations are used throughout this thesis. These will be
summarized here.

\begin{itemize}
\item{
Euclidean momentum vectors are denoted by a capital letter, that is $P
= (p_0, \vec p\,)$.  The length of a momentum vector $\vec p$ is denoted as $p =
\vert \vec p\,\vert$.  }
\item{
The integral over Euclidean momenta, the bosonic sum-integral 
and the fermio\-nic sum-integral are respectively defined as
\begin{equation}
  \int_P \equiv \int \frac{\mathrm{d}^{d+1} P}{(2\pi)^{d+1}} \;,
 \;\;\;\;
  \sumint_P \equiv T \!\!\!\! \sum_{p_0 = 2\pi n T} 
 \int \frac{\mathrm{d}^{d}p}{(2\pi)^{d}} \;,
\;\;\;\;
  \sumint_{\{P\}}\!\! \equiv T \!\!\!\!\!\! \sum_{p_0 = 2\pi (n+1/2) T + i \mu} 
 \int \frac{\mathrm{d}^{d}p}{(2\pi)^{d}} \;.
\end{equation}}
\item{
The difference of a sum-integral and an integral are for
bosonic fields and fermi\-onic fields respectively defined as
\begin{equation}
  \sumdiff_P f(P) \equiv \sumint_P - \int_P \;,
 \;\;\;\;\;\;\;\;\;
  \sumdiff_{\{P \}} \equiv \sumint_{\{P\}} - \int_P \;.
\end{equation}}
\item{
Integration over space(-time) will be depending on the context written as
\begin{equation}
 \int_x \equiv \int \mathrm{d} x_0 \int \mathrm{d}^d x \;\;\;\;\;
\mathrm{or} \;\;\;\;
 \int_x \equiv \int_0^\beta \mathrm{d} x_0 \int \mathrm{d}^d x \;.
\end{equation}
}
\end{itemize}

 %intro
\chapter{Finite temperature and density field theory}

In a system with a large number of particles like for example a gas,
it is very cumbersome, if not impossible, to calculate the
trajectories of individual particles.  On the other hand collective
properties, like
 the pressure and number densities, characterize the
system of particles as a whole and are therefore in many cases much
more interesting than the behavior of individual particles.  These
collective properties can really be calculated using statistical
methods.

In this chapter first the basics of classical statistical physics will
be summarized. Then, by using path-integrals in Euclidean space-time,
classical statistical physics will be cast in a form suitable for
quantum field theories, called finite temperature and density field
theory. As an illustration the pressure of a free scalar and of a free
fermion field theory are obtained. Both expressions for the pressure are
explicitly evaluated after a general explanation of how frequency sums,
that naturally arise in finite temperature calculations, can be
computed. At the end of this chapter further techniques for the
evaluation of frequency sums are developed, which are useful for the
numerical computation of more complicated sums that arise in
interacting field theories.

A more extensive introduction to finite temperature and density
field theory can be found in the books by \citet{Kapusta1989}
and \citet{LeBellac2000} and the review article of
\cite{Landsman1987}.

%---------------------------------------------------------------------
\section{Classical statistical physics}
Consider a box of volume $V$, having total energy $E$ and filled with
$N$ particles. This box can be divided into regions 1 and 2. Clearly,
it holds that $V = V_1 + V_2$, $E = E_1 + E_2$ and $N = N_1 +
N_2$. Let $\Omega(E, V, N)$ be the number of ways in which the total
energy $E$ can be distributed over $N$ particles in a volume $V$. This
quantity $\Omega$ is called the number of micro-states. Now the two
postulates of statistical physics are,
\begin{enumerate}
\item{All micro-states are equally likely to occur.}
\item{In equilibrium the system will choose the state that is the most likely
to occur.}
\end{enumerate}
Combining postulate 1 and 2 gives that the equilibrium state is the
one with the highest number of micro-states.  The number of
micro-states of the complete box can be written as the product of the
micro-states of the two different regions, $\Omega = \Omega_1
\Omega_2$, where $\Omega_{i} = \Omega_{i}(E_{i}, V_{i}, N_{i})$. It is
convenient to turn this equality into an additive relation by
introducing a quantity called the entropy $S$ which is defined as $S =
\log \Omega$. Then the total entropy of the box $S$ is equal to the sum of the
entropies of the different regions, $S = S_1 + S_2$.  The two
postulates can be translated into the condition that the total entropy
is maximal in equilibrium. If the entropy is maximal it holds that
\begin{equation}
  0 = \pdercol{S}{E_1}_{V, N} = \pdercol{S_1}{E_1}_{V_1, N_1}
  + \pdercol{S_2}{E_1}_{V_2, N_2} = 
  \pdercol{S_1}{E_1}_{V_1, N_1} - \pdercol{S_2}{E_2}_{V_2, N_2} \;,
\end{equation}
where it was used that the total energy $E=E_1+E_2$ is constant. It
follows that in equilibrium
\begin{equation}
  \pdercol{S_1}{E_1}_{V_1, N_1} = \pdercol{S_2}{E_2}_{V_2, N_2} \;.
\end{equation}
In equilibrium the temperatures $T_i$ of the two regions $1$ and $2$
should be equal, that is $T_1 = T_2$. Hence $\partial S_1 / \partial
E_1$ should be some function of temperature. The correct definition of 
temperature turns out to be
\begin{equation}
  \frac{1}{T_1} \equiv \pdercol{S_1}{E_1}_{V_1, N_1} \;,
\end{equation} 
because in that way it is possible to derive the experimentally verified
ideal gas law, $P V = N T$. Similarly since $N=N_1+N_2$ is constant
it holds that in equilibrium
\begin{equation}
 \pdercol{S_1}{N_1}_{E_1, V_1} = \pdercol{S_2}{N_2}_{E_2, V_2} \;.
\end{equation}
In equilibrium it take as much energy to transfer one particle from
region 1 to region 2 as to do the opposite. This energy is called
chemical potential, so in equilibrium the chemical potentials should
be equal, that is $\mu_1 = \mu_2$. As a result $\partial S_1 /
\partial N_1$ should be some function of chemical potential. It turns
out that the correct definition is
\begin{equation}
  \mu_1 = - T_1 \pdercol{S_1}{N_1}_{E_1, V_1} \;.
\end{equation}
because it gives rise to the correct distribution function for
fermions, Eq.~(\ref{eq:fermidiracdistr}).

Now consider a box of fixed volume $V$ placed in a very large heat
bath of constant temperature $T$ and constant chemical potential
$\mu$. The box is allowed to exchange energy and particles with the
heat bath.  The total system of heat bath and box together has energy
$E_0$ and contains $N_0$ particles. The probability $p_r$ that the box
has energy $E_r$ and contains $N_r$ particles is equal to the
probability that the heat bath has energy $E_0 - E_r$ and contains
$N_0 - N_r$ particles. So it follows that $p_r$ is proportional to
$\Omega(E_0 - E_r, N_0 - N_r)$, the number of micro-states of the heat
bath. In terms of entropy one has that
\begin{equation}
  p_r = C^{-1} \exp \left[S(E_0 - E_r, N_0 - N_r) \right] \;,
\end{equation}
where $C$ is a normalization factor.
Assuming the heat bath is large implies that $E_0 \gg E_r$ and
$N_0 \gg N_r$. Hence it is possible to expand the entropy
of the heat bath around $E_0$ and $N_0$,
\begin{equation}
  S(E_0 - E_r, N_0 - N_r) =  
  S(E_0, N_0) - E_r \pderst{S(E,N)}{E}_{E_0, N_0}
  - N_r \pderst{S(E, N)}{N}_{E_0, N_0} 
+ \ldots \;.
\end{equation}
In higher orders of the expansion one gets terms like $\partial^2 S /
\partial E^2 \vert_{E_0, N_0} \approx \partial (1/T) / \partial E$,
which reflects the change of the heat bath temperature when energy is
transferred into the box. Because it is assumed that the heat bath has
constant temperature $\partial (1/T) / \partial E = 0$, so this higher
order term can be neglected. Another higher order term of the
expansion is $\partial^2 S / \partial N^2 \vert_{E_0, N_0} \approx -
\partial (\mu/T) / \partial N$ which reflects the change of chemical
potential divided by temperature when particles are transferred into
the box. Again because it is assumed the heat bath has constant
temperature and chemical potential, this higher order term can be
neglected too. For the same reason one can assume that $\partial^2 S
/ \partial E\, \partial N |_{E_0, N_0}$ vanishes as well.

As a result, the probability $p_r$ that the box has energy $E_r$ and
contains $N_r$ particles is equal to
\begin{equation}
  p_r = Z^{-1} \exp \left[-\beta (E_r - \mu N_r) \right] \;, 
  \label{eq:probfunc}
\end{equation}
where $Z$ is a normalization factor different from $C$ and $\beta = 1 /
T$. 

The normalization factor $Z$ which is also called the partition
function, is equal to the sum of all probabilities,
\begin{equation}
  Z = \sum_r \exp \left[-\beta (E_r - \mu N_r) \right] \;.
\end{equation}
The partition function contains all information of the collective or
macroscopic behavior of the thermodynamic system. Strictly speaking
this is already a quantum mechanical equation, since the energies are
assumed to be discrete.  In the classical case, one has to replace the
sum over states by an integral. Using the partition function one can
calculate thermodynamical quantities, like the energy density of the
box,
\begin{equation}
 \mathcal{E}  = \frac{1}{V} \sum_r p_r E_r = - \frac{1}{V}\pder{\log Z}{\beta} \;,
\end{equation} 
the number density of particles in the box,
\begin{equation}
 n = \frac{1}{V} \sum_r p_r N_r 
   = \frac{1}{\beta V} \pder{\log Z}{\mu} \;,
\end{equation}
and the entropy density,
\begin{equation}
 \mathcal{S} = \frac{1}{V} \sum_r p_r \log p_r
   = - \frac{\beta}{V} \pder{\log Z}{\beta} \;.
\end{equation}
Using the definition of the pressure, $\mathrm{d} E_r = - P_r \mathrm{d}V$ which is
valid at constant temperature and number density, it follows that the
pressure is given by
\begin{equation}
 \mathcal{P} = - \sum_r p_r \frac{\partial E_r}{\partial V} =
\frac{1}{\beta} \pder{\log Z}{V} \;.
\end{equation}
Typically, the width $L$ of a system is much larger than the inverse
temperature, (i.e.\ $L \gg 2\pi \beta$), such that one can use the
infinite volume limit to describe the thermodynamics of a finite
volume to good approximation. The advantage of the infinite volume
limit is that field theoretic calculations simplify.  In all
calculations performed in this thesis, this infinite volume limit is
taken. Then it turns out that $\log Z$ becomes proportional to $V$,
such that the pressure becomes
\begin{equation}
 \mathcal{P} = \frac{1}{\beta V} \log Z \;.
 \label{eq:pressurefieldtheory}
\end{equation}
Instead of calculating $\log Z / \beta V$ directly, in this thesis the
pressure will be calculated via the effective potential (see
Chapter~3) which in its minimum equals $\log Z / \beta V$.

%---------------------------------------------------------------------
\section{Quantum statistical physics}
Of many physical systems one does not know the energies $E_r$ and the
number densities $N_r$ exactly. Most often only the Hamiltonian
$\oper{H}$ and a corresponding number operator $\oper{Q}$ which
commutes with $\oper{H}$ is known. Denoting the eigenstates of
$\oper{H}$ and $\oper{Q}$ by $\left \vert r \right >$ , the partition
function expressed in terms of the Hamiltonian and number operator
becomes
\begin{equation}
  Z = \sum_r \matr{r}{e^{- \beta\left(\oper{H} - \mu \oper{N}\right)}}{r} =
  \trace \, e^{-\beta \left(\oper{H} - \mu \oper{N}\right) } \;.
\end{equation}
The thermal expectation value of an operator
$\oper{A}$ can also be expressed in terms of a trace,
\begin{equation}
  \expval{\oper{A}} = \frac{1}{Z}\,\trace \! \left[ 
\oper{A} 
  e^{-\beta 
\left(\oper{H}  - \mu \oper{N}\right)
  } \right] \;,
\end{equation}
An expectation value is independent of the choice of basis due to the
cyclic property of the trace.

As an example of the formalism, it will be shown how to derive the
Bose-Einstein distribution function.  This distribution function gives
the number of states as a function of energy and temperature for a
non-interacting bosonic system.  The Bose-Einstein distribution
function is the expectation value of the number
operator. Non-interacting bosons obey the following harmonic
oscillator Hamiltonian
\begin{equation}
 \oper{H} = \tfrac{1}{2} \sum_k \omega_k 
 \left( 
   \danni{a}{k} \dcrea{a}{k} +
   \dcrea{a}{k} \danni{a}{k}
 \right)\;,
\end{equation}
where $\omega_k$ is the energy of the state with momentum $k$
and $\danni{a}{k}$ and $\dcrea{a}{k}$ are respectively annihilation and
creation operators which satisfy the usual commutation relation
for bosonic operators,
$[ \danni{a}{k},\, \dcrea{a}{l}] = \delta_{kl}$ and
$[ \danni{a}{k},\, \danni{a}{l}] = 
  [\dcrea{a}{k},\, \dcrea{a}{l}] = 0$.
The bosonic number operator is given by
$\oper{N}_k = \dcrea{a}{k} \danni{a}{k}$.  
The expectation value of the number operator is
\begin{equation}
   \expval{\oper{N}_k} 
  = \frac{1}{Z}\, \trace \left ( e^{-\beta \left( \oper{H} - \mu \oper{N} \right)}
    \dcrea{a}{k} \danni{a}{k} \right) 
  = \frac{1}{Z}\, \trace \left ( \danni{a}{k} e^{-\beta \left( \oper{H} - \mu \oper{N} \right)}
    \dcrea{a}{k} e^{\beta \left( \oper{H} - \mu \oper{N} \right)} e^{-\beta \left( \oper{H} - \mu \oper{N} \right)} \right) \;,
\end{equation}
With use of the following equality
\begin{equation}
  e^B A \, e^{-B} = A + \comm{B}{A} + \tfrac{1}{2} \comm{B}{\comm{B}{A}} + \ldots \;,
\end{equation} 
and the fact that $\comm{\oper{H} - \mu \oper{N}}{\dcrea{a}{k}} =
\left( \omega_k - \mu \right) \dcrea{a}{k}$ it can be shown that
\begin{equation}
  \expval{\oper{N}_k} = \frac{1}{Z} \trace \left ( \danni{a}{k}
    \dcrea{a}{k} e^{-\beta (\omega_k - \mu ) } e^{-\beta \left( \oper{H} - \mu \oper{N} \right)} \right) =
  \expval{1 - \oper{N}_k} e^{-\beta(\omega_k - \mu)} \;.
\label{eq:boseeinsteinexpval}
\end{equation}
The Bose-Einstein distribution function $n(\omega_k) =
\expval{\oper{N}_k}$ follows from the last equation,
\begin{equation}
  n(\omega_k) = \frac{1}{e^{\beta (\omega_k - \mu)} - 1} \;.
\end{equation}

In a similar way it is possible to derive the Fermi-Dirac distribution,
which is the expectation value of the fermionic number density. The fermion creation
and annihilation operators satisfy anti-commutation relations. As a
result one picks up a minus sign in \eqref{eq:boseeinsteinexpval} when
swapping the annihilation and creation operators.  
The Fermi-Dirac distribution is given by 
\begin{equation}
  \tilde n(\omega_k) = \frac{1}{e^{\beta (\omega_k - \mu)} + 1} \;.
 \label{eq:fermidiracdistr} 
\end{equation}

\section{Statistical field theory}
Using the path integral formalism it is possible to obtain the
partition function of a field theory.  Consider a bosonic field $\hat
\phi(t, \vec x\,)$. The thermal expectation value of a product of two bosonic
fields in equilibrium with a heat bath of temperature $T$ is given by
\begin{equation}
  \expval{\oper{\phi}(t_1, \vec x_1) \oper{\phi}(t_2, \vec x_2)} =
  \frac{1}{Z} \trace \bigl [ 
  \oper{\phi}(t_2, \vec x_2) e^{-\beta \oper{H}}
  \oper{\phi}(t_1, \vec x_1) e^{\beta \oper{H}} 
   e^{-\beta \oper{H}} \bigr ] \;.
\end{equation}
The dynamics of a field $\oper{\phi}(t, \vec x)$ is entirely described by
its Hamiltonian $\oper{H}$, which can be used to determine the time evolution
of the fields,
\begin{equation}
  \oper{\phi}(t, \vec x\,) = e^{i t \oper{H}} \oper{\phi}(0, \vec x\,)
  e^{-i t \oper{H}} \;.
\end{equation}
By identifying $t = i \beta$ a connection between inverse temperature
and imaginary time is found. As a result
\begin{equation}
  \expval{\oper{\phi}(t_1, \vec x_1) \oper{\phi}(t_2, \vec x_2)} =
  \expval{\oper{\phi}(t_2, \vec x_2) \oper{\phi}(t_1 + i \beta, \vec x_1)}
  \;.
  \label{eq:imtimetranslation}
\end{equation}
Using the relation between inverse temperature and imaginary time
the partition function can be written in terms of a path integral.
For this, consider a transition matrix element between an initial
bosonic state $\phi_i$ and a final state $\phi_f$ in ordinary field
theory. Such a transition element in terms of a path integral is given
by the following expression
\begin{equation}
  \left < \phi_f \left \vert \exp[-i(t_f-t_i) \hat H] \right \vert
  \phi_i \right>
 =  \int \mathcal{D}' \phi \; 
       \exp \left [i {\int_{t_i}^{t_f} \measure{t}}
       { \int \measure{^d x}} 
         \mathcal{L[\phi]} \right] \;,
\end{equation} 
where the prime (') on the measure indicates that the path integral is taken
over fields which satisfy the following boundary condition
$\phi(t_{i}, x) = \langle \phi_{i} \vert \hat \phi(t_i, x) \vert  \phi_{i} \rangle$ and
$\phi(t_{f}, x) = \langle \phi_{f} \vert \hat \phi(t_f, x) \vert  \phi_{f} \rangle$.  If one makes
the identification $t = -i \tau$ and if one chooses $t_i = 0$ and $t_f
= -i \beta$ one finds
\begin{equation}
  \left < \phi_f \left \vert \exp[-\beta \hat H] \right \vert
  \phi_i \right>
 =  \int \mathcal{D}' \phi \; 
       \exp \left [- {\int_{0}^{\beta} \measure{\tau}}
       { \int \measure{^d x}} 
         \mathcal{L[\phi]} \right]  \;.
\end{equation}
The last equation enables one to write the partition function
in terms of a path integral,
\begin{equation}
  Z = \sum_{\phi_n} \left< \phi_n \left \vert \exp[-\beta \hat H] \right \vert
  \phi_n \right>
 =  \int 
    \mathcal{D} \phi \; 
       \exp \left [- \int_{0}^{\beta} \measure{\tau}
      \int \measure{^d x} 
         \mathcal{L[\phi]} \right] \;,
\end{equation}
where the integration is implicitly over all fields which obey the
condition $\phi(\tau = 0, \vec x\,) = \pm \phi(\tau = \beta, \vec
x\,)$.  Since $\vert \phi\rangle$ and $-\vert \phi\rangle$ describe
the same physical state the sign of the boundary conditions on the
bosonic fields cannot be determined in this way.  However, this can be
done by considering a two-point function.

For bosonic fields the two-point function evaluated at $\tau = 0$ and
$\tau$, where $\tau$ is between 0 and $\beta$, is given by
\begin{equation}
  \expval{T_\tau  \hat \phi( 0)
   \hat \phi(\tau) }
    =
    \expval{\hat \phi(\tau)  \hat \phi( 0)}
    =
    \expval{\hat \phi(\beta ) \hat \phi(\tau)}
    = 
    \expval{T_\tau \hat \phi(\tau) \hat \phi( \beta)} \;,
   \label{eq:bosongreensfunction}
\end{equation}
where \eqref{eq:imtimetranslation} was used. Here $T_\tau$ indicates
time ordering in imaginary time. By choosing $\tau = \beta$ it follows
from \eqref{eq:bosongreensfunction} that the boundary
condition on bosonic fields has a $+$ sign,
\begin{equation}
  \phi(\tau = 0, \vec x\,) = \phi(\tau = \beta, \vec x\,) \;.
\end{equation}
For fermions similar arguments can be used, but since time
ordering for fermions requires an additional minus sign, one finds an
anti-periodicity condition for fermionic fields
\begin{equation}
  \psi(\tau = 0, \vec x\,) = - \psi(\tau = \beta, \vec x\,) \;. 
\end{equation}

So thermal field theory is in essence a Euclidean field theory
where one dimension ($\tau$) is compactified to a circle.
As a consequence of this, the Fourier transform of a bosonic field 
becomes a sum over so-called Matsubara frequencies, 
\begin{equation}
\phi(\tau, \vec x\,) = {\frac{1}{\beta} \sum_{n}}
     {\int \spaceint{d}{k}} 
    e^{i \omega_n \tau + i \vec k \cdot \vec x} \tilde \phi(K)
     \equiv \sumint_K  
e^{i \omega_n \tau + i \vec k \cdot \vec x} \tilde \phi(K)\;,
\label{eq:momentumdecomscalarfield}
\end{equation} 
where the Matsubara frequencies are $\omega_n = 2\pi n T$. The capital
$K$ is a momentum vector in Euclidean space, $K = (\omega_n, \vec k\,)$.
The symbol $\isumint_K$ denotes a so-called sum-integral, where the
sum is over bosonic modes, and will arise often in finite temperature
calculations. The momentum representation for fermions is given by
\begin{equation}
  \psi(t, \vec x\,) = 
  \frac{1}{\beta} \sum_n 
  \int \spaceint{d}{k} e^{i \tilde \omega_n + i \vec k \cdot \vec x}
   \tilde \psi(K) \equiv \sumint_{ \{ K \} } 
e^{i \tilde \omega_n \tau + i \vec k \cdot \vec x}
  \tilde \psi(K) \;,
\end{equation} 
where the Matsubara frequencies for fermions are $\tilde \omega_n =
(2n + 1) \pi T$. The symbol $\isumint_{\{ K\} }$ denotes a
sum-integral, where the sum is over fermionic Matsubara modes.

Two different integration contours are often used in equilibrium
finite temperature field theory.  In the derivation above, the
Matsubara contour was used. This is a contour starting at $t=0$
straight down the imaginary axis to $t=-i\beta$, which gives rise to
the so-called imaginary-time formulation of thermal field
theory. Another possibility is the Keldysh contour which starts at
$t_i = -\infty$, goes along the real axis to $t_1 = \infty$, down to
$t_2 = t_1 - i \epsilon$, back under the real axis to $t_3 = t_i - i
\epsilon$ and finally to $t_f = t_i - i\beta$.  This Keldysh contour
gives rise to the so-called real-time formalism. The real-time
formalism is favored over the imaginary-time formalism when quantities
have to be obtained in Minkowskian space-time at finite temperature as
is for example the case for spectral densities. To calculate such a
spectral density in the imaginary-time formalism one has to make an
analytic continuation, which can be avoided by using the real-time
formalism.  All calculations in this thesis will be performed using
the imaginary-time formalism.

As an example of finite temperature field theory, the pressure of a
free scalar theory will be calculated. The Lagrangian density of this
theory in Euclidean space is given by
\begin{equation}
  \mathcal{L} = \frac{1}{2} \partial_\mu \phi \partial_\mu \phi +
  \frac{1}{2} m^2 \phi^2 \;,
\end{equation}
where $m$ is the mass of the scalar field. The action of a free field
theory is quadratic in the fields, hence the Gaussian path integral
can be computed exactly (see for example \citet{Weinberg1995},
Chapter~9). One finds using the fact that $\log \mathrm{Det}\, A =
\mathrm{Tr} \log A$
\begin{equation}
  \log Z = 
\parbox{1cm}{\includegraphics{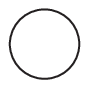}}
=-\frac{1}{2} \mathrm{Tr}
  \, \log \left(-\partial^2 + m^2 \right) \;,
 \label{eq:partfuncscalarfields}
\end{equation}
where the single closed loop denotes the corresponding Feynman diagram
of this contribution to $\log Z$. In contrast to for example a
two-point function, $\log Z$ contains no external vertices, so all its
diagrams are necessarily closed. In an interacting field theory more
complicated closed loop diagrams contribute to the pressure next to
the single closed loop, for examples see Figs.~\ref{fig:bubbles} and
\ref{fig:chains}. The Feynman rules needed to evaluate these kind of 
loop diagrams at finite temperature can be found in for example
\citet{Kapusta1989}.  The functional trace in
Eq.~(\ref{eq:partfuncscalarfields}) is over a complete set of
functions that satisfy the periodic boundary conditions in imaginary
time for scalar fields. The trace can be evaluated by going to
momentum space. As a result
\begin{equation}
 \log Z  = - \frac{\beta V}{2} \sumint_P \log(P^2 + m^2) \;,
 \label{eq:partfuncscalar}
\end{equation}
were the sum-integral is defined in
Eq.~(\ref{eq:momentumdecomscalarfield}). The pressure can now be
calculated by applying Eq.~(\ref{eq:pressurefieldtheory}).  Since it
is only possible to measure pressure differences, it is convenient to
normalize the pressure at zero temperature to zero. In order to do
this the contribution at zero temperature which is
\begin{equation}
-\frac{1}{2} \int_P \log(P^2+m^2) \label{eq:zerotcontr}\;,
\end{equation}
will be subtracted from the contribution at finite temperature
which is
\begin{equation}
 -\frac{1}{2} \sumint_P \log(P^2 + m^2) \label{eq:finitetcontr} \;.
\end{equation}
The zero temperature contribution, Eq.~(\ref{eq:zerotcontr}) is
clearly ultraviolet divergent. It can be evaluated by applying an
ultraviolet momentum cut-off or using dimensional regularization. The
finite temperature contribution, Eq.~(\ref{eq:finitetcontr}) is
ultraviolet divergent as well. Because high-momentum modes at finite
temperature are exponentially suppressed by a Bose-Einstein
distribution function (as will be shown in the next section) and since
Eq.~(\ref{eq:finitetcontr}) becomes equal to Eq.~(\ref{eq:zerotcontr})
in the limit of zero temperature, the divergences of
Eq.~(\ref{eq:zerotcontr}) and Eq.~(\ref{eq:finitetcontr}) are the
same. Hence the difference between those equations, which is the
normalized pressure is ultraviolet finite.  One then finds that the
pressure of a free scalar field in $d+1$ dimensions is given by
\begin{equation}
 \mathcal{P} = - \frac{1}{2} \left[ 
  \sumint_P \log(P^2 + m^2) - \int_P \log(P^2+m^2) \right]
  \equiv
  -\frac{1}{2} \sumdiff_P \log(P^2 + m^2) 
 \label{eq:bosonpres}
 \;.
\end{equation}
In the following section it will be explained how this expression can
be computed.

To obtain the pressure of a free fermion field theory consider
its Lagrangian density in Minkowskian
space 
\begin{equation}
  \mathcal{L} = \bar \psi \left( i \gamma^\mu \partial_\mu 
- m + \mu \gamma_0 \right) \psi \;,
\end{equation}
where $\mu$ is a chemical potential for the fermion particle minus
antiparticle number $\int \psi^\dagger \psi$. In Euclidean
space this Lagrangian density becomes
\begin{equation}
  \mathcal{L} = \bar \psi \left( -\gamma_0 \partial_0 
+ i \gamma_i \partial_i - m + \mu \gamma_0 \right) \psi \;.
\end{equation}
Performing the Gaussian path integral gives that
\begin{equation}
  \log Z = 
\parbox{1cm}{\includegraphics{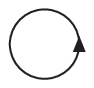}}
= 
\log \mathrm{Det}\, \left( -\gamma^0 \partial_0 + i
  \gamma_i \partial_i - m + \mu \gamma_0 \right) \;,
\end{equation}
where the determinant is over the Dirac indices and a complete set of
functions that satisfy anti-periodic boundary conditions in imaginary
time. After going to momentum space it follows that
\begin{equation}
  \log Z = \beta V \sumint_{\{ P \}} \log \mathrm{det}\, \left(i\gamma_0 \tilde
  \omega_n + 
  \gamma_i p_i - m + \mu \gamma_0 \right) \;.
 \label{eq:presfermtomomentum}
\end{equation}
Evaluating the determinant over the Dirac indices and subtracting
the divergent zero temperature contribution one finds that the
pressure of a free fermion field in 4 dimensions is given by
\begin{equation}
  \mathcal{P} = 2 \sumdiff_{ \{P\} } \log \left(P^2 + m^2 \right) \;,
 \label{eq:fermionpres}
\end{equation}
where $p_0 = \tilde \omega_n + i \mu$. Like in the bosonic case
discussed in the previous paragraph, this pressure is finite.  In the
following section it will be explained how this pressure can be
calculated.

\section{Analytic calculation of sum-integrals}
As was discussed in the previous section one often has to evaluate
sum-integrals in finite temperature field theory. In this section a
method to perform these sum-integrals analytically will be
discussed. The following section is devoted to the numerical
evaluation of sum-integrals.

\begin{figure}[t]
\begin{center}
  \input{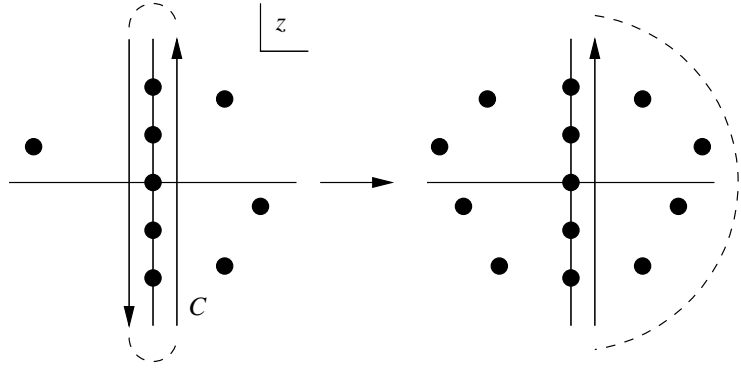} 
  \caption{Contour for summation formula. The black dots on the imaginary
           axis denote the poles of $\coth(\beta z / 2)$, while the other dots
           are possible poles of $f(z)$ (in the left figure) and of
           $f(z) + f(-z)$ (in the right figure).}
  \label{fig:contour}
\end{center}
\end{figure}

In order to calculate a sum-integral, one has to perform an infinite
sum over Matsubara modes, after which the integration over momenta has
to be done. Such a sum over Matsubara modes can be obtained by using
contour integration. Consider a particular sum
\begin{equation}
 \frac{1}{\beta} \sum_{n=-\infty}^{\infty} f(z = i \omega_n) \;,
\end{equation}
where $\omega_n = 2 \pi n T$ as for bosonic fields.  This expression
can viewed as a sum over residues of some function which has simple
poles located at $z= i\omega_n$. A sum over residues is equivalent to
an integration around all the poles, which is sketched in the
left-hand part of Fig.~\ref{fig:contour}.  Consider the function
$\coth(\beta z / 2)$.  It has only simple poles at $z = i \omega_n$,
which all have residue $2 / \beta$. So assuming $f(z)$ has no poles on
the imaginary axis, $g(z) = \coth(\beta z / 2) f(z)$ has simple poles
at $z= i\omega_n$ with residue $2 f(i\omega_n) / \beta$.  This allows
one to write the sum as the following integral
\begin{equation}
  \frac{1}{\beta} \sum_{n=-\infty}^{\infty} f(z = i \omega_n) =
  \frac{1}{2} \frac{1}{2\pi i}
  \int_C \measure{z} f(z) \coth \left (\frac{\beta z}{2} \right) \;,
\end{equation} 
where the contour $C$ is depicted in the left part of Fig.~\ref{fig:contour}. 

Now one can use that $\coth(\beta z / 2) = 1 + 2 n(z) = -1 - 2 n(-z)$,
where $n(z) = 1 / [\exp(\beta z) - 1]$ is the Bose-Einstein distribution
function. If $\lim_{z \rightarrow \infty} z g(z) = 0$, one can split
$C$ in two pieces along the imaginary axis and bring them
together
\begin{equation}
  \int_C \measure{z} g(z) = 
	\int_{-i \infty + \epsilon}^{i \infty + \epsilon}
	\measure{z} \left[ g(z) - g(-z) \right] \;.
\end{equation}
The last equation can be used to write the sum as
\begin{equation}
  \frac{1}{\beta} \sum_{n=-\infty}^{\infty} f(z = i \omega_n) =
  \frac{1}{2} \frac{1}{2\pi i}
  \int_{-i \infty + \epsilon}^{i\infty + \epsilon}
  \measure{z} \left[f(z) + f(-z) \right] \left[ 1 + 2 n(z) \right] \;.
\end{equation}
If $f(z)$ falls off rapidly enough at $z \rightarrow \pm \infty$ it is
possible to close the contour as is done in the right-hand part of
Fig.~\ref{fig:contour}. The integral can now be calculated
straightforwardly by summing over the residues. One should be aware
that the contour in the right part Fig.~\ref{fig:contour} goes
clockwise, so one picks up an additional minus sign when
applying the residue theorem to calculate the integral.

A frequently arising sum (see for example the gap equations calculated
in Chapters 4, 5 and 6) is the one over the propagator which using
$f(z) = 1 / (-z^2 + \omega_p^2)$ with $\omega_p^2 = p^2 + m^2$
results in,
\begin{equation}
  \frac{1}{\beta} \sum_n \frac{1}{P^2 + m^2} = \frac{1}{2\omega_p} 
  \left[ 1 + 2 n (\omega_p) \right] \;,
  \label{eq:bosonpropsum}
\end{equation}
where $P = (p_0, \vec p\,)$  with $p_0 = \omega_n$.
After integrating over spatial momenta one finds the following
important result
\begin{equation}
  \sumint_P \frac{1}{P^2 + m^2}
  = \int \frac{\mathrm{d}^d p}{(2\pi)^d} \frac{1}{2 \omega_p}
 \left[ 1 + 2 n(\omega_p) \right]
 \;.
\end{equation}
By integrating
Eq.~(\ref{eq:bosonpropsum}) over $\omega_p$ the sum-integral of a
logarithmic function can be obtained. This sum-integral arises in the
calculation of the pressure of a bosonic field theory (see for example
Eq.~(\ref{eq:finitetcontr}) of the previous section and Chapters 4, 5
and 6). One finds
\begin{equation}
   \sumint_P \log \left(P^2 + m^2 \right)
   = \int
  \frac{\mathrm{d}^d p}{(2\pi)^d}
\left[\omega_p + 2 T \log
\left(1 - e^{-\beta \omega_p} \right)
\right] + C\;,
\label{eq:sumintpres}
\end{equation}
where $C$ is an infinite constant which is independent of $\omega_p$
and temperature. Equation (\ref{eq:sumintpres}) is ultraviolet divergent;
all divergences arise from the integral over $\omega_p$ and the
constant $C$. The high-momentum modes which depend on temperature are
exponentially suppressed so they do not give rise to divergences.
In the limit of zero temperature a sum over
Matsubara modes changes into an integration over $p_0$, hence 
Eq.~(\ref{eq:sumintpres}) becomes in the limit of zero temperature
\begin{equation}
  \int_P \log \left(P^2 + m^2 \right) = \int
  \frac{\mathrm{d}^d p}{(2\pi)^d} \omega_p + C \;,
\end{equation}
which shows that all ultraviolet divergences of
Eq.~(\ref{eq:sumintpres}) are contained in the zero-temperature
contribution. The pressure of a free bosonic field as is defined in
Eq.~(\ref{eq:bosonpres}) is hence ultraviolet finite and given by
\begin{equation}
 \mathcal{P} = -\frac{1}{2} \sumdiff_P \log(P^2 + m^2)
 = 
  -T \int \frac{\mathrm{d}^d p}{(2\pi)^d} \log
\left(1 - e^{-\beta \omega_p} \right)\;.
\end{equation}

In Chapter 7 a theory with fermions in the presence of a chemical
potential will be discussed. In that case $p_0 = 2\pi (n + 1/2)T + i
\mu = \omega_n + \pi T + i\mu$. Using the formalism presented above it
can be shown that
\begin{equation}
   \frac{1}{\beta} \sum_n \frac{1}{P^2 + m^2} = \frac{1}{2\omega_p}
  \left[1 + \tilde n_+(\omega_p) + \tilde n_-(\omega_p) \right]
 \label{eq:sumintfermprop}
 \;,
\end{equation}
where $\tilde n_\pm(\omega_p) = 1 / ( \exp {\beta (\omega_p \pm \mu)}
+ 1 )$. After integrating Eq.\ (\ref{eq:sumintfermprop}) over
$\omega_p$ the fermionic sum-integral of a logarithmic function can be
obtained. This sum-integral arises in the calculation of the pressure
of a fermionic field theory (see for example
Eq.~(\ref{eq:fermionpres}) in the previous section and Chapter 7). One
finds
\begin{equation}
   \sumint_{ \{ P\}} \log \left(P^2 + m^2 \right)
 = \int  \frac{\mathrm{d}^d p}{(2\pi)^d}
\left[ \omega_p 
+ T \sum_{\pm} \log 
\left(1 + e^{-\beta (\omega_p \pm \mu)} \right)
\right] + C\;,
\label{eq:sumintfermpres}
\end{equation}
where $C$ is a divergent constant which is independent of temperature
and $\omega_p$. Like in the bosonic case, the high-momentum modes of
the term which depends on temperature is exponentially suppressed an
hence does not give rise to an ultraviolet divergence. All divergences
are contained in the zero-temperature contribution, as a result the
pressure of a free fermionic field in 3+1 dimensions (defined in
Eq.~(\ref{eq:fermionpres})) is finite and given by
\begin{equation}
 \mathcal{P} = 2 T \sum_\pm \int \frac{\mathrm{d}^3 p}{(2\pi)^3}
 \log \left(1 + e^{-\beta (\omega_p \pm \mu)} \right) \;.
\end{equation}

If $m=0$ and $\mu =0$ the temperature-dependent parts of the sum-integrals
can be obtained exactly. The integrals which appear after summing over
Matsubara frequencies can be evaluated with use of the
following identities with $\eta = \pm 1$ and $n>1$
\begin{multline}
  \int_0^\infty \mathrm{d}x \frac{x^{n-1}}{e^x - \eta} =
  \int_0^\infty \mathrm{d}x x^{n-1} e^{-x} \sum_{s=0}^{\infty} 
  \eta^s e^{-sx} = 
  \eta \Gamma(n) \sum_{s=1}^{\infty} \frac{\eta^n}{s^n} \\= 
  \left \{ \begin{array}{cc} 
   \Gamma(n) \zeta(n) & \mathrm{if} \; \eta = + 1 \\
   (1 - 2^{1-n}) \Gamma(n) \zeta(n) & \mathrm{if} \; \eta = -1
  \end{array} \right. \;,
  \label{eq:finint}
\end{multline} 
where $\Gamma(n)$ is the gamma function which obeys: $\Gamma(n+1) = n
\Gamma(n) $ and $\zeta(n)$ is the Riemann zeta function. Some useful
values of the Riemann zeta function are: $\zeta(2) = \pi^2 / 6$,
$\zeta(3) \approx 1.202$ and $\zeta(4) = \pi^4 / 90$. Combining Eqs.\ 
(\ref{eq:bosonpres}), (\ref{eq:sumintpres}) and (\ref{eq:finint}) gives
the following result for the pressure of a massless bosonic spin-0
degree of freedom in 3+1 dimensions
\begin{equation}
 \mathcal{P}_b = -\frac{1}{2} \sumdiff_P \log P^2 = \frac{\pi^2}{90}T^4 
 \;.
 \label{eq:presbosondof}
\end{equation}
By combining Eqs.~(\ref{eq:fermionpres}), (\ref{eq:sumintfermpres}) and
(\ref{eq:finint}) the pressure of a massless spin-$1/2$ fermionic field
in the absence of a chemical potential in 3+1 dimensions is obtained,
which reads
\begin{equation}
 \mathcal{P}_f = 2 \sumdiff_{\{P \}} \log P^2 = 
 \frac{7 \pi^2}{180}T^4\;.
  \label{eq:presfermdof}
\end{equation}

The last two equations can be used to obtain the pressure of QCD in
the limit of infinite temperature. Due to asymptotic freedom the
quarks and the gluons are effectively non-interacting in this limit.
Moreover since the temperature is then much larger than the quark
masses, the quarks are effectively massless.  So in the limit of
infinite temperature the pressure of QCD is the sum of the pressure of
eight massless gluons and $N_f$ massless quarks. Since all eight
gluons have two transverse polarizations their contribution to the QCD
pressure in the infinite temperature limit is $16 \mathcal{P}_b$. The
$N_f$ quarks all carry three colors, hence their contribution to the
QCD pressure in the infinite temperature limit is $3 N_f
\mathcal{P}_f$. Adding the gluonic and quark contribution gives the
QCD pressure in the infinite temperature limit which is $\mathcal{P} =
37 \pi^2 T^4 / 90$ for $N_f=2$ and $\mathcal{P} = 19 \pi^2 T^4 / 36$
for $N_f=3$. In the confined phase at low temperatures one expects the
QCD pressure to be dominated by a gas of particles with lowest mass,
which are the pions. Since there are three pions which are spin-0
particles, the pressure of QCD at low temperatures is something in the
order of $\mathcal{P} = \pi^2 T^4 / 30$ (if one takes into account
that the pions have a mass this pressure even becomes smaller, see
Chapter~6). Hence the pressure divided by $T^4$ at low temperatures is
much smaller than in the infinite temperature limit as also can be
seen from Fig.\
\ref{fig:fullQCDpres}, which display the results of lattice
calculations of the pressure of QCD. In this figure, the infinite
temperature limits calculated in this paragraph are drawn as well.

\section{Numerical computation of sum-integrals}
\label{sec:abelplana}
As was discussed in the previous sections, partition functions can be
obtained by calculating sum-integrals. It was shown that these
sum-integrals are typically ultraviolet divergent, however the
difference between a sum-integral and an integral is finite because of
the exponential suppression of high momentum modes. In the examples
discussed in the previous section, the sum-integrals were obtained
relatively straightforwardly because the sum over Matsubara modes
could be performed analytically. However in more complicated cases,
which for example arise in Chapters~4, 5 and 6, an analytic result for
this sum can no longer be obtained. In this section a method will be
developed, which can be used to calculate the sum-integrals
numerically.

The sum-integrals which arise in calculating partition functions are
typically of the following form (see for example
Eq.~(\ref{eq:partfuncscalarfields}))
\begin{equation}
  \sumint_P f(P^2) \;.
  \label{eq:typicalsumint}
\end{equation}
As was argued in the previous section, such sum-integrals are most
often ultraviolet divergent. Hence, an immediate problem which arises
when calculating a sum-integral numerically in the brute-force way
(that is to perform the sum over Matsubara modes numerically and then
to do the integration over spatial momenta numerically as well) are
the ultraviolet divergences.  To treat these ultraviolet divergences
in a consistent way it is useful to split a sum-integral into a finite
part containing the difference between a sum and an integral and a
divergent part in the following way
\begin{multline}
\sumint_P f(P) = 
\sumdiff_P f(P) + \int_P f(P) 
\\
 = \int \!\! \spaceint{d}{p} \left[ \frac{1}{\beta} \!\! \sum_{p_0=2\pi nT} 
 \!\! f(P) - 
  \int \frac{\measure{p_0}}{2\pi} f(P) \right] +
  \int \spaceint{d+1}{P} f(P) 
\;.
 \label{eq:splitsumint}
\end{multline}
In the following two subsection it is subsequently discussed how the
integral and the difference between the sum-integral and the integral
can be computed numerically.

\subsection*{Computation of the integral}
The term $\int_P f(P)$ contains all possible ultraviolet divergences
of $\isumint_P f(P)$ and can be calculated for example by applying a
cut-off or dimensional regularization. Numerically a cut-off
regularization is the easiest, however dimensional regularization
is also possible numerically as was shown by \citet{Caravaglios2000}.
In the cases considered in this thesis,
the ultraviolet divergences $D$ of the term proportional to $\int_P
f(P)$ can always be extracted analytically by considering the
high-momentum behavior $g(P)$ of $f(P)$, i.e.\ $D = \int_P g(P)$ (see
also \citet{Blaizot2003}).  The term which contains the divergences
can now be written as
\begin{equation}
 \int_P f(P) = D + \int_P \left[ f(P) - g(P)
 \right] \;,
\end{equation}
where the last integral is finite and can easily be evaluated
numerically using standard techniques like Gauss-Legendre
integration. The rewriting of the divergences in terms of an integral
prevents subtraction of two large quantities which, due to the finite
machine precision, can give rise to huge numerical errors.  If
$f(P)$ depends explicitly on temperature (see Chapter 4, 5 and 6
for an example) $D$ can have a temperature dependence as well, giving
rise to temperature-dependent ultraviolet divergences. Such 
a divergence gives rise to renormalization problems, therefore
a careful analysis is required.

\subsection*{Computation of the difference between a sum-integral
and an integral} The difference between a sum-integral and an integral
is most often finite because it turns out that its high momentum modes
are exponentially suppressed (as was shown explicitly for two examples
in the previous section). Since both terms in the difference are
divergent, it is not easy to obtain this difference numerically by
using the expression in the second line of
Eq.~(\ref{eq:splitsumint}). However, using contour integration it is
possible to derive an expression which does not contain divergent
parts and hence will be suitable for numerical evaluation.  After
that, this expression has to be integrated over the spatial momenta
which can be done using standard numerical techniques like
Gauss-Legendre integration.

For simplicity $T = 1$ is taken in this section. Results can however
be easily generalized to any $T$ by rescaling.  Consider a function
$f(z)$ which is analytic for $2\pi (N - 1/2) \leq
\mathrm{Re}\, z \leq 2 \pi (M + 1/2)$, where $N$ and $M$ are integers. Using
that $\cot(z/2)$ has poles at $z=2 \pi n$ with residue $2$, it holds that
\begin{equation}
  \sum_{n=N}^{M} f(z = 2 \pi n) = \frac{1}{2\pi i} \frac{1}{2} \oint_C \measure{z}
 f(z) \cot(z/2) \;.
\end{equation}
Because of the analyticity requirement on $f(z)$ one is free to choose
the contour $C$ as long as it is closed, the points $z = 2 \pi n$ with
$N \leq n \leq M$ are included and other possible cuts and poles
of $f(z)$ are excluded.
\begin{figure}[t]
\begin{center}
  \input{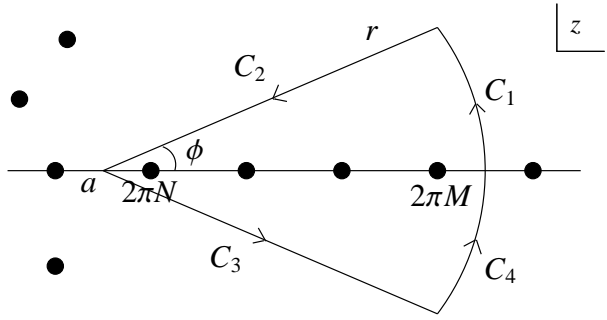} 
  \caption{Contour $C$ for summation formula. The black dots on the real
           axis denote the poles of $\cot(z / 2)$, while the other dots
           indicate possible poles and cuts of $f(z)$.}
  \label{fig:contourAP}
\end{center}
\end{figure}
The contour $C$ used in this section is displayed in Fig.\
\ref{fig:contourAP}, where $C_1$ goes from $a + r$ to $a + r e^{i
\phi}$, $C_2$ from $a + r e^{i \phi}$ to $a$, $C_3$ from $a$ to $a + r
e^{-i \phi}$ and $C_4$ from $a + r e^{-i \phi}$ to $a + r$. Here $a =
2 \pi (N - 1/2)$ and $r = 2\pi (M - N + 1)$.  Now it can be used that
\begin{equation}
  \cot (z/2) = - i \left( 1 + \frac{2}{e^{-iz} - 1} \right)  =
  i \left( 1 + \frac{2}{e^{iz} - 1} \right) \;,  
\end{equation}
to obtain
\begin{multline}
  \sum_{n=N}^{M} f(z = 2 \pi n) = 
	-\frac{1}{4\pi} \int_{C_1 \cup C_2} \measure{z} f(z)
	+\frac{1}{4\pi} \int_{C_3 \cup C_4} \measure{z} f(z) \\
	-\frac{1}{2\pi} \int_{C_1 \cup C_2} \measure{z} f(z) \frac{1}{e^{-iz} - 1}
	+\frac{1}{2\pi} \int_{C_3 \cup C_4} \measure{z} f(z) \frac{1}{e^{iz} - 1}
\;.
\end{multline}
Since it is assumed that $f(z)$ has no poles or cuts within the integration contour,  
\begin{eqnarray}
  \int_{C_1 \cup C_2} \measure{z} f(z) + \intab{a}{a+r} \measure{z} f(z) &=& 0 \;,\\
  \int_{C_3 \cup C_4} \measure{z} f(z) - \intab{a}{a+r} \measure{z} f(z) &=& 0 \;.
\end{eqnarray}
Combining the three equations above gives for the difference $\Delta_{N,M}$ of 
a sum and an integral
\begin{multline}
 \Delta_{N, M} \equiv \sum_{n=N}^{M} f(z = 2 \pi n)  - 
	\frac{1}{2\pi} \intab{2\pi(N-1/2)}{2\pi(M+1/2)}
	 \measure{z} f(z)
	= \\ \frac{1}{2\pi} \left[
	\int_{C_3 \cup C_4} \measure{z} f(z) \frac{1}{e^{iz} - 1}
	-\int_{C_1 \cup C_2} \measure{z} f(z) \frac{1}{e^{-iz} - 1}
	\right] \;.
\end{multline}
Now consider the limit $M \rightarrow \infty$. For functions $f(z)$ which
grow slower than an exponential in the limit $\vert z \vert\rightarrow \infty$, the
contribution coming from integration along $C_1$ and $C_4$ can be
neglected.  In that case 
\begin{multline}
 \Delta_{N, \infty}
	=  -\frac{1}{2\pi} \left[
	\intzepi \measure{\rho} e^{-i\phi} f(a + \rho e^{-i\phi}) \frac{1}{\exp(i\rho e^{-i\phi}) + 1} 
	\right. \\ \left.
	+ \intzepi \measure{\rho} e^{i\phi} f(a + \rho e^{i\phi}) \frac{1}{\exp(-i \rho e^{i \phi}) + 1}
	\right] \;.
\end{multline}
A convenient choice is $\phi = \pi/2$, which gives
\begin{equation}
  \Delta_{N,\infty} = \frac{i}{2 \pi} \intzepi \measure{\rho} \left[f(a - i \rho) - f(a + i\rho) \right]
	\frac{1}{e^\rho + 1} \;, 
  \label{eq:modabelplana}
\end{equation}
where $a = 2 \pi (N - 1/2)$ and $f(z)$ should be analytic for
$\mathrm{Re}\,z > a$. This formula is similar to the Abel-Plana
formula (where $a$ is taken to be $2\pi N$, see for example
\citet{Barton1981}). Eq.~(\ref{eq:modabelplana}) is, however, more convenient 
for numerical purposes due to the larger suppression factor of
$1/[\exp(\rho) + 1]$ as compared to $1/[\exp(\rho) -1]$ in the
original Abel-Plana formula. If $f(z) \in
\mathbb{R}$ for $z \in \mathbb{R}$, Eq.~(\ref{eq:modabelplana}) can be
simplified to
\begin{equation}
  \Delta_{N,\infty} = \frac{1}{\pi} \intzepi \measure{\rho} \mathrm{Im}\,f(a + i\rho)
	\frac{1}{e^\rho + 1} \;, 
 \label{eq:modabelplana2}
\end{equation}

The difference between a sum-integral and an integral can now be
calculated using the following expression
\begin{equation}
  \sumdiff_P f(P) = \int \frac{d^d p}{(2\pi)^d} \Delta_{-\infty, \infty} \;.
  \label{eq:sumdiffintegration}
\end{equation}
The term $\Delta_{-\infty, \infty}$ can be computed using
Eq.~(\ref{eq:modabelplana}) although not necessarily with $N=-\infty$,
since to apply that equation $f(P)$ should satisfy the analyticity
requirements discussed below Eq.~(\ref{eq:modabelplana}). To
illustrate the computation of Eq.~(\ref{eq:sumdiffintegration}) two
examples will be given below.

In order to obtain the pressure of a free scalar field theory (see
Eqs.~(\ref{eq:bosonpres}) and (\ref{eq:sumintpres})) the term
$\isumdiff_P \log(P^2+m^2)$ has to be computed. To evaluate this
expression numerically one takes $f(z) = \log(z^2 +
\omega_p^2)$.  Clearly the sum over all Matsubara modes ($z=2\pi n$)
diverges as does the integral over $z$. However, the difference
between the sum and integral is finite. To calculate this difference
it is not allowed to use
\eqref{eq:modabelplana} with $N=-\infty$, because $f(z)$ has a cut at
$\mathrm{Re} \,z = 0$.  One therefore has to split the difference into 
parts, for example in the following way
\begin{equation}
\Delta_{-\infty, \infty} = \Delta_{-\infty, -1} + \Delta_{-1,1} +
\Delta_{1, \infty} \;.
\end{equation}
Because $f(z)$ is even in $z$, $\Delta_{-\infty, -1} =
\Delta_{1,\infty}$.  The term $\Delta_{1, \infty}$ can be calculated numerically
using Eq.~(\ref{eq:modabelplana2}). The term $\Delta_{-1,1}$ can be
obtained numerically using the explicit expression for the difference
between a sum and integral. As a result
\begin{equation}
\Delta_{-\infty, \infty} = 2 \Delta_{1, \infty} + 
  \log(\omega_p^2) - \frac{1}{2\pi} \intab{-\pi}{\pi} 
 \measure{z} \log(z^2 + \omega_p^2) \;.
 \label{eq:diffboson}
\end{equation}
Now $\isumdiff_P \log(P^2+m^2)$ can be obtained by integrating
Eq.~(\ref{eq:diffboson}) over momenta using
Eq.~(\ref{eq:sumdiffintegration}).

In Chapters 4, 5, and 6 of this thesis, next-to-leading $1/N$
corrections to the pressure are investigated. To obtain these
corrections one has to calculate a sum-integral of the following form
\begin{equation}
 F = \sumdiff_P f(P)
  \label{eq:suminttocalc}\;,
\end{equation}
where $f(P) = \log\left[c + I(P,m) \right]$, $c$ is some constant and
\begin{equation}
  I(P,m) = \sumint_Q \frac{1}{Q^2+m^2} \frac{1}{(P+Q)^2 + m^2} \;.
\end{equation}
The function $f(P)$ is even in $p_0$, has a cut for $\mathrm{Re}(p_0)
= 0$ and depends explicitly on temperature. In order to use
Eq.~(\ref{eq:modabelplana2}) to calculate the difference between a sum
and an integral one has to split $\Delta_{-\infty, \infty}$ as follows
\begin{equation}
  \Delta_{-\infty,\infty} = 2 \Delta_{N,\infty} + \sum_{n =
  -N+1}^{N-1} f(p_0 = 2 \pi n) - \frac{1}{2\pi} \int^{2\pi(N-1/2)}_{-2\pi(N-1/2)}
  \mathrm{d} p_0 f(p_0)
 \label{eq:sumintcalc} \;,
\end{equation}
where $N \geq 1$ because of the cut in $f(P)$. The term
$\Delta_{N,\infty}$ can be obtained numerically using
Eq.~(\ref{eq:modabelplana2}).  It was checked numerically for several
examples that changing $N$ has no effects on $\Delta_{-\infty,
\infty}$ as expected.  After calculating $\Delta_{-\infty,\infty}$ the
integration over $p$ can be done straightforwardly to obtain $F$ using
Eq.~(\ref{eq:sumdiffintegration}). It was observed numerically that
the difference of a sum and an integral is dominated by the
low-momentum modes which can be understood as being due to the
exponential suppression of the high-momentum modes at finite
temperature. Hence after integration over momenta $p$ one obtains a
finite result for $F$ as expected. This conclusion applies to
difference of sum-integrals and integrals of more general $f(P)$
as well.
 %tft
\chapter{The effective potential and the $\boldsymbol{1/N}$ expansion}
\markboth{The effective potential and the $1/N$ expansion}{}

As was discussed in the previous chapter, thermodynamical quantities
can be derived by calculating the partition function. In this chapter,
it will be made clear how such a partition function can be obtained by
locating the extremum of a so-called effective potential. The
derivation of this effective potential will be reviewed in the first
section of this chapter. Then the large-$N$ approximation, which can
be used to investigate non-perturbative phenomena, will be discussed.
Thereafter, the auxiliary field method will be studied. Using the
auxiliary field method the large-$N$ approximation of the models
examined in this thesis can be obtained systematically. Finally it
will be explained why some effective potentials expressed in terms of
auxiliary fields can have temperature-dependent ultraviolet
divergences outside the minimum.

\section{The 1PI effective action and the effective potential}
\label{sec:formdefeffaction}
The one particle irreducible (1PI) effective action is a useful tool
for calculating partition functions. It is especially preferred if a
certain field can obtain a vacuum expectation value.  This happens for
example in a theory with spontaneous symmetry breaking.  In ordinary
perturbation theory one expands around the trivial vacuum, for which
the vacuum expectation values of fields vanish, say $\left <\phi
\right> = 0$. In the 1PI effective action approach however, one
expands around the true vacuum. This means that the 1PI method allows
for a nonzero vacuum expectation value of $\phi$, $\left<\phi \right>
= \bar \phi$.  This vacuum expectation value can be found by
minimizing the 1PI effective action. Because in the 1PI method one
perturbs around the true vacuum it is more advantageous to use this
method over ordinary perturbation theory. Moreover it allows one to
resum whole classes of Feynman diagrams and to investigate
non-perturbative physics. More details concerning the derivation of
the effective action given below can be found in quantum field theory
textbooks, for example \citet{Weinberg1995}, \citet{Peskin1995} and
\citet{ZinnJustin1996}.

\subsection*{Formal definition of the effective action}
Consider a general scalar field theory with Lagrangian density
$\mathcal{L}[\phi]$ in Euclidean space-time. The partition function of
this theory in the presence of a source term $J$ is given by
\begin{equation}
  Z \left[ J \right] 
	= 
  \int \mathcal{D} 
  \phi \;
  \exp \left[ - S\left[\phi \right] - \int \measure{^d x} J(x) \phi(x)
	\right ]\;,
 \label{eq:partfuncscalar3}
\end{equation}
here $S[\phi] = \int \mathrm{d}^d x \mathcal{L}[\phi]$ denotes the
classical action. From this partition function one can compute the
generating functional for connected Green's functions which is 
\begin{equation}
  W \left [J \right] = - \log Z \left[ J\right] \;.
\end{equation}
These Green's functions are obtained by functional differentiating
$W[J]$ with respect to $J$.
For example the vacuum expectation value of
$\phi$ of the theory in the presence of a source term is given by
\begin{equation}
 \frac{\delta W[J]}{\delta J(x)}
  = \expval{\phi(x)}_J  
  \equiv
  \bar \phi(x) \;. 
 \label{eq:scalarvev}
\end{equation}
Similarly, the connected two-point correlation function can be found
as follows
\begin{equation}
  \frac{\delta^2 W[J]}{\delta J(x) \delta J(y)} =
   \expval{T \phi(x)\phi(y)}_J - \expval{\phi(x)}_J \expval{\phi(y)}_J 
  \equiv D(x, y) \;.
\end{equation}

The 1PI effective action $\Gamma[\bar \phi]$ is the generating
functional of 1PI diagrams. An 1PI diagram is a diagram that is still
connected after cutting one internal line. The 1PI effective action is
the Legendre transform of $W[J]$,
\begin{equation}
  \Gamma[\bar \phi] = W[J] - \int \measure{^d x} J(x) \bar \phi(x) \;.
 \label{eq:effactiondef} 
\end{equation}
The effective action is a function of $\bar \phi$ only and not of
$J$. The source term $J$ in Eq.~(\ref{eq:effactiondef}) is to be
chosen in such a way that the vacuum expectation value of $\phi$ in
the theory with a source term will become equal to $\bar \phi$. The
1PI effective action is, unlike the classical action, an
action which contains all contributions arising from quantum
fluctuations. Extremizing the effective action with respect to $\bar
\phi$ gives
\begin{equation}
  \frac{\delta \Gamma[\bar \phi]}{\delta \bar \phi(x)} = \int
  \measure{^d y} \frac{\delta W[J]}{\delta J(y)} \frac{\delta
  J(y)}{\delta \bar \phi(x)} - \int \measure{^d y} \frac{\delta
  J(y)}{\delta \bar \phi(x)} \bar \phi(y) - J(x) = -J(x) \;.
\end{equation}
Hence at an extremum of the effective action the source term has to
vanish, in other words $J = 0$. The extremum will be denoted by $\varphi$(x),
\begin{equation}
   \left. \frac{\delta \Gamma[\bar \phi]}
  {\delta \bar \phi(x)} \right \vert_{\bar \phi = \varphi} = 0 \;.
\label{eq:ext1pi}
\end{equation}
From the definition of the effective action,
Eq.~(\ref{eq:effactiondef}), it follows that the effective action at
the extremal value is equal to
\begin{equation}
  \Gamma[\varphi] = - \log Z[0] \;.
\end{equation}
This important equation shows that the partition function can be
obtained by calculating the minimal value of the effective action.  

It often happens that the vacuum solution of Eq.~(\ref{eq:ext1pi}) is translational invariant,
which implies that $\varphi(x)$ is space(-time) independent. However,
translational non-invariant solutions for $\varphi(x)$ are sometimes
possible. Instantons which for example arise in the
$\mathbb{C}P^{N-1}$ model discussed in Chapter~5, are translational
non-invariant solutions. But due to their dependence on space(-time),
these translational non-invariant solutions often have a larger action
than the translational invariant solution.  If $\varphi(x)$ is
translational invariant it is useful to define the effective potential
$\mathcal{V}$ which is minus the effective action divided by the
volume $V$ of the space
\begin{equation}
 \mathcal{V}\left( \bar \phi \right) = -\frac{1}{V} \Gamma \left[ \bar \phi
 \right] \;,
\end{equation}
where the sign of the effective potential is chosen in such a way that
the extremal value of the finite temperature effective potential
becomes equal to the pressure. The extremal value of the effective
potential is given by
\begin{equation}
 \mathcal{V}\left( \varphi \right) = \frac{1}{V} \log Z[0] \;. 
  \label{eq:mineffpot}
\end{equation}
At finite temperature the volume of the space $S \times
\mathbb{R}^d$ is $\beta V$. Therefore at finite temperature the
extremal value of the effective potential is given by $\log Z[0] /
\beta V$ which is equal to the pressure. In the rest of the thesis
this equation will be used to calculate the thermodynamical
quantities.

Next to the 1PI effective action, a 2PI effective action
\citep{Cornwall1974} and even more general nPI effective actions exist
as well. The 2PI effective action is the Legendre transform of the
generating functional for connected Green's functions in the presence
of a source term $J$ for the field $\phi$ and a source term $K$ for
the two-point function. The 2PI formalism is very useful for
out-of-equilibrium quantum field theory calculations (see for example
\citet{Berges2004}). This is because unlike the 1PI method, the 2PI
formalism lacks the so-called secularity problem, which causes the
perturbation series of a time-dependent quantity to diverge at late
times (see for example \citet{Arrizabalaga2004a}). Moreover, it is very
natural to introduce Gaussian initial density matrices in the 2PI
formalism. Since in this thesis all results are obtained for
equilibrium situations, the 2PI formalism will not be discussed
further. Although it is of course also possible to perform equilibrium
calculations using the 2PI formalism.

\subsection*{Perturbative calculation of the effective action}
In practice, the effective action has to be calculated in
perturbation theory. Following the method of \citet{Jackiw1974}, the
field $\phi(x)$ is replaced by the sum of its vacuum expectation value
$\bar \phi(x)$ and a quantum fluctuating field $\chi(x)$. In this way it is
possible to perturb around the true vacuum $\bar \phi(x)$. Taylor
expanding the action and the source term around $\bar \phi(x)$ gives,
\begin{equation}
\begin{split}
  S[\bar \phi + \chi] & + \int \measure{^d x} J(x)
  \left[ \chi(x) + \bar \phi(x) \right ]
 = S[\bar \phi] + \int \measure{^d x} J(x) \bar \phi(x) 
  \\  
& + 
  \int \measure{^d x} 
  \left[\left. \frac{\delta S[\phi]}{\delta \phi(x)} \right 
     \vert_{\phi=\bar \phi} + J(x)\right]
  \chi(x)
  + \frac{1}{2!} \int \measure{^d x} \measure{^d y}
  \chi(x) G(x, y)^{-1} \chi(y) 
  \\  & +
  \frac{1}{3!} \int \measure{^d w} \measure{^d x} \measure{^d y}
  \left. \frac{\delta^3 S[\phi]}{\delta 
\phi(w) \delta \phi(x) \delta \phi(y)}\right \vert_{\phi = \bar \phi} 
  \chi (w) \chi(x) \chi(y) + \ldots \;.
\end{split}
\label{eq:actionexp}
\end{equation}
where the bare inverse propagator is given by
\begin{equation}
  G(x, y)^{-1} = \left. \frac{\delta^2 S[\phi]}{\delta \phi(x) \delta
  \phi(y)}\right \vert_{\phi = \bar \phi} \;.
\end{equation}
Using the expansion of the action, Eq.\ (\ref{eq:actionexp}), and the
definition of the effective action, Eq.\ (\ref{eq:effactiondef}), it follows 
that the effective action obeys
\begin{equation}
  \Gamma[\bar \phi] = S[\bar \phi] + \Gamma_2[\bar \phi] \;.
\end{equation}
So the effective action $\Gamma[\bar \phi]$ is equal to the sum of 
the classical action $S[\bar \phi]$ and the quantum corrections
$\Gamma_2[\bar \phi]$ which are given by
\begin{equation}
  \Gamma_2[\bar \phi] = - \log
  \int \mathcal{D} 
  \chi \;
  \exp \left[ - S'\left[\chi \right] - \int \measure{^d x} J'(x)
  \chi(x) \right ] \;.
\end{equation}
In the expression for $\Gamma_2$ the action $S'[\chi]$ is equal to 
\begin{equation}
\begin{split}
  S'[\chi] = & 
  \frac{1}{2!} \int \measure{^d x} \measure{^d y}
  \chi(x) G(x, y)^{-1} \chi(y) 
  \\& +
  \frac{1}{3!} \int \measure{^d w} \measure{^d x} \measure{^d y}
  \left. 
  \frac{\delta^3 S[\phi]}{\delta \phi(w) 
  \delta \phi(x) \delta \phi(y)}\right \vert_{\phi = \bar \phi} 
  \chi (w) \chi(x) \chi(y) 
  + \ldots \;,
\end{split}
\label{eq:gamma2}
\end{equation}
and the current $J'$ is equal to
\begin{equation}
  J'(x) 
   = \left. \frac{\delta S[\phi]}{\delta \phi(x)} \right 
     \vert_{\phi=\bar \phi} 
- \frac{\delta \Gamma[\bar \phi]}{\delta \bar \phi(x)} 
  = - \frac{\delta \Gamma_2[\bar \phi] }{\delta \bar \phi(x)}  \;.
 \label{eq:currentjprime}
\end{equation}

The quantum corrections to the effective action can be obtained by
calculating $\Gamma_2[\bar \phi]$ in perturbation theory by summing
Feynman diagrams. In order to find out which kind of diagrams
contribute to $\Gamma_2[\bar \phi]$, consider $Z[0]$. In perturbation
theory, $Z[0]$ is equal to 1 plus the sum of all closed loop Feynman
diagrams times the contribution of the single loop (see also
Sec.~2.3). The quantity $W[0] = -\log Z[0]$ is equal to the sum all
connected closed loop Feynman diagrams. And $\Gamma[\bar \phi = 0] =
-W[J']$ is equal to the sum of all one particle irreducible Feynman
diagrams, for a proof see for example \citet{ZinnJustin1996}. Here
$J'$ is a source term that is chosen in such a way that the corresponding
vacuum expectation value $\bar
\phi$ vanishes.
As one might expect the source term $J'$ of
Eq.~(\ref{eq:currentjprime}) is tuned automatically in such a way that
the vacuum expectation value $\bar \chi$ of the fluctuating $\chi$
field vanishes, for a proof see \citet{Jackiw1974}. 
Hence $\Gamma_2$ is equal to the
sum of all closed loop 1PI diagrams with bare propagator $G$ and vertices
as given by the shifted action $S'$. It is important to realize that
although a tadpole term arises in the shifted action, tadpole diagrams
do not contribute to $\Gamma_2$. The tadpoles form part of the current
$J'$, see Eq.\ (\ref{eq:currentjprime}), which forces the vacuum
expectation value of the $\chi$ field to vanish. 

If the contribution of higher order 1PI diagrams to $\Gamma_2$ is
suppressed, for instance due to some small coupling constant, the main
contribution to $\Gamma_2$ will arise from the single closed loop. In
such a case the effective action can be approximated by
\begin{equation}
 \Gamma[\bar \phi] \approx S[\bar \phi] +
   \frac{1}{2} \log \mathrm{Det}\, G \;,
 \label{eq:effapr}
\end{equation}
where the $\log \mathrm{Det}$ term arises from the Gaussian
integration over the quantum fluctuations $\chi$.  As will be argued
in the following sections, Eq.~(\ref{eq:effapr}) can be used to obtain
thermodynamical quantities to next-to-leading order in $1/N$, where $N$
is the number of fields in the theory.

An explicit calculation of the effective potential in a scalar
$\lambda \phi^4$ field theory has been performed by
\citet{Coleman1974} in an expansion in small $\lambda$. This work was
generalized to finite temperature by \citet{Dolan1974}. Although the
discussion of the effective potential in this section was limited to
scalar fields only, the effective potential for a theory with gauge
fields and fermions is obtained analogously. In this thesis finite
temperature effective potentials will be calculated not in a
perturbative expansion in the coupling constant, but rather in an
expansion in $1/N$. This $1/N$ expansion will be explained next.

\section{The $1/N$ expansion}\label{sec:largeNexp}
To calculate a quantity in an interacting field theory it is often
necessary to make some kind of approximation. Expanding in the
coupling constant $\lambda$ is a widely used method. This however has
the drawback that it only works for small couplings. Moreover,
contributions which are non-analytic in $\lambda$, like
$\exp(-1/\lambda)$, will not be found in an expansion around $\lambda
= 0$.  If the coupling constant is large, which happens for example to
be the case in QCD at low energies, perturbation theory is not
reliable. Hence in this non-perturbative regime one has to use other
approaches like for example the $1/N$ expansion, lattice
discretization, Dyson-Schwinger or renormalization group methods.

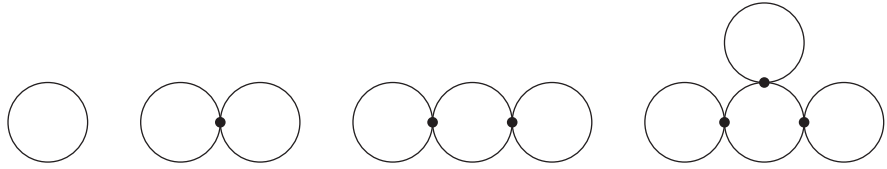
\begin{figure}[t]
\begin{center}
\begin{picture}(330,70)(0,0)
\CArc(20,20)(15,-180, 180)

\Vertex(85,20){2}
\CArc(70,20)(15,-180,180)
\CArc(100,20)(15,-180,180)

\Vertex(165,20){2}
\CArc(150,20)(15,-180,180)
\CArc(180,20)(15,-180,180)
\Vertex(195,20){2}
\CArc(210,20)(15,-180,180)

\Vertex(275,20){2}
\CArc(260,20)(15,-180,180)
\CArc(290,20)(15,-180,180)
\Vertex(305,20){2}
\CArc(320,20)(15,-180,180)
\CArc(290,50)(15,-180,180)
\Vertex(290,35){2}

\end{picture}
\end{center}
\caption{Some examples of bubble diagrams in the $\mathrm{O}(N)$ linear sigma model.
An infinite number of these types of diagrams with all possible
insertions of bubbles contribute at leading order in $1/N$. The lines
in these diagrams denote the bare propagators, which can be read off
directly from the Lagrangian density of the $\mathrm{O}(N)$ linear
sigma model, Eq.~(\ref{eq:laglinearsigma}).}
\label{fig:bubbles}
\end{figure}

\begin{figure}[t]
\begin{center}
\begin{picture}(285,60)(0,0)

\CArc(30,25)(20,-180,180)
\Oval(30,25)(10,20)(0)
\Vertex(10,25){2}
\Vertex(50,25){2}

\Oval(100,39)(5,17)(0)
\Oval(90,22)(5,17)(120)
\Oval(110,22)(5,17)(240)
\Vertex(100,7){2}
\Vertex(117,38){2}
\Vertex(82,38){2}

\Vertex(155,40){2}
\Vertex(185,40){2}
\Vertex(155,10){2}
\Vertex(185,10){2}
\Oval(170,10)(5,15)(0)
\Oval(170,40)(5,15)(0)
\Oval(155,25)(5,15)(90)
\Oval(185,25)(5,15)(90)

\Vertex(220,40){2}
\Vertex(250,40){2}
\Vertex(220,10){2}
\Vertex(250,10){2}
\Vertex(255,25){2}
\Oval(235,10)(5,15)(0)
\Oval(235,40)(5,15)(0)
\Oval(220,25)(5,15)(90)
\Oval(250,25)(5,15)(90)
\CArc(265,25)(10,-180,180)

\end{picture}
\end{center}
\caption{Some examples of chain diagrams in the $\mathrm{O}(N)$ 
linear sigma model.  An infinite number of these types of diagrams
with all possible insertions of bubbles contribute at next-to-leading
order in $1/N$.}
\label{fig:chains}
\end{figure}
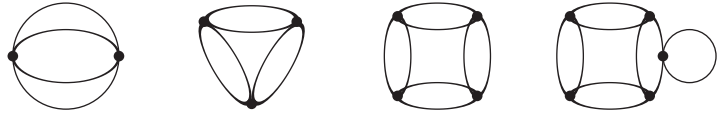

In the large-$N$ method the expansion parameter is the number of
fields $N$. In order to perform this expansion one has to change the
symmetry group of the theory under consideration. To illustrate this
point, consider the $\mathrm{O}(4)$ linear sigma model (which is
discussed in more detail in Chapter \ref{chap:onmodel}). This 
model has the following Lagrangian density
\begin{equation} 
  \mathcal{L} = \frac{1}{2} \left(\partial_\mu 
  \phi_i\right)^2 + \frac{\lambda_b}{32} \left (
  \phi_i \phi_i\!-\!4 f_{\pi,b}^2 \right)^2  \;, \;\;\;\;\; 
  i = 1 \ldots 4 \;,
 \label{eq:laglinearsigma3}
\end{equation}
where $\lambda_b$ is a bare 4-point coupling constant and $f_{\pi, b}$
is the bare ``pion decay'' constant.  To apply the $1/N$ expansion in the
$\mathrm{O}(4)$ linear sigma model it has to be generalized to the
$\mathrm{O}(N)$ linear sigma model. The Lagrangian density of the
$\mathrm{O}(N)$ linear sigma model is given by
\begin{equation} 
  \mathcal{L} = \frac{1}{2} \left(\partial_\mu 
  \phi_i\right)^2 + \frac{\lambda_b}{8 N} \left (
  \phi_i \phi_i\!-\!N f_{\pi,b}^2 \right)^2  \;, \;\;\;\;\; 
  i = 1 \ldots N \;.
 \label{eq:laglinearsigma}
\end{equation}
As one can see from Eq.~(\ref{eq:laglinearsigma}), the coupling
constants $\lambda_b$ and $f_{\pi,b}$ have been rescaled by factors
of $N$. This redefinition of the coupling constants is important when
applying a $1/N$ approximation, since it assures that the relative
strength of the interactions is not changed when varying $N$. The
coupling constants are rescaled in such a way that the action
naturally scales with $N$. In the $\mathrm{O}(N)$ linear sigma model
the diagrams which are contributing to the pressure at leading order
are the bubble diagrams and all possible insertions of bubbles, see
for example \citet{Jackiw1974}. These diagrams are called daisy and
superdaisy diagrams as well and are displayed in
Fig.~\ref{fig:bubbles}. The chain diagrams which are displayed in
Fig.~\ref{fig:chains} contribute to next-to-leading order in
$1/N$. All these diagrams can be resummed using the auxiliary field
method which will be discussed in the next section. In this way it is
possible to obtain an expansion which is really in $1/N$ and not for
example in $\lambda / N$. This allows one to investigate the
non-perturbative large $\lambda$ behavior of the model.

Unfortunately the situation in QCD is not that simple. The obvious
generalization is to consider an $\mathrm{SU}(N_c)$ gauge theory and
to take the number of colors, $N_c$, as an expansion parameter. The
coupling constant $g^2$ has then to be rescaled to $g^2 / N_c$. As was
shown by \citet{tHooft1973}, only planar diagrams contribute to
leading order in the $1/N_c$ expansion. Non-planar diagrams with only one
handle contribute at next-to-leading order. Just like in the
$\mathrm{O}(N)$ linear sigma model an infinite number of diagrams
contribute to the QCD pressure at leading and next-to-leading order in
$N_c$. But unfortunately until now no one has found a way to resum all
these diagrams, the auxiliary field method discussed in the next
section for example does not work for QCD. 

Another possibility is to give all quark flavors the same mass and
expand in the number of flavors $N_f$. Using this method
\citet{Moore2002} calculated for zero quark mass the full
non-perturbative large $N_f$ contribution to the QCD pressure up to
next-to-leading order. \citet{Ipp2003} extended this study to non-zero
chemical potential. Unfortunately this approach gives only limited
insight in QCD, since large $N_f$ QCD is not asymptotically free and
hence behaves completely differently from real QCD.

A hint that in the future methods may become available which can be used
to perform analytic calculations in the non-perturbative regime of QCD
comes from the anti-de-Sitter conformal field theory (AdS/CFT)
correspondence. \citet{Maldacena1997} conjectured that $\mathcal{N} =
4$ supersymmetric $\mathrm{U}(N)$ Yang-Mills theory in 4 dimensions
(which is a conformal gauge field theory) is equivalent to a
10-dimensional type IIB string theory on $\mathrm{AdS}_5 \times
S^5$. It turns out that if the supersymmetric Yang-Mills theory is
strongly coupled, the corresponding string theory is weakly
interacting. So by applying perturbation theory to this 10-dimensional
string theory one can obtain insight in the non-perturbative behavior
of the corresponding 4-dimensional gauge theory. One is of course eager
to find an analogous result for real QCD in 4 dimensions.

\section{The auxiliary field method}
The $1/N$ expansion can in some theories be systematically performed
by introduction of auxiliary fields. These auxiliary fields will have
no dynamics so they will not influence the physical content of the
theory. They will be shifted by a constant in
order to obtain a Lagrangian density which is quadratic in the $N$
original fields. This allows for Gaussian integration, after
which the $1/N$ expansion follows naturally.

As an example consider the $\mathrm{O}(N)$ model defined in Eq.\
(\ref{eq:laglinearsigma}). Its partition function is given by
Eq.\ (\ref{eq:partfuncscalar3}). 
An auxiliary field $\alpha$ can be added to this model
by changing the Lagrangian density as follows
\begin{equation}
 \mathcal{L} \rightarrow \mathcal{L} + \frac{N}{2 \lambda_b} \alpha^2 \;.
\end{equation}
The original partition function times an infinite constant (which
drops out when calculating physical quantities) is recovered after
integration over the $\alpha$ field. Since a field is nothing else than
an integration variable of the path-integral, it is always possible
to shift this field by a constant. If the $\alpha$ field is shifted
according to
\begin{equation}
\alpha \rightarrow \alpha - 
    \frac{i \lambda_b}{2 N} \left(\phi_i \phi_i -  N f_{\pi,b}^2
    \right) \;,
\end{equation}
the Lagrangian density becomes quadratic in the $\phi$ fields,
\begin{equation}
    \mathcal{L} = \frac{1}{2} \left( \partial_\mu \phi_i \right)^2 
 - \frac{i}{2} \alpha
  \left(\phi_i \phi_i - N f_{\pi,b}^2 \right) 
+ 
  \frac{N}{2 \lambda_b} \alpha^2 \;.
\end{equation}
It is now possible to perform the integration over the scalar fields,
resulting in the following effective action (not to be confused with
the 1PI effective action, here $\alpha$ is still a field which should
be integrated over)
\begin{equation}
    S_{\mathrm{eff}}[\alpha] =
  \frac{N}{2}  
  \mathrm{Tr} \log\left( -\partial^2 - i \alpha\right)  
   +  \frac{N}{2} \int_{0}^{\beta} \! \mathrm{d}\tau\int \! \mathrm{d}^3 x
     \left(
             i f_{\pi,b}^2 \alpha 
         +  \frac{\alpha^2}{\lambda_b} 
   \right) \;.
 \label{eq:effaction37}
\end{equation}
The auxiliary field can obtain a vacuum expectation value $\bar
\alpha$ which is
assumed to be translational invariant. Hence to obtain for example the
pressure, it is necessary to calculate the effective potential and
minimize it with respect to $\bar \alpha$. The leading order term of
this effective potential is just the classical effective action,
$S_\mathrm{eff}(\bar
\alpha)$ and is proportional to $N$. By expanding $S_\mathrm{eff}$
around its vacuum expectation value the quantum corrections to the
effective potential can be obtained.  As will be shown in detail in
Chapters~4, 5 and 6, the leading term of these corrections is the
logarithm of a determinant of the propagator of the quantum
fluctuations which is proportional to $N^0$, see also 
Eq.~(\ref{eq:effapr}).
The contributions which
arise from the 1PI diagrams of the quantum fluctuations are of order
$1/N$ and higher because the vertices of these diagrams turn out to
have $1/\sqrt{N}$ suppressing factors.

\section[Temperature-dependent ultraviolet
 divergences]{Temperature-dependent ultraviolet divergences\footnote{This
section is based on: {\it Thermodynamics of the O($N$) non-linear
sigma model through next-to-leading order in $1/N$}, H.J. Warringa,
proceedings of the SEWM 2004 meeting, World Scientific (2005).}}
\label{sec:tdepdivs}

The finite temperature effective potential expressed in terms of the
vacuum expectation value of the auxiliary field discussed in the
previous section is calculated explicitly for different models in
Chapters 4, 5 and 6. In all these models it turns out that the
effective potential contains temperature-dependent ultraviolet
divergences. However, as follows from the calculations, these
divergences become temperature-independent at the minimum. These
temperature-dependent divergences make renormalization of the
effective potential outside the minimum impossible because it will
require a renormalization prescription which is different for every
temperature. Such a prescription would lead to arbitrary
temperature dependence and hence is meaningless.

At first sight it might seem surprising that these
temperature-dependent ultraviolet divergences arise at all, since one
expects that temperature cannot influence the ultraviolet behavior of
a theory due to the exponential suppression of high-momentum modes of
finite temperature contributions. To see why it is nevertheless
possible that an effective potential can have temperature-dependent
divergences outside its minimum, consider as an example the
$\mathrm{O}(N)$ linear sigma model defined in
Eq.~(\ref{eq:laglinearsigma}). For other models the argument is
similar. The effective action of the $\mathrm{O}(N)$ linear sigma
model expressed in terms of the auxiliary field,
$S_\mathrm{eff}[\alpha]$ is given in Eq.~(\ref{eq:effaction37}).  The
effective action in the presence of a space-time independent source $i
J$ for $\alpha$ is defined by
\begin{equation}
  S_\mathrm{eff}[J, \alpha] = S_\mathrm{eff}[\alpha] +  
  i J \int_0^\beta \measure{\tau} \int \measure{^3 x} \alpha(x) \;.
\end{equation}
The quantity $Z[J]_T$ is defined here as the partition function of the
theory with action $S_\mathrm{eff}[J, \alpha]$ at temperature $T$. Now
$\log Z[J=0]_T / \beta V$ is equal to the pressure of the linear sigma
model. One therefore expects that this quantity should be well-defined
and that it contains no temperature-dependent divergences.  So one can
assume that $\log Z[0]_T$ does not contain temperature-dependent
divergences.  As a result, $\log Z[J]_T$, contains also no
temperature-dependent divergences as long as $J$ is independent of
temperature.  The reason for this is that since both $J$ and
$f^2_{\pi,b}$ are proportional to $\alpha$ in the Lagrangian density,
any source $J$ can be incorporated in $f^2_{\pi,b}$ by the redefinition
$N f^2_{\pi,b} / 2 + J \rightarrow N f^2_{\pi,b} / 2$.

The effective potential in the minimum is equal to $\log Z[0]_T$.  On
dimensional grounds, the divergences of the minimum of the effective
potential should depend on $f_{\pi,b}$. For example in Chapter 6 it is
found that the effective potential contains terms proportional to the
divergent factor $\Lambda^2 f_{\pi,b}^2$, where $\Lambda$ is an
ultraviolet momentum cut-off.
Since any $J$ can be incorporated in $f^2_{\pi,b}$, varying $J$ will
change the divergences of $\log Z[J]_T$. Hence the divergences
of $\log Z[J]_T$ will depend on $J$.

The effective potential is given by $\mathcal{V}(m^2) = \log Z[J]_T /
\beta V- J m^2$, where $\bar \alpha = i m^2 $ is the vacuum expectation value
of the $\alpha$ field. That this vacuum expectation value is purely
imaginary is proved in Chapter 4. In the expression for the effective
potential, $J$ is defined implicitly, since $J$ is the source term
which gives the $\alpha$ field the vacuum expectation value $i m^2$.
If the temperature is varied, the implicitly defined $J$ has to change
in order to keep the vacuum expectation value of the $\alpha$ field
fixed to $im^2$. Therefore the implicitly defined $J$ has to depend on
temperature. In the previous paragraph it was argued that the
divergences of $\log Z[J]_T$ depend on $J$. Since $\log Z[J]_T$ is
part of the effective potential, the divergences of the effective
potential will depend on temperature for any non-zero $J$.

As is shown in Eq.~(\ref{eq:mineffpot}), the source term $J$ vanishes
at an extremum of the effective potential. This holds for any
temperature. So using the arguments in the previous paragraphs
temperature-dependent divergences do not arise at the minimum of the
effective potential, but only outside the minimum.  Since physical
quantities like the pressure, entropy density and energy density are
to be calculated at the minimum of the effective potential they do not
suffer from the renormalization problems caused by
temperature-dependent divergences.

%For the same reason, it could very well be that the full QCD effective
%action expressed in terms contains temperature dependent divergences. 

 %1/n auxfield
\chapter{Thermodynamics of the nonlinear sigma model in d=1+1}

The thermodynamics of the $\mathrm{O}(N)$ nonlinear sigma model is
studied in this chapter in the $1/N$ approximation. The effective
potential, from which one can derive thermodynamical quantities, will
be calculated to next-to-leading order in $1/N$. It is found that the
effective potential contains temperature-dependent divergences which
become independent of temperature at the minimum. As a result it turns
out that the effective potential can only be renormalized meaningfully
at the minimum, hence thermodynamical quantities which are obtained at
the minimum, can be rendered finite consistently. The renormalized
pressure will be calculated to next-to-leading order in $1/N$.
Furthermore in intermediate steps of the calculation, thermal infrared
renormalons are encountered which potentially can give rise to
ambiguities in the end result.  It will be shown that the correctly
renormalized pressure is free from these ambiguities.  This chapter is
based on: {\it Thermodynamics of the O($N$) nonlinear sigma model in
(1+1)-dimensions,} J.O.\ Andersen, D.\ Boer and H.J.\ Warringa, Phys.\
Rev.\ {\bf D69} 076006, (2004).

%=====================================================================
\section{Introduction}

The $\mathrm{O}(N)$ nonlinear sigma model in $1+1$ dimensions has been
studied extensively at zero temperature as a toy model for QCD. It is
a scalar field theory which classically is scale invariant. Due to
renormalization of the quantum corrections a scale is introduced.  As
was shown by \citet{Polyakov1975} the nonlinear sigma model is
asymptotically free like QCD. It is renormalizable both perturbatively
and in the $1/N$ expansion. Moreover, as is discussed in more detail
in Chapter~5, it contains instanton solutions for $N=3$ only. In that
chapter it is also shown that the $\mathrm{O}(3)$ nonlinear sigma
model is equivalent to the $\mathbb{C}P^{1}$ model.

Unlike the nonlinear sigma model in more than two dimensions, where
the theory is no longer renormalizable, there is no spontaneous
symmetry breaking of the global $\mathrm{O}(N)$ symmetry for any value
of the coupling constant. This reflects the theorem of
\citet{Mermin1966} and \citet{Coleman1973}, 
which forbids spontaneous breakdown of a continuous symmetry in a
homogeneous system in one spatial dimension at any temperature.  The
nonlinear sigma model suffers from infrared divergences in
perturbation theory, because the scalar fields remain massless in that
case.  It was conjectured by \citet{Elitzur1983}\footnote{This article
was submitted in 1978.} and shown by \citet{David1981} that the
infrared divergences cancel in $\mathrm{O}(N)$-invariant correlation
functions.  However, like in QCD, a non-perturbative mass-gap is
generated dynamically due to the interactions. In contrast to QCD
where the spontaneous breakdown of chiral symmetry breakdown is
responsible for non-zero masses, the non-vanishing mass of the scalar
fields in the nonlinear sigma model is not related to symmetry
breaking.  In the large-$N$ limit, which is equivalent to summing all
so-called daisy and super-daisy graphs (see Chapter~3), it turns out
that the mass at zero temperature is given by
$m=\mu\exp{(-2\pi/g^2)}$, where $g$ is the coupling constant and $\mu$
is the renormalization scale.  The mass is non-analytic in $g$ which
explains that it vanishes in perturbation theory.

\citet{Dine1981} have investigated the nonlinear sigma model in $1+1$
dimensions at finite temperature. They calculated the free energy in
perturbation theory and to leading order in the large-$N$ limit.  In
the weak-coupling expansion, they showed that the two-loop
contribution to the ideal gas vanishes and that the three-loop
contribution is infrared finite. 
In this chapter the analysis of \citet{Dine1981} will be extended to
next-to-leading-order (NLO) in the $1/N$ expansion.  To obtain the
thermodynamical quantities the effective potential will be calculated.
At zero temperature, the effective potential (or equivalently the
Gibbs free energy) has been investigated at this order by
\citet{Root1974} and \citet{Biscari1990}. 

There are several reasons to calculate the $1/N$ corrections to the
thermodynamics. Since the nonlinear sigma model is one of the few
field theories in which non-perturbative calculations can be done
(partly) analytically, it is very useful to check whether the $1/N$
expansion is converging. Furthermore renormalization at finite
temperature is non-trivial, since next to the renormalization scale an
additional scale is introduced. Normally one expects that the
ultraviolet behavior of a theory will not depend on temperature
because the high-momentum modes of thermal contributions are
exponentially suppressed due to the Bose-Einstein distribution
factor. However, it is found that the next-to-leading order effective
potential contains temperature dependent divergences. Fortunately it
turns out that the effective potential can still be renormalized at
the minimum, but it then contains an ambiguity which is related to a
so-called infrared renormalon
\citep{David1982}. It will be shown that this ambiguity vanishes when
calculating the pressure.

It was shown by \citet{David1982} that the $1+1$ dimensional nonlinear
sigma model contains infrared renormalons. Renormalons (see
\citet{Beneke1999} for a review) give rise to ambiguities in
resummming a perturbative series in the coupling constant $g$. The
size of these ambiguities may give insight in the importance of
non-perturbative corrections.  In QCD these infrared renormalons
appear as well, which in principle might teach us about the size of
non-perturbative physics in QCD. Because the $\mathrm{O}(N)$ nonlinear
sigma model can be accessed both non-perturbatively (via the $1/N$
expansion) and perturbatively (in an expansion in $g$) the nonlinear
sigma model is widely used
\citep{David1984, David1986, Beneke1998} to study these
infrared renormalons. In this chapter such ambiguities are encountered
when calculating the effective potential. However it turns out that
they vanish when calculating the pressure. Furthermore it is
investigated how these renormalons depend on temperature. It is found
that thermal infrared renormalons in the nonlinear sigma model are
independent of temperature at the minimum of the effective potential,
however outside the minimum the renormalon residue become
temperature-dependent while the position of the renormalon pole is not
affected by temperature. Thermal renormalons have been studied before
by
\citet{Loewe2000} in $\phi^4$ theory in 3+1 dimensions. In this
theory, one deals with ultraviolet renormalons only and thus it
resembles QED rather than QCD.  They found that the residues of the
ultraviolet renormalon poles in the Borel plane are
temperature-dependent while the position of the poles are not.

The Gross-Neveu model, which is a 1+1 dimensional toy model which
includes also fermions was recently studied by \citet{Blaizot2003} at
finite temperature to next-to-leading order in the $1/N$ expansion as
well.  While there are similarities between this model and the
nonlinear sigma model, such as dynamical mass generation and
asymptotic freedom, no problems related to renormalization of the
effective potential was encountered by \citet{Blaizot2003}.

This chapter is organized as follows. In Sec.~4.2 the Lagrangian
density of the nonlinear sigma model will be given. Thereafter, in
Sec.~4.3 it will be discussed how the pressure could in principle be
obtained by a weak-coupling constant expansion. Sec.~4.4 is devoted to
the calculation of the effective potential to next-to-leading order in
$1/N$. The zero-temperature case will be discussed first, after which
the derivation of the finite temperature effective potential will
follow. In Sec.~4.5 the pressure for the $\mathrm{O}(N)$ nonlinear
sigma model will be presented. Some approximations for the pressure at
high temperatures will be discussed in Sec.~4.6. Thermal infrared
renormalons are investigated in Sec.~4.7. A summary and conclusions are
given in Sec.~4.8.

%=====================================================================
\section{The nonlinear sigma model}
The nonlinear sigma model is a scalar field theory with an
$\mathrm{O}(N)$ symmetry. It is described by a Lagrangian density
which only consists of a kinetic term and a constraint
\begin{equation}
   \mathcal{L} = \frac{1}{2} \partial_\mu \phi_i \partial^\mu \phi_i 
\;,\;\;\;\;\;\;\;
   \phi_i(x) \phi_i(x) = N/g_b^2 
\;,\;\;\;\;\;\;\; 
i = 1 \ldots N \;,
\label{eq:lagran_non}
\end{equation}
where $g_b$ is the bare coupling constant. The constraint forces the
fields to live on a $N-1$-dimensional hypersphere, which causes the
interactions between the fields. Since $g_b$ has no dimension in
$d=1+1$, the nonlinear sigma model is renormalizable. As one can see
from Eq.~(\ref{eq:lagran_non}), the classical $\mathrm{O}(N)$
nonlinear sigma model is scale invariant.  Due to renormalization of
the quantum corrections a scale $\mu$ is introduced in the theory. As
will be shown explicitly in Sec.~4.4, the $\mathrm{O}(N)$ nonlinear
sigma model is asymptotically free.  For $N=2$ the nonlinear sigma
model is a free field theory. This can be seen by choosing $\phi_1 =
\sqrt{2} \cos(\theta) / g_b$ and $\phi_2 =
\sqrt{2} \sin(\theta) / g_b$. The constraint is satisfied automatically by
this choice, and the Lagrangian density turns into
\begin{equation}
 \mathcal{L} = \frac{1}{g_b^2} (\partial_\mu \theta)^2 \;,
\end{equation}
which is the Lagrangian density of a free field theory.

Thermodynamical quantities of the nonlinear sigma model can be derived
from the partition function which is given by
\begin{equation}
 Z =  \int \prod_{i=1}^{N} \mathcal{D} \phi_i 
  \prod_x \delta \left(N/g_b^2 - \phi_i \phi_i\right)
  \exp \left [- \int_0^{\beta} \measure{\tau} \int \measure{x} 
  \mathcal{L} \right] \;,
\end{equation}
In the following section thermodynamics in the weak-coupling expansion
will be discussed. The remaining sections of this chapter will deal with the
non-perturbative $1/N$ approach.

%=====================================================================
\section{Thermodynamics in the weak-coupling expansion}
\label{sec:pertexp}
The weak-coupling expansion can be performed by writing the scalar
fields as $\phi = (\pi_1, \pi_2, \ldots, \pi_{N-1}, \sigma)$.  Due to
the constraint it holds that $\sigma^2 = N/g_b^2 - \pi_i \pi_i$ with
$i =1 \ldots N -1$.  Integrating over the $\sigma$ field results in
the following partition function
\begin{equation}
  Z =  \int \prod_{i=1}^{N-1} \mathcal{D} \pi_i 
  \prod_x \theta\left(N/g_b^2 - \pi_i \pi_i\right)
  \exp \left [- \int_0^{\beta} \measure{\tau} \int \measure{x} 
  \mathcal{L}_{\mathrm{eff}}(\pi) \right] \;,
\end{equation}
where $\theta(x)$ is the step function. This step function reflects
the fact that due the original constraint on the $\phi$ fields, $\pi_i
\pi_i$ never can be larger than $N/g_b^2$. The effective Lagrangian
density $\mathcal{L}_{\mathrm{eff}}$ can be found by inserting the
expression for $\sigma$ in terms of $\pi$ in the original Lagrangian
density. As a result one finds
\begin{equation}
 \mathcal{L}_{\mathrm{eff}}(\pi) = \frac{1}{2} \partial_\mu \pi_i 
        \partial^\mu \pi_i
  + \frac{g_b^2}{2} \frac{(\pi_i \partial_\mu \pi_i)^2}{N  - g_b^2  \pi_i 
        \pi_i} - \frac{1}{2} \delta^2(0)\log
   \left(N/g_b^2 -  \pi_i \pi_i \right)\;,
  \label{eq:efflagrnon}
\end{equation}
where the term proportional to $\delta^2(0)$ arises from integration
over the delta function $\delta(N/g_b^2 - \phi_i \phi_i)$.

For small values of $g_b^2$ the $\theta(x)$ function is only
vanishing when $\pi(x)$ is large. Since large values of $\pi$
give a small contribution to the partition function one can approximate 
$ \theta(N / g_b^2 - \pi_i \pi_i) \approx 1$ which gives 
\begin{equation}
  Z \approx \int \prod_{i=1}^{N-1} \mathcal{D} \pi_i 
  \exp \left [- \int_0^{\beta} \measure{\tau} \int \measure{x} 
  \mathcal{L}_{\mathrm{eff}}(\pi) \right] \;.
\end{equation} 
If $g_b^2=0$ it can be seen from $\mathcal{L}_{\mathrm{eff}}$ that the
nonlinear sigma model contains $N-1$ non-interacting $\pi$
fields. Hence the leading contribution to the pressure of the
nonlinear sigma model is that of a gas of $N-1$ non-interacting scalar
fields which is
\begin{equation}
   \mathcal{P} = -\frac{N-1}{2} \sumdiff_K \log(K^2)
= (N-1)\frac{\pi}{6}T^2 \;.
\label{eq:nonlinpertpres}
\end{equation}

Using the effective Lagrangian density Eq.~(\ref{eq:efflagrnon}),
\cite{Dine1981} showed that the order $g_b^2$ correction to the pressure
vanishes. In their calculation, effects of the step-function were not
taken into account. However, using the the $1/N$ expansion, one does
find corrections to the free pressure as will be discussed in detail
in the following sections.

%=====================================================================
\section{The effective potential}\label{sec:non1overn}
By introducing an auxiliary scalar field $\alpha$, the constraint on
the $\phi$ fields can be expressed in terms of a path integral over
the $\alpha$ fields in the following way
\begin{equation}
 \prod_{x} \delta \left( \phi_i \phi_i - N / g_b^2 \right)
 = C \int \mathcal{D} \alpha 
  \exp \left[\int_x \frac{i}{2} \alpha \left( \phi_i \phi_i - N / g_b^2
 \right) \right] \;,
\end{equation}
where $C$ is a temperature and $g_b$ independent normalization
constant. As a result one finds the following Lagrangian density which
is equivalent to the original nonlinear sigma model Lagrangian density
\begin{equation}
\mathcal{L}=\frac{1}{2} \partial_{\mu}\phi_i \partial^{\mu}\phi_i 
- \frac{i}{2}\alpha(\phi_i \phi_i - N / g_b^2) + J_i \phi_i\;,
\label{eq:lagrang_non}
\end{equation}
where $J_i$ is a source term for the scalar field. It is useful to
keep this source term for a moment in order to determine the scalar
field propagator. By calculating this propagator to next-to-leading
order in $1/N$ one can determine the physical mass of the scalar field
to next-to-leading order. Any renormalization prescription should give
rise to a finite physical mass. The expression for the physical mass
will be used later in order to renormalize the so-called gap
equation to next-to-leading order in $1/N$.

Using the auxiliary field method the Lagrangian density becomes
quadratic in the scalar fields. Therefore one can perform the integral
over these fields. This yields the following exact expression for the
partition function of the nonlinear sigma model
\begin{equation}
 Z = \int \mathcal{D} \alpha \exp \left( - S_{\mathrm{eff}}
 \right) \;,
\end{equation}
where the effective action is given by \citep{Novikov1984}
\begin{equation}
  S_\mathrm{eff} = 
  \frac{N}{2} \mathrm{Tr} \log \left(-\partial^2 - i \alpha \right)
  + i \frac{N}{2 g_b^2} \int_x\,
  \alpha 
  + \frac{1}{2} \int_x \int_y J_i(x) \langle x  \vert
  \frac{1}{-\partial^2 - i \alpha} \vert y \rangle J_i(y)
  \label{eq:effactionsource} 
  \;. 
\end{equation}
In the last expression the trace is over a complete set of functions
that satisfy the boundary conditions for scalar fields at finite
temperature.

The vacuum expectation value of the $\alpha$ field, $\left < \alpha
\right >$ is purely imaginary since from $S_{\mathrm{eff}}(\alpha)^* =
S_{\mathrm{eff}}(-\alpha)$ and changing the integration over $\alpha$
to $-\alpha$ it follows that
\begin{equation}
  \left < \alpha \right >^* =
 \int \mathcal{D} \alpha\, \alpha \exp\left[-S_\mathrm{eff}(-\alpha) \right]
  = 
  -\int \mathcal{D} \alpha\, \alpha \exp\left[-S_\mathrm{eff}(\alpha) \right]
  = 
- \left < \alpha \right > 
   \;.
 \label{eq:alphaimag}
\end{equation}
It is very likely that (in absence of a source term) the vacuum
expectation value of $\alpha$ is space-time independent because the
Euclidean space is homogeneous. The auxiliary field $\alpha$ can be
written as the sum of its complex vacuum expectation value $i m^2$ and a quantum
fluctuating field $\tilde{\alpha}$
\begin{equation}
\alpha=i m^2+\tilde{\alpha} / \sqrt{N} \;.
\label{eq:defalpha}
\end{equation}
Putting $J = 0$ and expanding the effective action around the vacuum
expectation value of $\alpha$ gives
\begin{multline} 
  S_\mathrm{eff} = \frac{N}{2} \mathrm{Tr} \log ( -\partial^2 + m^2 )
        - \frac{N m^2}{2 g_b^2} \beta V - \frac{i \sqrt{N}}{2} \mathrm{Tr} \left (
   \frac{1} { -\partial^2 + m^2} \tilde \alpha \right)
 \\ + \frac{1}{4} \mathrm{Tr} 
  \left(\frac{1} {-\partial^2 + m^2 } 
  \tilde \alpha \right)^2
  + \mathcal{O}( 1 / \sqrt{N} ) \;.
\label{eq:effactionexpansion}
\end{multline}
The last equation shows that $\tilde \alpha$ n-point vertices are
suppressed by powers of $1/\sqrt{N}$. This is of course due to the
definition of $\tilde \alpha$ in Eq.~(\ref{eq:defalpha}).  This
definition is convenient since now one immediately sees from the
vertices that higher loop diagrams are suppressed. One could of course
choose another definition, for example, $\alpha = i m^2 + \tilde
\alpha$. Then the same conclusions would follow, since in that case the
$\tilde \alpha$ propagator has an additional suppressing factor of
$1/N$.  Inserting a plane wave basis in
Eq.~(\ref{eq:effactionexpansion}) results in
\begin{multline}
    \frac{\mathrm{S}_{\mathrm{eff}}}{\beta V} = 
\frac{N}{2}
\sumint_P \log \left (P^2 + m^2 \right)
        - \frac{N m^2}{2 g_b^2} 
      + \sqrt{N} \times \mathrm{terms\;linear\;in\;}
\tilde \alpha
 \\  +  \frac{1}{2} \sumint_P 
\tilde \alpha(P)
\Pi(P, m) 
\tilde \alpha(P)^* + \mathcal{O}(1 / \sqrt{N} )
  \label{eq:expandedeffactionmomentum} \;,
\end{multline}
where the terms linear in $\alpha$ were not written down since these
terms (tadpoles) do not contribute to the effective potential as 
explained in Chapter~3. The inverse $\tilde \alpha$ propagator is
equal to
\begin{equation}
\Pi(P,m) =  \frac{1}{2}\sumint_Q
\frac{1}{Q^2+m^2}\frac{1}{(P+Q)^2+m^2} \;.
\label{Pit}
\end{equation}
As discussed in more detail in Chapter~3, the effective potential
$\mathcal{V}(m^2)$ can be obtained by integrating over the quantum
fluctuations $\tilde
\alpha$. The leading term of this effective potential proportional to
$N$ stems from the classical action. The next-to-leading order term is
originating from the Gaussian integral over the quantum fluctuations
$\tilde \alpha$. The remaining contributions are of order $1/N$, since
one needs at least two three-point vertices (which are proportional to
$1/\sqrt{N}$) to form a closed diagram.  Choosing the sign of the
effective potential in such a way that the pressure is equal to the
minimum of the effective potential one obtains
\begin{equation}
\mathcal{V} = 
\frac{N m^2}{2 g_b^2}- \frac{N}{2} \sumint_P \log \left(P^2+m^2\right)
-\frac{1}{2} \sumint_{P}\log \Pi(P,m) + \mathcal{O}(1/N) \;.
\label{P-NLO1}
\end{equation}

The $\phi$ field propagator can be found by differentiating
Eq.~(\ref{eq:effactionsource}) twice with respect to the sources $J$
and expanding $\alpha$ around its vacuum expectation value. If one
then keeps terms to next-to-leading order in $1/N$ and inserts the
wave function renormalization factor $Z_\phi$ one finds that the
scalar propagator in momentum space equals
\begin{equation}
  D_\phi(P, m) = \frac{Z_\phi}{P^2 + m^2 - \frac{1}{N} \Sigma(P,
  m)}\;,
  \label{eq:scalarfieldprop}
\end{equation}
where the self-energy is equal to
\begin{equation}
  \Sigma(P, m) = \sumint_Q \frac{1}{(P+Q)^2 + m^2} \frac{1}{\Pi(Q, m)} \;.
  \label{eq:scalarselfenergy}
\end{equation}

It is important to keep in mind that $m^2$ is the vacuum expectation
value of the $\alpha$ field. This vacuum expectation value can be
found by minimizing the effective potential with respect to
$m^2$. Since the effective potential is ultraviolet divergent and the
renormalization of the nonlinear sigma model is non-trivial, the zero
temperature case will be discussed separately in the following
subsection. Thereafter, the finite temperature effective potential
will be investigated.

%---------------------------------------------------------------------
\subsection*{The effective potential at zero temperature}
At zero temperature one can obtain analytic results for the effective
potential. The sum over Matsubara modes changes into an integration over
momentum. The effective potential $\mathcal{V}$ to next-to-leading order in the
$1/N$ expansion is at zero temperature given by
\begin{equation}
\mathcal{V} = \frac{N m^2}{2g_b^{2}} - \frac{N}{2}
\int\frac{\mathrm{d}^2 P}{(2\pi)^2} 
\log \left(P^2+m^2\right)
-\frac{1}{2}\int \frac{\mathrm{d}^2 P}{(2\pi)^2}
 \log \Pi(P,m)\;.
\label{V0}
\end{equation}
It follows after integration over momenta that $\Pi(P,m)$, which is
the inverse $\tilde \alpha$ propagator, at zero temperature equals
(see for example \citet{Novikov1984})
\begin{equation}
\Pi(P,m) =  \frac{1}{4\pi \sqrt{P^2} \sqrt{P^2 + 4 m^2}} \log
\left(\frac{\sqrt{P^2 + 4m^2} + \sqrt{P^2}}{\sqrt{P^2+4m^2} -\sqrt{P^2}} \right)\;.
\end{equation}
The integrals integrals in Eq.~(\ref{V0}) will be evaluated using an ultraviolet
momentum cutoff $\Lambda$. For that one needs the following integral
\begin{multline}
\int \frac{\mathrm{d}^2 P}{(2\pi)^2}
\log \log
\left(\frac{\sqrt{P^2 + 4m^2} + \sqrt{P^2}}{\sqrt{P^2+4m^2} -\sqrt{P^2}} \right)
= \frac{m^2}{4\pi} \int_1^X \mathrm{d} t \left(1 -  \frac{1}{t^2} \right)
 \log \log (t) 
 \\
 = \frac{m^2}{4\pi}
 \left( \left. \left [ t \log \log (t) - \li (t)  \right]\right \vert^X_1 
  + \left. \left[ t \log \log(1/t) - \li(t)  \right]\right \vert^{1/X}_1 \right)\;,
\end{multline}
where 
\begin{equation}
X = \frac{\sqrt{\Lambda^2 + 4m^2} + \sqrt{\Lambda^2}}{
\sqrt{\Lambda^2+4m^2} -\sqrt{\Lambda^2}} \;,
\end{equation}
and
$\mathrm{li}(x)$ is the logarithmic integral
\begin{equation}
  \mathrm{li}(x) = \mathcal{P} \int^x_0 \measure{t} \frac{1}{\log(t)}
  \;.
\end{equation}
Here $\mathcal{P}$ indicates a principal-value prescription for the
integral. Since the integrand of the logarithmic integral has a pole
at $t=1$, a prescription to evaluate the integral is necessary. This
prescription is the source of the renormalon ambiguity discussed in
Sec.~4.7. For example the logarithmic integral in the $\pm i \epsilon$
prescription
\begin{equation}
  \li_\pm(x) = \int_0^x \mathrm{d}t\, \frac{1}{\log(t) \pm i \epsilon} \;,
\end{equation}
differs from the logarithmic integral using the principal value
prescription, that is $\li(x) = \li_\pm(x) \pm i \pi$.

Using that $\lim_{x\rightarrow 1} [\li(x) - x \log \log(x)] =
\gamma_E$, where $\gamma_E$ is the Euler-Mascheroni constant, and 
after dropping divergences that are $m$-independent in the
limit $\Lambda^2 \to \infty$, one obtains
\begin{multline}
\mathcal{V}  =  \frac{N m^2}{2 g_b^2} - 
\frac{N m^2}{8\pi}\left(1 + \log \frac{\Lambda^2}{m^2} \right)
- \frac{1}{8\pi} 
\left[\left(\Lambda^2 + 2 m^2\right) \log \log
\left(\frac{\Lambda^2}{m^2}  \right)
- m^2 \mathrm{li} \left(\frac{\Lambda^2}{m^2}\right) 
\right. \\+\left.  2 m^2 \left(\gamma_E -1 -
\log \left (\frac{\Lambda^2}{4m^2} \right) \right) \right]\;.
\label{eq:effpotzerot}
\end{multline}
As one can see from Eq.~(\ref{eq:effpotzerot}), the effective
potential is highly ultraviolet divergent. Its divergences can be
removed by renormalization. One way to do this is by adding or
subtracting infinite, $m$-independent constants. This merely shifts
the whole effective potential, but since relative energy differences
will not change this does not have physical effects. This subtraction
is already performed in Eq.\ (\ref{eq:effpotzerot}) but does not
remove all divergences. Another possibility to remove divergences is
to absorb $m$-independent infinite constants in a redefinition of the
coupling constant $g_b$.  However, this still leaves two divergences
in Eq.~(\ref{eq:effpotzerot}). The only possibility to renormalize
the term $\Lambda^2 \log \log (\Lambda^2 /m^2)$ would be to
subtract an $m$-dependent infinite constant since
$\lim_{\Lambda \rightarrow
\infty} \Lambda^2 \log \log ( \Lambda^2 /m^2) - \Lambda^2 \log \log
(\Lambda^2 /\mu^2)$ is still divergent if $\mu^2 \neq m^2$.  The
$m^2\mathrm{li}(\Lambda^2/m^2)$ term has the same problem. This
divergence should be absorbed into the coupling constant since it is
proportional to $m^2$. But there does not exist a single
$m$-independent function which can be subtracted from
$\mathrm{li}(\Lambda^2/m^2)$ to obtain a finite result in the limit
$\Lambda \to \infty$.  So it seems that the effective potential is not
renormalizable, unless $m$-dependent and potentially $T$-dependent
counter terms are introduced.

Moreover, subtracting the $\Lambda^2 \log
\log(\Lambda^2/m^2)$ $- m^2 \li (\Lambda^2 /m^2)$ term as is
done for instance by \citet{Biscari1990} for $T=0$, gives rise to
another problem. This term is called the perturbative tail and does not
arise in dimensional regularization suggesting that this subtraction
is the correct thing to do. But if one would calculate the effective
potential using dimensional regularization one would find a $\pm i
\pi$ ambiguity instead \citep{David1984} depending on whether the
limit $\epsilon \downarrow 0$ or $\epsilon \uparrow 0$ is taken. This
ambiguity is the same as the renormalon ambiguity mentioned above due
to the choice of prescription in the logarithmic integral. In Sec.~4.5
it is shown that the perturbative tail and its accompanying ambiguity
cancel when calculating the pressure which is the minimum of the
effective potential at finite temperature minus the minimum of the
effective potential at zero temperature.

To obtain physical quantities like the mass of the scalar fields and
the pressure, one has to evaluate the effective potential at its
minimum. The condition for the minimum is given by the equation
\begin{equation}
\frac{\partial \mathcal{V}} {\partial m^2} = 0\;.
\label{gap}
\end{equation}
Equation (\ref{gap}) is often referred to as a gap equation, since
solving this equation determines in leading order the gap in the
excitation spectrum at zero spatial momentum or equivalently the mass
of the scalar particles.  Differentiating the effective potential with
respect to $m^2$, Eq.~(\ref{eq:effpotzerot}), one obtains the gap
equation which reads
\begin{equation}
 \frac{4 \pi}{g_b^2} =
\left(1 - \frac{2}{N} \right) \log \left( \frac{\Lambda^2}{m^2} \right)
 + \frac{1}{N} \left[2 \log \log 
\left( \frac{\Lambda^2}{m^2} \right)
-  \mathrm{li}\left(\frac{\Lambda^2}{m^2}\right) + 2 \gamma_E + 4 \log 2
                       \right]\;. 
\label{gapintermsofm}
\end{equation}
This gap equation still contains the problematic divergence
proportional to $\mathrm{li}(\Lambda^2/m^2)$. But unlike the effective
potential, the gap equation {\it can} be renormalized. For that the
gap equation has to be expressed in terms of the finite
physical mass $m_\phi$. 

The physical mass $m_\phi$ of the scalar fields follows from the pole
of the $\phi$ propagator, Eq.~(\ref{eq:scalarfieldprop}), at zero
spatial momentum in Minkowski space. To calculate this
propagator to next-to-leading order the self-energy,
Eq.~(\ref{eq:scalarselfenergy}), has to be evaluated. At zero
temperature this self-energy can be simplified by integrating over
angles, resulting in
\begin{equation}
 \Sigma(P, m) = \frac{1}{2 \pi} \int_0^\Lambda \mathrm{d} Q\,
  \frac{1}{[ (P^2 + Q^2 + m^2)^2 - 4 P^2 Q^2]^{1/2}} 
  \frac{1}{\Pi(Q,m)}
\label{eq:selfenergyintegr} \;.
\end{equation}
Since the expression for the propagator,
Eq.~(\ref{eq:scalarfieldprop}), is in Euclidean space-time, one has to
analytically continue the propagator to Minkowski space to
obtain the physical mass.  Because the self-energy only depends on
$P^2$ as one can see from Eq.\ (\ref{eq:selfenergyintegr}),
continuation to Minkowski space boils down to replacing the Euclidean
two-vector $P^2$ by the Minkowskian two-vector $-p^2$. Neglecting
terms of order $1/N^2$, the physical mass $m_\phi$
yields~\citep{Flyvbjerg1990},
\begin{equation}
  m_\phi^2 = m^2 - \frac{1}{N} \Sigma(P, m)\vert_{P^2 = -m_\phi^2} =
 m^2 + \frac{m^2}{N} \li\left(\frac{\Lambda^2}{m^2}\right)\;.
\label{massshift}
\end{equation}
Solving Eq.~(\ref{massshift}) for $m^2$ results in
\begin{equation}
m^2 = m_\phi^2 - \frac{m_\phi^2}{N}\li\left(\frac{\Lambda^2}{m_\phi^2}
\right)\;.
\label{massshift2}
\end{equation}
Since the physical mass $m_\phi$ should be finite, it follows from
Eq.\ (\ref{massshift2}) that $m^2$, which is the vacuum
expectation value of the $\alpha$ field, receives divergent
contributions due to the quantum fluctuations at next-to-leading order
in $1/N$. 

Expressing the gap equation~(\ref{gapintermsofm}) in terms of
the finite physical mass $m_\phi^2$, using Eq.~(\ref{massshift2})
gives
\begin{equation}
 \frac{4 \pi}{g_b^2}  =  \left(1 - \frac{2}{N} \right) \log
\left (\frac{\Lambda^2}{m_\phi^2} \right)
     + \frac{2}{N} \left[ \log \log \left(\frac{\Lambda^2}{m_\phi^2}\right)
  + \gamma_E + \log 4 \right]\;.
\label{gapintermsofmphi}
\end{equation}
Because $\lim_{\Lambda \rightarrow \infty} [\log \log(\Lambda^2 / m_\phi^2) - \log \log
(\Lambda^2 / \mu^2)] = 0$, the gap
equation can now be renormalized.
Making the substitution $g_b^2 \rightarrow Z_{g^2} g^2$ with $g^2 =
g^2(\mu)$ and
\begin{equation}
Z_{g^2}^{-1} =  1+\frac{g^2}{4\pi}
\left(1 - \frac{2}{N} \right)
\log \frac{\Lambda^2}{\mu^2}+
\frac{1}{N} \frac{g^2}{2\pi}\log \log
\frac{\Lambda^2}{\mu^2} \;,
\label{Z1}
\end{equation}
one obtains the renormalized gap equation which reads
\begin{equation}
\frac{4\pi}{g^2(\mu)} = 
\left(1 - \frac{2}{N}\right)
\log \frac{\mu^2}{m_{\phi}^2} + \frac{2}{N}
\left(\gamma_E + \log 4 \right)\;.
\label{eq:gapren}
\end{equation}
The expression Eq.~(\ref{Z1}) for $Z_{g^2}^{-1}$ is exact in
$g^2(\mu)$ up to order $1/N^2$ corrections and results in the known
next-to-leading order $\beta$-functions as calculated by
\citet{Rim1984}, \citet{Orloff1986} and \citet{Biscari1990}
\begin{eqnarray}
  \beta(g_b^2) & = &  
  \Lambda \frac{\mathrm{d} g_b^2}{\mathrm{d}\Lambda} 
= -\left(1-\frac{2}{N}\right)\frac{g_b^4}{2\pi} \left(1 +
\frac{1}{N}\frac{g_b^2}{2\pi} \right)\;, \\
  \beta(g^2)  & = & 
  \mu \frac{\mathrm{d} g^2}{\mathrm{d}\mu} 
= - \frac{g^4}{2\pi} \left(1 - \frac{2}{N} \right) \;.
\label{eq:nonbetafuncs}
\end{eqnarray}
The negative sign in front of the $\beta$-functions shows that the
nonlinear sigma model is indeed asymptotically free. The
next-to-leading order beta function gives already the correct result
for $N=2$ because for a free theory the beta function vanishes. This is
an indication that the $1/N$ expansion in the nonlinear sigma model
might work well down to low values of $N$.
Solving the renormalized gap equation (\ref{eq:gapren}) for the
physical mass $m_\phi$ gives
\begin{equation}
  m^2_\phi = \mu^2 \exp \left \{ - \left [ \frac{4\pi}{g^2} -
 \frac{2}{N} \left(\gamma_E + \log 4 \right) \right] / \left (1 -
 \frac{2}{N} \right) \right \} \;.
 \label{eq:physmass}
\end{equation}
This equation illustrates the need for the non-perturbative approach,
since the physical mass turns out to be non-analytic in the coupling
constant. For $N=2$ this gives a zero physical mass, reflecting the
fact that the $\mathrm{O}(2)$ nonlinear sigma model is a free field
theory.

Using the gap equation, one can obtain the value of the effective
potential $\mathcal{V}(m^2)$ at the minimum which will be denoted by
$\mathcal{V}^{T=0}(m_0^2)$, where $m_0 = m_\phi$ given by
Eq.~(\ref{eq:physmass}). In terms of bare quantities, one finds
\begin{equation}
\mathcal{V}^{T=0}(m_0^2)
=  -\left(N - 2 \right)\frac{m_0^2}{8\pi}
     - \frac{1}{8 \pi} \Lambda^2 \log \frac{4\pi}{g_b^2} \;.
\label{pressure0}
\end{equation}
In contrast to the effective potential as a function of $m^2$, this
equation shows that the effective potential at the minimum can be
renormalized. It is expressed in terms of a finite mass which can be
found by solving the renormalized gap equations. The quadratic
divergence that is left in the minimum of the effective potential can
be removed by a subtraction. However, as explained before this term is
ambiguous, it would differ by a constant $i\pi$ if a different
prescription is used.  The expression for the minimum of the effective
potential at zero temperature will be subtracted from the minimum of
the effective potential at finite temperature in order to obtain a
finite pressure which is moreover unambiguous. As is shown in Sec.~4.5
this ambiguity can be removed without renormalizing the effective
potential as well.

For completeness, the wave-function renormalization factor of the
scalar fields will be derived in this paragraph. The wave-function
renormalization constant $Z_\phi$ follows from requiring that the pole
of the propagator has a residue equal to unity. This implies that
\begin{equation}
Z_\phi = 1 - \frac{1}{N} \left. \frac{\mathrm{d} \Sigma(P, m)}
{\mathrm{d} P^2} \right \vert_{P^2=-m_\phi^2} = 
1 + \frac{1}{N} \log \log \left(\frac{\Lambda^2}{\mu^2}\right)\;.
\label{wf1}
\end{equation}
The wave-function renormalization Eq.~(\ref{wf1}) is in accordance with
that obtained by \citet{Flyvbjerg1990}.
\cite{Rim1984} have calculated the wave-function renormalization
constant in dimensional regularization. Their result also
agrees with Eq.~(\ref{wf1}) upon identifying $\log
(\Lambda^2/\mu^2) \to2/\epsilon$ where $d=2-\epsilon$.

%---------------------------------------------------------------------
\subsection*{The effective potential at finite temperature}
The results at zero temperature are obtained analytically, but at
finite temperature this is in general not possible. Therefore,
numerical methods will be applied to calculate thermodynamical
quantities.  However, it is still possible to isolate the ultraviolet
divergences analytically and discuss the renormalization at finite
temperature without the need for numerical evaluation.

The effective potential through next-to-leading order in $1/N$ 
can be written as $\mathcal{V}(m^2) = N
\mathcal{V}_\mathrm{LO}(m^2) + \mathcal{V}_\mathrm{NLO}(m^2)$ where
\begin{equation}
\mathcal{V}_{\mathrm{LO}}=
\frac{m^2}{2g_b^2}-\frac{1}{2} \sumint_P\log\left(P^2+m^2\right)
\;,\;\;\;\;\;\;\;
\mathcal{V}_{\mathrm{NLO}} =
-\frac{1}{2}\sumint_{P}\log \Pi(P,m) \;.
\label{P-NLO}
\end{equation}
After summing over Matsubara modes and subtracting divergent $m$-
and $T$-independent terms one obtains in the limit $\Lambda
\rightarrow \infty$ for the leading order effective potential
\begin{equation}
\mathcal{V}_\mathrm{LO} = 
\frac{m^2}{2 g_b^2} - 
\frac{m^2}{8\pi}\left(1 + \log \frac{\Lambda^2}{m^2} \right)
 + \frac{1}{8 \pi} T^2 J_0(\beta m) \;,
\end{equation}
where
\begin{equation}
J_0(\beta m) = \frac{8}{T^2} \int_0^\infty \frac{\mathrm{d}q\,
q^2}{\omega_q} 
 n(\omega_q)\;.
\end{equation}
Here $\omega_q=\sqrt{q^2+m^2}$ and $n(x)=[\exp(\beta x)-1]^{-1}$ is
the Bose-Einstein distribution function.

To calculate the next-to-leading order contribution to the effective
potential one needs to know the inverse $\tilde \alpha$ propagator
$\Pi(P,m)$ at finite temperature which is defined in Eq.~(\ref{Pit}).
Summing over Matsubara frequencies and averaging over angles,
$\Pi(P,m)$ reduces to
\begin{equation}
\Pi(P,m) =  \frac{1}{4\pi \sqrt{P^2} \sqrt{P^2 + 4 m^2}} \log
\left(\frac{\sqrt{P^2 + 4m^2} + \sqrt{P^2}}{\sqrt{P^2+4m^2}
-\sqrt{P^2}} \right)
+ \Pi_T(P,m)
\;.
\end{equation}
where
\begin{equation}
\Pi_T(P,m) =
\frac{1}{2 \pi} \int_{-\infty}^{\infty}\frac{\mathrm{d} q}{\omega_q} 
\frac{P^2 + 2pq}{ (P^2 + 2pq)^2 + 4 p_0^2 \omega_q^2}
n(\omega_q)\;.
\label{eq:nonlinpit}
\end{equation}
This expression cannot be simplified further. It shows that the
inverse $\tilde \alpha$ propagator $\Pi(P,m)$ is temperature-dependent
and ultraviolet finite. The next-to-leading order contribution to
the effective potential is not finite though.

\begin{figure}[t]
\includegraphics{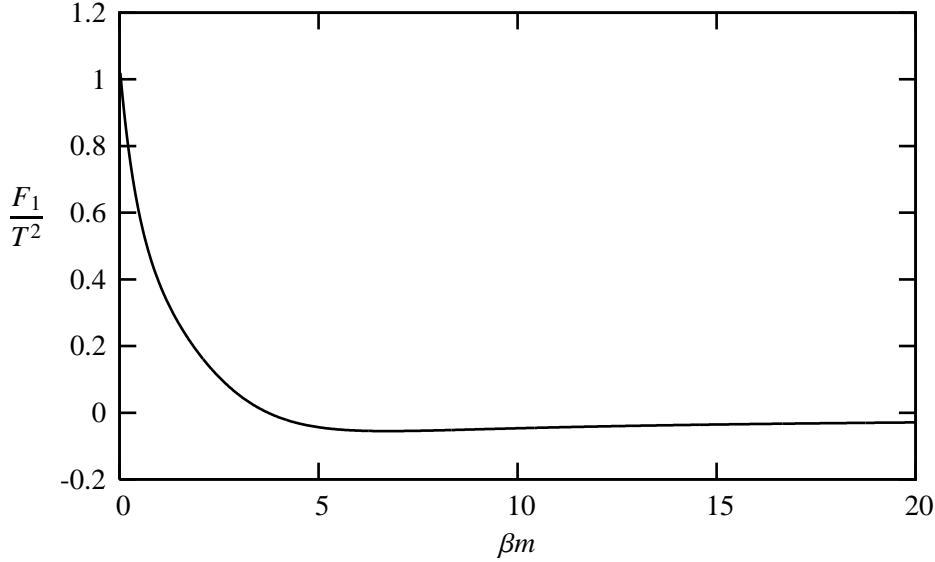}
\caption{The function $F_1 / T^2$ as a function of $\beta m$.
 This function contributes to the next-to-leading order finite
temperature effective potential.}
\label{fig:nonf1}
\end{figure}

The next-to-leading order contribution to the effective potential can
be written as
\begin{equation}
  \mathcal{V}_{\mathrm{NLO}} = -\frac{1}{2} \left(D_1 + D_2 + F_1 + F_2
  \right) \;,
\end{equation} 
where $D_1$ and $D_2$ contain divergences and $F_1$ and $F_2$ are
finite terms. As explained in Sec.~\ref{sec:abelplana} and
as verified numerically, the quantity
\begin{equation}
F_1 \equiv  \sumdiff_P \log \Pi(P, m),
\end{equation}
is finite since its integrand is suppressed by a Bose-Einstein like
distribution factor. To calculate $F_1$ a modified Abel-Plana formula
(see Sec.~\ref{sec:abelplana}) was used. The function $F_1 / T^2$
depends on $\beta m$ only and is displayed in Fig.~\ref{fig:nonf1}. As
is discussed in more detail in Sec.~\ref{sec:hightappr}
$\lim_{m\rightarrow 0} F_1(m) = \pi /3 T^2$ which can also be seen
from Fig.~\ref{fig:nonf1}. The limit of $\beta m \to \infty$ is
equivalent to fixing $m$ and taking $T \to 0$. In this limit $F_1$
should vanish because it is the difference of a finite temperature and
zero temperature contribution.
 
Because $F_1$ is finite all divergences of the next-to-leading order
effective potential arise from
\begin{equation}
 \int \frac{\mathrm{d}^2 P}{(2\pi)^2}
 \log \Pi(P,m) \;.
 \label{eq:deff1}
\end{equation}
The nonlinear sigma model only contains ultraviolet divergences. It is
infrared finite since the mass $m$ is non-vanishing. Therefore, in
order to isolate the divergences in Eq.~(\ref{eq:deff1}) one has to
consider the limit of momenta large compared to $T$. Due to the
suppression by the Bose-Einstein distribution function of high
momentum modes, only internal momenta $q$'s up to order $T$ contribute
to $\Pi_T(P,m)$. Hence, if the external momenta $P$ are much larger
than $T$ one can neglect factors of $q$ compared to factors of $P$ in
the high-momentum limit. In this way one can approximate the inverse
$\tilde \alpha$ propagator as follows
\begin{equation} 
\Pi_T(P,m) \approx \frac{1}{4 \pi} \frac{P^2}{P^4 + 4 m^2
p_0^2} J_1 \;,
\label{PiHE}
\end{equation}
where
\begin{equation}
  J_1({\beta m}) = 4 \int_0^\infty 
  \frac{\mathrm{d}q\,}{\omega_q}\, n(\omega_q)\;.
\end{equation}
In order to split off the prefactor in front of the logarithm of the zero
temperature part of $\Pi(P,m)$, it is useful to write $\Pi(P,m)$ as
\begin{equation}
\Pi(P, m) \equiv \frac{1}{4\pi} \frac{1}{\sqrt{P^2 \left(P^2 + 4m^2 \right)}}
 \tilde \Pi(P,m) \;.
\end{equation}
The factor $\sqrt{P^2(P^2+4m^2)}$ gives the following divergent 
contribution to the free energy
\begin{equation}
 D_1 = -\frac{1}{2} \int \spaceint{2}{P} \log \left(P^2 + 4m^2\right) 
  =  
 -\frac{m^2}{2 \pi} \left[1 + \log 
 \left( \frac{\Lambda^2}{4 m^2} \right) \right] \;.
\end{equation}
The divergent term $D_1$ is temperature independent. Hence it is
also  contained in the effective potential at zero temperature as
can be verified from Eq.~(\ref{eq:effpotzerot}).
The other divergent term $D_2$ and the finite term $F_2$ can be found
by calculating
\begin{equation}
\int \frac{\mathrm{d}^2 P}{(2\pi)^2} 
\log \tilde \Pi(P,m) = D_2 + F_2 \;.
\label{eq:findivd2f2}
\end{equation}
In order to isolate the divergent term $D_2$, one needs the large-$P$
behavior of $\tilde\Pi(P,m)$ which by using Eq.~(\ref{PiHE}) is found
to be
\begin{equation}
\tilde \Pi(P,m) \approx 
\tilde \Pi_\mathrm{HM}(P,m) = \log \left( \frac{P^2}{m^2} \right) 
 + \frac{2m^2}{P^2}
\left(1 + J_1 \right)
 - \frac{4 m^2 p_0^2}{P^4} J_1 + \mathcal{O}
   \left(\frac{m^4}{P^4}\right),
\end{equation}
where HM stands for high-momentum approximation. By integrating
$\log \tilde \Pi_\mathrm{HM}(P,m)$ over momentum it is found
that the divergent term $D_2$ is given by
\begin{equation}
D_2  = 
\frac{1}{4 \pi} \left[\Lambda^2 \log \log 
 \left( \frac{\Lambda^2}{\bar m^2} \right)
 - \bar m^2 \, \li 
  \left( \frac{\Lambda^2}{\bar m^2} \right) \right]
+ \frac{m^2}{2 \pi} \log \log \left (\frac{\Lambda^2}{\bar m^2} \right) \;,
\label{D2}
\end{equation}
here $\bar{m}^2 \equiv m^2 \exp[-J_1(\beta m)]$. In the limit of zero
temperature it can be seen that the divergence $D_2$ coincides with
the divergences found in the effective potential at zero temperature,
Eq.~(\ref{eq:effpotzerot}).
Since $\bar m^2$ depends explicitly on temperature, Eq.~(\ref{D2})
shows that the next-to-leading order effective potential contains
temperature dependent divergences. This seems surprising since one
expects that scales at which finite temperature are important are much
smaller than the ultraviolet cutoff. As a result, ultraviolet
divergences should be independent of temperature. In
Sec.~\ref{sec:tdepdivs} a general argument is given why these
$T$-dependent divergences arise. Hence apart from problems at zero
temperature discussed in the previous subsection, it even becomes 
impossible to renormalize the effective potential in a temperature
independent way. But, as will be discussed in the next section, these
problems disappear at the minimum of the effective potential.

\begin{figure}[t]
\includegraphics{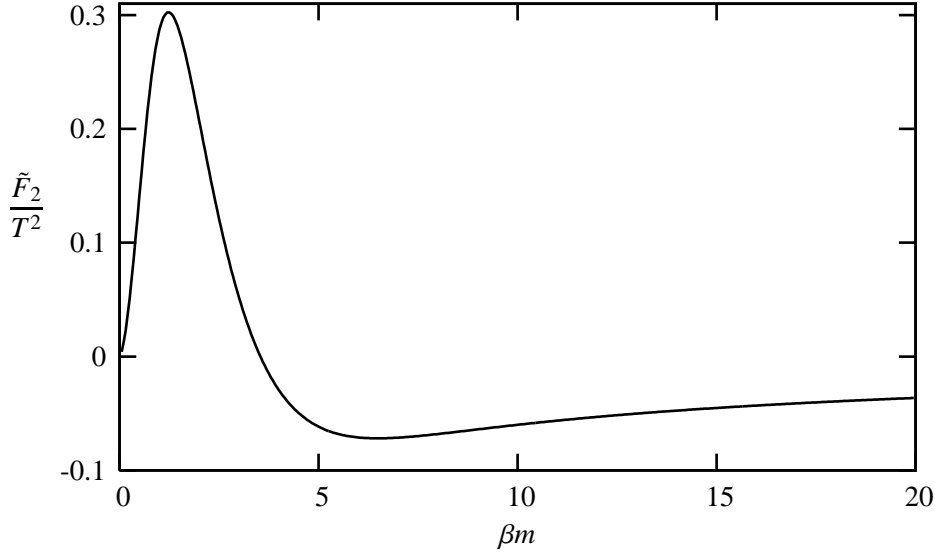}
\caption{The function $\tilde F_2 / T^2$ as a function of $\beta m$. This function
contributes to the next-to-leading order finite temperature effective
potential.}
\label{fig:nonf2}
\end{figure}

The finite function $F_2$ defined in Eq.~(\ref{eq:findivd2f2}) is the
difference of two divergent quantities. Due to the finite machine
precision the maximal reachable accuracy of the difference would be
rather small if the divergent quantities are calculated separately. It
is therefore better to write $D_2$ partly in terms of an integral in
the following way
\begin{equation}
  D_2 = \mathcal{P} \int \frac{\mathrm{d}^2 P}{(2\pi)^2} \log \log
  \left( \frac{P^2}{\bar m^2} \right) + \frac{m^2}{2\pi}
 \log \log \left( \frac{\Lambda^2}{\bar m^2} \right) \;.
\end{equation}
The last equation can be used to write $F_2$ in terms of an
integral and a very slowly diverging term.
\begin{equation}
  F_2 = \mathcal{P} \int \spaceint{2}{P} \log \left[ 
  \frac{\tilde \Pi(P,m)}
  {\log \left( P^2 / \bar m^2 \right)} \right]
 -  \frac{m^2}{2 \pi} \log \log 
  \left( \frac{\Lambda^2}{\bar m^2} \right) \;. 
\label{eq:intf2}
\end{equation}
To calculate $F_2$ numerically, the integral in Eq.~(\ref{eq:intf2})
was evaluated up to a cutoff $\Lambda$. Then for increasing values of
$\Lambda$, $F_2$ was calculated until convergence to a finite number
was found. This convergence indicates that by the method discussed
above indeed all divergences were isolated. In the limit of zero
temperature it can be seen from Eq.~(\ref{eq:effpotzerot}) that $F_2 =
m^2 \gamma_E / 2\pi$. It is convenient to subtract this zero
temperature contribution from $F_2$ in order to define $\tilde F_2$ as
\begin{equation}
 \tilde F_2 = F_2 - \frac{m^2}{2 \pi} \gamma_E \;.
\end{equation}
The function $\tilde F_2 / T^2$ is a function of $\beta m$ and is
displayed in Fig.~\ref{fig:nonf2}. In Sec.~\ref{sec:hightappr} it is
shown that for small $\beta m$, $F_2 \to 0$. Consequently, $\tilde F_2
\to 0$ in agreement with Fig.~\ref{fig:nonf2}. 
For large values of $\beta m$, which implies low temperatures, $\tilde
F_2 \to 0$. This is because in the definition of $\tilde F_2$ the zero
temperature contribution is subtracted.

\begin{figure}[t]
\includegraphics{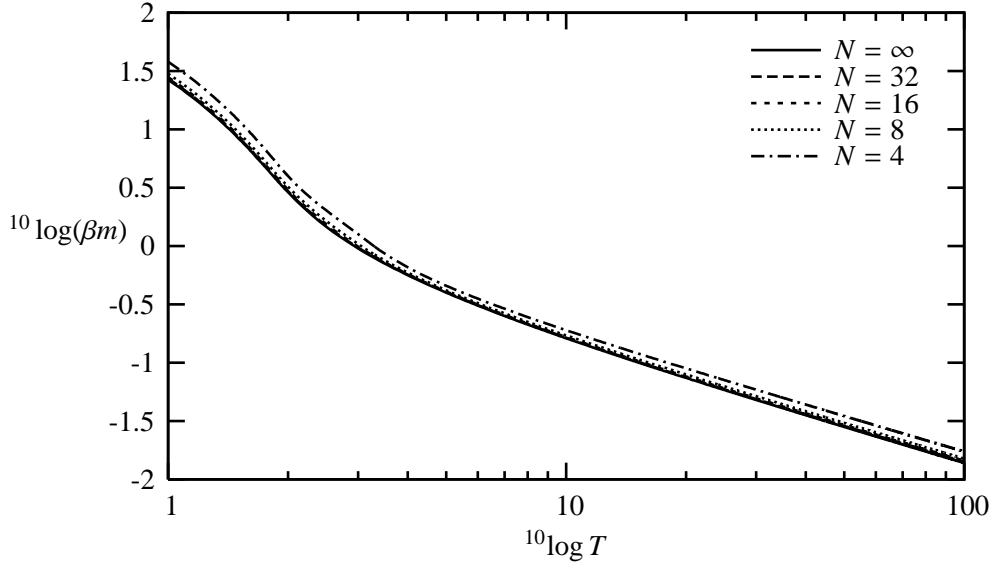}
\caption{The function 
$^{10}\log(\beta m_\phi)$ as a function of $T$, for different values
of $N$.}
\label{fig:solgap}
\end{figure}

Putting everything together, the finite temperature effective
potential becomes
\begin{equation}
\mathcal{V} = \frac{N m^2}{2 g_b^2}
-\frac{N m^2}{8\pi}\left[1 + \log \left(\frac{\Lambda^2}{m^2}  \right) 
\right]
  + \frac{N}{8\pi} T^2 J_0
- \frac{1}{2} \left(F_1 + D_1 + F_2 + D_2 \right) \;.
\label{eq:finaleffpot}
\end{equation}
The gap equation~(\ref{gap}) at nonzero temperature is
\begin{multline}
  \frac{4 \pi}{g_b^2} =  \log \left( \frac{\Lambda^2}{\bar m^2}\right)            
  + \frac{1}{N} \left[2 \log \log 
  \left( \frac{\Lambda^2}{\bar m^2} \right)
          -  \frac{\mathrm{d} \bar m^2}{\mathrm{d} m^2}
      \li 
  \left( \frac{\Lambda^2}{\bar m^2} \right)
\right. \\ \left. 
  - 2 \log \left( \frac{\Lambda^2}{4 m^2}  \right) + 
          4 \pi \frac{\mathrm{d}(F_1+F_2)}{\mathrm{d} m^2} \right] \;.
\label{g2}
\end{multline}
From the fact that $g_b^2$ is temperature independent, one can
conclude that $\bar{m}^2$ is also temperature independent {\it at
leading order} in the $1/N$ expansion, when it is a solution to the
gap equation. This will be used later on to conclude that the pressure
can be renormalized in a temperature-independent way. The
renormalization of the gap-equation at finite temperature proceeds
analogously to the zero-temperature case discussed in the
previous subsection. To remove the problematic li divergence the gap
equation will, like was done in the zero temperature case, be
expressed in terms of the physical mass $m_\phi$ at finite
temperature. This mass can be obtained by finding the pole of the
finite temperature $\phi$ field propagator,
Eq.~(\ref{eq:scalarfieldprop}), analytically continued to Minkowski
space. Here the finite temperature physical mass will be defined
as the pole of the propagator at $p_0 =0$ and $p^2 = -m_\phi^2$. 
It is found that
\begin{equation}
  m^2 = m_\phi^2 - \frac{\bar m^2_\phi}{N} \left[ \li 
\left( \frac{\Lambda^2}{\bar m_\phi^2}\right) + F_3\right]\;,
  \label{massshiftnonzerotemp}
\end{equation}
where $\bar m_\phi = m_\phi \exp[-J_1(\beta m_\phi)]$ and $F_3$ is a
finite function that depends on the temperature as well as $m_\phi$.
Since $m_\phi$ is merely used as a way to express the renormalized gap
equation in terms of finite quantities, any choice of $F_3$ will do
and we choose $F_3=0$. It was checked numerically that other choices
indeed do not alter the final result for the pressure. Using
Eq.~(\ref{massshiftnonzerotemp}), the gap equation expressed in the
physical mass becomes
\begin{equation}
  \frac{4 \pi}{g_b^2} = \log 
 \left (\frac{\Lambda^2}{\bar m_\phi^2} \right)           
  + \frac{1}{N} \left[2 \log \log 
  \left( \frac{\Lambda^2}{\bar m_\phi^2} \right)
      - 2 \log \left (\frac{\Lambda^2}{4 m_\phi^2} \right) +
          4 \pi \frac{\mathrm{d}(F_1+F_2)}{\mathrm{d} m_\phi^2} \right] \;.
\end{equation}
The gap equation can be rendered finite by the substitution $g_b^2
\rightarrow Z_{g^2} g^2(\mu)$, where $Z_{g^2}^{-1}$ is the zero
temperature renormalization constant given by Eq.~(\ref{Z1}). Hence
unlike the effective potential, the gap equation can be renormalized
$T$-independently without problems. The renormalized gap equation becomes
\begin{equation}
  \frac{4 \pi}{g^2(\mu)} = \left(1-\frac{2}{N} \right) 
\log \frac{\mu^2}{\bar m_\phi^2}
+ \frac{2}{N} \left[J_1(\beta m_\phi)+ \log 4 + 2 \pi 
\frac{\mathrm{d}(F_1+F_2)}{\mathrm{d} m_\phi^2} \right] \;.
\label{RenGapEqNonzeroT}
\end{equation}
In Fig.~\ref{fig:solgap} the solution to this renormalized gap
equation for the arbitrary choice $g^2 (\mu =500) = 10$ is displayed
as a function of temperature and for different values of $N$.

%=====================================================================
\section{Pressure}\label{sec:nlsmpres}
\begin{figure}[t]
\includegraphics{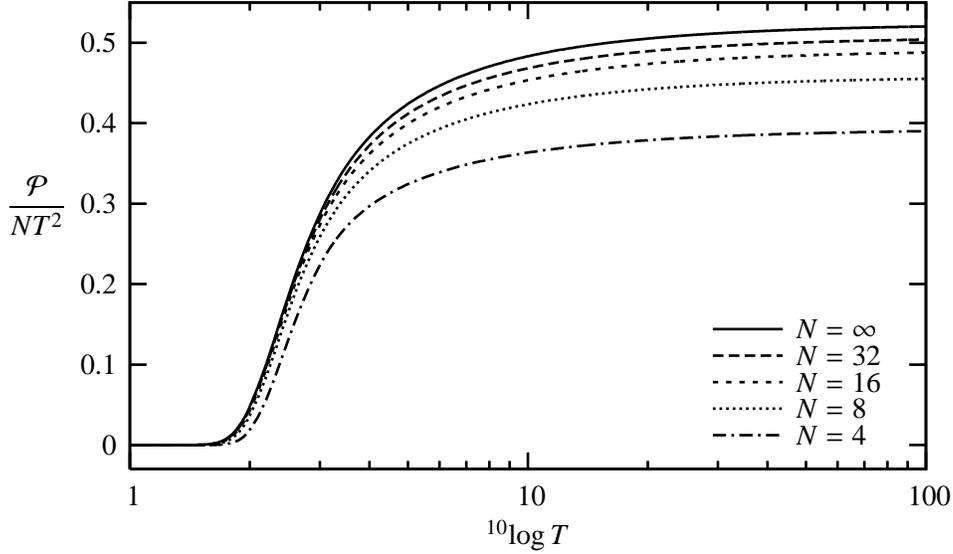}
\caption{Pressure of the $\mathrm{O}(N)$ nonlinear sigma model 
to next-to-leading order in $1/N$, normalized to $N T^2$, as a
function of temperature, for different values of $N$.}
\label{fig:nonpres}
\end{figure}

By solving the renormalized gap equation (\ref{RenGapEqNonzeroT}) one
can find the minimum value of the effective potential which is equal
to the pressure. The physical mass at this minimum will be denoted as
$m_\phi^2 = m_\phi^2(T)$ and $m_{\phi}^2(0) = m_\phi^2(T=0)$.  By
expressing the effective potential in terms of the physical mass using
Eq.~(\ref{massshiftnonzerotemp}) the pressure can be calculated. For
this one has to expand the $J_0$ and $J_1$ functions in $1/N$. Furthermore, one
should also take into account the zero temperature results since the
pressure is the difference of the minimum of the effective potential
at finite temperature and at zero temperature (given in
Eq.~(\ref{pressure0})).  As a result due to this difference the
problematic divergences drop out and one obtains
\begin{multline}
\mathcal{P} \equiv \mathcal{P}^{T} - \mathcal{P}^{T=0} 
 =  \frac{N-2}{8 \pi} \left[ m_{\phi}^2(0)
        -m_\phi^2 \right] 
 +  \frac{N}{8\pi} \left[ T^2 J_0(\beta m_\phi) +  m_\phi^2 
J_1(\beta m_\phi) \right] \\
 + \frac{1}{2} \left[m^2_\phi \frac{\mathrm{d}(F_1+F_2)}{\mathrm{d}
m_\phi^2} - F_1 - F_2 \right] \;.
\label{eq:finitepressure}
\end{multline}
This finite pressure was evaluated numerically, after solving
Eq.~(\ref{RenGapEqNonzeroT}) for $m_\phi(T)$.  The result for
different values of $N$ is shown in Fig.~\ref{fig:nonpres}, for the arbitrary
choice $g^2(\mu=500) = 10$, hence $T$ is given in the same units as
$\mu$.
As can be seen in the figure, $\mathcal{P}/N T^2$ approaches an
$N$-dependent constant (to be evaluated in the next section) at large
temperatures. This constant is exactly the pressure of a
non-interacting gas of $N-1$ bosons divided by $N$, which is
$\mathcal{P}_{\mathrm{SB}} / N = (N-1) \pi T^2 /6 N$.  This is an
immediate consequence of the asymptotic freedom and the fact that one
degree of freedom is effectively removed due to the constraint.  This
expression coincides with the perturbative result,
Eq.~(\ref{eq:nonlinpertpres}), as expected.  In the limit of zero
temperature the pressure divided by $T^2$ approaches zero
exponentially as it effectively becomes a gas of heavy quasi-particles,
see Fig.~\ref{fig:solgap}. 
The figure shows that corrections are really of order
$1/N$, which indicates that the $1/N$ is a good expansion for the
nonlinear sigma model.

\begin{figure}[t]
\includegraphics{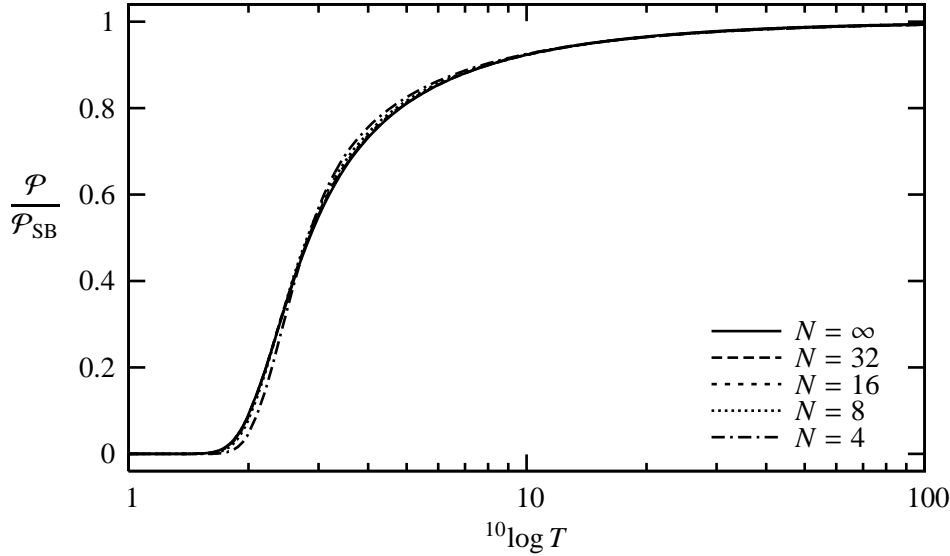}
\caption{Pressure of the $\mathrm{O}(N)$ nonlinear sigma model 
normalized its value at $T\rightarrow \infty$, as a function of
temperature for different values of $N$.}
\label{fig:presdivfree}
\end{figure}

\begin{figure}[t]
\begin{center}
\vspace{4.7cm}
\scalebox{0.4}{\includegraphics{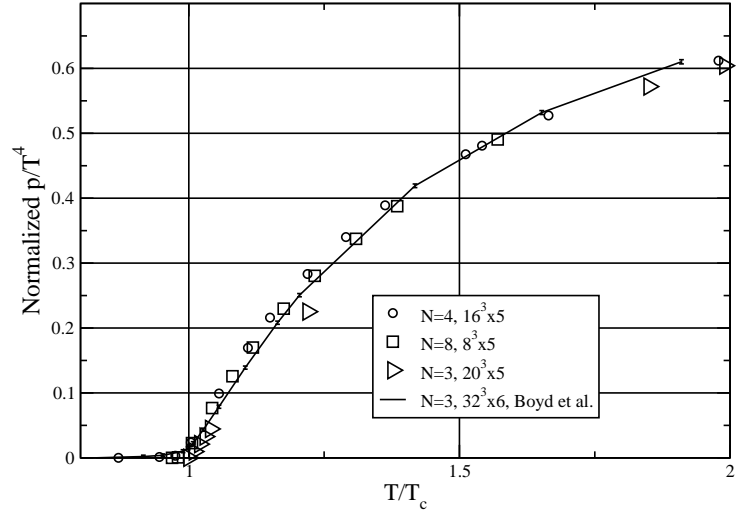}}
\end{center}
\caption{Results of lattice calculations by \citet{Bringoltz2005} of the 
pressure of pure SU($N$) Yang-Mills theory normalized to the pressure
of a free gas for $N=3$ (triangles), $N=4$ (circles) and $N=8$
(squares). The solid line is the $N=3$ result of \citet{Boyd1996}.}
\label{fig:pressu348}
\end{figure}

If the pressure, for a given value of $N$, is normalized to its value
at $T = \infty$, the normalized pressure has a very small dependence
on $N$ as is displayed in Fig.\ \ref{fig:presdivfree}. Similar
behavior is found in lattice calculations of the normalized pressure
of the pure SU$(N_c)$ gauge theory in 4-dimensions
\citep{Bringoltz2005}, see Fig.\ \ref{fig:pressu348}.  Also lattice
calculations performed by \citet{Karsch2000} show that for full QCD
the normalized pressure turns out to be rather insensitive to the
number of flavors.

To obtain the pressure, the next-to-leading order gap equation was
used. But as was argued by \citet{Root1974} one in principle only
needs the leading order gap equation in order to obtain the value of
the effective potential at the minimum.  Writing the solution to the
gap equations as
\begin{equation}
m^2 = m^2_\mathrm{LO} + m^2_{\mathrm{NLO}} / N \;.
\end{equation}
and Taylor expanding the effective
potential, one obtains (up to $\mathcal{O}(1/N)$ corrections)
\begin{equation}
  \mathcal{V}(m^2) =  
  N \mathcal{V}_{\mathrm{LO}}(m^2_{\mathrm{LO}}) 
   + \mathcal{V}_{\mathrm{NLO}}(m_{\mathrm{LO}}^2)  
  + m^2_{\mathrm{NLO}}
  \left. \frac{\partial \mathcal{V}_{\mathrm{LO}}(m^2)} 
    {\partial m^2} \right \vert_{m^2 = m^2_\mathrm{LO}} 
\label{eq:gapexp2}
\end{equation}
The last term of Eq.~(\ref{eq:gapexp2}) vanishes by using the leading-order 
gap equation. The pressure $\mathcal{P}$ can now be written as
\begin{equation}
\mathcal{P} \equiv N
\mathcal{P}_{\mathrm{LO}} + \mathcal{P}_{\mathrm{NLO}} 
\end{equation}
From the discussion above it follows that
\begin{equation}
  \mathcal{P}_{\mathrm{LO}}  = 
     \mathcal{V}^{T}_{\mathrm{LO}}(m^2_T) 
   - \mathcal{V}^{T=0}_{\mathrm{LO}}(m^2_0)  
  \;, \;\;\;\;\;\;\; 
  \mathcal{P}_{\mathrm{NLO}} =
     \mathcal{V}^{T}_{\mathrm{NLO}}(m^2_T) 
   - \mathcal{V}^{T=0}_{\mathrm{NLO}}(m^2_0) \;,
\end{equation}
where $m^2_T$ is the solution of the leading-order gap equation, at
temperature $T$.  By using the leading order gap equation it can be
shown that at the minimum, part of $D_2$ becomes
\begin{multline}
 \frac{1}{4 \pi} \left[\Lambda^2 \log \log 
 \left( \frac{\Lambda^2}{\bar m^2} \right)
 - \bar m^2 \, \mathrm{li}
  \left( \frac{\Lambda^2}{\bar m^2} \right) \right]
= 
\\ 
 \frac{\Lambda^2}{4\pi} 
 \left[ \log \left(\frac{4 \pi}{g_b^2} \right) 
 -  \exp \left(-\frac{4\pi}{g_b^2} \right)
  \mathrm{li}\, \exp \left( \frac{4 \pi}{g_b^2} \right)
 \right] \;.
\label{eq:d2divsnon}
\end{multline}
Since $D_2$ contains the logarithmic integral it is dependent on the
prescription and has an ambiguity.  From Eq.~(\ref{eq:d2divsnon}) it
follows that the problematic $T$-dependent divergences become
$T$-independent at the minimum of the effective potential, and vanish
in the calculation of the pressure due to the subtraction of the zero
temperature contribution. As a result the pressure is
unambiguous, i.e.\ independent of the choice of prescription.

%=====================================================================
\section{High-temperature approximations}\label{sec:hightappr}
\citet{Bochkarev1996} consider the nonlinear sigma model in 3+1 dimensions,
at next-to-leading order in the $1/N$ expansion. Since their result
for the pressure cannot be obtained analytically, they resort to a
``high-energy approximation''. In this section the same approximation
for the $1+1$-dimensional case will be made to compare it with
the exact numerical results obtained in Sec.~\ref{sec:nlsmpres}.

The essence of the high-energy approximation of \citet{Bochkarev1996} is
that one neglects the zero temperature part of $\Pi(P,m)$ and
approximates it by its temperature-dependent part $\Pi_T(P,m)$.
According to \citet{Bochkarev1996} one can then approximate 
$\Pi_T(P,m)$ by its high-energy behavior. In the present case of
the nonlinear sigma model in $1+1$ dimensions, these
approximations amount to
\begin{equation}
\Pi(P,m) \approx \Pi_T(P,m) \approx \frac{1}{4 \pi} \frac{P^2}{(p_0^2
+\omega_+^2) (p_0^2 + \omega_-^2)} J_1\;,
\end{equation}
where $\omega_{\pm}=\sqrt{p^2+m^2}\pm m$.  This expression is
identical to the high-momentum approximation to $\Pi_T(P,m)$,
Eq.~(\ref{PiHE}).  In the high-energy approximation the effective
potential is given by
\begin{multline}
\mathcal{V}_\mathrm{HEA} = 
\frac{N m^2}{2g_b^{2}}- \frac{N}{2} \sumint_{P}\log\left(P^2+m^2\right)
-\frac{1}{2}\sumint_{P}\log P^2
\\
+\frac{1}{2} \sumint_{P}\log\left(p_0^2+\omega_{+}^2\right)
+\frac{1}{2} \sumint_{P}\log\left(p_0^2+\omega_{-}^2\right)
+ \mathcal{O}(1/N)
\;.
\label{pnr}
\end{multline}
The resulting expression for the gap equation is
\begin{equation}
\frac{N}{g_b^2} = N\sumint_{P}\frac{1}{P^2+m^2}
-\sumint_{P}\frac{\omega_{+}^2}{mE_p}\frac{1}{p_0^2+\omega_{+}^2}
+\sumint_{P}\frac{\omega_{-}^2}{mE_p}\frac{1}{p_0^2+\omega_{-}^2}
\;.
\label{gnlo}
\end{equation}
Again, the gap equation requires coupling constant renormalization. In
this approximation, it is found that the renormalization constant is
\begin{equation}
Z_{g^2}^{-1}=1+\frac{g^2}{4\pi} \left(1-\frac{2}{N}\right) 
\log \left( \frac{\Lambda^2}{\mu^2} \right)\;,
\end{equation}
which differs slightly from the exact renormalization constant to
next-to-leading order in $1/N$, Eq.~(\ref{Z1}). It is in fact equal to
the perturbative one as for instance is discussed in
\citet{Peskin1995}. Making the substitution $g_b^2\rightarrow
Z_{g^2}g^2$, one obtains
\begin{equation}
 1 = \frac{g^2}{4\pi N}\left[NJ_1-K_1^{+}-K_1^{-}
+2(N-2)\log \frac{\mu}{m}
\right]\;.
\end{equation}
where the functions $K_1^{\pm}$ are defined as
\begin{equation}
K_1^{\pm} = \pm 4\int_0^{\infty}
 \frac{\mathrm{d}p\;\omega_{\pm}}{m E_p}
n(\omega_{\pm})\;. 
\end{equation}
By substituting the gap equation~(\ref{gnlo}) into the expression for
the effective potential in the high-energy approximation,
Eq.~(\ref{pnr}), the pressure 
becomes
\begin{multline}
\mathcal{P}_{\mathrm{HEA}} = 
\frac{N}{8\pi}\left[J_0T^2+(J_1-1)m^2
\right]+\frac{\pi}{6}T^2
\\
-\frac{1}{8\pi}\left[(K_0^{+}+K_0^{-})T^2+(K_1^{+}+K_1^- -2)m^2\right]
\;,
\label{pnlo}
\end{multline}
where the functions $K_0^{\pm}$ are given by
\begin{equation}
K_0^{\pm}  = \frac{8}{T^2}
\int_0^{\infty} 
\frac{\mathrm{d}p\;p^{2}}{E_p}n(\omega_{\pm})\;.
\end{equation}

\begin{figure}[t]
\includegraphics{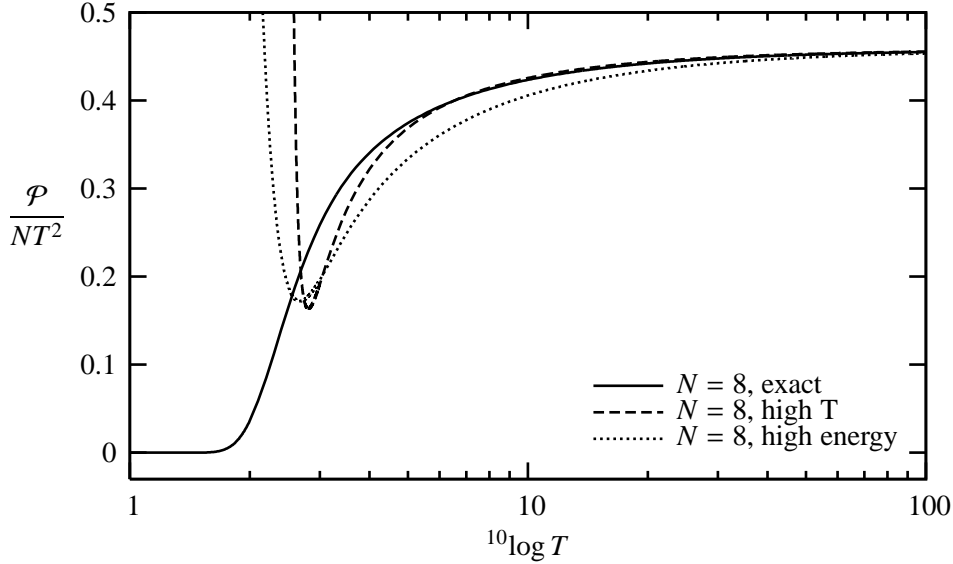}
\caption{Exact pressure as a function of temperature at
next-to-leading order for $N=8$ compared with the
high-energy and high-temperature approximations.}
\label{s13}
\end{figure}

From Fig.~\ref{s13}, one can see that the high-energy approximation
underestimates the pressure compared to the exact result.  The
advantage of an approximation like the high-energy approximation is
that the analytic calculations are simpler and that it is easier to
implement numerically. However, it is not a satisfactory approximation
because one cannot really justify that the temperature-independent
part of $\Pi(P,m)$ can be left out. It may even be considered
surprising that this approximation works. In Chapter~6 it will be
shown that the high-energy approximation in the nonlinear sigma model
in $3+1$ dimensions actually leads to wrong results.

It is possible to derive a different approximation, which is better
than the high-energy approximation. The inverse $\tilde
\alpha$-propagator $\Pi$ at finite temperature can be obtained by
first integrating over the momentum which gives 
\begin{equation}
  \Pi^T(P, m)  =   \frac{1}{2 \beta} 
\sum_{q_0=2n\pi T}
  \frac{1}{\sqrt{m^2 + q_0^2}} \frac{P^2 + 2 q_0 p_0}
  {P^4 + 4 q_0 (q_0 + p_0) P^2 + 4m^2 p^2} \;.
\label{pipi}
\end{equation}
Since it follows from the leading order gap equation that for high
temperatures and for all values of the coupling constant, $m \ll T$
see Fig.~\ref{fig:solgap}, $\Pi(P, m)$ can at high temperatures (HT)
be approximated by keeping only the $q_0 = 0$
mode in the sum Eq.~(\ref{pipi}) as follows
\begin{equation}
\Pi(P,m) \approx \Pi_{\mathrm{HT}}(P,m) = 
\frac{1}{2} \frac{1}{\beta m} \frac{P^2}{P^4 + 4 m^2 p^2}\;.
\label{eq:nonlinpiht}
\end{equation}
By using that $P^4 + 4m^2 p^2$ can be written as $[p_0^2 + (p+ im)^2 +
m^2][p_0^2 + (p-im)^2 +m^2]$ and shifting $p \rightarrow p \pm im$
after taking the logarithm, it is found that
\begin{equation}
  \sumdiff_P \log (P^4 + 4m^2 p^2) = 2 \sumdiff_P \log(P^2 + m^2) \;.
\end{equation}
Hence the function $F_1$ can be
approximated by
\begin{equation}
  F_1 \approx  \sumdiff_P \log \Pi_{\mathrm{HT}}(P,m) 
= \frac{1}{2\pi} T^2 J_0(\beta m) - \frac{\pi}{3}T^2 \approx
\frac{\pi}{3} T^2\;,
\label{ApproxofFs}
\end{equation}
The function $F_2 \approx 0$ in the high-temperature approximation
because $\int_P \log \Pi_{HT}$ does not contain a finite term that
depends on temperature. The part proportional to $\log \beta m$
gives rise to a divergence which by definition is not part of $F_2$.
These high temperature limits agree with the numerical calculations
displayed in Figs.~\ref{fig:nonf1} and \ref{fig:nonf2}.

The numerical calculation of $F_1 + F_2$ shows that for $m/T
\simordertwo 0.1$ this approximation has an error smaller than $10$
percent. To obtain the pressure in the high-temperature approximation,
Eq.~(\ref{eq:finitepressure}) was applied using the approximations to
$F_1$ and $F_2$.  The result for the pressure in the high-temperature
approximation is shown for comparison in Fig.~\ref{s13} (again for
$g^2(\mu=500) = 10$). It can be seen that the high-temperature
approximation is much better than the high-energy approximation.

One can approximate the pressure even further by expanding the
functions $J_0$ and $J_1$ in the limit $\beta m\rightarrow 0$:
\begin{eqnarray}
J_0 &=& \frac{4\pi^2}{3}-4\pi\beta m
-2\left(\log \frac{\beta m}{4\pi}+\gamma_E-\frac{1}{2}\right)(\beta m)^2 +
\mathcal{O}\left((\beta m)^4\right) \;, 
\label{ApproxofJ0}
\\
J_1 &=& 
\frac{2\pi}{\beta m} +2\left(\log \frac{\beta m}{4\pi}+\gamma_E\right) 
+ \mathcal{O}\left((\beta m)^2\right)\;.
\label{ApproxofJ1}
\end{eqnarray}
Inserting the approximations given in
Eqs.~(\ref{ApproxofFs}), (\ref{ApproxofJ0}) and (\ref{ApproxofJ1}) into
the gap equation~(\ref{RenGapEqNonzeroT}), one obtains
\begin{equation}
\beta m \approx \pi \left[ \left(\frac{2\pi}{g^2(\mu)}-\frac{\log
4}{N}\right)\left(1+\frac{2}{N} \right)-\gamma_E - \log \frac{\mu\beta}{4\pi}
\right]^{-1}\;,
\end{equation}
which indicates that $\beta m \sim 1/\log T$ for large $T$.  In the
limit $m/T \rightarrow 0$, the high-temperature
approximation of the pressure is
\begin{equation}
  \frac{\mathcal{P}}{N T^2} \approx \frac{\pi}{6} \left(1 - \frac{1}{N}\right)
  - \left(1-\frac{2}{N}\right) \frac{m}{4 T}\;, 
\end{equation}
where the first term is the pressure of a gas of free massless 
particles with $N-1$ degrees of freedom.

%=====================================================================
\section{Thermal infrared renormalons}
An observable $O$ in a quantum field theory like QCD is often only
known in terms of an expansion in the coupling constant
$\alpha$. Because negative $\alpha$ gives rise to unphysical behavior
as was argued by \citet{Dyson1952} for QED, it is expected that this
series expansion has zero radius of convergence. Hence one should view
it at best as an asymptotic series. One might wonder whether it is
possible by resumming the series expansion to learn about the
non-perturbative physics. In specific cases this is possible, but more
often one finds so-called renormalons, which give rise to ambiguities
in the relation between $O$ and its resummed version. These
ambiguities may give an insight in the size of the non-perturbative
corrections.

Since one can apply the $1/N$ expansion to the
nonlinear sigma model, observables can be calculated exactly in terms
of the coupling constant. These exact expressions can be used to study
the renormalons in the nonlinear sigma model as was done at zero
temperature by \citet{David1982,David1984,David1986} and
\citet{Beneke1998}. In this section this analysis is extended to
finite temperature.
 
This section consists of two parts. In the first part a short
introduction into renormalons will be given. In the second part the
nonlinear sigma model will be used to investigate the temperature
dependence of the renormalons and how they affect the possible
definitions of the renormalized effective potential, pressure and gap
equation. For a more extensive review on renormalons see
\citet{Altarelli1995} and \citet{Beneke1999}.

%---------------------------------------------------------------------
\subsection*{Renormalons}
An observable $O$ can in general be expressed in
terms of a perturbation series in $\alpha = g^2$ in the following way
\begin{equation}
 O(\alpha) \sim \sum_{n=0}^{\infty} a_n \alpha^n 
 \label{eq:oalphaseries}\;.
\end{equation}
It often happens that this series is an asymptotic series which is not
convergent. Such an asymptotic series can sometimes be resummed by the Borel
transform method, although in general this resummation is not
unique.  The Borel transform of $O(\alpha)$ is defined as
\begin{equation}
 B(t) = \sum_{n=0}^{\infty} \frac{a_n}{n!} t^{n} \;.
\end{equation}
The observable $O$ can now be expressed in terms of an integral over its Borel transform
\begin{equation}
 O(\alpha) = \frac{1}{\alpha} \int_{0}^{\infty} \mathrm{d}t\,
  e^{-t/\alpha} B(t) \;,
\end{equation}
which has the series expansion of Eq.~(\ref{eq:oalphaseries}).
For example if one takes $a_n = (-1)^n n!$, the series expansion for
$O$ is clearly divergent.  The Borel transform however is convergent
for $\vert t \vert < 1$ and equal to $B(t) = 1 / (1+t)$. To
resum $O(\alpha)$ it is assumed that $B(t) = 1 / (1+t)$ for any value
of $t$. The resummed expression for $O$ now yields
\begin{equation}
  O(\alpha) = \frac{1}{\alpha} \int_{0}^{\infty}
 \mathrm{d}t\, e^{-t/\alpha} \frac{1}{1+t} \;.
  \label{eq:integralo}
\end{equation}
The integral in Eq.~(\ref{eq:integralo}) is convergent and, moreover,
well-defined since there are no singularities along the integration
contour. So in this example the asymptotic series has been resummed
unambiguously. However, sometimes this method gives rise to
ambiguities. Take for example $a_n = n!$ in which case $B(t) = 1 /
(1-t)$ and
\begin{equation}
  O(\alpha) = \frac{1}{\alpha} \int_{0}^{\infty}
 \mathrm{d}t\, e^{-t/\alpha} \frac{1}{1-t} \;.
  \label{eq:integralo2}
\end{equation}
In this case the integral is not well defined since there is a
singularity at $t=1$ on the integration contour. This singularity is
called the renormalon pole. One can still give a
meaning to this resummed series by introducing a prescription how to integrate
around the pole
\begin{equation}
  O(\alpha)_\pm = \frac{1}{\alpha} \int_{0}^{\infty}
 \mathrm{d}t\, e^{-t/\alpha} \frac{1}{1-t \pm i \epsilon} \;.
  \label{eq:integralo23}
\end{equation}
This resummation gives rise to ambiguities, since one does not know
whether to take the $+$, $-$ or a combination like the principal
value prescription. But as long as the differences between the
prescriptions are not that big, it does not really matter. The
difference is equal to the residue at $t=1$, which is called the
renormalon residue. In this case it is equal to
\begin{equation}
 O(\alpha)_+ - O(\alpha)_- = \frac{2 \pi i }{\alpha} \exp \left(-1/\alpha \right) \;.
\end{equation}
So in this example the series can be resummed up to ambiguities of
order $\exp(-1 / \alpha ) / \alpha$. These ambiguities are tiny when
$\alpha$ is small. However for larger values of $\alpha$ they can
become important. These ambiguities indicate that the non-perturbative
corrections to Eq.~(\ref{eq:oalphaseries}) could be in this case of order
$\exp(-1 / \alpha ) / \alpha$.

\subsection*{Renormalons in the nonlinear sigma model}
Renormalons appear in the vacuum expectation value of the $\alpha$
field. A calculation shows that at zero temperature one can write
\citep{David1986}
\begin{equation}
\left<\alpha \right>  =  m_{\mathrm{LO}}^2 + \frac{4\pi m_{\mathrm{LO}}^2}{N} 
\int \frac{\mathrm{d}^2P}{(2\pi)^2} \frac{1}{\Pi} \frac{\partial \Pi}{\partial 
m_{\mathrm{LO}}^2} + \mathcal{O} \left(\frac{1}{N^2}
\right)\;,
\label{vevalpha}
\end{equation}
where $m_{\mathrm{LO}}^2\equiv \Lambda^2 \exp (-4\pi/g_b^2)$.
This equation is in agreement with the gap
equation~(\ref{gapintermsofm}) if $m^2$ is split as follows $m^2 = m_{\mathrm{LO}}^2 +
m_{\mathrm{NLO}}^2/N$.
The part of the integral in Eq.\ (\ref{vevalpha}) that has the infrared
renormalon pole in the Borel plane is in fact the contribution from
the integrand in the limit $P^2/m^2 \to \infty$ (its ``perturbative tail'')
\begin{equation}
\int^\Lambda \frac{\mathrm{d}^2 P}{(2\pi)^2} \frac{1}{\Pi_\infty} \frac{\partial
\Pi_\infty}{\partial m^2}  =  - \frac{1}{4\pi} 
{\mathrm{li}}\left(\frac{\Lambda^2}{m^2}\right)
 =  
- \frac{1}{4\pi} \frac{\Lambda^2}{m^2} e^{-x} {\mathrm{Ei}} (x)\;,
\label{renormalon}
\end{equation}
where $x=\log(\Lambda^2/m^2)$ and $\mathrm{Ei}(x)$ the exponential integral
which is defined below. In the limit $x\to \infty$, the logarithmic
integral has the asymptotic expansion
\begin{equation}
e^{-x} {\mathrm{Ei}} (x)  =  \sum_{n=0}^\infty \frac{n!}{x^{n+1}} \mp i\pi
e^{-x} 
 =  \int_0^\infty \mathrm{d}b \, \frac{e^{-bx}}{1-b} \mp i\pi
e^{-x} \;,
\label{rp}
\end{equation}
where $\arg(b) = \pm \varepsilon$.  From Eq.~(\ref{rp}), it is clear
that there is a renormalon pole at $b=1$.  This shows that when
$\Lambda \to \infty$ the value of $\left<\alpha\right>$ is
inherently ambiguous at next-to-leading order, due to the freedom in
the choice of prescription. \citet{David1984} has shown that this ambiguity also
arises in dimensional regularization.

The same problem appears in the calculation of the effective potential
even at its minimum, but {\it not} in the gap equation. The latter can
be seen from the last term of Eq.\ (\ref{V0}) which contributes to the
gap equation as follows
\begin{equation}
\frac{1}{2} \frac{\partial}{\partial m^2} \int \frac{\mathrm{d}^2
P}{(2\pi)^2}\log \Pi  = \frac{1}{2} \int \frac{\mathrm{d}^2
P}{(2\pi)^2} \frac{1}{\Pi} \frac{\partial \Pi}{\partial m^2} \;.
\end{equation}
The ambiguity that would arise from this term when removing its perturbative
tail (cf.\ Eq.\ (\ref{renormalon})) cancels in the gap equation
(\ref{gapintermsofm}) against the one arising in 
$m^2$ (cf.\ Eq.\ (\ref{vevalpha})).

The perturbative tail of the effective potential, i.e.\ the first two
terms of $D_2$ defined in Eq.~(\ref{D2}), corresponds to poles in the
Borel plane at $b=0$ and $b=1$, respectively.  Since $\bar m^2$ is
only temperature independent at the minimum (at LO only, but that is
sufficient because one is working at NLO), the subtraction of the
perturbative tail will become temperature dependent, {\em except} at
the minimum. Since subtracting temperature-dependent divergences
renders the remaining temperature-dependent terms meaningless, it is
impossible to define a finite effective potential at finite
temperature. In order to avoid any renormalon ambiguity, it also not 
possible to obtain a finite effective potential or even an unambiguous
minimum at zero temperature.  However, as was shown in Sec.~4.5 the
quantity $\mathcal{P}^{T} - \mathcal{P}^{T=0}$ is free from renormalon
ambiguities and is finite after temperature-independent coupling
constant renormalization.

Finally, using Eq.~(\ref{vevalpha}) one can investigate the finite
temperature dependence of renormalon contributions to
$\langle\alpha\rangle$ and the effective potential.  One can show that
Eq.~(\ref{vevalpha}) at finite temperature has exactly the same
renormalon contribution, i.e.\ neither the pole nor the residue become
temperature dependent. Secondly, the perturbative tail of the
effective potential which is given by the first two terms of $D_2$,
corresponds to poles in the Borel plane at $b=0$ and $b=1$.  The
positions of the renormalon poles are not affected by temperature.
Only the residues become temperature dependent, except at the minimum
of the potential, as we concluded earlier.  The fact that renormalon
pole positions are not affected by temperature, but residues are, is
also the case for the thermal ultraviolet renormalons in $\phi^4$ in
$3+1$ dimensions studied by \citet{Loewe2000}.

\section{Summary and Conclusions}
 
To summarize, the pressure in the nonlinear sigma model was calculated
at finite temperature to next-to-leading order in the $1/N$
expansion. The main result is that one can obtain an unambiguous,
finite pressure, by subtracting the zero-temperature value of the
pressure and renormalization of the coupling constant in a
temperature-independent way.  This procedure cannot be carried out
away from the minimum of the effective potential and it was argued
that defining a finite, effective potential by the subtraction of the
so-called perturbative tail, leads to ambiguities associated with
infrared renormalons.  In general, these become temperature dependent,
and this casts doubt on the usefulness of defining a finite effective
potential outside the minimum. Since it turns out that, as it should, 
physical quantities are independent of the choice of prescription, one
can apply any prescription for the logarithmic integral
without worrying about the possible ambiguities. This is what will 
be done in the remaining chapters of this thesis.

The expression for the pressure was calculated numerically at finite
temperature. This calculation shows that the $1/N$ expansion is a
meaningful expansion for all temperatures.  The high-energy
approximation that was originally applied to the nonlinear sigma model in 3+1
dimensions by \citet{Bochkarev1996} was also studied.  In 1+1 dimensions,
where one can compare with exact numerical results, it was found that
it underestimates the pressure for all temperatures. An improved
approximation was suggested, the so-called high-temperature
approximation. This approximation has the advantage that it is quite
easy to produce numerical results and agrees better with the exact
results.  At asymptotically high temperatures the pressure approaches
that of a gas of $N-1$ free massless particles. It was found that
the pressure divided by the pressure at infinite temperatures has
a very weak dependence on $N$. Similarly behavior is found in
lattice calculations as a function of $N_f$ and $N_c$.

 %non
\chapter{Thermodynamics of the $\boldsymbol{\mathbb{C}P^{N-1}}$ model}
\markboth{Thermodynamics of the $\mathbb{C}P^{N-1}$ model}{}

Using the $1/N$ expansion, the influence of quantum instantons to
the thermodynamics of the $\mathbb{C}P^{N-1}$ model in 1+1 dimensions
is studied. The auxiliary field effective potential is calculated to
next-to-leading order in $1/N$ and turns out to have
temperature-dependent ultraviolet divergences. These divergences can only be
renormalized at its minimum just like in the nonlinear sigma model
discussed in Chapter~4. By using that the $\mathbb{C}P^1$ model is
equivalent to the $\mathrm{O}(3)$ nonlinear sigma model it is argued
that the pressure for intermediate temperatures is dominated by the
effect of quantum instantons. This chapter is based on: {\it The
effect of quantum instantons on the thermodynamics of the
$\mathbb{C}P^{N-1}$ model}, J.O.~Andersen, D.~Boer and H.J.~Warringa,
hep-th/0602082.

\section{Introduction}  
It was discovered by \citet{Belavin1975} that the classical equations
of motion of Euclidean QCD have finite action solutions with nontrivial
topology. These solutions are called instantons. They are stationary
points of the classical action. In perturbation theory one most often
only takes into account the trivial vacuum and small fluctuations
around it.  However, in the non-perturbative regime the instanton
solutions and the fluctuations around it can really contribute to
physical quantities as was first observed by \citet{tHooft1976}. He
showed that instantons are giving rise to an additional source of
$\mathrm{U}(1)_A$ symmetry breaking in QCD, which is necessary to
explain the relatively large mass of the $\eta$ meson.

In this chapter it is investigated whether instantons can have an
influence on the thermodynamical quantities. To answer this question,
the $\mathbb{C}P^{N-1}$ model in $1{+}1$ dimensions will be studied since it
admits instanton solutions for all $N$.
The $\mathbb{C}P^{N-1}$ model was introduced by
\citet{Eichenherr1978}. It is possible to investigate its
non-perturbative regime using the $1/N$ expansion \citep{DAdda1978}.
An attractive feature of the $\mathbb{C}P^{N-1}$ model in $1{+}1$
dimensions are its similarities with QCD in $3{+}1$ dimensions. Like
QCD it is an asymptotically free theory. Moreover, classically the
$\mathbb{C}P^{N-1}$ model is scale-invariant. Due to renormalization
of the quantum corrections a mass scale is introduced in the
model. Furthermore, a mass gap for the scalar fields is generated
dynamically by the interactions. The local $\mathrm{U}(1)$ symmetry of
the $\mathbb{C}P^{N-1}$ model generates a long range interaction
between the scalar fields. Since the Coulomb potential in ${1+1}$
dimensions is proportional to the distance between the charges, this
model has confinement \citep{DAdda1978}. Finally as alluded to above,
the classical equations of motion of the $\mathbb{C}P^{N-1}$ model
admit instanton solutions \citep{Golo1978}. \citet{Lazarides1979}
investigated the $\mathbb{C}P^{N-1}$ model at finite temperature for
the first time and showed that at non-zero temperature the model is no
longer confining. For a review on the $\mathbb{C}P^{N-1}$ model at
finite temperature, see \citet{Actor1985}. Instantons in QCD at finite
temperature are discussed by \citet{Gross1980} and in the large $N_c$
limit by \citet{Schafer2004}.

As is explained in Chapter 3, thermodynamical quantities like the
pressure can be obtained by minimizing the finite temperature
effective potential. In this chapter the finite temperature effective
potential will be calculated to next-to-leading order in $1/N$.  Its
zero-temperature counterpart was calculated by
\citet{Campostrini1992}.  It turns out that the effective
potential contains temperature-dependent ultraviolet divergences.
These divergences become temperature-independent at the
minimum. Thermodynamic quantities like the pressure are defined at the
minimum of the effective potential and hence can be calculated without
such renormalization problems. The same phenomenon is also found in
the $\mathrm{O}(N)$ nonlinear sigma model in $1{+}1$ dimensions
(Chapter 4) and the $\mathrm{O}(N)$ linear sigma model in $3{+}1$
dimensions (Chapter 6). A general explanation of how these
temperature divergences arise is given in Sec.~3.4.

Instantons are characterized by a quantized winding number $Q$, which
is a topological invariant. The instantons give a contribution
proportional to $\exp(-\pi N Q / g_b^2)$ to the partition function,
where the non-analyticity in $g_b^2$ shows that the contribution is
non-perturbative. This contribution also indicates that instantons
effects disappear in the limit $N \rightarrow \infty$ as was argued by
\citet{Witten1978}. As a result one has to perform calculations to
next-to-leading order in $1/N$ to investigate the effects of
instantons in the $1/N$ expansion. It was argued by
\citet{Witten1978} using the $1/N$ expansion at zero temperature
that quantum corrections let the instanton configurations to
disappear. However,
\citet{Jevicki1979} showed that classical instantons still are present in the
quantum effective action, not longer as stationary solutions but as
poles. However, at finite temperature,
\citet{Affleck1980a, Affleck1980b, Affleck1980c} found that the
large-$N$ quantum effective action does contain stationary solutions
with quantized topological charge, called quantum instantons. Hence,
at finite temperature configurations with non-trivial topology can
contribute to physical quantities. In this chapter the same conclusion
will be drawn. The pressure is calculated for topological
configurations with zero winding number. The contribution from other
configurations will be ignored since they are difficult to
calculate. It is found that for intermediate temperatures, where the
leading order contribution to the pressure divided by $T^2$ displays a
sharp increase, the next-to-leading $1/N$ contributions gives rise to
a negative pressure. This is unphysical because it results in a
negative entropy. Therefore, the negative pressure indicates that an
important contribution, most probably from the configurations with
non-zero winding number, is left out. This
conclusion is further strengthened by using a correspondence
discovered by
\citet{Eichenherr1978} between the $\mathbb{C}P^1$ model and the
$\mathrm{O}(3)$ nonlinear sigma model. In the $1/N$ approximation
of the $\mathrm{O}(3)$ nonlinear sigma model one implicitly includes
all instanton configurations. Hence the difference between the
pressure of the $Q=0$ sector of the $\mathbb{C}P^1$ model and the
pressure of the $\mathrm{O}(3)$ nonlinear sigma model gives the
contribution of the topological configurations with $Q \neq 0$ to the
pressure of the $\mathbb{C}P^1$ model. In this way a strong hint is
found that the topological configurations with $Q \neq 0$ give a large
contribution to the pressure for intermediate temperatures.

This introduction will be ended by mentioning some related studies of
the $\mathbb{C}P^{N-1}$ model.
\citet{Schwab1982}, \citet{Munster1982} and
\citet{Munster1983} have investigated the $\mathbb{C}P^{N-1}$ model 
on the sphere $S^2$. \citet{Munster1983} remarked that it is naive to
 expand only around the saddle point $A_\mu = 0$ because it leads
to an improper treatment of zero modes.
Furthermore, the $\mathbb{C}P^{N-1}$ model has been studied on the
lattice as well, see for example \citet{Campostrini1992b, Campostrini1992c}
and \citet{Vicari1993}. A way to investigate the relevance of the
topological solutions is by adding a term $\theta Q$ to the
action. \citet{Olejnik1994} and \citet{Schierholz1995} calculated for
$N=4$ the free energy as a function of $\theta$. They found that
depending on the size of the coupling constant there is a phase
transition from the confined to the deconfined phase for $\theta
\leq
\pi$.  \citet{Azcoiti2004} found that for $N=10$ the CP symmetry is
spontaneously broken for $\theta = \pi$.
The $\mathbb{C}P^{N-1}$ model also has applications in condensed
matter physics. \citet{Pruisken2003} and \citet{Pruisken2005} have
applied this model to investigate the quantum Hall
effect. \citet{Ichinose1990} studied antiferromagnetism using the
$\mathbb{C}P^{N-1}$ model at finite temperature.

This chapter is organized as follows. In Sec.\ 5.2 the
$\mathbb{C}P^{N-1}$ model is introduced. The calculation of the
effective action and effective potential are performed in Secs.\ 5.3
and 5.4. In Sec.\ 5.5 the results of the calculation of the pressure
are discussed. Finally a summary and conclusions are given in Sec.\
5.6.

\section{The $\mathbb{C}P^{N-1}$ model}
The $\mathbb{C}P^{N-1}$ model is described by the following Lagrangian
density
\begin{equation}
  \mathcal{L} = \frac{1}{2} \partial_\mu \phi_i^* \partial^{\mu}
   \phi_i + \mathcal{L}_{\mathrm{int}}
   \;,\;\;\;\;\;\;\; \phi_i^* \phi_i = N / {g_b^2} \;,
      \;\;\;\;\;\;\;
  i = 1 \ldots N \;,
 \label{eq:cpnlagrandens}
\end{equation}
where $\phi(x)$ is a complex scalar field and $g_b$ is the bare coupling
constant. Under an $\mathrm{U}(1)$ transformation which is
parametrized by $\sigma(x)$, $\phi(x) \rightarrow
\exp[ i \sigma(x) ] \phi(x)$. 
By requiring that the Lagrangian density is invariant under local
$\mathrm{U}(1)$ transformations, the interaction term
$\mathcal{L}_{\mathrm{int}}$ can be determined from the transformation
rule of the Lagrangian density which reads
\begin{equation}
  \delta \mathcal{L} = i (\partial_\mu \sigma) 
  \left(\partial^\mu \phi_i^* \right) \phi_i
  + \delta \mathcal{L}_{\mathrm{int}} = 0 \;.
 \label{eq:lagrantrans}
\end{equation}
Hence the interaction term should contain derivatives. 
If the interaction Lagrangian density is chosen as follows
\begin{equation}
  \mathcal{L}_{\mathrm{int}} = \frac{g_b^2}{2 N}  
   \left( \phi_i^* \partial_\mu \phi_i \right) 
  \left( \phi_j^* \partial^\mu \phi_j \right) \;,
\label{eq:cpnint}
\end{equation}
then Eq.~(\ref{eq:lagrantrans}) is satisfied.  As a result the model
becomes invariant under global $\mathrm{SU}(N)$ transformations as well.

The Lagrangian density can also be written in terms of gauge fields
$A_{\mu}$ which transform under a $\mathrm{U}(1)$ gauge transformation
as
\begin{equation}
  A_\mu(x) \rightarrow A'_\mu(x) = A_\mu(x) - \partial_\mu \sigma(x)
\;.
\end{equation}
The Lagrangian density expressed in gauge fields is given by
\begin{equation}
  \mathcal{L} = \frac{1}{2} 
  \left \vert \mathcal{D}_{\mu} \phi_i \right \vert^2
\;,\;\;\;\;\;\;\; \phi_i^* \phi_i = N / {g_b^2} \;.
  \label{eq:cpnlagrgaugefields}
\end{equation}
where the covariant derivative $\mathcal{D}_\mu = \partial_\mu + i A_\mu$.
By solving the equations of motions of the $A_\mu$ fields one finds
\begin{equation}
  A_{\mu} =  i \frac{g_b^2}{N} \phi_i^* \partial_{\mu} \phi_i^{\phantom{*}} \;.
 \label{eq:amutrans}
\end{equation}
If this expression is inserted into
\eqref{eq:cpnlagrgaugefields}
the original Lagrangian density \eqref{eq:cpnlagrandens} and
Eq.~(\ref{eq:cpnint}) is recovered.

\subsection*{Coset models}
The $\mathbb{C}P^{N-1}$ and the $\mathrm{O}(N)$ nonlinear sigma model
are examples of coset models. A coset model is a theory of fields
$\phi(x)$ which take values in the coset space $G / H$.  Here $G$
denotes a compact Lie group and $H$ a closed subgroup of $G$. The
coset space $G/ H$ is a manifold though only a Lie group if $H$ is an
invariant subgroup of $G$. The coset space has dimension
$\mathrm{dim}\, G / H = \mathrm{dim}\, G - \mathrm{dim}\,H$. In
2-dimensional Euclidean space, $\phi$ is a map from $\mathbb{R}^2
\rightarrow G/H$, whereas at finite temperature this map becomes $S
\times \mathbb{R} \rightarrow G / H$. In a coset model the fields
transform as $\phi(x) \rightarrow g \phi(x)$ where $g \in G$. The
Lagrangian density of a coset model is given by 
\begin{equation}
 \mathcal{L}  = \frac{1}{2} \eta^{\mu \nu}
 g^{ij}(\phi) \partial_\mu \phi^*_i \partial_\nu \phi^{\phantom{*}}_j
 \;,
\label{eq:lagrancoset}
\end{equation}
where $\eta_{\mu \nu}$ is the space-time metric ($\delta_{\mu \nu}$ in
Euclidean space) and $g_{ij}(\phi)$ is a metric on the coset space
$G/H$. The metric $g_{ij}$ is to be chosen in such a way that the
Lagrangian density is invariant under transformations $G$.

Coset models have an interesting connection to the topology of
space-time. If one identifies all points at space-time infinity, the
fields $\phi(x)$ become a map from $S^2 \rightarrow G / H$. These maps
can be topologically nontrivial, which implies that there are maps
$\phi(x)$ which cannot be deformed continuously into each other. Such
maps can be classified according to the second homotopy group of
$G/H$, $\pi_2(G/H)$.  In the case that the second homotopy group is
nontrivial, the action of the coset model has more than one stationary
solution, which in Euclidean space-time are called instantons. At
finite temperature all points at $x_1 = \pm \infty$ can be identified,
as a result space-time becomes $S^2$ as well. Hence finite
temperature instantons (also called calorons) do exist, but unlike
instantons at zero temperature they satisfy the boundary conditions
for scalar fields in imaginary time \citep{Lazarides1979,
Affleck1980a, Gross1980, Bruckmann2005}.

The $\mathrm{O}(N)$ nonlinear sigma model is an $\mathrm{O}(N) /
\mathrm{O}(N-1) \cong S^{N-1}$ coset model with real scalar fields. 
Since in a coset model, the fields take their value in the coset
space, it follows that $\phi_i \phi_i = N/g_b^2$, where $i=1\ldots N$
and $N/g_b^2$ is the radius of $S^{N-1}$. If the metric is chosen to
be $g_{ij} =
\delta_{ij}$ the Lagrangian density becomes invariant under G. 
Furthermore the Lagrangian density discussed in Chapter~4, 
Eq.~(\ref{eq:lagran_non}), is recovered from Eq.~(\ref{eq:lagrancoset})
in this way. Since $\pi_2(S^{N-1})
\neq 0$ only for $N=3$, the $\mathrm{O}(N)$ nonlinear sigma model has
instanton solutions for $N=3$ only.

The $\mathbb{C}P^{N-1}$ model is an $\mathrm{SU}(N) / \mathrm{U}(N-1)
\cong \mathbb{C}P^{N-1}$ coset model. The space $\mathbb{C}P^{N-1}$
is the so-called complex projective space, it is the space of
$N$-dimensional complex vectors $z$ satisfying the equivalence
relation $z \sim z'$ if $z = \lambda z'$ where $\lambda \in
\mathbb{C}$ and the relation $z^\dagger z = c$.
Hence the $N$ scalar fields
$\phi_i(x)$ of the $\mathbb{C}P^{N-1}$ model satisfy $\phi_i^*(x)
\phi_i(x) = N/g_b^2$. The equivalence relation together with the constraint
directly translates into the requirement of $\mathrm{U}(1)$ gauge
invariance. The corresponding metric on the coset space can be read
off from Eq.\ (\ref{eq:cpnint}), $g_{ij} = \delta_{ij} + g_b^2
\phi^{\phantom{*}}_i \phi^*_j / N$. It turns out that
 $\pi_2( \mathbb{C}P^{N-1} ) =
\mathbb{Z}$, hence the $\mathbb{C}P^{N-1}$ model has instanton
solutions for any $N$.

\subsubsection*{Instantons}
Classical instantons are solutions to the classical equations of motion in
Euclidean space-time which have finite action. They
can be characterized by an integer topological charge $Q$ which
is given by
\begin{equation}
 Q = \frac{1}{2 \pi i} \frac{g_b^2}{N} \int \mathrm{d}^2 x\,
   \epsilon_{\mu \nu} \partial_\mu \left(
 \phi_i^*  \partial_\nu \phi_i^{\phantom{*}}
\right) \;.
\end{equation}
Using Eq.~(\ref{eq:amutrans}) the expression for $Q$ can be written in
terms of the gauge field as follows
\begin{equation}
Q = \frac{1}{2\pi} \int \mathrm{d}^2 x\, \epsilon_{\mu \nu} 
  \partial_\mu A_\nu\;.
 \label{eq:topchargegf}
\end{equation}
To show that $Q$ is an integer, consider a configuration with finite
action.  In that case the fields $\phi$ should go to a constant times
a phase factor (arising from gauge invariance) at infinity,
$\lim_{x\rightarrow \infty} \phi(x) \equiv \phi_\infty \exp[i
\sigma(x)]$. Since $Q$ is the integral over a total divergence, it can
be written as a surface integral at infinity. As a result one finds
\begin{equation}
  Q = \frac{1}{2 \pi} \int_0^{2\pi}  \mathrm{d} \theta \,
  \frac{\mathrm{d} \sigma({R=\infty}, \theta)}{\mathrm{d} \theta}
  = \frac{1}{2 \pi} \Delta \sigma \;,
  \label{eq:topologicalcharge}
\end{equation}
where $\Delta \sigma$ denotes the change in $\sigma$ by going along a
circle at infinity. Since the fields with finite action should be
continuous functions, $\sigma$ is a multiple of $2\pi$. Hence $Q$ can
only take integer values.

Like any
vector, $C_{\mu}^i = \mathcal{D}_\mu \phi^i$ satisfies 
$\vert C_\mu^i \pm i \epsilon_{\mu \nu} C_\nu^i \vert^2
\geq 0$. Working out this equation gives a lower bound on the Lagrangian
density of the $\mathbb{C}P^{N-1}$ model
\begin{equation}
  \mathcal{L} \geq \pm \tfrac{1}{2} i \epsilon_{\mu \nu}
  \partial_\mu \left( \phi^*_i \partial_\nu \phi_i^{\phantom{*}} \right) 
  = \pm \frac{N}{2g_b^2} \epsilon_{\mu \nu} \partial_\mu A_\nu
 \;.
\end{equation}
Integrating this equation over $x$ results in the classical action $S$
which satisfies the following lower bound
\begin{equation}
  S \geq \frac{\pi N}{g_b^2} \left \vert Q \right \vert \;.
\end{equation}
For instantons this lower bound turns into an equality. The explicit form
of the instantons can be found by solving the self-duality equation
\begin{equation}
  \mathcal{D}_\mu \phi_i = \pm i \epsilon_{\mu \nu} \mathcal{D}_\nu \phi_i \;.
\end{equation}
The solutions of this equation, the instantons, are of the following
form \citep{Golo1978}
\begin{equation}
  \phi_i(x) = \left({N}/{g_b^2} \right)^{1/2} e^{i \sigma(x)}
  \frac{w_i(x_0 \pm i x_1)}{\left \vert w(x_0 \pm i x_1) \right
  \vert}\;,
\end{equation} 
where $w(x_0 \pm i x_1)$ is a smooth $N$-dimensional complex
vector function. The one instanton (with $Q=1$) for example can be
written as \citep{Golo1978, DAdda1978} 
\begin{equation}
\phi_i(x) = 
 \frac{\lambda u_i + \left[(x_0 - a_0) - i(x_1-a_1) \right]v_i}
{ \left(\lambda^2 + (x-a)^2 \right)^{1/2} } \;,
\end{equation}
where $a_\mu$ is the position of the instanton in space and $\lambda$
the size of the instanton. The constants $u$ and $v$ obey the
following relations, $\vert u \vert^2 = N/g_b^2$, $\vert v \vert^2 =
N/g_b^2$ and $u^* \cdot v =0$.

These results can be generalized to finite temperature. If all points
at $x_1 = \pm \infty$ are identified the space changes from $S \times
\mathbb{R}$ into $S^2$. To obtain a finite action at non-zero
temperature, the fields $\phi$ still have to go to a constant,
$\lim_{x\rightarrow \infty} \phi(x) \equiv \phi_\infty \exp[i
\sigma(x)]$. Due to the finite temperature boundary condition
$\sigma(x_0, x_1) = \sigma(x_0+\beta, x_1)+2 \pi N$.  The topological
charge is still quantized at finite temperature
\citep{Affleck1980a}. The action has the same bound as at zero
temperature.  The explicit form of the instanton at finite temperature
(also called calorons) follow from solving the self-duality equations
using the boundary condition on $x_0$, see \citet{Affleck1980b}.

\subsubsection*{A correspondence between $\boldsymbol{\mathbb{C}P^{1}}$ and 
$\boldsymbol{\mathrm{O}(3)}$} 

The $\mathbb{C}P^{1}$ model is equivalent to the $\mathrm{O}(3)$
nonlinear sigma model, which was shown by \citet{Eichenherr1978}, see
also \citet{Banerjee1994}. This means that there is a one-to-one
correspondence between those two models. Such an equivalence for
example implies that one can obtain the pressure of the
$\mathbb{C}P^{1}$ model by calculating the pressure of the
$\mathrm{O}(3)$ nonlinear sigma model.  The correspondence can be made
explicit by writing the $\mathrm{O}(3)$ nonlinear sigma fields
$\chi(x)$ as follows
\begin{equation}
  \chi_a(x) = \left (g_b^2 / N \right)^{1/2} \phi^*_i(x) 
  (\sigma_a)^{\phantom{**}}_{ij} \phi^{\phantom{*}}_j(x)
  \;,\;\;\;\;\;\;\;\;a = 1 \ldots 3\;,
  \label{eq:duality}
\end{equation}
where $\sigma_a$ are the three $2 \times 2$ Pauli matrices satisfying
$\{ \sigma_a,
\sigma_b \}= 2 \delta_{ab}$. Using Eq.~(\ref{eq:duality}) the
$\mathrm{O}(3)$ nonlinear sigma Lagrangian density,
Eq.~(\ref{eq:lagran_non}), with constraint turns into the
$\mathbb{C}P^{1}$ Lagrangian density, \eqref{eq:cpnlagrandens}, with
the corresponding constraint.

\section{Effective action}
To obtain the effective action, the constraint from
\eqref{eq:cpnlagrandens} can be implemented by introducing an auxiliary
Lagrange multiplier field $\alpha$. This gives rise to the following
Lagrangian density which is equivalent to the original Lagrangian
density Eq.~(\ref{eq:cpnlagrandens})
\begin{equation}
   \mathcal{L} = \frac{1}{2} 
  \left \vert \mathcal{D}_{\mu} \phi_i \right \vert^2
  - \frac{i}{2}  \alpha \left(\phi_i^* \phi^{\phantom{*}}_i
- N / g_b^2\right)
\end{equation}
Since the Lagrangian density is now quadratic in the $\phi$ fields, the
Gaussian integration over these fields can be performed. This results
in the following effective action 
\begin{equation}
  S_{\mathrm{eff}} = N \mathrm{Tr}\, \log \left (-\mathcal{D}_{\mu} 
    \mathcal{D}^{\mu} - i \alpha \right)  
 + i \frac{N}{2g_b^2} \int \mathrm{d}^2 x\, \alpha(x) 
\;.
   \label{eq:effectiveactionCPN}
\end{equation}
The covariant derivative $\mathcal{D}_{\mu}$ transforms as $U^\dagger
\mathcal{D}_{\mu} U$. Hence, as expected this effective action is
invariant under local $\mathrm{U}(1)$ transformations. In order to
obtain the effective potential, $S_{\mathrm{eff}}$ has to be expanded
around the vacuum expectation values of the $\alpha$ field. The vacuum
expectation value of the $\alpha$ field is purely imaginary as is
proved in Eq.~(\ref{eq:alphaimag}). Therefore $\alpha$ can be
expressed in terms of the sum of its vacuum expectation value $im^2$
and a quantum fluctuating field $\tilde \alpha$ as follows, $\alpha= i
m^2 + \tilde \alpha / \sqrt{N}$.  For convenience the gauge fields are
rescaled by a factor $\sqrt{N}$. This rescaling with factors of
$\sqrt{N}$ does not have any effect on the final results since the
$\mathrm{\tilde \alpha}$ and $A_\mu$ fields are integration
variables of the path integral. It is just a convenient way to
organize the $1/N$ expansion. Working out the square of the covariant
derivative and inserting the vacuum expectation value gives the
following effective action
\begin{multline}
  S_{\mathrm{eff}} = N \Trace \log \left (-\partial^2 + m^2 - 
  \frac{i}{\sqrt{N}} 
 \left \{ \partial_{\mu}, A^{\mu} \right \}
  +  \frac{1}{N} A_{\mu} A^{\mu} - i
  \frac{\tilde \alpha}{\sqrt{N}}
   \right) \\
  - \frac{N}{2 g_b^2} \int \mathrm{d}^2 x
  \left[m^2 - \frac{i}{\sqrt{N}} \tilde \alpha(x) \right] 
 \label{eq:effaction} \;.
\end{multline}

Now as shown by \citet{Affleck1980a} $S_\mathrm{eff}$ contains
stationary solutions $A_\mu$ at finite temperature, which vanish at
zero temperature according to \citet{Witten1978}.  These solutions
have a quantized topological charge given by
Eq.~(\ref{eq:topchargegf}).  Since these solutions are
stationary points of an action in which quantum effects are
incorporated, they are called `quantum instantons'. In the limit of
high temperature, the exact form of these quantum instantons can be
found \citep{Affleck1980a}. Such instantons need to be considered in a
full calculation of the pressure.  As a first step to investigate the
relevance of the quantum instantons to the pressure, only the
fluctuations around the trivial configuration $A_\mu = 0$ are taken
into account in this chapter. The result for the pressure of the
$\mathbb{C}P^{1}$ model with $Q=0$ will be compared to the pressure of
the $\mathrm{O}(3)$ nonlinear sigma model in which the contribution of
all quantum instantons is take into account. Because the
$\mathbb{C}P^{1}$ model is equivalent to the $\mathrm{O}(3)$ nonlinear
sigma model this comparison should therefore give the contribution to
the pressure of the quantum instantons.

To obtain the pressure for the $Q=0$ configuration, $S_{\mathrm{eff}}$
has to be expanded around the vacuum expectation value $im^2$ and the
trivial configuration $A_\mu =0$ (with $Q=0$). Expanding
$S_{\mathrm{eff}}$ gives
\begin{multline}
  S_\mathrm{eff} = N \Trace \log  \left (-\partial^2 + m^2 \right)
  + N \Trace \sum_{k=1}^{\infty} \frac{(-1)^{k+1}}{k} \left( 
  \frac{\frac{1}{N} A_{\mu}A^{\mu} + 
  \frac{1}{\sqrt{N}} \tilde \alpha - \frac{i}{\sqrt{N}}
   \left \{ \partial_{\mu}, A^{\mu} \right \} }
  { -\partial^2 + m^2 } \right)^k 
\\
 - \frac{N}{2 g_b^2} \int \mathrm{d}^2 x
  \left[m^2 - \frac{i}{\sqrt{N}} \tilde \alpha(x) \right] 
\;.
\label{eq:cpneffact}
\end{multline}
With help of the following relations
\begin{eqnarray}
\mathrm{Tr}\, O &=& \int_x \langle x \left \vert O \right \vert x \rangle \;,
\\
 \langle x \left \vert 
\frac{1}{
-\partial^2 + m^2} \right \vert
  y \rangle &=&
  \int_P e^{i P(x-y)} \frac{1}{P^2 + m^2} 
\;,
\\
 - i\langle x \left \vert A^{\mu}  \partial_{\mu} \right \vert
  y \rangle  &=&
 \int_P e^{ i P(x-y)} P_{\mu} A^{\mu}(x) \;,
\\
 -i \langle x \left \vert \partial_{\mu} A^{\mu} \right \vert
  y \rangle
  &=&
  \int_P e^{ i P(x-y)} P_{\mu} A^{\mu}(y) \;, 
\end{eqnarray}
the traces in $S_\mathrm{eff}$ can be evaluated. To next-to-leading order in $1/N$ 
these relations lead to
\begin{eqnarray}
\Trace \left(
   \frac{\tilde \alpha}{ -\partial^2 + m^2} \right)
&=&
 \int_X \tilde \alpha(x) \int_P \frac{1}{P^2 + m^2} \;,
\\
-i \Trace
\left(
\frac{\{\partial_\mu, A^{\mu} \}}{-\partial^2 + m^2} 
\right)
&=& 2 \int_X A^{\mu} (x) \int_P \frac{P_\mu}{P^2 + m^2} = 0 \;,
\\
  \Trace \left( \frac{A_{\mu}A^{\mu}}{-\partial^2 + m^2} \right) 
&=& 
  \int_{X, Y, P} e^{iP(x-y)} A^{\mu}(x) A^{\nu}(y) 
  \frac{\delta_{\mu \nu}}{P^2 + m^2} \;,
\end{eqnarray}
\begin{multline}
 \Trace \left(
   \frac{\tilde \alpha}{ -\partial^2 + m^2}
 \frac{\tilde \alpha}{ -\partial^2 + m^2}
\right) 
 \\ =
\int_{X,Y,P} e^{iP(x-y)} \tilde \alpha(x) \tilde \alpha(y) 
  \int_Q \frac{1}{(P+Q)^2 + m^2} \frac{1}{Q^2 + m^2}\;,
\end{multline}
\begin{multline}
  i \Trace \left(
   \frac{\tilde \alpha}{ -\partial^2 + m^2}
   \frac{\{\partial_\mu, A^{\mu} \}}{-\partial^2 + m^2}
    +
   \frac{\{\partial_\mu, A^{\mu} \}}{-\partial^2 + m^2}
   \frac{\tilde \alpha}{ -\partial^2 + m^2}
  \right) \\=
  2 \int_{X,Y,P} e^{iP(x-y)} \tilde \alpha(x) A^{\mu}(y) 
  \int_Q \frac{P_\mu + 2 Q_\mu }{\left[(P+Q)^2 + m^2\right] 
\left[Q^2 + m^2\right]} = 0 \;,
\end{multline} 
\begin{multline}
\Trace \left(
  \frac{\{\partial_\mu, A^{\mu} \}}{-\partial^2 + m^2}
   \frac{\{\partial_\nu, A^{\nu} \}}{-\partial^2 + m^2}
 \right) \\
 =
 - \int_{X, Y, P} e^{iP(x-y)}
  A^{\mu}(x) A^{\nu}(y)
\int_Q
\frac{(P_\mu + 2 Q_\mu)(P_\nu + 2 Q_\nu)}
 {\left[(P+Q)^2 + m^2\right] \left[Q^2 + m^2\right]} \;.
\end{multline}
Using the results above one gets the following effective action up to
corrections of order $1 / \sqrt{N}$ \citep{DAdda1978}
\begin{multline}
  S_{\mathrm{eff}} = N \Trace \log  \left (-\partial^2 + m^2 \right)
  -\frac{N m^2}{2 g^2} \beta V
  + i \frac{\sqrt{N}}{2} 
  \int_X \tilde \alpha (x) \left (\frac{1}{g_b^2} -  \int_P \frac{1}{P^2 + m^2} \right)
  \\
+ \frac{1}{2} \int_{X, Y} \tilde \alpha(x) \Gamma(x - y) \tilde \alpha(y)
  + \frac{1}{2} \int_{X, Y} A^{\mu}(x) \Delta_{\mu \nu}(x - y)  A^{\nu}(y)
  + \mathcal{O} \left (\frac{1}{\sqrt{N}} \right),
 \label{eq:effaction2}
\end{multline}
where the inverse $\tilde \alpha$ and gauge field propagators are given by
\citep{DAdda1978}
\begin{multline}
 \Gamma(P) = \int_Q \frac{1}{(P+Q)^2 + m^2} \frac{1}{Q^2 + m^2}
  \\=  \frac{1}{2 \pi} \frac{1}{\sqrt{P^2 (P^2 + 4m^2)}} \log \left(
  \frac{\sqrt{P^2 + 4 m^2} + \sqrt{P^2}}{\sqrt{P^2 + 4m^2} - \sqrt{P^2}}
  \right) \;,
\end{multline}
\begin{multline}
 \Delta_{\mu \nu}(P) =
  \int_Q \frac{2 \delta_{\mu \nu}}{Q^2 + m^2} -
 \int_Q \frac{(P_\mu + 2 Q_\mu)(P_\nu + 2 Q_\nu)}
 {\left[(P+Q)^2 + m^2\right] \left[Q^2 + m^2\right]}
 \\ 
  = \left(\delta_{\mu \nu} - \frac{P_{\mu} P_{\nu}}{P^2} \right) 
  \Delta_{\mu}^{\mu}(P) \;,
\end{multline}
\begin{multline}
\Delta_{\mu}^{\mu}(P) = \int_Q \frac{2 - 2 \epsilon}
 {Q^2 + m^2} - \int_Q \frac{2}{(P+Q)^2 + m^2}
 + (P^2 + 4m^2) \int_Q \frac{1}{(P+Q)^2 + m^2} \frac{1}{Q^2 + m^2}
\\ = 
  \frac{1}{2 \pi} \left[\sqrt{\frac{P^2 + 4m^2}{P^2}} \log \left(
  \frac{\sqrt{P^2 + 4 m^2} + \sqrt{P^2}}{\sqrt{P^2 + 4m^2} - \sqrt{P^2}}\right)
 - 2
  \right] \;.
\end{multline}
The trace of the inverse gauge field propagator was evaluated using
dimensional regularization in $d=2-\epsilon$ dimensions, in that case
$\delta_\mu^\mu = 2-\epsilon$. Of course using a Pauli-Villars
regulator one obtains the same result \citep{DAdda1978}, but one has
to be careful in applying this regularization technique in this case
because it explicitly breaks gauge invariance. 

Equation (\ref{eq:effaction2}) shows that although a kinetic term for
the gauge fields was absent from the classical action, it follows that
such a term is generated by the quantum corrections. The Lorentz
structure of the inverse gauge field propagator follows from gauge
invariance which requires the inverse propagator to be transverse,
that is $P_\mu \Delta^{\mu \nu} = 0$.

The nonzero temperature results can be obtained by changing the
integrals over momenta into sum-integrals. Due to gauge invariance,
the inverse gauge field propagator is still transverse at nonzero
temperature. Its tensor structure at finite temperature is the same
as at zero temperature (this is typical for $1+1$ dimensions, in more
dimensions Lorentz symmetry breaking terms can appear at finite
temperature). At finite temperature one finds \citep{Lazarides1979}
\begin{equation}
\Gamma(P) = 
  \frac{1}{2 \pi} \frac{1}{\sqrt{P^2 (P^2 + 4m^2)}} \log \left(
  \frac{\sqrt{P^2 + 4 m^2} + \sqrt{P^2}}{\sqrt{P^2 + 4m^2} - \sqrt{P^2}}
  \right) + \Gamma_T(P) \;,
\end{equation}
\begin{equation}
\Delta_{\mu}^{\mu}(P) =
  \frac{1}{2 \pi} \left[\sqrt{\frac{P^2 + 4m^2}{P^2}} \log \left(
  \frac{\sqrt{P^2 + 4 m^2} + \sqrt{P^2}}{\sqrt{P^2 + 4m^2} - \sqrt{P^2}}\right)
 - 2
  \right] + (P^2 + 4m^2) \Gamma_T(P) \;,
\end{equation}
where $\Gamma_T(P)$ is up to a factor 2 equal to Eq.~(\ref{eq:nonlinpit})
\begin{equation}
 \Gamma_T(P) = 
\frac{1}{\pi} \int_{-\infty}^{\infty}\frac{\mathrm{d} q}{\omega_q} 
\frac{P^2 + 2pq}{ (P^2 + 2pq)^2 + 4 p_0^2 \omega_q^2}
n(\omega_q)\;.
\end{equation}

In the low-momentum limit at zero temperature the inverse gauge field
propagator becomes
\begin{equation}
  \Delta_{\mu \nu} \approx \frac{1}{12 \pi m^2}
  \left( P^2 \delta_{\mu \nu} - P_\mu P_\nu \right) \;.
\end{equation}
This is expression is up to a constant (which could be absorbed in the
gauge fields by a redefinition) equal to the ordinary inverse photon
propagator which gives rise to a Coulomb potential between two
charges.  In $1+1$ dimensions, this potential is proportional to the
distance between the charged particles.  Therefore the
$\mathbb{C}P^{N-1}$ model has confinement at zero temperature
\citep{DAdda1978,Samuel1983}.  However, confinement disappears at
non-zero temperature as was shown by
\citet{Lazarides1979}, see also \citet{Actor1985}.

\section{Effective potential}
By integrating the effective action $S_\mathrm{eff}$ over the quantum
fluctuations $\tilde \alpha$ and $A_\mu$ one obtains the effective
potential. The leading term of the effective potential can be read off
directly from Eq.~(\ref{eq:effaction2}). In order to obtain the
next-to-leading order corrections a Gaussian integration has to be
performed. For the contribution arising from the $\tilde \alpha$
fluctuations this is straightforward. But to calculate the gauge field
contribution to the effective potential one has to fix the gauge,
because otherwise it is not possible to invert the gauge field inverse
propagator. In the generalized Lorenz gauge this boils down to adding
the gauge fixing term
\begin{equation}
\frac{1}{2 \xi} \int_x \left( \partial^{\mu} A_\mu
\right)^2\;,
\end{equation}
to the effective action. As a consequence the ghost fields give 
a contribution to the effective potential which is $\isumint_P \log
P^2$.  After subtracting $T$-independent constants it is found that the gauge field
and ghost contribution to the effective potential together is given by
\begin{multline}
  \mathcal{V}_{\mathrm{gauge}}(m^2) 
=
\sumint_P \log P^2 - \frac{1}{2} \sumint_P \log \mathrm{det}\,
 \left ( \Delta_{\mu \nu} + \frac{1}{\xi} P_\mu P_\nu \right) \\
=   
\frac{1}{2} \sumint_P \log P^2 
-\frac{1}{2} \sumint_P \log \Delta_{\mu}^{\mu}
\;.
\end{multline}
which is independent of $\xi$ as expected.  The same result could of
course also be obtained in any another gauge, for example the $A_0 =
0$ gauge. In that gauge the ghost fields give a contribution of
$\frac{1}{2} \isumint \log( P_0^2 )$, whereas the $A_1$ gauge field
gives a contribution of $-\frac{1}{2} \isumint \log [ (1 - P_1^2 / P^2)
\Delta_{\mu}^{\mu}]$.  Together, this leads of course to the same
result as in the generalized Lorenz gauge.

From \eqref{eq:effaction2} and the results above, one obtains the
complete finite temperature effective potential
\begin{equation}
 \mathcal{V}(m^2) = N \mathcal{V}_\mathrm{LO}(m^2) +
 \mathcal{V}_\mathrm{NLO}(m^2) \;,
\end{equation}
where the leading order contribution is given by
\begin{equation}
 \mathcal{V}_\mathrm{LO}(m^2) = 
  \frac{m^2}{2 g_b^2} - \sumint_P \log(P^2 + m^2)\;, 
\end{equation}
and the next-to-leading order contribution by
\begin{equation}
 \mathcal{V}_\mathrm{NLO}(m^2) =
  - \frac{1}{2} \sumint_P \log \Gamma(P)
  - \frac{1}{2} \sumint_P \log \Delta_{\mu}^{\mu}(P)
  + \frac{1}{2} \sumint_P \log P^2  \;.
\end{equation}

The effective potential is ultraviolet divergent. To regulate
the divergences an ultraviolet momentum cutoff $\Lambda$ is
introduced. This yields
\begin{equation}
 \mathcal{V}_\mathrm{LO}(m^2) =
  \frac{m^2}{2 g_b^2} - \frac{m^2}{4\pi} 
   \left[1 + \log \left(\frac{\Lambda^2}{m^2} \right) \right]
     + \frac{1}{4\pi} T^2 J_0 (\beta m) \;. 
\end{equation}
Here
\begin{equation}
  J_0(\beta m) = \frac{8}{T^2} \int_0^\infty \frac{\mathrm{d}q\,
  q^2}{\omega_q} n(\omega_q)\;.
\end{equation}
The minimum of the leading order effective potential obeys
the following gap equation
\begin{equation}
  \frac{1}{g_b^2} = \frac{1}{2 \pi} \log 
\left( \frac{\Lambda^2}{m^2} \right)
  + \frac{1}{2\pi} J_1(\beta m) \;,
 \label{eq:logapeqcpn}
\end{equation}
where
\begin{equation}
J_1({\beta m}) = 4 \int_0^\infty \frac{\mathrm{d}q\,}{\omega_q}\, n(\omega_q)\;.
\end{equation}
The gap equation can be rendered finite by the substitution $g_b^2 \rightarrow
Z_g^2 g^2(\mu)$, where
\begin{equation}
\frac{1}{Z_g^2} = 1+ \frac{g^2}{2\pi} 
 \log \left ( \frac{\Lambda^2}{\mu^2} \right) \;.
\end{equation}
From this renormalization prescription it follows that the leading
order beta-function is given by
\begin{equation}
  \beta(g^2) \equiv
  \mu \frac{\mathrm{d} g^2(\mu)}{\mathrm{d} \mu} = -\frac{g^4}{\pi} +
   \mathcal{O}\left(\frac{1}{N} \right) \;.
\end{equation}
The negative sign shows that the $\mathbb{C}P^{N-1}$ model is
asymptotically free.

In order to calculate the next-to-leading order contribution
to the effective potential, $\mathcal{V}_\mathrm{NLO}$ is
written as a sum of divergent ($D$) and finite parts ($F$) in
the following way
\begin{equation}
 \mathcal{V}_\mathrm{NLO}(m^2) = 
     - \frac{1}{2} \left( D_1 + D_2 + F_1 + F_2 + F_3 + F_4 \right)
   - \frac{\pi}{3}T^2 \;.
  \label{eq:effpot1}
\end{equation}
where the divergent and finite quantities are defined through the
following relations
\begin{equation}
\begin{split}
   D_1 + F_1 = \int_P \log \tilde \Gamma(P) \;, & \;\;\;\;\;\;
    F_3 = \sumdiff_P \log \tilde \Gamma(P) 
   \;, \\
 D_2 + F_2 = \int_P \log \tilde \Delta_{\mu}^{\mu}(P)
  \;, & \;\;\;\;\;\;
       F_4 = \sumdiff_P \log \tilde \Delta_{\mu}^{\mu}(P)\;.
\end{split}
\end{equation} 
Here $\tilde \Gamma(P) \equiv 2 \pi \sqrt{P^2(P^2 + 4m^2)} \Gamma(P)$ and $\tilde
\Delta^{\mu}_{\mu}(P) \equiv 2\pi \sqrt{P^2 / (P^2 + 4m^2)}
\Delta_{\mu}^{\mu}(P)$.

The functions $D_1$ and $D_2$ are functions which contain all
divergences of the next-to-leading order effective potential. In order
to obtain these ultraviolet divergences, the high-momentum limits of
$\tilde \Gamma(P)$ and $\tilde \Delta_\mu^\mu(P)$ are needed.
In the high-momentum approximation ($\vert \vec p \vert \gg T$)
one finds using Eq.~(\ref{PiHE})
\begin{eqnarray}
  \tilde \Gamma(P)
   &\approx& 
   \log\left(\frac{P^2}{\bar m^2} \right) + \frac{2m^2}{P^2}
         + \frac{2 m^2 J_1(\beta m)}{P^2}\left(1 - \frac{2 p_0^2}{P^2} \right) +
 \mathcal{O}\left (\frac{m^4}{P^4} \right) \;, 
 \label{eq:gammaExp}
 \\
\tilde \Delta_{\mu}^{\mu}(P)
 &\approx& 
\log\left(\frac{P^2}{\bar m_e^2} \right) + \frac{6m^2}{P^2}
         + \frac{2 m^2 J_1 (\beta m)}{P^2} \left(1 - \frac{2 p_0^2}{P^2} \right) +
 \mathcal{O}\left (\frac{m^4}{P^4} \right) \;,
  \label{eq:deltaExp}
\end{eqnarray}
where $\bar m^2 = m^2 \exp[- J_1(\beta m) ]$ and $\bar m^2_e = m^2
\exp[2 - J_1(\beta m)]$.

The divergences $D_1$ and $D_2$ can be obtained by integrating
\eqref{eq:gammaExp} and
\eqref{eq:deltaExp} over momenta. It is found that
\begin{eqnarray}
  D_1 &=& \frac{1}{4\pi} \left[
  \Lambda^2 \log \log \left( \frac{\Lambda^2}{\bar m^2} \right)
  - \bar m^2 \mathrm{li}\, \left( \frac{\Lambda^2}{\bar m^2} \right)
  + 2 m^2 \log \log \left( \frac{\Lambda^2}{\bar m^2} \right)
      \right ] \;,
\\ 
   D_2 &=& \frac{1}{4\pi} \left[
     \Lambda^2 \log \log \left( \frac{\Lambda^2}{\bar m_e^2} \right)
 - \bar m_e^2 \mathrm{li}\, \left( \frac{\Lambda^2}{\bar m_e^2} \right)
    + 6 m^2 \log \log \left( \frac{\Lambda^2}{\bar m_e^2} \right)
\right] \;.
\end{eqnarray}
From these two equations it can be seen that (through $\bar m^2$ and
$\bar m_e^2$) the effective potential contains temperature dependent
divergences. They cannot be renormalized in a temperature-independent
way. However, by using the leading order gap equation
(\ref{eq:logapeqcpn}) one can show that $\bar m^2$ and $\bar m_e^2$
become temperature-independent at the minimum of the effective
potential. Therefore the quadratic divergence and the divergence
proportional to the li function become temperature-independent at the
minimum. As a result these divergences can be renormalized at the
minimum, as will be done explicitly in the following section.

\begin{figure}[t]
\begin{center}
\input{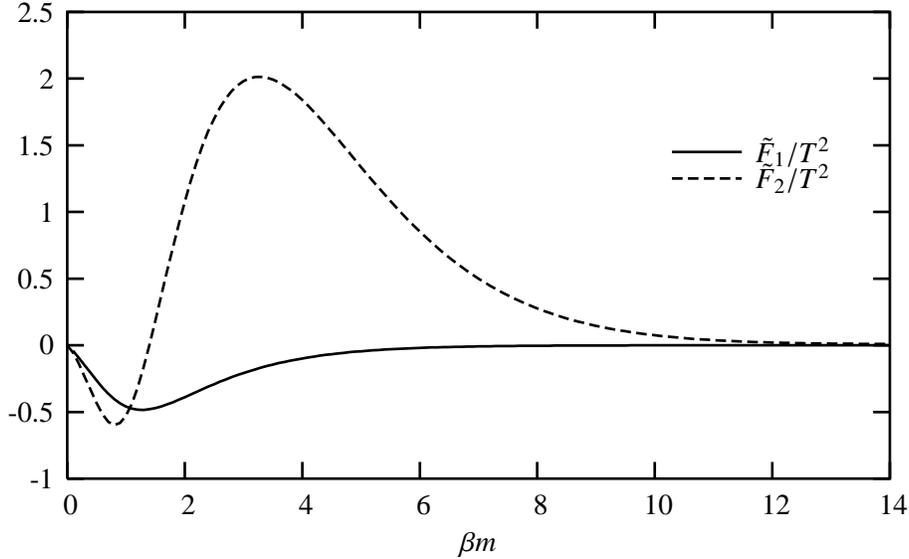}
\caption{The finite functions $\tilde F_1(\beta m) /T^2$ and 
$\tilde F_2(\beta m)/T^2$. These functions contribute to the next-to-leading
order effective potential.}
\label{fig:f1f2}
\end{center}
\end{figure}

The finite functions $F_1$ and $F_2$ will be obtained numerically. In
order to calculate these functions, the divergences will be written
partly in terms of an integral as is discussed in
Sec.~\ref{sec:abelplana}. This prevents subtracting large quantities which
can give rise to big numerical errors. The functions $F_1$ and $F_2$
are calculated using the following expressions
\begin{eqnarray}
  F_1 &=& \mathcal{P} \int_P \log \left [\frac{\tilde \Gamma(P)}{\log
       \left(P^2 / \bar m^2\right)} \right] - 
    2 m^2 \log \log \left(
       \frac{\Lambda^2}{\bar m^2} \right)
   \\ F_2 &=& \mathcal{P}
       \int_P \log
 \left[
   \frac{\tilde \Delta_\mu^\mu(P)}{\log \left(P^2 / \bar m_e^2\right)}
 \right]
 - 6 m^2 \log \log \left( \frac{\Lambda^2}{\bar m_e^2} \right) \;,
\end{eqnarray}
here $\mathcal{P}$ denotes the principal value integral. At zero
temperature it is found that $F_1 \approx \frac{m^2}{2\pi} \gamma_E$
and $F_2 \approx \frac{m^2}{2\pi} c_1$, where $c_1 \approx 0.611 671
457 \ldots$. This is in agreement with the zero temperature
calculations of \citet{Campostrini1992}. For convenience the finite
temperature parts of $F_1$ and $F_2$ are defined as $\tilde F_{1} =
F_1 - m^2 \gamma_E / (2\pi)$ and $\tilde F_{2} = F_2 - m^2 c_1 /
(2\pi)$. These functions divided by $T^2$ depend on $\beta m$ only and are
displayed in Fig.~\ref{fig:f1f2}. For large $\beta
m$ these functions go to zero because that is equivalent to taking the
limit of zero temperature. For small $\beta m$ they go to zero as
well, for the same reason as discussed in Sec.~4.6 for the function $F_2$.

\begin{figure}[t]
\begin{center}
\input{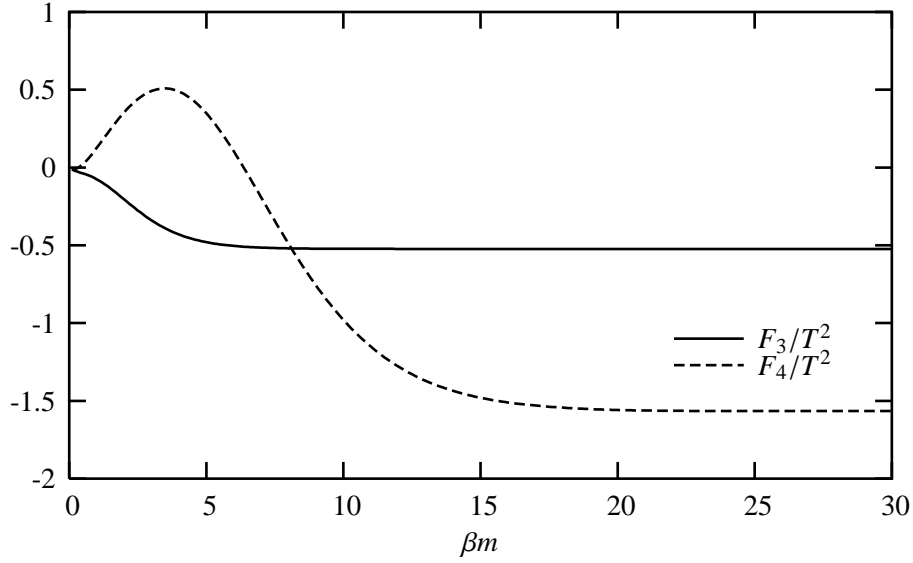}
\caption{The functions $F_3(\beta m)/T^2$ and $F_4(\beta m)/T^2$
as a function of $\beta m$. These functions contribute to the
next-to-leading order effective potential.}
\label{fig:f3f4}
\end{center}
\end{figure}

The finite functions $F_3/T^2$ and $F_4/T^2$ were calculated using the
method explained in Sec.~\ref{sec:abelplana}. They are displayed in
Fig.~\ref{fig:f3f4}.
The $\beta m$ large limit of $F_3$ and $F_4$ can be obtained by noting
that for large $\beta m$ the temperature dependent part of the inverse
propagator does not contribute to $F_3$ and $F_4$. Furthermore, the
dominant contribution to the difference of a sum-integral and an
integral arises from the low momentum modes. So the large $m$ behavior of
the zero temperature inverse propagators can be used to obtain a large
$\beta m$ approximation for $F_3$ and $F_4$. It is found that
\begin{equation} 
F_3 \approx \frac{1}{2} \sumdiff_P \log P^2 
= - \frac{\pi}{6} T^2 \;,\;\;\;\;\;\;
F_4 \approx \frac{3}{2} \sumdiff_P \log P^2 
 = - \frac{\pi}{2} T^2 \;.
\end{equation}
As one can see in Fig.\ \ref{fig:f3f4} this is in agreement with the
numerical calculations.

The small $\beta m$ limit of $F_3/T^2$ and $F_4/T^2$ can be obtained
as well. For that one needs the small $\beta m$ limit of
$\Gamma(P)$. That limit can be found by first performing the momentum
integration and taking the zero modes as is explained around
Eq.~(\ref{eq:nonlinpiht}).  In this limit
\begin{equation}
  \Gamma(P) \approx \frac{1}{\beta m} \frac{P^2}{P^4 + 4m^2 p^2} \;.
\end{equation}
This results in
\begin{eqnarray}
  F_3 &\approx& \frac{1}{2} \sumdiff_P \log P^2 
+ \frac{1}{2}\sumdiff_P \log (P^2 + 4m^2) 
 - \sumdiff_P \log(P^2) \approx 0\;,
\\
  F_4 &\approx&  \frac{1}{2} \sumdiff_P \log P^2
  -  \frac{1}{2}\sumdiff_P \log (P^2 + 4m^2) 
  + \sumdiff_P \log \left( 
  \frac{2 \pi}{\beta m} - 2
   \right )
 \approx 0 \;.
\end{eqnarray}
These limits are also in agreement with the numerical calculations as
can be seen from Fig.~\ref{fig:f3f4}.

\section{Contribution of quantum instantons to the pressure}
In the previous section the effective potential was evaluated. It was
found that the effective potential contains temperature-dependent
ultraviolet divergences. At the minimum these temperature-dependent
divergences will disappear as will be discussed now. To calculate the
effective potential at the minimum one only needs to solve the leading
order gap equation (\ref{eq:logapeqcpn}) as was shown by
\citet{Root1974} and as discussed in Sec.~\ref{sec:nlsmpres}. 
%To show this the solution to the gap equation is
%written as 
%\begin{equation}
%m^2 = m^2_\mathrm{LO} + m^2_{\mathrm{NLO}} / N \;.
%\end{equation}
%By Taylor expanding the effective
%potential one obtains (up to ${\cal O}(1/N)$ corrections)
%\begin{equation}
%  \mathcal{V}(m^2) =  
%  N \mathcal{V}_{\mathrm{LO}}(m^2_{\mathrm{LO}}) 
%   + \mathcal{V}_{\mathrm{NLO}}(m_{\mathrm{LO}}^2)  
%  + m^2_{\mathrm{NLO}}
%  \left. \frac{\partial \mathcal{V}_{\mathrm{LO}}(m^2)} 
%    {\partial m^2} \right \vert_{m^2 = m^2_\mathrm{LO}} 
%\label{eq:cpngapexp2}
%\end{equation}
%The last term of Eq.~(\ref{eq:cpngapexp2}) vanishes by using the leading 
%order gap equation. The pressure ${\cal P}$ can now be written as
%\begin{equation}
%\mathcal{P} \equiv N
%\mathcal{P}_{\mathrm{LO}} + \mathcal{P}_{\mathrm{NLO}} 
%\end{equation}
As a result the leading and next-to-leading order contributions to
the pressure are given by
\begin{equation}
  \mathcal{P}_{\mathrm{LO}}  = 
     \mathcal{V}^{T}_{\mathrm{LO}}(m^2_T) 
   - \mathcal{V}^{T=0}_{\mathrm{LO}}(m^2_0)  
  \;, \;\;\;\;\;\;\; 
  \mathcal{P}_{\mathrm{NLO}} =
     \mathcal{V}^{T}_{\mathrm{NLO}}(m^2_T) 
   - \mathcal{V}^{T=0}_{\mathrm{NLO}}(m^2_0) \;,
\end{equation}
where $m^2_T$ is the solution of the leading-order gap equation
(\ref{eq:logapeqcpn}) at temperature $T$. By using the leading order gap
equation (\ref{eq:logapeqcpn}) it can be shown that at the minimum
the divergent terms $D_1$ and $D_2$ become
\begin{eqnarray}
 \!\!\!\! D_1 \!\!&=& \!\! \frac{\Lambda^2}{4\pi} \left[
 \log \left( \frac{2\pi}{g_b^2} \right)
 - \exp \left( - \frac{2 \pi}{g_b^2} \right) 
  \mathrm{li} \exp \left( \frac{2 \pi}{g_b^2} \right)
 \right]
  + \frac{2 m_T^2}{4\pi} 
\log \log \left( \frac{\Lambda^2}{\bar m_T^2} \right)
\;, \\
 \!\!\!\! D_2 \!\!&=& \!\!
\frac{\Lambda^2}{4\pi} \left[
 \log \left( \frac{2\pi}{g_b^2}{-}2 \right)
 - \exp \left(2{-} \frac{2 \pi}{g_b^2} \right) 
  \mathrm{li} \exp \left( \frac{2 \pi}{g_b^2} {-}2\right)
 \right]
  + \frac{6 m_T^2}{4\pi} 
\log \log \left( \frac{\Lambda^2}{\bar m_{eT}^2} \right)
\;.
\end{eqnarray}
Hence the temperature-dependent quadratic divergence and the
divergence proportional to the li function become
temperature-independent at the minimum of the effective
potential. They cancel in the calculation of the pressure due to the
subtraction of the zero temperature contribution.  The divergences
proportional to the $\log \log$ function can be absorbed into the
coupling constant $g_b^2$. The renormalization factor $Z_g^2$ to
next-to-leading order becomes in this way
\begin{equation}
\frac{1}{Z_g^2} = 1+ \frac{g^2}{2\pi} 
 \log \left ( \frac{\Lambda^2}{\mu^2} \right) + \frac{2}{N}
  \frac{g^2}{\pi} \log \log \left( \frac{\Lambda^2}{\mu^2}\right)
+ \mathcal{O} \left( \frac{1}{N^2} \right) \;.
\end{equation}
From this renormalization prescription it follows that the 
beta-function to next-to-leading order in $1/N$ is given by
\begin{equation}
  \beta(g^2) \equiv
  \mu \frac{\mathrm{d} g^2(\mu)}{\mathrm{d} \mu} = -\frac{g^4}{\pi} +
   \mathcal{O}\left(\frac{1}{N^2} \right) \;.
\end{equation}

Using the results above it follows that the leading and
next-to-leading order contributions to the pressure are given by
\begin{eqnarray}
  \mathcal{P}_\mathrm{LO} &=& \frac{m^2_T}{2 g^2} - \frac{m^2_T}{4\pi}
 \left[1 + \log \left( \frac{\mu^2}{m_T^2} \right) \right] +
 \frac{T^2}{4\pi} J_0(\beta m_T) + \frac{m^2_0}{4\pi} \;,
\\
 \mathcal{P}_\mathrm{NLO} &=&
- \frac{1}{2} \left[ \tilde F_1(m_T) + \tilde F_2(m_T) + F_3(m_T) +
F_4(m_T) \right]
\nonumber \\
& &
 + \frac{1}{4\pi} (\gamma_E + c_1) (m_0^2 - m_T^2) 
   - \frac{\pi}{3}T^2 
\;.
\label{eq:cpnpress}
\end{eqnarray}

\begin{figure}[t]
\includegraphics{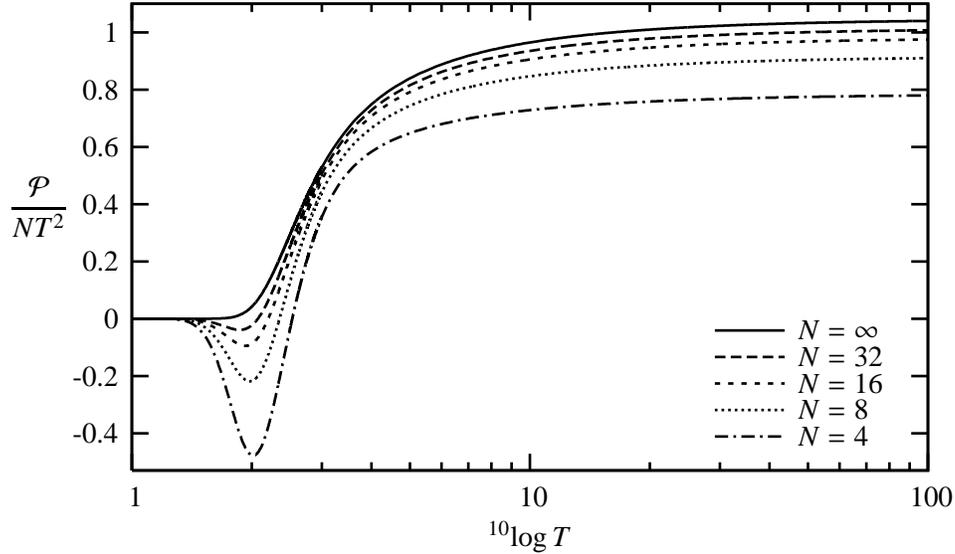}
\caption{Contribution of the zero winding number configurations to the 
pressure $\mathcal{P}$ of the $\mathbb{C}P^{N-1}$ model normalized to
$NT^2$, as a function of temperature, and for different values of $N$.}
\label{fig:cpnpressure}
\end{figure}

The results of the calculation of the pressure are displayed in
Fig.~\ref{fig:cpnpressure} for the arbitrary choice $g^2(\mu = 500) =
10$, for different values of $N$. As one can see, for low temperatures
and all finite values of $N$ the pressure first decreases for
increasing values of $T$. A decreasing pressure is unphysical because
it implies that the entropy which is the derivative of the pressure
with respect to temperature becomes negative. But this is in
disagreement with the third law of thermodynamics that states that the
entropy is minimal at zero temperature.  If one
believes that the $1/N$ expansion is a good expansion (which is for
example shown in the $\mathrm{O}(N)$ nonlinear sigma model discussed
in Chapter 3) it is likely that the reason for this negative pressure
is that the effective action, Eq.~(\ref{eq:cpneffact}) was only
expanded around the vacuum $A_\mu = 0$ solution with zero winding
number. The contribution of other vacua (quantum instantons) with
nonzero winding number was left out of the calculations. As one can
see from the figure, the problem of the negative pressure becomes less
severe if $N$ becomes larger. This is in agreement with the fact that
the instanton contribution vanishes in the $N \rightarrow \infty$ limit
\citep{Witten1978} as is discussed in Sec.~5.1. Moreover, that this problem
arises at low temperatures is also reasonable because instantons are
non-perturbative in $g$. They become less important at small
couplings, so they should vanish at high temperatures due to
asymptotic freedom.

\begin{figure}[t]
\scalebox{1.0}{\includegraphics{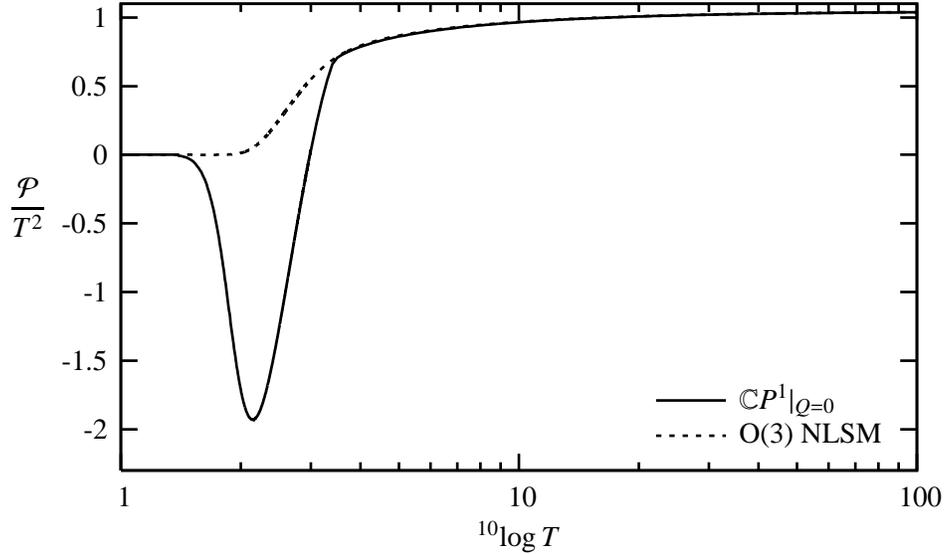}}
\caption{Pressure $\mathcal{P}$ of the $\mathrm{O}(N)$ 
nonlinear sigma model to NLO in $1/N$ for $N=3$ compared to the
contribution to the pressure of the configuration with zero winding
number $Q$ of the $\mathbb{C}P^{1}$ model.  The pressures are
normalized to $T^2$ and displayed as a function of temperature, for
the $g^2 (\mu = 500) = 10$.}
\label{fig:cp1o3pres}
\end{figure}

\begin{figure}[t]
\includegraphics{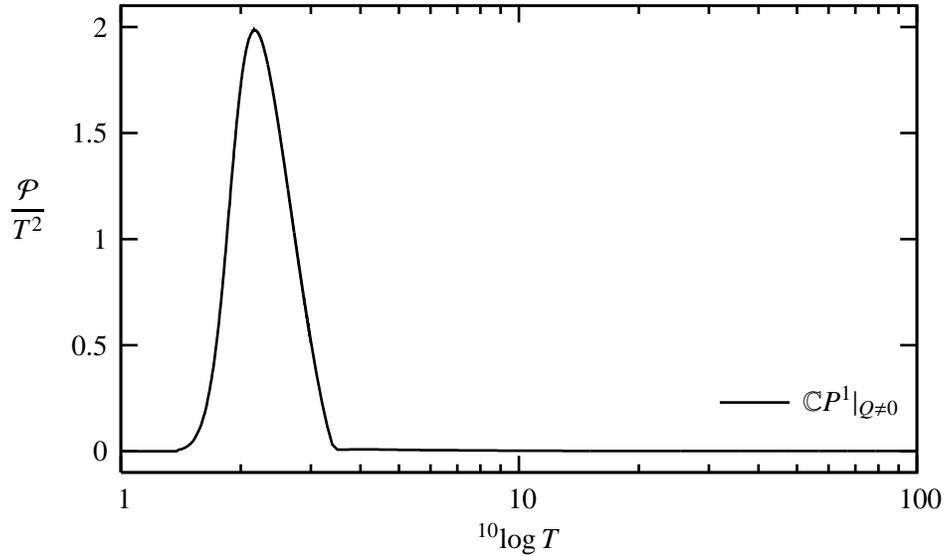}
\caption{Contribution of the configurations with nonzero winding number $Q$ to the 
pressure $\mathcal{P}$ of the $\mathbb{C}P^{1}$ model. The pressure
is normalized to $T^2$ and displayed as a function of temperature.}
\label{fig:cpnpresinstcontr}
\end{figure}

Using the equivalence between the $\mathrm{O}(3)$ nonlinear sigma
model and the $\mathbb{C}P^{1}$ model (see Sec.~5.2) the exact
contribution of the quantum instantons with nonzero winding number to
the pressure can be found. Since there are no gauge fields in the
$\mathrm{O}(3)$ nonlinear sigma model and because the integration over
the scalar fields can be done exactly the effects of all quantum
instantons are automatically included its effective potential (see
Chapter 4).  Due to the equivalence, the pressure of the
$\mathbb{C}P^{1}$ model should be exactly equal to that of the
$\mathrm{O}(3)$ nonlinear sigma model. In Fig.~\ref{fig:cp1o3pres} the
result of the NLO calculation of the pressure of the $\mathrm{O}(N)$
nonlinear sigma model for $N=3$ is compared to the contribution to the
pressure of the $\mathbb{C}P^{1}$ model with zero winding number. It
can be seen that for very low and high temperatures both pressures
coincide. For intermediate temperatures these pressures differ. This
difference is displayed in Fig.~\ref{fig:cpnpresinstcontr}.  This is a
strong indication that quantum instantons give a sizable contribution
to the pressure and other thermodynamical quantities at where the
pressure (divided by $T^2$) increases considerably.

\section{Summary and Conclusions}
In this chapter the effect of quantum instantons on the
thermodynamical quantities was investigated. For that the effective
potential of the $\mathbb{C}P^{N-1}$ model expanded around a
background with zero winding number was calculated to next-to-leading
order in $1/N$. It was shown that the effective potential contains
temperature-dependent divergences which only can be renormalized at
the minimum of the effective potential. Hence thermodynamical
quantities (which are all defined at the minimum of this effective
potential) can be rendered finite like in the (non)linear sigma
models discussed in Chapters~4 and 6. 

It was found that for all values of $N$ the contribution of the vacuum
with zero winding number gives rise to a negative pressure, which is
large for intermediate temperatures where the leading order pressure
divided by $T^2$ increases strongly. Since this is unphysical, it
indicates that the quantum instantons must give a considerable
contribution to the pressure for intermediate temperatures. In
agreement with the vanishing of the instantons in the $N \rightarrow
\infty$ limit this problem of the negative pressure becomes less
severe for large values of $N$.

For $N=2$ the exact contribution of the quantum instantons was found
using an equivalence between the $\mathrm{O}(3)$ nonlinear sigma model
and the $\mathbb{C}P^1$ model. In the $1/N$ approximation to the
$\mathrm{O}(3)$ nonlinear sigma model one implicitly integrates over
all quantum instantons and one finds a finite pressure in
next-to-leading order in $1/N$. Comparing this result to the pressure
of the $\mathbb{C}P^1$ model with zero winding number gives the exact
contribution of the quantum instantons for $N=2$. They give a large
contribution for intermediate temperatures where the pressure divided
by $T^2$ raises quickly.

A possible extension of this work would be to add a term $\theta Q$ to
the effective action and investigate the dependence of the
thermodynamical quantities on $\theta$. Moreover one could try to take
the fluctuations around the quantum instantons with nonzero winding
number into account explicitly and see whether they will reduce the
problem of the negative pressure, but this is not attempted in this
thesis.
 %cpn
\chapter{Thermodynamics of $\boldsymbol{\mathrm{O}(N)}$ sigma models in d=3+1}
\markboth{Thermodynamics of $\mathrm{O}(N)$ sigma models in d=3+1}{}\label{chap:onmodel}

The $\mathrm{O}(N)$ linear and nonlinear sigma model in $3+1$
dimensions are low-energy effective theories for QCD.  The pressure of
these two low-energy effective models will be computed up to
next-to-leading order in $1/N$ by minimizing the effective
potential. It is found that the finite temperature effective potential
contains temperature dependent divergences which become
temperature-independent at the minimum.  The calculated pressure can
serve as a prediction for the pressure of QCD at low temperatures and
will be compared to approximations discussed previously in the
literature. Finally a mass bound on the sigma meson is presented. This
chapter is based on: {\it Thermodynamics of O$(N)$ sigma models: $1/N$
corrections}, J.O.\ Andersen, D.\ Boer and H.J.\ Warringa, Phys.\
Rev.\ {\bf D70} 116007, (2004).

\section{Introduction}

Although the QCD Lagrangian possesses a chiral symmetry in the limit
of zero quark masses, the true QCD ground state does not respect this
symmetry. The chiral symmetry is spontaneously broken by quantum
effects.  To be specific, QCD with $N_f$ massless
quarks has a global $\mathrm{SU}(N_f)_L\times
\mathrm{SU}(N_f)_R$ symmetry, which for the ground state at low
temperatures is broken down to an $\mathrm{SU}(N_f)_V$
symmetry. According to Goldstone's theorem, there is a massless,
spinless particle for each generator of a broken global continuous
symmetry. In this case this implies the occurrence of $N_f^2-1$
Goldstone bosons. In phenomenological applications $N_f$ is either two
or three, and one also has to take into account the explicit symmetry
breaking due to the nonzero quark masses. Both the spontaneous and the
explicit chiral symmetry breaking are apparent in the low-energy
hadronic particle spectrum, where the expected number of relatively
light mesons is observed (e.g.\ the three pions for $N_f=2$). At
sufficiently high temperatures one expects the chiral symmetry to be
restored and lattice simulations of QCD suggest that this happens at a
temperature of approximately 170 MeV depending on the number of quarks
and their masses.  

In the case of two flavors, the situation is the simplest, since one
can exploit the fact that the $\mathrm{SU}(2)_L\times
\mathrm{SU}(2)_R$ symmetry is locally isomorphic to $\mathrm{SO}(4)$.
If baryons (nucleons in this case) and the $\rho$ and $\omega$ meson
are not included the simpler $\mathrm{O}(4)$ linear sigma model can be
used as a low-energy effective theory for describing the dynamics of
three pion fields and one sigma field. These four fields form a
four-dimensional vector $\phi$ in the fundamental representation of
$\mathrm{O}(4)$.  At low temperature, the $\mathrm{O}(4)$ symmetry is
spontaneously broken down to $\mathrm{O}(3)$, where the sigma field
acquires a vacuum expectation value and the three pions are
interpreted as the Goldstone bosons.  At high temperatures the vacuum
expectation value of the sigma field vanishes. According to the
calculations performed in this chapter, the symmetry is restored via a
second-order phase transition in agreement with other calculations in
the literature.  If the symmetry is restored, the pion and the sigma
field have the same mass.

For $N_f >2$ there is no connection between the
$\mathrm{SU}(N_f)_L\times \mathrm{SU}(N_f)_R$ model and the
$\mathrm{O}(N)$ linear sigma model since the symmetries differ.  The
$\mathrm{O}(N)$ sigma models have besides their relevance to
low-energy QCD also other applications. For example the Higgs field in
the Standard Model is described by an $\mathrm{O}(4)$ linear sigma
model. In that case the scalar fields are also coupled to the
electroweak gauge fields and the fermions. The spontaneous symmetry
breakdown of the $\mathrm{O}(4)$ symmetry causes the Higgs field to
acquire a vacuum expectation value. As a result the weak vector bosons
and the fermions (except for the neutrinos) become massive. Another
application is in cosmology. It is believed that during early times
the universe went through a period of rapid expansion, called
inflation. The field which controls the inflation, called the inflaton
can be described by a linear sigma model (see for
example \citet{Kolb1990}). In condensed matter physics the
$\mathrm{O}(N)$ sigma models are used to model spin-spin interactions (see
for example \citet{Itzykson1995}) .

In this chapter both the $\mathrm{O}(N)$ linear sigma model and
$\mathrm{O}(N)$ nonlinear sigma model in 3+1 dimensions will be
studied at finite temperature and to next-to-leading order in the
$1/N$ expansion.  At zero temperature, the $1/N$ expansion was applied
to $\mathrm{O}(N)$ sigma models a long time ago at leading order (LO)
by \citet{Coleman1974} and at next-to-leading order (NLO) by
\citet{Root1974}.  At finite temperature the LO $1/N$ contribution has been
studied by \citet{Meyers-Ortmanns1993}. In that case, the effective
potential is that of an ideal gas and thus straightforward to compute.
The NLO $1/N$ corrections are less trivial, since they involve a
momentum-dependent self-energy and cannot be evaluated analytically.
A high-temperature expansion was performed by
\citet{Jain1993} to obtain purely analytical results for the linear
sigma model. Similarly
\citet{Bochkarev1996} resorted to a ``high-energy'' approximation,
which makes the calculations manageable. However, this approximation
is uncontrolled and it is difficult to assess how reasonable it is,
unless one calculates the full NLO $1/N$ corrections. These
corrections will be calculated in this chapter and compared to the
``high-energy'' approximation.

The $\mathrm{O}(N)$ sigma models have also been studied in detail at
finite temperature using various other approaches.  A systematic study
has been carried out by \citet{Chiku1998} using optimized perturbation
theory.  The method was used to calculate spectral functions,
properties of the effective potential, and dilepton emission rates.
The 2PI formalism \citep{Cornwall1974} (see
Sec.~\ref{sec:formdefeffaction}) has also been used to examine various
properties of the $\mathrm{O}(N)$ linear sigma models at finite
temperature~\citep{Baym1977,Amelino-Camelia1993,Amelino-Camelia1997,
Roh1998,Petropoulos1999, Lenaghan2000,Nemoto2000, Aarts2004,
Arrizabalaga2005c, Roder2005}, see \citet{Petropoulos2004} for a recent
review. For example, in several papers the temperature dependence of
the pion and sigma masses, and of the vacuum expectation value of the
sigma field, have been investigated. The calculations of the scalar
field effective potential as a function of temperature have been
carried out in the Hartree approximation and the large-$N$
limit~\citep{Amelino-Camelia1993,Amelino-Camelia1997,Roh1998,Petropoulos1999,Lenaghan2000,
Nemoto2000}. In these cases, the gap equations for the propagators are
easy to solve since the self-energy reduces to a local mass term. In
the Hartree approximation (which implies ignoring momentum-dependent
self-energies), the result has been shown to be problematic (and a
first-order phase transition occurs), which has been remedied by
including more diagrams in the truncation
\citep{Verschelde2002,Baacke2003}, resulting in a second-order phase
transition.  If one goes beyond the Hartree approximation or includes
the full NLO contributions in the $1/N$ expansion, the gap equations
become nonlocal and very difficult to solve.

The advantage of using the 1PI formalism over the 2PI formalism is
that one does not have to make a Hartree or other approximation to
calculate the full $1/N$ corrections to the thermodynamical
quantities. However, to describe out of equilibrium phenomena, the 2PI
formalism is favored as is explained in Sec.~\ref{sec:formdefeffaction}.

The approach followed in this chapter is similar to the study of the
nonlinear sigma model in 1+1 dimensions (see Chapter~4) and to that of
the $\mathbb{C}P^{N-1}$ model in 1+1 dimensions (see Chapter~5).
Similar conclusions about the renormalization of the effective
potential in 3+1 dimensions as in 1+1 dimensions will be drawn. It
turns out that at NLO, temperature-independent renormalization is only
possible at the minimum of the auxiliary field effective potential.
This aspect of the $1/N$ expansion was missed in previous work by
\citet{Jain1993} and \citet{Bochkarev1996}, since the renormalization is considerably
simplified or even ignored in the various approximations.

Since explicit chiral symmetry breaking plays a very important role in
the actual hadron spectrum at low energy, the case of explicit
symmetry breaking will also be considered in this chapter. The results
change considerably and moreover, a critical temperature cannot be
determined in that case, since the second-order phase transition turns
into a smooth cross-over.
 
The nonlinear sigma model in 3+1 dimensions is non-renormalizable and
should be viewed as an effective theory, which is valid up to a
certain energy scale where new physics enters. Strictly speaking the
linear sigma model is renormalizable, but since it becomes a trivial theory in the
limit where the cutoff goes to infinity (see for example
\citet{Amelino-Camelia1997}), it will be treated as a theory with a finite
cutoff.  Given a finite cutoff, terms are called divergences when they
are increasing in magnitude without bound as the cutoff is
increased. The low-energy physics should be independent of such terms
(decoupling) and they can be subtracted in the renormalization
procedure in order to avoid increasing sensitivity to the ultraviolet
cutoff as it grows. On general grounds, one expects the temperature
dependence to be insensitive to an increasing cutoff due to the
exponential suppression provided by the Bose-Einstein
distribution function. Therefore one expects the renormalization to be possible
in a temperature-independent way.

This chapter is organized as follows. In Sec.~6.2, the effective
action of the linear and nonlinear sigma model in the $1/N$ expansion
are discussed. In Sec.~6.3, the effective potential and gap equations
to next-to-leading order are calculated. In Sec.~6.4, the results for
the pressure at next-to-leading order for general $N$ and for the
special case of $N=4$ are presented. Also, the so-called high-energy
approximation is discussed and compared with exact numerical
results. Sec.~6.5, is devoted to the choice of parameters for $N=4$,
in order to make contact with low-energy QCD phenomenology. A bound on
the scalar sigma meson is derived too in this section. In Sec.~6.6, a
summary and conclusions are given.

%=====================================================================
\section{Effective actions}
The Euclidean Lagrangian density of the $\mathrm{O}(N)$-symmetric
linear sigma model with a symmetry breaking term proportional to $H$
is given by
\begin{equation} 
   \mathcal{L} = \frac{1}{2} \left(\partial_\mu 
  \phi_i\right)^2 + \frac{\lambda_b}{8 N} \left (
  \phi_i \phi_i\!-\!N f_{\pi,b}^2 \right)^2 - \sqrt{N} H \phi_N\;,
\label{eq:lagrangian_lin_sigma} 
\end{equation}
where $i = 1 \ldots N$. Summation over repeated indices is implicitly
understood. The subscript $b$ denotes a bare quantity.
The coupling constants are rescaled with factors of $N$ 
in such a way that for large $N$ the action naturally scales as $N$
as is explained in Sec.~\ref{sec:largeNexp}. 

It is possible to eliminate the quartic interaction term from
Eq.~(\ref{eq:lagrangian_lin_sigma}) by introducing an auxiliary field
which is denoted by $\alpha$, in order to allow for Gaussian integration. 
To this end one adds to the Lagrangian density 
Eq.~(\ref{eq:lagrangian_lin_sigma}) the term
\begin{equation}
    \mathcal{L}_\alpha = \frac{N}{2 \lambda_b} \left[\alpha - 
    \frac{i \lambda_b}{2 N} \left(\phi_i \phi_i -  
    N f_{\pi,b}^2 \right) \right]^2 \;,
    \label{eq:lagrangian_alpha}
\end{equation}
such that one has
\begin{equation}
    \mathcal{L} = \frac{1}{2} \left( \partial_\mu \phi_i \right)^2 
 - \frac{i}{2} \alpha
  \left(\phi_i \phi_i - N f_{\pi,b}^2 \right) 
+ 
  \frac{N}{2 \lambda_b} \alpha^2
   -\sqrt{N} H \phi_N\;.
  \label{eq:lagrangian_auxfield}
\end{equation}
If one integrates over the $\alpha$ fields, the original Lagrangian in
Eq.\ (\ref{eq:lagrangian_lin_sigma}) is recovered. Hence Eq.\
(\ref{eq:lagrangian_auxfield}) is equivalent to the original
Lagrangian density of the linear sigma model. In the limit
$\lambda_b\rightarrow\infty$ the term quadratic in $\alpha$ vanishes
and one obtains the Lagrangian of the nonlinear sigma model
Eq.~(\ref{eq:lagran_non}) with a symmetry breaking term.

If explicit symmetry breaking is absent ($H=0$), the field $\phi$
acquires a vacuum expectation value by spontaneously breaking the
symmetry. Because of the residual $\mathrm{O}(N-1)$ symmetry, it is
possible to write $\phi = (\pi_1, \pi_2, \ldots, \pi_{N-1}, \sigma)$,
such that only $\phi_N=\sigma$ has a nonzero expectation value.  For
$H > 0$ the same argument applies, because the action is minimal when
the $\sigma$ field is the only one that acquires an expectation value.

%Equation (\ref{eq:lagrangian_auxfield}) is quadratic in the scalar fields. 
%It is therefore possible to integrate immediately over all scalar fields to obtain an
%effective action. However, if one proceeds, one has to expand the $\alpha$ field around
%its vacuum expectation value. In this expansion it is automatically assumed that

Integrating over the $\pi$'s gives the following effective action
\begin{multline}
    \mathrm{S}_{\mathrm{eff}} =
\frac{1}{2}(N-1)  
\mathrm{Tr} \log\left( -\partial^2 - i \alpha\right)  
   + 
        \int_{0}^{\beta} \! \mathrm{d}\tau\int \! \mathrm{d}^3 x \left[
       \frac{1}{2} \left( \partial_\mu \sigma \right)^2
  - \frac{i}{2} \alpha\, \sigma^2
  \right.  
\\  
\left.
         + \frac{i}{2} N f_{\pi,b}^2 \alpha 
         +  \frac{N}{2 \lambda_b} \alpha^2 
      - \sqrt{N} H \sigma   
   \right] \;.
\label{sigalp}
\end{multline}
The scalar fields $\sigma$ and $\alpha$ can be written as a sum of
space-time independent vacuum expectation values $im^2$ and
$\bar{\sigma}$, and quantum fluctuating fields $\tilde{\alpha}$ and
$\tilde{\sigma}$
\begin{equation}
  \alpha = i m^2 +\frac{\tilde{\alpha}}{\sqrt{N}}\;,
  \;\;\;\;\;\;\;\;\;\;\;\;
  \sigma = \sqrt{N}\bar{\sigma} + \tilde{\sigma}
\label{eq:s12}
\;.
\end{equation}
Equation~(\ref{eq:lagrangian_alpha}) can be used to show that the vacuum
expectation value of $\alpha$ is purely imaginary (for a proof, see
Sec.\ \ref{sec:non1overn}). The vacuum expectation value of $\sigma$ is
proportional to $\sqrt{N}$, which follows from
Eq.~(\ref{eq:lagrangian_lin_sigma}). Substituting Eqs.~(\ref{eq:s12})
into Eq.~(\ref{sigalp}), the effective action $S_\mathrm{\!eff}$ can be
written as
\begin{eqnarray}
  \nonumber
    S_{\!\mathrm{eff}} &=& 
\frac{1}{2}(N-1)\mathrm{Tr} 
       \log\left( -\partial^2 + m^2 - \frac{i \tilde \alpha}{\sqrt{N}}\right) 
        - \beta V N H \bar \sigma  
\\ && \nonumber
+  \int_{0}^{\beta}\mathrm{d} \tau \int
\mathrm{d}^3x 
   \bigg[
       \frac{1}{2} \left ( \partial_\mu \tilde \sigma \right)^2
     + \frac{1}{2} \left(m^2\!-\! 
        \frac{i \tilde \alpha}{\sqrt{N}}\right) 
        \left(\sqrt{N} \bar \sigma\!+\!\tilde \sigma\right)^2
    \\ &&  
        -  \frac{N}{2} f_{\pi,b}^2 
\left(m^2\!-\!\frac{i \tilde \alpha}{\sqrt{N}} 
        \right) - \frac{N}{2 \lambda_b} 
        \left(m^2\!-\!\frac{i \tilde \alpha}{\sqrt{N}}\right)^2 
 - \sqrt{N} H \tilde \sigma
\bigg] \;.
  \label{eq:shiftedeffaction}
\end{eqnarray}
Expanding Eq.~(\ref{eq:shiftedeffaction}) in powers of $1/\sqrt{N}$ up to
corrections of order $1/\sqrt{N}$, 
one finds
\begin{eqnarray}
    \frac{\mathrm{S}_{\!\mathrm{eff}}}{\beta V} &=& 
\frac{1}{2}(N-1)
\sumint_P \log \left (P^2 + m^2 \right)
        - \frac{N m^2}{2} \left (f^2_{\pi, b} - \bar \sigma^2 \right)
        \nonumber \\
  && - \frac{N m^4}{2 \lambda_b}
        - N H \bar \sigma 
 + \sqrt{N} \times \mathrm{terms\;linear\;in\;}
\tilde \alpha\;\mathrm{and}\;\tilde \sigma 
    \nonumber \\
&&   +  \frac{1}{2} \sumint_P 
\chi^\mathrm{T}
\left(
\begin{array}{cc} 
\tfrac{1}{2} \Pi(P, m) + \frac{1}{\lambda_b}  
  & -i \bar \sigma \\
  -i \bar \sigma & P^2 + m^2  
\end{array}
\right)
\chi^*
  \label{eq:expandedeffaction} \;,
\end{eqnarray}
where $\chi^T = (\tilde \alpha(P)$,\, $\tilde \sigma(P))$ is a vector
containing the Fourier transforms of $\tilde \alpha$ and $\tilde
\sigma$, and the function $\Pi(P,m)$ is given by
\begin{equation}
 \Pi(P, m) = \sumint_Q \frac{1}{Q^2+m^2} \frac{1}{(P+Q)^2+m^2} \;.  
\end{equation}

%=====================================================================
\section{Effective potential and gap equations}

One can obtain the effective potential to next-to-leading order
in the $1/N$ expansion from Eq.~(\ref{eq:expandedeffaction})
by performing the Gaussian integral over the fluctuating fields $\tilde
\alpha$ and $\tilde \sigma$. 
Up to corrections of order $1/N$, the effective potential can be written
as
\begin{equation}
{\cal V}(m^2, \bar \sigma) = 
  N \mathcal{V}_{\mathrm{LO}}(m^2,\bar \sigma) 
+ \mathcal{V}_{\mathrm{NLO}}(m^2, \bar \sigma)\;,
\label{nloeq}
\end{equation}
where
\begin{eqnarray}
  \mathcal{V}_{\mathrm{LO}}\left(m^2, \bar \sigma \right)
 &=& \frac{m^2}{2} \left ( f_{\pi,b}^2 - \bar \sigma^2 \right) + 
        \frac{m^4}{2 \lambda_b} + H \bar \sigma
  -\frac{1}{2} \sumint_P \log(P^2 + m^2 )  
  \label{eq:effpotlo}
   \;, \\ 
   \mathcal{V}_{\mathrm{NLO}}\left(m^2, \bar \sigma\right) 
  &=& -\frac{1}{2} \sumint_P \log I(P,m) 
\;.
\end{eqnarray}
Here,
\beq
  I(P,m) = 16 \pi^2 \Pi(P, m) + \frac{32 \pi^2}{\lambda_b} 
+ \frac{32 \pi^2 \bar \sigma^2}
      {P^2 + m^2} \;.
\label{Ipm}
\eeq
To derive the effective potential, divergent constants which are
independent of $\bar\sigma$, $m$ and the temperature were subtracted.
Equivalently, these terms can be removed by adding a vacuum
counterterm to the effective potential. In the
following, such terms are simply dropped.

In thermodynamic equilibrium, the system will be in the state that
extremizes the effective potential with respect to $m^2$ and $\bar
\sigma$.  This extremum can be found by differentiating the effective
potential with respect to $m^2$ and $\bar{\sigma}$, which gives
\begin{eqnarray}
  \sumint_P \frac{1}{P^2 + m^2}
   - \frac{2 m^2}{\lambda_b} +
  \frac{1}{N} \sumint_P 
   \frac{ \frac{\mathrm{d} \Pi(P, m)}{\mathrm{d} m^2}
  - \frac{2 \bar \sigma^2}{ \left( P^2 + m^2 \right)^2}}
  {\Pi(P, m) + \frac{2}{\lambda_b}+ \frac{2\bar \sigma^2}
      {P^2 + m^2}}
 & =&
   \left( f_{\pi,b}^2 - \bar \sigma^2\right) \;, 
  \label{eq:gapmsq} 
  \\
  \left(m^2 + \frac{2}{N} \sumint_P \frac{1}{P^2 + m^2} 
  \frac{1}{\Pi(P,m) + \frac{2}{\lambda_b} + 
  \frac{2 \bar \sigma^2}{P^2 + m^2}} \right)\bar \sigma 
   &=& H \;. 
  \label{eq:gapsigma}
\end{eqnarray}
These equations are often referred to as gap equations.  Solving the
gap equations gives $m$ and $\bar \sigma$ as a function of the
parameters $f_\pi$, $H$ and $\lambda$, and of the temperature. The 
solution of the gap equation is needed to calculate thermodynamical
quantities like the pressure.

The inverse $\tilde \alpha$ and $\tilde \sigma$ propagators can also
be obtained from Eq.~(\ref{eq:expandedeffaction}). For this
the $2 \times 2$ matrix in Eq.~(\ref{eq:expandedeffaction}) has to be 
inverted. As a result one finds
\begin{eqnarray}
  D^{-1}_{\tilde \alpha} (P, m)  &=& \frac{1}{2} \Pi(P,m) + 
  \frac{1}{\lambda_b} + \frac{\bar \sigma^2}{P^2 + m^2} \;, 
\label{eq:inversetildealphaprop}
  \\
  D^{-1}_{\tilde \sigma} (P, m)  &=& P^2 + m^2 + 
  \frac{2 \bar \sigma^2}{\Pi(P, m) + 2 / \lambda_b}\;.
\label{inversesigmaprop}
\end{eqnarray}
In these equations the values for $m^2$ and $\bar \sigma$ are
determined by solving the gap equations. 
The $\pi$ propagator is equal to (see Eq.~(\ref{eq:effactionsource}))
\begin{equation}
  D_{\pi} = \left< \frac{1}{-\partial^2 -i \alpha} \right> \;.
\end{equation}
Expanding $\alpha$ around its vacuum expectation value $im^2$ in the
previous equation, and using the expression for $\tilde \alpha$
propagator, Eq.~(\ref{eq:inversetildealphaprop}), it follows that the
inverse $\pi$ propagator in momentum space is given by
\begin{equation}
 D^{-1}_{\pi} (P, m) = P^2 + m^2 + 
  \frac{2}{N} \sumint_Q  \frac{1}{(P+Q)^2 + m^2} 
  \frac{1}{\Pi(Q, m) +
\frac{2}{\lambda_b} + \frac{2 \bar \sigma^2}{Q^2 + m^2}} \;.
\end{equation}
Gap equation (\ref{eq:gapsigma}) can be used to show that in the
broken phase where $\bar \sigma \neq 0$ (for $H=0$) the pion
propagator has a pole at $P^2=0$. This implies that also to
next-to-leading order in $1/N$ the pions are massless, in accordance
with Goldstone's theorem.

From Eq.\ (\ref{inversesigmaprop}) it follows that in the unbroken
phase, the $\sigma$ mass becomes equal to the leading order mass of
the $\pi$ field, which is $m^2$. It may appear therefore that the
$\sigma$ and $\pi$ masses are not equal at next-to-leading order in
the unbroken phase, but this is not a correct conclusion.  This is
because the $\sigma$ field only starts to propagate at next-to-leading
order, so its $1/N$ mass corrections require a next-to-next-to-leading
order calculation.  In the calculation of the pressure to
next-to-leading order one only needs the leading order masses as will
be explicitly shown below in Eq.~(\ref{gapexp}).

The leading-order and next-to-leading order contributions to the
effective potential in $3+1$ dimensions will be explicitly calculated in
the next subsections. The integrals over momentum are regulated using
an ultraviolet momentum cutoff $\Lambda$. Furthermore it is assumed
that the cutoff is large compared to the other scales in the problem,
that is $\Lambda \gg m, 2 \pi T$.

%---------------------------------------------------------------------
\subsection*{Leading-order contribution}
\label{sub1}
The leading-order contribution to the effective potential is
\begin{eqnarray}
{\cal V}_{\rm LO} &=&
  \frac{m^2}{2} \left(f^2_{\pi,b}- \frac{\Lambda^2}{16\pi^2} - 
  \bar{\sigma}^2\right)
+ \frac{T^4}{64\pi^2} J_0(\beta m)
\nonumber \\
&& +
\frac{m^4}{64\pi^2}\left[
  \frac{32 \pi^2} {\lambda_b} + 
  \log \left ( \frac{\Lambda^2}{m^2} \right)
+ \frac{1}{2}  
\right] + H \bar \sigma\;,
\label{LO}
\end{eqnarray}
where the function $J_0(\beta m)$ is
\begin{equation}
  J_0(\beta m) = \frac{32}{3 T^4} \int_0^{\infty}\mathrm{d}p\;\frac{p^4}{\omega_p}
  n(\omega_p)\;.
\label{Jzero}
\end{equation}
Here, $n(\omega_p) = [\exp(\beta \omega_p) - 1]^{-1}$ is the
Bose-Einstein distribution function. Since a finite effective momentum
cutoff $\Lambda$ was introduced in the theory, one makes an error in
the evaluation of $J_0$ by summing over all Matsubara modes and
integrating up to infinite momenta instead of up to
$\Lambda$. However, this error is negligible as long as $\Lambda
\gg m, 2 \pi T$. This remark also applies to the functions $J_1$,
$K_0^\pm$ and $K_1^\pm$ defined below. For an investigation of how one
could apply a finite cut-off in the calculation of sum-integrals see
\citet{Amte1993}.

Equation (\ref{LO}) contains ultraviolet divergences in the sense
explained in Sec.~6.1.  These divergences can be dealt with by
defining the renormalized parameters $f_{\pi}^2$ and $\lambda$ as
\bqa
f^2_\pi &=& f^2_{\pi,b} -
\Lambda^2 / 16 \pi^2 \;,
\\ 
  \frac{32\pi^2}{\lambda} &= &\log
  \left(\frac{\Lambda^2}{\mu^2} \right) + \frac{32 \pi^2}{\lambda_b} \;,
  \label{eq:lambdarenorm}
\eqa
where $\lambda = \lambda(\mu)$. 
The renormalization group equation for the running coupling $\lambda$
that follows from Eq.~(\ref{eq:lambdarenorm}) is
\begin{equation} 
  \beta(\lambda) = \mu \frac{\mathrm{d}
  \lambda}{\mathrm{d} \mu} = \frac{\lambda^2}{16 \pi^2} \;.
\label{beta1}
\end{equation} 
The $\beta$-function is exact to all orders in $\lambda^2$ in the
large-$N$ limit, but differs from the perturbative one obtained at one
loop, see \citet{Peskin1995}.  However, at next-to-leading order they agree as
will be shown in the next subsection.  After this renormalization, the
leading-order effective potential becomes
\begin{equation}
  \mathcal{V}_{\mathrm{LO}}
  = \frac{m^2}{2}\left(
  f_{\pi}^2 - \bar \sigma^2 \right) 
    +\frac{m^4}{64 \pi^2} 
  \left[\frac{32 \pi^2}{\lambda} + 
  \log \left ( \frac{\mu^2}{m^2} \right)
   + \frac{1}{2} \right]
 + 
  \frac{T^4}{64 \pi^2} J_0(\beta m) + H \bar \sigma\;.
  \label{eq:effpotlo3d}
\end{equation}

Since the potential term in the Lagrangian should always have a
minimum in order to have a stable theory, $\lambda_b$ must be positive
(cf.\ \citet{Amelino-Camelia1997} for a detailed discussion). 
From Eq.\ (\ref{eq:lambdarenorm}) it immediately 
follows that there is a maximal value for the cutoff given by
\begin{equation}
  \Lambda_\mathrm{max} = \mu \exp \left( \frac{16 \pi^2}{\lambda}
  \right) \;.
\label{cutmax}
\end{equation}
Therefore the linear sigma model should be viewed as an effective
theory, which is at most valid up to the cutoff given by
Eq.~(\ref{cutmax}).  Taking the cutoff to infinity is equivalent to
taking $\lambda$ to zero, which implies that the theory is trivial.
One should keep in mind that the renormalized leading-order effective
potential does not depend explicitly on $\Lambda$, but is only valid
for $m$ and $T$ much smaller than $\Lambda_{\mathrm{max}}$.  When
$\Lambda = \Lambda_{\mathrm{max}}$ the linear sigma model reduces to
the nonlinear sigma model, since in this case $\lambda_b = \infty$.

The leading-order renormalized gap equations follow from
differentiating Eq.\ (\ref{eq:effpotlo3d}) with respect to $m^2$ and
$\bar{\sigma}$ and are given by
\begin{eqnarray}
  G & = & 16 \pi^2 f^2_\pi \;,
 \label{eq:logapeqm}
  \\
  H &=& m^2 \bar \sigma  \;,
 \label{eq:logapeqsigma}
\end{eqnarray}
where
\beq
G  =
T^2 J_1(\beta m) + 16 \pi^2 \bar \sigma^2 
  - m^2 \log \left( \frac{\mu^2}{m^2} \right) 
    -  \frac{32\pi^2 m^2}{\lambda} 
\;.
\eeq
Here, the function $J_1(\beta m)$ is defined as
\begin{equation}
 J_1(\beta m) =  \frac{8}{T^2} \int_0^{\infty}
\mathrm{d}p\;\frac{p^2}{\omega_p} n(\omega_p) \;.
\label{J1}
\end{equation}

\begin{figure}[t]
\begin{center}
\includegraphics{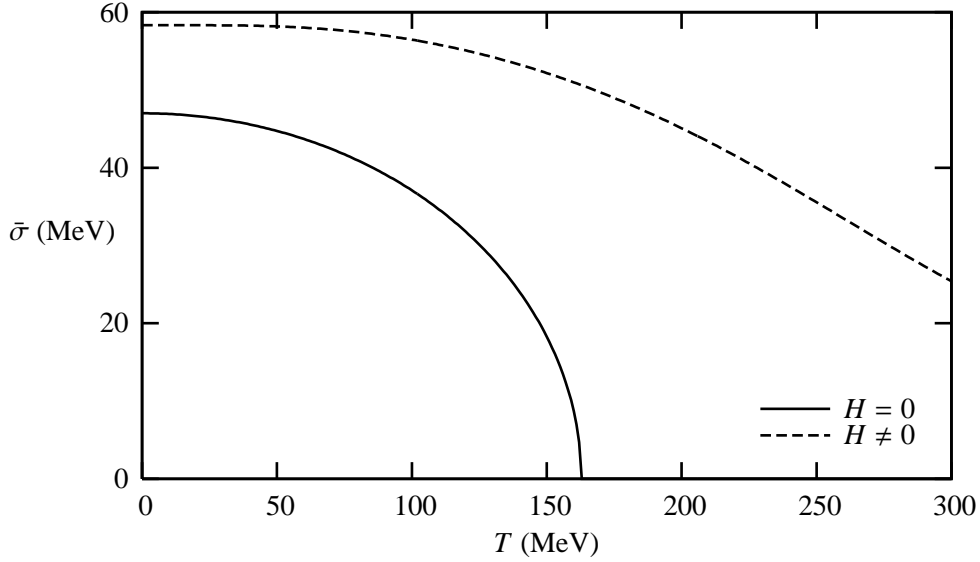}
\caption{Leading order sigma condensate for $H=0$ (solid line) and $H \neq 0$ (dashed line).
The parameter choice is discussed in Sec.~6.5.}
\label{fig:barsigma}
\end{center}
\end{figure}

\begin{figure}[t]
\begin{center}
\includegraphics{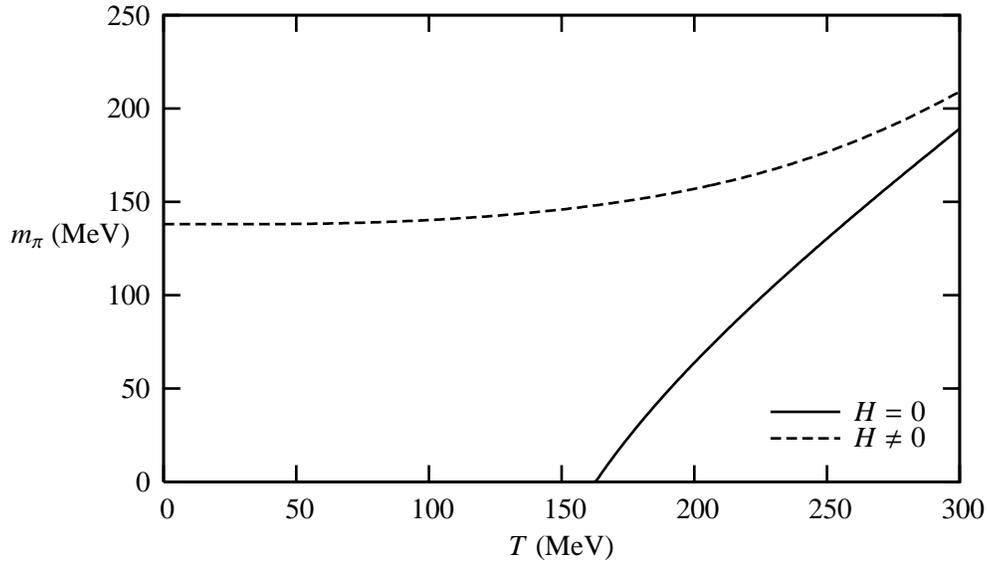}
\caption{Leading order pion mass for $H=0$ (solid line) 
and $H \neq 0$ (dashed line). The parameter choice is discussed in
Sec.\ 6.5}
\label{fig:mpi}
\end{center}
\end{figure}

In Figs.~\ref{fig:barsigma} and \ref{fig:mpi} the vacuum expectation
value of the sigma field and the pion mass are respectively displayed
as a function of temperature. The cases with ($H \neq 0$) and without
($H=0$) explicit symmetry breaking are considered. Both figures were
calculated using the parameter set discussed in Sec.\ 6.5.

If $H=0$, one can show by using the gap equation (\ref{eq:logapeqsigma}) 
that either $m=0$ or $\bar \sigma = 0$. From the gap equation
(\ref{eq:logapeqm}) it follows that for $m=0$ the 
expectation value of $\sigma$ has the temperature dependence
\begin{equation}
  \bar \sigma = \sqrt{f_\pi^2 - \frac{T^2}{12}}\;.
\end{equation}
The order parameter for symmetry breaking, $\bar \sigma$ vanishes
continuously. Hence at $T = T_c \equiv \sqrt{12} f_\pi$ there is a
second-order phase transition. Below $T_c$ the $\mathrm{O}(N)$ symmetry is
broken spontaneously to $\mathrm{O}(N-1)$ since $\bar
\sigma \neq 0$. Above $T_c$ the $\mathrm{O}(N)$ symmetry is restored and one
has $\bar \sigma = 0$ and $m \neq 0$.

%---------------------------------------------------------------------
\subsection*{Next-to-leading order contribution}
\label{sub2}
In this subsection it will be shown that it is not possible to
renormalize the next-to-leading order effective potential in a
temperature-independent way. It turns out that one can only
renormalize the effective potential at the minimum, since the
temperature-dependent divergences become temperature independent by
using the leading order gap equations. Hence physical quantities like
the pressure, which can be obtained from the minimum of the effective
potential can be renormalized consistently.  To show this, the
divergent parts of the effective potential will be extracted. This can
be done analytically.  A perhaps more familiar example in which the
effective potential is meaningless outside the minimum arises at zero
temperature in gauge theories. In this case, the effective potential
depends on the gauge-fixing condition except at the
minimum~\citep{Jackiw1974,Nielsen1975,Kobes1991}.

In order to isolate all divergences, one in principle needs to
evaluate $\Pi(P,m)$ including corrections of order $m^2 / \Lambda^2$,
since such terms can also give rise to divergences in the effective
potential. However, since the linear sigma model is an effective
theory, Eq.~(\ref{eq:lagrangian_lin_sigma}) should be viewed as the
part containing only the so-called relevant operators, see
\citet{Polchinski1984}.  For instance, irrelevant operators of
mass-dimension six are not included in the Lagrangian density. Such
operators also contribute to $\Pi(P,m)$ at order
$1/\Lambda^2$. Therefore, for consistency with
Eq.~(\ref{eq:lagrangian_lin_sigma}), order $1/\Lambda^2$ terms in
$\Pi(P,m)$ are not considered.  Although it is possible to obtain an
exact analytic expression for the zero-temperature part of $\Pi(P,m)$,
here it will only be given up to order $1 / \Lambda^2$ since this
expression is much less complicated.  As a result one finds
\begin{multline}
\Pi(P,m)
= \frac{1}{16 \pi^2} \left[ \log \left( \frac{\Lambda^2}{m^2} \right)
  + 1 
   + \sqrt{\frac{P^2+4m^2}{P^2}} 
  \log \left( \frac{\sqrt{P^2+4m^2} - \sqrt{P^2}}{\sqrt{P^2+4m^2} + 
  \sqrt{P^2}} \right) \right] \\ 
+ \Pi_T(P,m) \;,
\end{multline}
where the temperature-dependent part of $\Pi(P,m)$ equals
\citep{Bochkarev1996}
\begin{equation}
\Pi_T(P,m) =
   \frac{1}{8\pi^2p}\int_0^{\infty}\mathrm{d}q\;\frac{q}{\omega_q}
  \log \left( \frac{q^2 + p q + A^2}{q^2 - p q + A^2} \right)
  n(\omega_q) \;.
\end{equation}
Here
\begin{equation}
  A^2 = \frac{P^4 + 4 m^2 p_0^2}{4 P^2} \;.
\label{Akwadraat}
\end{equation}
In the limit $P \gg m, T$, $\Pi_T(P,m)$ can be approximated by
\begin{equation}
  \Pi_T(P,m) \approx 
 \frac{1}{8 \pi^2} \left[
  \frac{T^2}{P^2} J_1(\beta m) - 
  \frac{4m^2 T^2 p_0^2}{P^6} J_1(\beta m) -
  \frac{\left(3 P^2 - 4 p^2\right)T^4}{P^6} J_0(\beta m)
  \right] \;.
\end{equation}

The next-to-leading-order effective potential has only ultraviolet
divergences. Using the leading order renormalization of $\lambda_b$,
it is easily seen that $I(P,m)$ (defined in Eq.~(\ref{Ipm})) becomes
finite. Also, the difference
\begin{equation}
\sumdiff_P \log I(P,m) 
\;,
\end{equation}
is finite (cf.\ Sec.~\ref{sec:abelplana}).  Therefore, all possible
divergences of $\mathcal{V}_{\mathrm{NLO}}$ can be isolated by
calculating
\begin{equation}
  -\frac{1}{2} \int_P \log I_{\mathrm{HM}}(P, m) \;,
\end{equation}
where $I_{\mathrm{HM}}(P,m)$ is the high-momentum (HM) approximation
to $I(P,m)$. It gives the large-$P$ behavior of $I(P,m)$.
After averaging over angles, one finds 
\begin{equation}
  \log I_{\mathrm{HM}} =
  \log C_1 + \frac{1}{P^2} \frac{C_2}{C_1} 
- \frac{1}{2 P^4} 
  \left( \frac{C_2}{C_1}
  \right)^2 
  + \frac{1}{P^4} \frac{C_3}{C_1} \;,
\end{equation}
where
\begin{eqnarray}
  C_1 &=& \log \left( \frac{\mu^2}{P^2} \right)+ 1 +
   \frac{32\pi^2}{\lambda} \;,
\\
  C_2 &=& - 2 m^2 \left[1 + \log \left(\frac{P^2}{m^2} \right) \right]
  + 32\pi^2 \bar \sigma^2 + 2 T^2 J_1(\beta m)\;,
\\
  C_3 &=& 
+ m^4 \left[2 \log \left( \frac{P^2}{m^2} \right) - 1 \right]
   - m^2 \left[ 32 \pi^2 \bar \sigma^2 + 2 T^2 J_1(\beta m) \right] \;.
\end{eqnarray}
By integrating the function $\log I_{\mathrm{HM}}$ over $P$, all the
divergences of the NLO effective potential can be found.  The
logarithmic and power divergences are given by the quantity $D$, which
is
\begin{eqnarray}
  D &=&
\frac{1}{16\pi^2} 
\biggl \{ \biggr. 
   \Lambda^2  e^{1+32\pi^2 / \lambda_b} 
  \,\mathrm{li} \left( \frac{1}{e^{1+32\pi^2 / \lambda_b}} \right) G 
  + 2 m^4 \log \left(\frac{\Lambda^2}{m^2} \right) 
\nonumber \\ 
   & & 
  \phantom{ \frac{1}{16\pi^2} 
\biggl \{ \biggr.}
  - m^2 \Lambda^2 \left[1 +
 2 e^{1+32\pi^2/\lambda_b}
  \, \mathrm{li} 
  \left( \frac{1}{e^{1+32\pi^2 / \lambda_b}} \right) 
  \right]
  \biggl. \biggr \} \;,
\label{eq:nlodivs} 
\end{eqnarray}
while the terms that have a small cutoff dependence through their
dependence on $\lambda_b$, are given by the quantity $E$, which is
\begin{equation}
E =
  \frac{1}{16\pi^2} 
  \Biggl [ \Biggr. 
  3 m^2 \left(-G + \tfrac{3}{2} m^2 \right) \log
   \left(
         1 + \frac{32\pi^2}{\lambda_b} \right)
  + \left(G - 2 m^2 \right)^2 \frac{1}{ 1 + \frac{32\pi^2}{\lambda_b}}
  \Biggl. \Biggr ] \;.
  \label{eq:slowdiv}
\end{equation}
Since $G$ depends explicitly on the temperature, it is impossible to
renormalize the next-to-leading-order effective potential in a
temperature-independent way. However, at the minimum, the
leading-order gap equation~(\ref{eq:logapeqm}) can be used to show
that $G = 16\pi^2 f_\pi^2$. Hence, the divergences become independent
of the temperature at the minimum and renormalization can be carried
out in a temperature-independent manner. This will be discussed in
more detail next.

The divergence proportional to $G$ in Eq.~(\ref{eq:nlodivs}) is
independent of $m$ in the minimum even though it depends on $f_\pi$.
This divergence can be removed by vacuum renormalization.  The
divergent terms which are proportional to $m^2$ can be removed by
defining the renormalized parameter $f_\pi$ as
\begin{equation}
  f_\pi^2 = f^2_{\pi, b} - \left(1 + \frac{2}{N} \right)
  \frac{\Lambda^2}{16\pi^2} 
  -
  \frac{1}{N} \frac{\Lambda^2}{4\pi^2}
  \bigg[ 
  e^{1+32\pi^2/\lambda_b} \, \mathrm{li}  
 \left( \frac{1}{e^{1+32\pi^2 / \lambda_b}} \right) \bigg]
  \;.
  \label{eq:fpirenormnlo}
\end{equation}
The remaining divergence is proportional to $m^4$
and is removed by renormalizing
$\lambda_b$ as follows
\begin{equation} 
  \frac{32\pi^2}{\lambda} = 
   \frac{32 \pi^2}{\lambda_b} + 
\left(1 + \frac{8}{N} \right)
 \log
\left(\frac{\Lambda^2}{\mu^2} \right)
  \label{eq:lambdarenormnlo} \;.
\end{equation} 
From Eq.~(\ref{eq:lambdarenormnlo}), the $\beta$-function 
governing the running of $\lambda$ can be obtained
\begin{equation}
\beta(\lambda)
= \frac{\lambda^2}{16 \pi^2} \left( 1+ \frac{8}{N} \right)\;,
\end{equation}
which coincides with the standard one-loop $\beta$-function in
perturbation theory, see \citet{Peskin1995}. One could argue from this
renormalization that one can only trust the $1/N$ expansion for $N \gg
8$.  Although the $1/N$ correction to the $\beta$ function indeed has
a large coefficient, this is not the case for the effective potential
itself as will be shown below.

As mentioned, the terms in $E$ have a small cutoff dependence through
their dependence on $\lambda_b$. These terms will not be renormalized,
since they do not grow without bound with increasing cutoff and are
not strictly speaking divergences.  The effective potential does not
become increasingly sensitive to them with increasing cutoff. The term
in the first line of Eq.~(\ref{eq:slowdiv}) becomes smaller if
$\Lambda$ is increased and the absolute value of the other term from
$E$ increases as a function of $\Lambda$, but is bounded by a finite
number which is independent of $\lambda_b$ and $\Lambda$. Moreover,
renormalizing these terms would invalidate the $1/N$ expansion,
because of their magnitude. This is similar to ordinary perturbation
theory, where one is only allowed to do finite renormalizations that
do not invalidate the perturbative expansion.  A final reason for not
renormalizing these terms is the connection with the nonlinear sigma
model ($\lambda_b = \infty$). In that case, the terms from $E$ are not
divergent and $f_\pi$ will be renormalized just as in
Eq.~(\ref{eq:fpirenormnlo}) with $\lambda_b =
\infty$.

The next-to-leading order correction changes the critical temperature
$T_c$. Since the next-to-leading order gap equations are complicated,
it is not possible to obtain an analytical expression for the critical
temperature at next-to-leading order. However, in the limit of small
$\lambda_b$ and $H=0$ the gap equations simplify to
\bqa
 && \sumint_P \frac{1}{P^2 + m^2}
   - \frac{2 m^2}{\lambda_b} 
  =
   \left( f_{\pi,b}^2 - \bar \sigma^2\right) \;, 
   \hspace{1cm}
  \label{eq:gapmsqsmalllambda} 
  \\
 && \left(m^2 + \frac{\lambda_b}{N} \sumint_P \frac{1}{P^2 + m^2}  
  \right) \bar \sigma
   = 0 \;.
  \label{eq:gapsigmasmalllambda}
\eqa
From the gap equations it follows that the critical temperature at NLO is 
\beq
T_c = \sqrt{\frac{12}{1+2/N}} f_\pi \;.
\eeq
This result is the same as obtained by \citet{Jain1993} and
\citet{Bochkarev1996}.  It is probably only correct in the
weak-coupling limit and $T_c$ may depend on $\lambda$ at NLO in $1/N$.
In the next section it can be seen that the transition at NLO remains
second order.

%---------------------------------------------------------------------
%=====================================================================
\section{Pressure}
Like in Chapters 4 and 5, the pressure ${\cal P}(T)$ is defined as the
value of the effective potential at the minimum at temperature $T$
minus its value at the minimum at zero temperature. As was shown in
the previous section, it is possible to renormalize the effective
potential at the minimum. The pressure is therefore a well-defined
quantity.  In order to determine the value of the next-to-leading
order effective potential in the minimum, only the gap equation to
leading order~\cite{Root1974} is needed. Writing the solutions to the
gap equations as
\begin{eqnarray}
m^2 &=
&m^2_\mathrm{LO} + m^2_{\mathrm{NLO}} / N \;, \\
\bar \sigma &=& \bar \sigma_{\mathrm{LO}} + \bar \sigma_{\mathrm{NLO}}
/ N \;,
\end{eqnarray}
and Taylor expanding the effective
potential~(\ref{nloeq}), one obtains (up to ${\cal O}(1/N)$ corrections)
\begin{eqnarray}
\nonumber
  \mathcal{V}(m^2, \bar \sigma) &=& 
  N \mathcal{V}_{\mathrm{LO}}(m^2_{\mathrm{LO}}, \bar
  \sigma_\mathrm{LO}) 
   + \mathcal{V}_{\mathrm{NLO}}(m_{\mathrm{LO}}^2, \bar
\sigma_{\mathrm{LO}})  
  \\ && \nonumber
  + m^2_{\mathrm{NLO}}
  \left. \frac{\partial \mathcal{V}_{\mathrm{LO}}(m^2)} 
    {\partial m^2} \right \vert_{m^2 = m^2_\mathrm{LO}} 
+  \bar \sigma_{\mathrm{NLO}}
  \left. \frac{\partial \mathcal{V}_{\mathrm{LO}}(\bar \sigma)} 
    {\partial  \bar \sigma} \right \vert_{\bar \sigma = \bar
  \sigma_\mathrm{LO}}\;.
\label{gapexp}
\end{eqnarray}
The last two lines of Eq.~(\ref{gapexp}) vanish by using the leading-order 
gap equations. 
In the following, the pressure ${\cal P}$ will be written as
\bqa
\mathcal{P} \equiv N
\mathcal{P}_{\mathrm{LO}} + \mathcal{P}_{\mathrm{NLO}} 
\;.
\eqa
From the discussion above, it follows that
\begin{eqnarray}
  \mathcal{P}_{\mathrm{LO}}  &=& 
     \mathcal{V}^{T}_{\mathrm{LO}}(m^2_T, 
\bar \sigma_T) 
   - \mathcal{V}^{T=0}_{\mathrm{LO}}(m^2_0, 
\bar \sigma_{0})  
  \;, \label{eq:lopres}
  \\
  \mathcal{P}_{\mathrm{NLO}} &=&
     \mathcal{V}^{T}_{\mathrm{NLO}}(m^2_T, 
  \bar \sigma_T) 
   - \mathcal{V}^{T=0}_{\mathrm{NLO}}(m^2_0, 
\bar \sigma_0) \;,
\end{eqnarray}
where $m^2_T$ and $\bar \sigma_T$ are the solutions of the leading-order 
gap equations~(\ref{eq:logapeqm}) 
and~(\ref{eq:logapeqsigma}), at temperature $T$. 

In the following, the results for the numerical evaluation of the
leading and the next-to-leading order contributions to the pressure
for general $N$ will be presented. Subsequently the $N=4$ case which
is of relevance for QCD phenomenology will be discussed.

As will be motivated in Sec.~\ref{sec:parameters}, the following
values for the parameters will be used: $\lambda (\mu =
100\;\mathrm{MeV}) = 30$, $f_\pi = 47\;\mathrm{MeV}$ (note that
$f_\pi$ as defined in Eq.~(\ref{eq:lagrangian_lin_sigma}) is $1/2$
times the more conventional definition) and if there is explicit
symmetry breaking $H = \left(104 \;\mathrm{MeV}
\right)^3$ in order to reproduce the physical value of the
pion mass at $T=0$.  A realistic choice of parameters would allow to
compare to lattice QCD simulations of the $N_f=2$ case for $T
\simordertwo T_c$, although this would require extrapolation of the
lattice results down to the actual pion masses.

%---------------------------------------------------------------------
\subsection*{Leading-order contribution to the pressure}

\begin{figure}[t]
\begin{center}
\scalebox{1.0}{\includegraphics{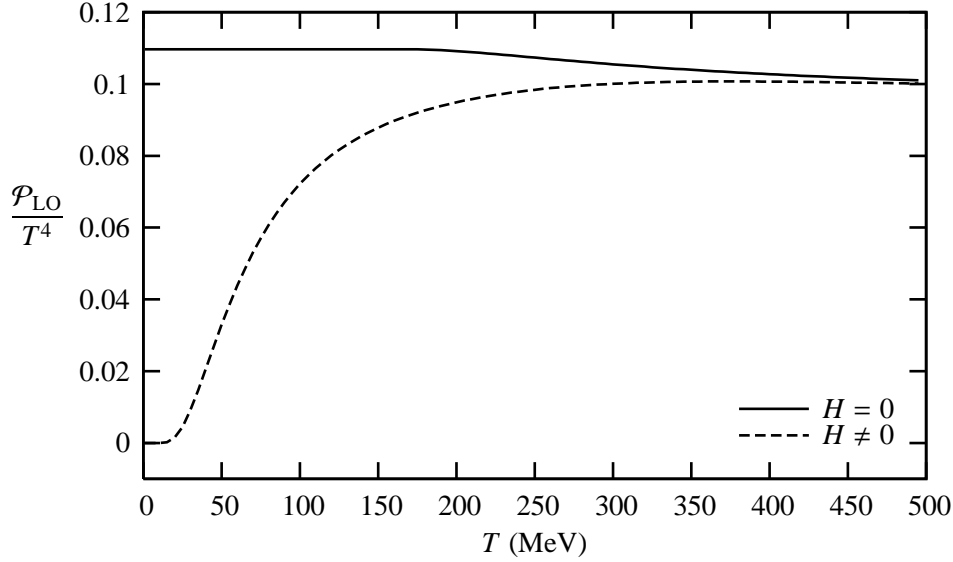}}
\caption{Leading-order pressure
${\cal P}_{\rm LO}$ normalized to $T^4$ as a function of
temperature without and with explicit symmetry breaking.}
\label{fig:leadingpressure}
\end{center}
\end{figure}

The leading order pressure can be obtained from
Eqs.~(\ref{eq:effpotlo3d}) and (\ref{eq:lopres}).  In
Fig.~\ref{fig:leadingpressure}, the leading-order pressure normalized
by $T^4$ is shown. If $H=0$, the pions are massless below $T_c$. Then
it can be see from the leading-order effective
potential~(\ref{eq:effpotlo3d}) that the pressure becomes equal to the
pressure of an ideal gas of massless particles: ${\cal P}_{\rm
LO}=\pi^2 T^4 / 90$. If $H \neq 0$, the pions are massive. In the
limit of zero temperature the $\mathcal{P}_\mathrm{LO}/T^4$ goes to
zero because of the Boltzmann suppressing factor $\exp(-m/T)$.

%---------------------------------------------------------------------
\subsection*{Next-to-leading order contribution to the pressure} 
To calculate the next-to-leading order contribution to the pressure,
$\mathcal{P}_{\mathrm{NLO}}$ is decomposed as follows
\begin{equation}
  \mathcal{P}_{\mathrm{NLO}} = D(m_T) - D(m_0) + F_1 + F_2 \;,
  \label{eq:nlopressplit}
\end{equation}
where $D(m)$ is the term containing logarithmic and power ultraviolet
divergences given in Eq.~(\ref{eq:nlodivs}),
and $F_1$ and $F_2$ are finite terms defined below. 

The term $F_1$ is defined by
\begin{multline}
  F_1 = -\frac{1}{2} \int_P \Bigg\{ \log \left [\Pi(P, m_T) + \frac{2}{\lambda_b} +
   \frac{2\bar \sigma_T^2}{P^2 + m_T^2}
 \right] \\
    -\log \left [\Pi(P, m_0) + \frac{2}{\lambda_b} +
   \frac{2\bar \sigma_0^2}{P^2 + m_0^2}
 \right ] \Bigg\} 
- D(m_T) + D(m_0) \;. 
  \label{eq:f1}
\end{multline}
Since the term $F_1$ contains the finite cutoff-dependent term $E$
(which is defined in Eq.~(\ref{eq:slowdiv})) it has a small dependence
on the cutoff as well.  The function $F_1$ was calculated numerically
by rewriting the terms involving $D$ as an integral like it was done
in Sec.~4.4. Then it is possible to subtract the integrands, instead
of the large values of the integral. In this way, it is easier to
avoid large numerical errors.

The function $F_2$ is defined by
\begin{equation}
   F_2 =- \frac{1}{2} 
\sumdiff_P \log \left [\Pi(P, m_T) + \frac{2}{\lambda_b} +
   \frac{2\bar \sigma_T^2}
      {P^2 + m_T^2} \right] \;.
 \label{eq:f2}
\end{equation}
In order to calculate the function $F_2$ a modified Abel-Plana formula
was used, see Sec.~(\ref{sec:abelplana}).  It turns out that due the
suppression of high momentum modes at finite temperature, low momentum
modes gives the main contribution to $F_2$.  This shows that the
``high-energy approximation'' of \citet{Bochkarev1996} (implicitly
defined as such a difference and directly related to $F_2$) is
invalid, since in that approximation the high-momentum modes are
assumed to give the main contribution.

\begin{figure}[t]
\begin{center}
\scalebox{1.0}{\includegraphics{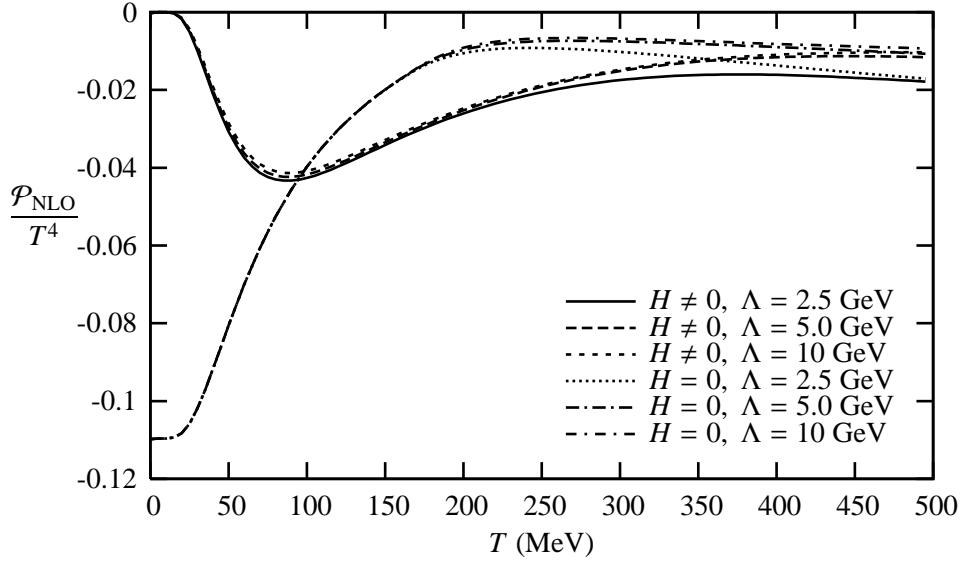}}
\caption{Next-to-leading order contribution to the pressure 
normalized to $T^4$, as function
of temperature for $H=0$ and $H = \left(104 \;\mathrm{MeV} \right)^3$,
for different values of the
cutoff $\Lambda$. }
\label{fig:nlopressure}
\end{center}
\end{figure}

After renormalization, it turns out that the next-to-leading order
contribution to the pressure is
\begin{equation}
  \mathcal{P}_{\mathrm{NLO}} = \frac{m_T^4}{8\pi^2} \log
  \left(\frac{\mu^2}{m_T^2} \right) 
  - \frac{m_0^4}{8\pi^2} \log
  \left(\frac{\mu^2}{m_0^2} \right)
 + F_1 + F_2
 \;,
\end{equation}
which is shown in Fig.~\ref{fig:nlopressure}. 
At $T=0$, $\mathcal{P}_{\mathrm{NLO}} / T^4$ can be calculated
exactly. This can be used as a check of the numerical calculations. At
$T=0$, clearly $F_1 = 0$, and hence $\mathcal{P}_{\mathrm{NLO}}/T^4 =
F_2 / T^4$.  At $T=0$, for low $P$, $I(P,m)$ is
dominated by $32\pi^2 \bar \sigma^2 / (P^2 + m^2)$. This gives
\beq
F_2 \approx \frac{1}{2} \sumdiff_P \log(P^2 + m^2) \;. 
\eeq
For $H=0$, the mass $m$ vanishes and so
$\mathcal{P}_{\mathrm{NLO}}/T^4 = -\pi^2 / 90$.  For $H =
(104\;\mathrm{MeV})^3$ it follows that $\mathcal{P}_{\mathrm{NLO}}/T^4
= 0$, since $T$ is in that case much smaller than $m$, such that the
pressure is exponentially suppressed.

The pressure for $H=0$ is approaching the $H \neq 0$ pressure at high
temperatures, indicating that the effects of the explicit
symmetry-breaking terms become smaller at higher temperatures. This is
because $H$ is a temperature-independent constant.

%---------------------------------------------------------------------
\subsection*{Pressure of the $\mathrm{O}(4)$ linear sigma model} 
\begin{figure}[htb]
\begin{center}
\scalebox{1.0}{\includegraphics{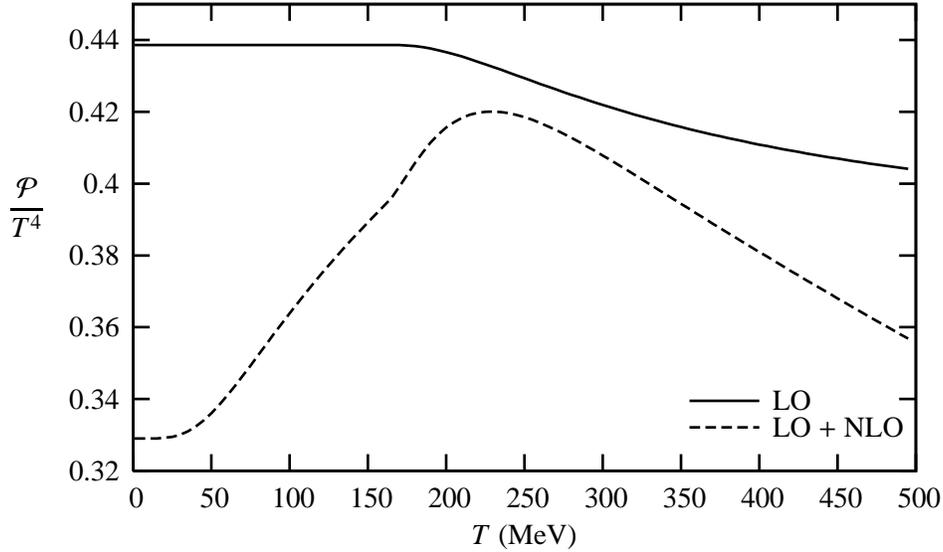}}
\caption{LO and NLO pressure normalized to $T^4$, for $N=4$ as
 a function of temperature, for $H=0$ and $\Lambda =5.0$ GeV.}
\label{fig:presN4h0}
\end{center}
\end{figure}

In order to make contact with two-flavor low-energy QCD the
$\mathrm{O}(4)$ model is studied in this subsection.  In
Fig.~\ref{fig:presN4h0}, the pressure normalized by $T^4$ for $N=4$
and $H=0$ to next-to-leading order as function of $T$ is shown.
The LO pressure below $T_c$ equals the pressure of a gas of four
massless non-interacting scalars.  This follows immediately from
Eqs.~(\ref{eq:effpotlo3d}) and~(\ref{Jzero}).  At NLO the sigma field
becomes massive. For temperatures much lower than $m_{\sigma}$, the
contribution to the pressure from the sigma is Boltzmann suppressed
and (to good approximation) it holds that ${\cal P}=\pi^2 T^4/30$,
which is the pressure of a gas of three massless non-interacting
scalars.  From the calculations presented here it can be concluded
that the transition to next-to-leading order is of second (or higher)
order since the derivative of the pressure is not diverging.

\begin{figure}[htb]
\begin{center}
\scalebox{1.0}{\includegraphics{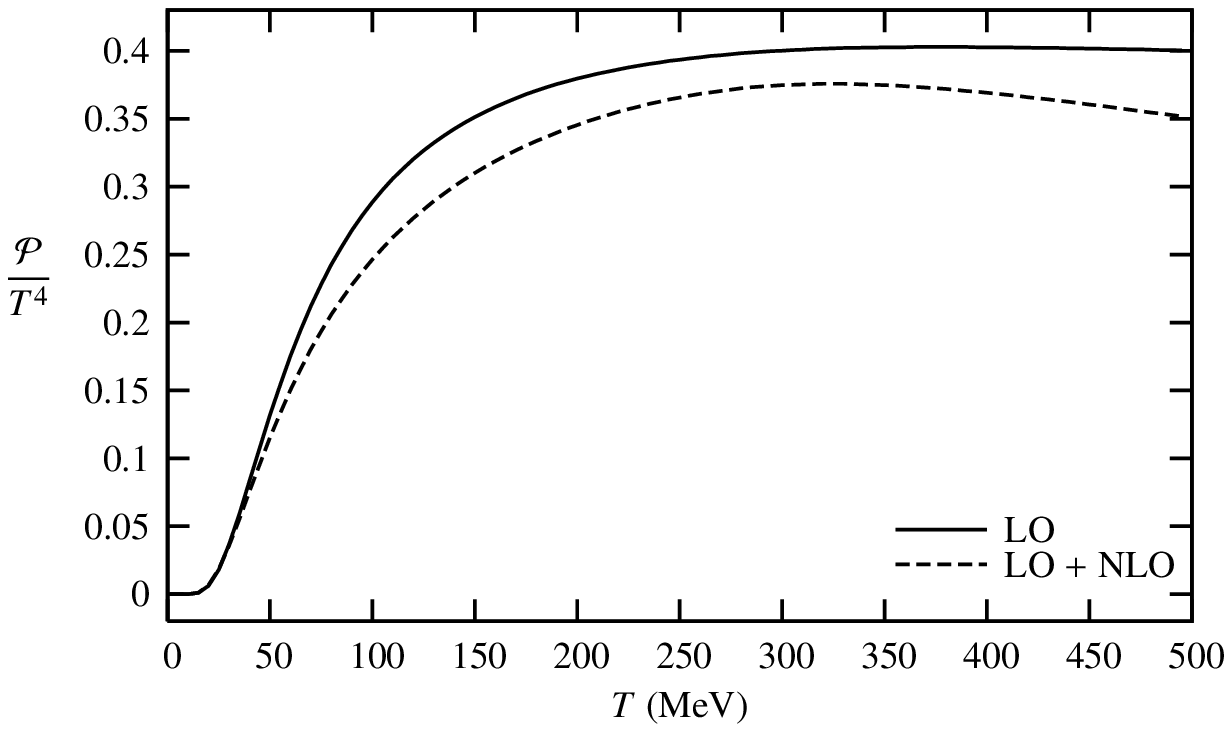}}
\caption{LO and NLO pressure for $N=4$ normalized to $T^4$, as
a function of temperature for $H = \left(104 \;\mathrm{MeV} \right)^3$  
and $\Lambda = 5.0$ GeV.}
\label{fig:presN4h1}
\end{center}
\end{figure}

In Fig$.$~\ref{fig:presN4h1}, the pressure normalized by $T^4$ for
$N=4$ and $H =
\left(104 \;\mathrm{MeV} \right)^3$ to next-to-leading order as
function of $T$ is shown.

In Figs.\ \ref{fig:presN4h0} and \ref{fig:presN4h1} the
cutoff $\Lambda =5.0$ GeV. To make contact with
low-energy QCD, a few comments on this choice are in order. For the
low-energy chiral Lagrangian, the cutoff is usually taken to be $8\pi
f_\pi$ (using our definition of $f_\pi$), which is around 1.2
GeV. However, for the present purpose this value would be at the limit
of applicability, since the critical temperature at which chiral
symmetry is (approximately) restored is only about a factor of 8
smaller and the requirement that $2 \pi T \ll
\Lambda$ should be satisfied. In this way one ensures that one sums over sufficient
Matsubara modes.  Therefore, the cutoff is taken to be considerably
larger to reduce the sensitivity of the results to the
cutoff. However, only for temperatures considerably lower than $T_c$
one expects that the result is of actual relevance for the QCD
pressure.

%---------------------------------------------------------------------
\subsection*{Pressure of the $\mathrm{O}(4)$ nonlinear sigma model} 
In the limit $\lambda_b = \infty$, the Lagrangian of the nonlinear
sigma model is obtained.  In the nonlinear sigma model there are no
counterterms available for logarithmic divergences.  Therefore only
$f_{\pi,b}$ will be renormalized as in Eq.~(\ref{eq:fpirenormnlo})
with $\lambda_b = \infty$. This implies that $F_2$ has a weak cutoff
dependence.

\begin{figure}[t]
\begin{center}
\scalebox{1.0}{\includegraphics{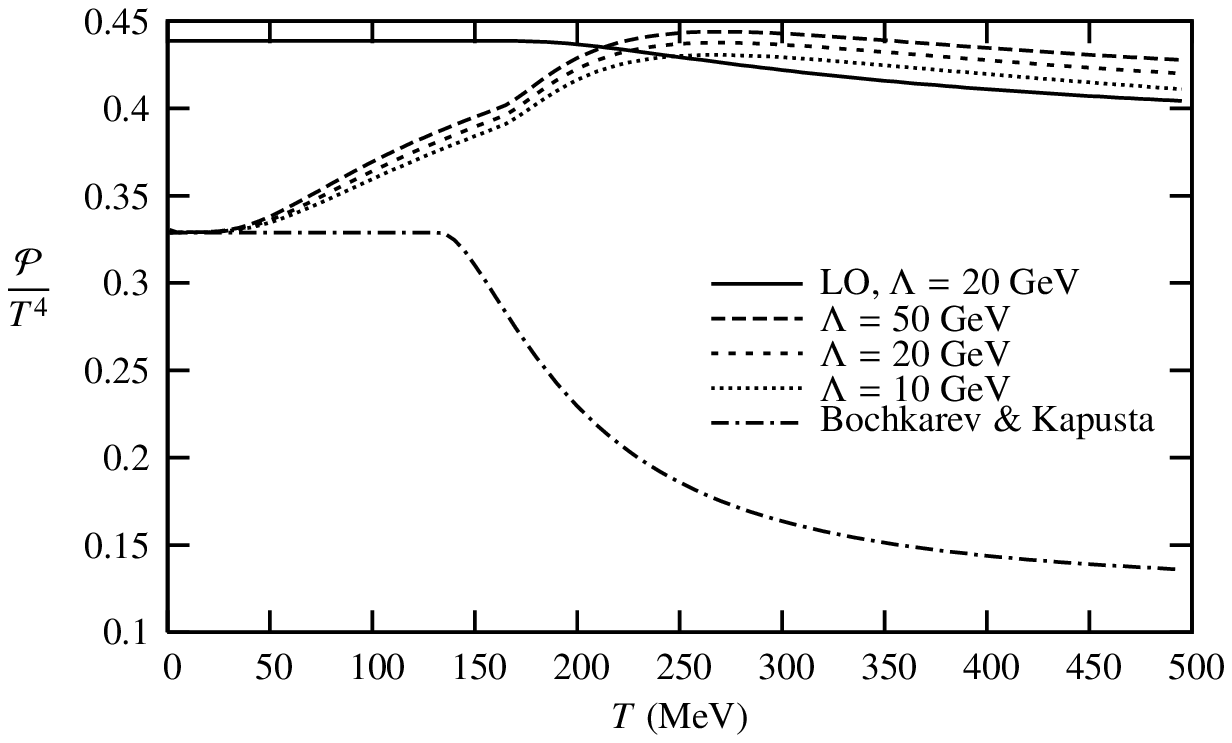}}
\caption{NLO pressure of the nonlinear sigma model for $N=4$ 
normalized to $T^4$, as a function of temperature for different values
of the cutoff $\Lambda$. For comparison the LO pressure and a curve
corresponding to the NLO pressure expression from
\citet{Bochkarev1996} (see their Eq.~(52)) is included.}
\label{fig:presnon}
\end{center}
\end{figure}

In Fig.~\ref{fig:presnon} the pressure of the $\mathrm{O}(4)$
nonlinear sigma model without explicit symmetry breaking ($H=0$),
through next-to-leading order in $1/N$ is shown. The pressure has been
calculated for different values of the cutoff. The LO result for
$\Lambda = 20\;\mathrm{GeV}$ is included.  For comparison, also the
pressure resulting from the ``high-energy approximation'' employed by
\citet{Bochkarev1996} is shown.  A considerable difference
between our results and those of the ``high-energy approximation'' is
observed.

In the approximations made by \citet{Bochkarev1996}, the $T=0$ part of
sum-integrals are omitted such that every $\isumint$ is replaced by
$\isumdiff$. Hence $\Pi(P,m)$ is simply replaced by $\Pi_T(P,m)$.  In the
high-energy approximation the latter is furthermore approximated by
\begin{equation}
\Pi_T(P,m) \approx \frac{T^2 J_1(\beta m)}{32 \pi^2 A^2},
\end{equation}
where $J_1$ and $A^2$ are given by Eqs.~(\ref{J1})
and~(\ref{Akwadraat}), respectively.  The term involving
$\bar{\sigma}^2$ in Eq.~(\ref{Ipm}) is omitted because
\citet{Bochkarev1996} incorrectly assumed that this term is $1/N$
suppressed with respect to the other contributions in Eq.~(\ref{Ipm}).
As a result the pressure in the ``high-energy approximation'' reduces to
\begin{equation}
  \mathcal{P} = \frac{N m^2}{2}\left( f_{\pi}^2 - \bar \sigma^2 \right) 
        -\frac{N}{2} \sumdiff_P \log(P^2 + m^2 )
  -\frac{1}{2} \sumdiff_P \log \frac{P^2}{P^4 + 4 m^2 p_0^2} \;.
\end{equation}
Defining the functions 
\begin{equation}
  K_0^{\pm}(\beta m) = \frac{32}{3 T^4} 
\int_0^{\infty}\mathrm{d}p\;\frac{p^4}{\omega_p}\;
  n(\omega_\pm)\;,
\end{equation}
where $\omega_{\pm}=\sqrt{p^2+m^2}\pm m$, the pressure becomes
\begin{equation}
  \mathcal{P} = \frac{N m^2}{2} \left( f_{\pi}^2 - \bar \sigma^2 \right) 
+ \frac{NT^4 J_0(\beta m)}{64 \pi^2} + \frac{\pi^2 T^4}{90}
   -\frac{T^4}{64 \pi^2} \left[K_0^+(\beta m) + K_0^-(\beta m) \right] \;.
\label{expnot}
\end{equation}
Expanding Eq.~(\ref{expnot}) in powers of $m/T$ and rescaling with
factors of $N$, Eq.~(52) of \citet{Bochkarev1996} is obtained.

For completeness, the gap equations in this approximation are given by
\begin{eqnarray}
16 \pi^2 f_\pi^2 &=& T^2 J_1(\beta m) + 16 \pi^2 \bar \sigma^2 
 - \frac{T^2}{N} \left[K_1^+(\beta m) + K_1^-(\beta m) \right]
\;,\label{BKgap}  \\
m^2 \bar \sigma &=& 0 \;,
\end{eqnarray}
where the functions $K_1^{\pm}$ are 
\begin{equation}
  K_1^{\pm}(\beta m) = \pm \frac{8}{T^2} 
\int_0^{\infty}\mathrm{d}p\;\frac{p^2}{\omega_p} \;\frac{\omega_\pm}{m} \; 
n(\omega_\pm)\;.
\end{equation}

There are several problems with the approach of \citet{Bochkarev1996}.
Firstly, it is incorrect to ignore zero-temperature contributions to
the pressure and it also obscures renormalization issues. Then, as was
argued below Eq.~(\ref{eq:f2}), the arguments for applying the
high-energy approximation are not valid. Furthermore, one cannot
neglect the term proportional to $\bar{\sigma}$ in $I(P,m)$ for
$T<T_c$. Fourthly, as the solutions to the gap equation (\ref{BKgap})
indicate, $m/T$ becomes significantly larger than one for $T>T_c$,
hence the $m/T$ expansion breaks down. If one were to use
Eq.~(\ref{expnot}) instead, one finds that the pressure even becomes
negative above $T \approx 300$ MeV.  Another problem is that for $T <
T_c$ their pressure is equal to that of a massless gas, which is
incorrect since the sigma meson is massive and included at NLO. Hence,
one expects a deviation from the ideal-gas pressure at $T <
T_c$. Finally, at high temperatures the next-to-leading order pressure
should become approximately equal to the leading order pressure
because chiral symmetry will be restored. This is not the case for the
pressure calculated by
\citet{Bochkarev1996} as can be seen from Fig.~\ref{fig:presnon}.

\citet{Jain1993} has calculated the thermodynamic potential
to NLO in the $\mathrm{O}(N)$ linear sigma model using a high-temperature
expansion. This approximation breaks down at low
temperatures. Therefore it is not useful to compare those results to
the results obtained in this chapter since they are strictly speaking 
only valid at low temperatures (much lower than the cutoff $\Lambda$).

%=====================================================================
\section{Choice of parameters}\label{sec:parameters}

The plots in the preceding sections are made using particular choices
of the parameters, namely, $\lambda (\mu = 100\;\mathrm{MeV}) = 30$,
$f_\pi = 47\;\mathrm{MeV}$ (note that this $f_\pi$ differs from the
more conventional definition by a factor of $1/2$) and if there is
explicit symmetry breaking, $H = \left(104\;\mathrm{MeV}
\right)^3$. In this section these choices are motivated. For
simplicity partly leading-order calculations are used for fixing the
parameters.

The values for $f_\pi$ and $m_\pi$ are chosen to be roughly equal to
their measured values: $f_\pi = 47\;\mathrm{MeV}$ and $m_{\pi}= 138$
MeV (the average of the measured masses of the $\pi^0, \pi^+$ and
$\pi^-$).  These values are used for determining the parameter $H$ as
follows.  Given a choice of $\lambda$ at some scale $\mu$ the LO
renormalized gap equations (\ref{eq:logapeqm}) and
(\ref{eq:logapeqsigma}) are solved for $\bar
\sigma$ and $m^2$, such that $m^2 = m_\pi^2$ (which is the correct
identification at leading order, see
Eq.~(\ref{inversesigmaprop})). For the choice of $\lambda(\mu =
100\;\mathrm{MeV}) = 30$, this results in $H = \left(104
\;\mathrm{MeV} \right)^3$.

The choice of $\lambda$ is motivated by considerations on the maximal
value of the cutoff and the sigma mass.
As explained below, the sigma mass turns out to be maximal if $\lambda(\mu =
100\;\mathrm{MeV}) = 80$. The problem with this choice of $\lambda$ is
that in that case the maximal value of the cutoff is 720 MeV. This is
very low and allows us only to do calculations up to around $T =
50\;\mathrm{MeV}$. Therefore, a lower value is chosen: $\lambda(\mu =
100\;\mathrm{MeV}) = 30$. Using that parameter choice
$\Lambda_{\mathrm{max}} = 19\;\mathrm{GeV}$ and the sigma mass is
equal to 256 MeV and 350 MeV in the case of $H = 0$ and $H =
(104\;\mathrm{MeV})^3$ respectively.

To obtain the mass of the sigma field, one has to find the poles of
the propagator in Minkowski space.  The physical mass
$m_{\mathrm{ph}}$ is often defined by the solution to the equation
\begin{equation}
  -m_{\mathrm{ph}}^2 + m^2 + \mathrm{Re}{\Sigma(p_0 = i m_\mathrm{ph}
   + \epsilon,\,p = 0,\, m)} = 0\;,
\end{equation}
where $\Sigma$ is the self-energy.  Using the expression for the
inverse $\sigma$ propagator in the $1/N$ expansion, Eq.\
(\ref{inversesigmaprop}), and choosing $\mu = m_\sigma$ (because all
results are independent of $\mu$, one can choose $\mu$ as one likes,
the choice $\mu = m_\sigma$ is just convenient), one finds that at
$T=0$ and for $H=0$
\begin{equation} 
  m^2_\sigma = \frac{32
\pi^2 f_\pi^2}{1 + \frac{32 \pi^2}{\lambda(m_\sigma)} + \frac{\pi^2}{1
+ 32\pi^2 / \lambda(m_\sigma)}} \;.  
 \label{eq:sigmamass}
\end{equation} 
Equation (\ref{eq:sigmamass}) can be maximized with respect to
$\lambda(m_\sigma)$.  This implies that $m_\sigma \leq \sqrt{16 \pi}
f_\pi \approx 333$ MeV, which is lower than the averaged measured
value of 400 - 800 MeV. An analogous bound was found using the
large-$N$ expansion by \citet{Einhorn1984} on the mass of the Higgs
boson. \citet{Patkos2002} show a figure in which the bound on the
sigma mass can be seen indirectly, although they do not comment on
this rather interesting fact.

For $H \neq 0$ a similar bound applies. In that case the maximal value
of the sigma mass can be found by maximizing
\begin{equation}
m^2_\sigma = m^2_{\pi} 
+ \mathrm{Re} \frac
{32 \pi^2 f_\pi^2
+ 2 m_\pi^2 \log \left( \frac{\Lambda^2}{m_\pi^2} \right)
+ \frac{64 \pi^2 m_\pi^2}{\lambda_b}
}{
  \log \left( \frac{\Lambda^2}{m_{\pi}^2} \right)
  + 1 + \sqrt{\frac{m^2_\sigma - 4 m_\pi^2}{m^2_\sigma}}
   \left[ \log \left( \frac{m_\sigma - \sqrt{m_\sigma^2 - 4 m_\pi^2}}
 {m_\sigma + \sqrt{m_\sigma^2 - 4 m_\pi^2}} \right) + i \pi \right]
  + \frac{32 \pi^2}{\lambda_b}
  }
\;,
\end{equation}
with respect to $\lambda_b$.
Finding the maximum of the previous equation requires solving the following
equation for $m_\sigma$ 
\begin{equation}
  m_\sigma^2 = \left[2 + \sqrt{1+ A^2(m_\sigma^2 /
  m_\pi^2)}\right] m_\pi^2 \;,
\end{equation}
where
\begin{equation}
  A(x) = \left(\frac{16\pi f_\pi^2}{m_\pi^2} - \frac{1}{\pi} \right)
  \frac{1}{\sqrt{1- 4 / x}} 
  - \frac{1}{\pi} \log \left( \frac{1 - \sqrt{1 - 4/x}}{1+ \sqrt{1 - 4 /x}}
\right) \;.
\end{equation}
By solving this equation it turns out that the maximal value of
$m_\sigma$ is equal to 433 MeV. This is on the low side of the
experimental measured values between 400 and 800 MeV.  The reason that
a rather low bound (which might be unrealistic) on the sigma meson
mass is found could be because possible essential three-flavor physics
was missed out in the calculation.

The bounds in the previous section were obtained using non-perturbative
large-$N$ expansion. So these bounds are valid for any value of
$\lambda$. In perturbation theory in small $\lambda$, the mass of the
sigma field is $m^2_\sigma = m^2 + \lambda_b \bar \sigma^2$.  This
result can be found by expanding the expression for the inverse
$\sigma$ propagator, Eq.\ (\ref{inversesigmaprop}). One might think
that this indicates that the sigma mass can grow to any value by
increasing $\lambda$. However, by increasing $\lambda$, the sigma
field will be coupled stronger to the pions. Since the sigma field is
not stable, but can decay into two pions, the decay width is increased
when $\lambda$ becomes bigger. As a result this decay shifts the mass
of the sigma meson to lower values.

\section{Summary and Conclusions}
In this chapter, the thermodynamics of the $\mathrm{O}(N)$ linear and
nonlinear sigma models to next-to-leading order in the $1/N$ expansion
was studied.

At next-to-leading order it was shown that one can renormalize the
effective potential in a temperature-independent manner only at the
minimum of the effective potential. By renormalizing the
next-to-leading order effective potential in the minimum the beta
function for $\lambda$ to next-to-leading order was found. This beta
function is consistent with the perturbative result.

The pressure for the linear and nonlinear sigma model to
next-to-leading order as a function of temperature was calculated
numerically. The results show that for the calculation of the pressure
$1/N$ is a good expansion, even if $N=4$. With a relatively realistic
choice of the parameters a prediction for the pressure of QCD for
temperatures below $T_c$ was made. The results for the
pressure disagree significantly with the calculations of those by
\citet{Bochkarev1996}. This is due to the fact that in the
calculations performed in this chapter the zero-temperature contributions
are not neglected and that the next-to-leading order contribution is
treated without resorting to a ``high-energy approximation''.

It was also found that in the linear sigma model the sigma mass has an
upper bound at zero temperature. This bound depends only on the
parameters $f_\pi$ and $m_\pi$. For a realistic choice of these
parameters, this implies that the mass of the sigma meson is smaller
than 433 MeV. This does not necessarily have consequences for the real
sigma meson, since the full three-flavor physics was not taken into
account. It would be interesting to investigate the temperature
dependence of this bound.

An interesting extension of these calculations would be the the
calculation of spectral functions at finite temperature. It would be
interesting to see how the bound on the sigma meson mass depends on
temperature. The methods developed in this chapter could also be
useful for the study of more complicated models incorporating
additional features of low-energy QCD.

 %on
\chapter{The phase diagram of the NJL model}

In order to obtain information about the structure of the QCD phase
diagram, one can investigate the phase diagram of low-energy effective
theories for QCD.  In this chapter phase diagrams of such an effective
theory, namely the Nambu--Jona-Lasinio (NJL) model, will be
presented. These phase diagrams are obtained numerically as a function
of temperature and of the up, down and strange quark chemical
potentials. Phases with broken chiral symmetry, color superconducting
phases and phases in which the pseudoscalar mesons condense can be
found in the calculated diagrams. It is shown numerically that color
superconducting and pseudoscalar condensed phases are separated by a
first order phase transition. This chapter is based on: \textit{Color
superconductivity versus pseudoscalar condensation in a three-flavor
NJL model}, H.J.~Warringa, D.~Boer and J.O.~Andersen, Phys.~Rev.~{\bf
D72} 014015, (2005).

\section{Introduction}

The NJL model \citep{Nambu1961}, the instanton liquid model
\citep{Shuryak1982} and random matrix models (see for example 
\citet{Verbaarschot2000})
 are examples of effective models which can be used to study the QCD
phase diagram at finite temperature and densities. Despite their
shortcomings, it is expected that these models do describe the
qualitative features of the QCD phase diagram in regions not
accessible to perturbative or lattice QCD. For example, 
a critical point at finite baryon chemical potential has been
predicted using effective models by
\citet{Halasz1998}. Furthermore, it is expected that due to the
existence of an attractive interaction, quarks can form
Cooper pairs. This is just as electrons do in ordinary
superconductivity due to the attractive phonon interaction between
electrons. As a result, at high baryon chemical potentials and low
temperatures, quark matter can for instance be in a two-flavor
color-superconducting phase (2SC) \citep{Bailin1984, Alford1998,
Rapp1998} in which two flavors pair, or in a color-flavor locked (CFL)
phase \citep{Alford1999, Alford1999q} in which three flavors pair. In
these color-superconducting phases gaps in the quasi-particle
excitation spectrum on the order of 100 MeV arise. These gaps are
large compared to the mass gaps of free quarks (the up and the down
current quark masses are about 5 MeV), indicating that the physics in
the color superconducting phase is very different from that in the
deconfined phase.

Many results on high density phase diagrams presented in the
literature have been obtained for equal up, down and strange quark
chemical potentials. However, this may not be directly relevant for
heavy-ion collisions or compact stars. For example to model compact
stars, one has to enforce electric and color neutrality and weak
equilibrium.  For this reason, different flavor and color chemical
potentials have been introduced in the NJL model~\citep{Alford2002,
Steiner2002}.  This gives rise to a more complicated phase diagram in
which one also finds new 2SC-like \citep{Neumann2003} and gapless
superconducting phases \citep{Alford1999b, Shovkovy2003, Alford2004a,
Ruster2004, Abuki2004, Ruster2005, Blaschke2005}. It is still an open
issue whether these phases are really stable.
Similarly, in heavy-ion collisions, a difference between the quark
chemical potentials arises if the number densities of the different
quark flavors are not the same. This difference can cause interesting
observable effects such as two critical endpoints~\citep{Klein2003,
Toublan2003}. However, instanton induced interactions tend to suppress
this effect as was shown by \citet{Frank2003}.

In addition to a more complicated structure of the superconducting
phases, different chemical potentials can also trigger pseudoscalar
condensation \citep{Son2001, Kogut2001}. This has been confirmed on
the lattice at zero baryon chemical potential ($\mu_B = \mu_u + \mu_d
+ \mu_s$) but finite isospin chemical potential ($\mu_I = \mu_u -
\mu_d$) by
\citet{Kogut2002}. In a pseudoscalar condensed phase, depending on the
flavors involved, the charged pion, neutral or charged kaon field
acquires a vacuum expectation value.  As a result parity
is broken spontaneously. 
Pseudoscalar condensation in the
two-flavor NJL model has been studied by \citet{Toublan2003} and
\citet{Barducci2004}, as a function of the different chemical
potentials at zero and finite temperature.  An extension to three
flavors was carried out by \citet{Barducci2005} as well.

The phase diagram of the three-flavor NJL model as a function of the
different chemical potentials including {\it both} pseudoscalar
condensation in the quark-antiquark channel and color
superconductivity had not yet been addressed. This is the subject of
this chapter. At zero temperature, pseudoscalar condensation is
possible if $\vert \mu_u - \mu_d \vert > m_\pi$, as was shown by
\citet{Son2001}, or if $\vert \mu_{u,d} - \mu_s \vert > m_K$
\citep{Kogut2001}. On the other hand, color superconducting phases
occur if the chemical potentials are large and approximately
equal. Therefore, one can imagine scenarios where for example $\mu_u
\approx \mu_d$ (the Fermi surfaces of the $u$ and $d$ quark should be
sufficiently close for Cooper pairing to occur) and $\mu_u \approx -
\mu_s$ (the Fermi surfaces of the $u$ and $\bar{s}$ should be
sufficiently close for kaon condensation to occur), with $\vert
\mu_{u,d} - \mu_s \vert > m_K$.  In such a case a 2SC phase is
competing against a phase in which kaons condense. Hence the phase
diagrams presented in this chapter are not a superposition of the
phase diagrams with only pseudoscalar condensation and with only color
superconductivity, which were calculated before in the literature. From
the calculations it follows that a coexistence phase of pseudoscalar
condensation and color superconductivity does not occur for the
parameters chosen and that these phases are separated by a first-order
transition. However, it is not excluded that other choices of
parameters may lead to such a coexistence phase, just as coexistence
of color superconductivity and chiral symmetry breaking may occur in
the NJL model for specific ranges of parameters as was found by
\cite{Blaschke2003}. Here a coexistence phase is to be understood as a
phase in which two condensates are nonzero simultaneously to
be distinguished from a mixed phase in which phases coexist.

In this chapter pseudoscalar condensation in the quark-antiquark
channel is studied. The pseudoscalar diquark interaction was not taken
into account.  This interaction is suppressed relative to the scalar
diquark interaction due to instantons~\citep{Rapp1998}. However, in
absence of instanton interactions, if one neutralizes the bulk matter
with respect to color and electric charges it is possible to have
pseudoscalar diquark condensation with rather large gaps as was shown
by \cite{Buballa2005b}.  According to \cite{Buballa2005b} pseudoscalar
diquark condensation in the NJL model is similar to pseudoscalar
condensation in the CFL phase studied with effective chiral models
studied by \citet{Casalbuoni1999}, \citet{Son2000},
\citet{Casalbuoni2000}, \citet{Schafer2000}, \citet{Bedaque2002a},
\citet{Kaplan2002} and \citet{Forbes2004}.

Overall charge neutrality conditions are not applied in this chapter,
this allows us to compare to the previous studies of \citet{Toublan2003},
\citet{Barducci2004} and \citet{Barducci2005}.
Imposing neutrality conditions is necessary to describe situations in
nature where the quark matter is realized at high densities such as in
the core of neutron stars. Neutrality conditions would qualitatively
affect the phase structure, leading for example to the observation
that in a macroscopic volume of electric neutral quark matter in
equilibrium with the weak interactions the 2SC phase is energetically
disfavored~\citep{Alford2002, Steiner2002}.

To obtain the phase diagrams an auxiliary field effective potential
will be minimized with respect to the vacuum expectation values. This
effective potential will be calculated in leading order in $1/N_c$
where $N_c$ is the number of colors. For simplicity next-to-leading
order $1/N_c$ corrections are not taken into account. However,
it is possible to calculate these corrections as was shown by
\citet{Hufner1994}. 
The leading
order $1/N_c$ expansion amounts to the so-called
mean field approximation.

This chapter is organized as follows. In Sec.~2, it will be argued how
the NJL model qualitatively arises from the QCD Lagrangian
density. The NJL model itself and the choice of parameters will be
treated in Sec.~3. In Sec.~4, the calculations and some of its
technical aspects are discussed. In Sec.~5, the phase diagrams are
presented and a summary and conclusions are given in Sec.~6.

%=====================================================================
\section{From QCD to the NJL model}
The NJL model can serve as a low-energy effective theory of QCD. In
this section it will be discussed qualitatively which approximations
one has to make in order to obtain the NJL model from QCD. A related
discussion about the connection between QCD and the NJL model can be
found in \citet{Bijnens1992} and \citet{Bijnens1995}.

The QCD Lagrangian density in Minkowki space, 
Eq.~(\ref{eq:QCDLagrangian}), can be written
as $\mathcal{L}_\mathrm{QCD} = \mathcal{L}_{\mathrm{f}} +
\mathcal{L}_{\mathrm{YM}} + g J_\mu^a A^\mu_a$, where the kinetic term
of the fermions is given by
\begin{equation}
  \mathcal{L}_{\mathrm{f}} 
  = \bar \psi \left( i \gamma^\mu \partial_\mu - M_0 + \mu \gamma_0
 \right) \psi \;, 
  \label{eq:lagrnjlfermkin}
\end{equation}
and the terms involving the gluon fields by
\begin{equation}
 \mathcal{L}_\mathrm{YM} = -\frac{1}{4} F^{\mu \nu}_a F^a_{\mu \nu} \;.
\end{equation}
The interaction between the gluons and the quarks is written in
terms of a fermion color current which is equal to $J_\mu^a = \bar
\psi T^a \gamma_\mu \psi$, where $T^a$ is a generator of $\mathrm{SU}(3)$. 
The partition function of QCD can be factored in the following way
\begin{equation}
 Z = \int  \mathcal{D}\bar \psi \mathcal{D} \psi \mathcal{D}A_\mu^a
  \exp \left(i S_\mathrm{QCD} \right) =
 \int \mathcal{D} \bar \psi \mathcal{D} \psi \exp \left(i S_\mathrm{f} +
 i S_\mathrm{int}[J_\mu^a]
\right)
 \;,
\end{equation}
where the quark interaction action $S_\mathrm{int}[J]$ is given by
\begin{equation}
  S_\mathrm{int}[J] = -i \log \int \mathcal{D}A_\mu^a 
 \exp \left(i S_\mathrm{YM} + i g \int_x 
 J_\mu^a A^\mu_a
 \right) \;.
\end{equation}
By adding a gauge fixing term and the
corresponding ghost fields one can perform
the integration over the gauge fields in such a way that the quark
interaction action can be written as
\begin{equation}
 S_{\mathrm{int}}[J] = 
S_\mathrm{int}[J=0] +
\frac{i g^2}{2} \int_x \int_y J_\mu^a(x) 
  D^{\mu \nu}_{ab}(x-y) J_\nu^b(y) 
+ \ldots \;.
\label{eq:sint}
\end{equation}
where for simplicity the 6-point and higher fermion interaction terms
will be neglected from now on (if $g$ is small this is certainly
allowed). In a more realistic low-energy effective theory the effect
of the 6-point and higher fermion interactions should be taken into
account.  In Eq.~(\ref{eq:sint}) $D^{\mu \nu}_{ab}(x-y)$ is the exact
gluon propagator of pure glue QCD in a specific gauge
\begin{equation}
  D^{\mu \nu}_{ab}(x-y) = \left< A^{\mu}_a(x) A^{\nu}_b(y) \right> \;.
\end{equation}
In the limit of high momenta the pure gluon propagator becomes (in the
Feynman ($\xi = 1$) gauge)
\begin{equation}
  D^{\mu \nu}_{ab}(p) = \frac{- i \delta_{ab} g^{\mu \nu} }{p^2} \;.
 \label{eq:gluonprop}
\end{equation}
At low momenta non-perturbative effects will modify this propagator.
Since the exact gluon propagator cannot be calculated analytically in
the low-momentum regime, one can try to make an ansatz for the exact
gluon propagator. Such an ansatz should reduce to
Eq.~(\ref{eq:gluonprop}) in the high-momentum limit.  One of the
non-perturbative effects is that gluons can form massive bound states
called glueballs (see for example \citet{Morningstar1999}). The effect
of these glueballs should somehow be reflected in the gluon
propagator. As an ansatz one could take
\begin{equation}
  D^{\mu \nu}_{ab}(p) =  \frac{- i \delta_{ab} g^{\mu \nu} }{p^2 - M^2} \;,
\end{equation}
where $M^2$ mimics the effect of a massive glue-ball. This mass should
be on the order of the scale at which non-perturbative physics starts
to play a role, which is roughly $1\; \mathrm{GeV}^2$.  Comparing this
ansatz to lattice calculations of the exact propagator for pure gluon
QCD in the Landau gauge ($\xi = 0$) \citep{Bernard1993, Bonnet2000,
Silva2004} shows it to be fairly realistic.

Since the objective is to obtain a low-energy effective theory for
QCD, the ansatz can at low-momenta be crudely approximated by
\begin{equation}
  D^{\mu \nu}_{ab}(p) = \left \{ \begin{array}{ll} 
 \frac{i}{M^2} \delta_{ab} g^{\mu \nu} & p^2 < \Lambda^2 \;,\\
  0 & p^2 > \Lambda^2 \;,
 \end{array}
 \right .
  \label{eq:gluonpropappr}
\end{equation}
where $\Lambda$ is an ultraviolet momentum cutoff. This ansatz,
Eq.~(\ref{eq:gluonpropappr}) will eventually lead to the NJL
model. The term $1/M^2$ acts as a sort of coupling constant and should
be in the order of $1\;\mathrm{GeV}^{-2}$. Because one has to fix a
gauge to obtain the gluon propagator, one should realize that after a
gauge transformation in principle the ansatz for the propagator should
transform as well. Since this is usually not done in the NJL model,
gauge invariance is lost, but a global $\mathrm{SU}(3)_c$ symmetry
is retained.

To obtain a more realistic low-energy effective theory for QCD one
should use a more general ansatz for the exact gluon propagator. Such
a better ansatz could be found by fitting it to lattice calculations
of the gluon propagator which were performed by \citet{Bernard1993},
\citet{Bonnet2000} and \citet{Silva2004}.  However, in most cases this
will lead to non-local interactions between the quarks which will be
more difficult to solve analytically in the end, see for example
\citet{Hashimoto2005}. In a first approximation the local interaction
could be sufficient. 

Now at finite temperatures and densities, the term
$S_{\mathrm{int}}[J=0]$ is the contribution of the pure gluon part of
QCD to the total pressure of QCD. This term does not depend on
chemical potentials but only on temperature. If one investigates
phase diagrams as a function of chemical potential at a fixed
temperature this term is just a constant added to the effective
potential. Hence to first approximation the effects of
$S_{\mathrm{int}}[J=0]$ can be neglected.  At finite temperature, the
exact gluon propagator depends on temperature, this could be taken
into account in a more realistic ansatz.

As a result the QCD partition function can be approximated by
\begin{equation}
  Z \approx C \int \mathcal{D}\bar \psi \mathcal{D}\psi \exp
 \left(i\int_x \mathcal{L}_{\mathrm{NJL}} \right)
\;,
\end{equation}
where the Lagrangian density of the NJL model is given by
\begin{equation}
 \mathcal{L}_{\mathrm{NJL}} = \mathcal{L}_{\mathrm{f}} 
  - \tilde g J_\mu^a J^\mu_a \;,
\end{equation}
here $\tilde g = g^2/2M^2$ is a coupling constant. The term $C$ contains the
gluonic energy density and does not depend on chemical potentials but
only on temperature. 

The qualitative arguments to obtain the NJL model will not play a role
in the following sections.  Below the NJL model will be taken as a
starting point. One must therefore be careful to extend the
conclusions obtained below to QCD without cautionary remarks,
especially at chemical potentials and temperatures close to the
cutoff. Nevertheless, there are certainly reasons to believe that the
qualitative features of the NJL model also extend to QCD. For instance
the meson spectrum is well described by the NJL model, see for example
\citet{Klevansky1992}. In the following section the NJL model will be
discussed in more detail.

%=====================================================================
\section{The NJL model}
The NJL model \citep{Nambu1961} was originally intended to describe
the interactions between nucleons. Nowadays, however, it is mainly used
as an effective model for QCD. In QCD the interaction between the
quarks is caused by gluon exchange. In the NJL model this gluon
exchange is approximated by a four-point color current-current
interaction. This is like the four-Fermi theory of the weak
interactions in which the low-energy exchange of a massive vector
boson is modeled by a four-point interaction. At energies much lower
than the mass of the vector bosons, the weak interactions are almost
independent of momentum. Therefore in that case the four-Fermi theory
works well. In the previous chapter it was argued that by making an
ansatz for the gluon propagator in the non-perturbative regime a
fermion four-point interaction arises too.  

There are many NJL-like models. In this chapter a three-flavor NJL
model with three colors is discussed.  The different NJL models are
based on a Lagrangian density which contains a kinetic term and a
four-point color current-current interaction
\begin{equation}
 \mathcal{L} = \bar \psi \left(i \gamma^\mu \partial_\mu - M_0
  + \mu \gamma_0 \right)\psi 
 - \tilde g \left( \bar \psi \gamma^\mu T_a \psi \right)^2 \;.
\end{equation}
The color, flavor and Dirac indices of the fermion fields $\psi$ are
suppressed in this equation for notational simplicity. The diagonal
mass matrix $M_0$ contains the bare quark masses $m_{0u}$, $m_{0d}$
and $m_{0s}$. The matrix $\mu$ is also diagonal and contains the quark
chemical potentials $\mu_u$, $\mu_d$ and $\mu_s$.  
The
matrices $T_a$ are the $8$ generators of $\mathrm{SU}(3)$ and act in
color space. They are normalized as $\mathrm{Tr}\, T_a T_b = 2
\delta_{a b}$. 

By applying several Fierz transformations to the current-current
interaction (see for example \citet{Buballa2005a}) and including only
terms which give rise to attractive $qq$ and $\bar{q}q$ channels, one
obtains the following Lagrangian density
\begin{equation}
  \mathcal{L} = \bar \psi \left(i \gamma^\mu \partial_\mu - M_0
  + \mu \gamma_0 \right)\psi 
  + \mathcal{L}_{\bar q q} + \mathcal{L}^{\mathrm{s}}_{qq} 
  + \mathcal{L}^{\mathrm{ps}}_{qq}
 \label{eq:lagrnjl} \;,
\end{equation}
where the quark-antiquark, the scalar diquark and the pseudoscalar
diquark interaction term are respectively given by
\begin{eqnarray}
 \mathcal{L}_{\bar q q} 
  &=& G \left[ \left(\bar \psi \lambda_a \psi \right)^2 + 
 \left(\bar \psi \lambda_a i \gamma_5 \psi \right)^2 \right] \;, 
\label{eq:lagrnjlqbarq}
\\
 \mathcal{L}^{\mathrm{s}}_{qq} &=&  
  \frac{3}{4} G \left(\bar \psi t_{A} \lambda_{B} C i \gamma_5 \bar \psi^T \right)
\left(\psi^T  t_{A} \lambda_{B} C i \gamma_5 \psi \right) \;, 
\label{eq:lagrnjldiq}
\\
\mathcal{L}^{\mathrm{ps}}_{qq} &=&  
  \frac{3}{4} G \left(\bar \psi t_{A} \lambda_{B} C \bar \psi^T \right)
\left(\psi^T  t_{A} \lambda_{B} C \psi \right)
 \;,
\end{eqnarray}
where $A, B \in \{2, 5, 7\}$, since only the interaction in the
antisymmetric color and flavor triplet channel is attractive.  The
matrices $\lambda_a$ are the $9$ generators of $\mathrm{U}(3)$ and act
in flavor space. They are normalized as $\mathrm{Tr}\, \lambda_a
\lambda_b = 2 \delta_{a b}$.  The matrices $t_a$ are the generators of
$\mathrm{U}(3)$ and act in color space.  Their normalization is
$\mathrm{Tr}\, t_a t_b = 2 \delta_{a b}$.  To remind the reader, the
antisymmetric flavor matrices $\lambda_2$, $\lambda_5$ and $\lambda_7$
couple up to down, up to strange and down to strange quarks,
respectively.  The charge conjugate of a field $\psi$ is denoted by
$\psi_c = C \bar \psi^T$ where $C = i \gamma_0 \gamma_2$.  The
coupling strength $3 G /4$ of the diquark interaction is fixed by the
Fierz transformation (see \citet{Buballa2005a}) and will be used in
this chapter. However, some authors discuss the NJL model with a
different diquark coupling constant (see for example
\citet{Ruster2005} for a comparison).

The NJL model has a symmetry structure similar to QCD, except for the
color symmetry. The NJL model is only invariant under global SU(3)
color transformations since it does not contain gauge fields.  In
absence of quark masses and chemical potentials, the Lagrangian
density of the NJL model has a global
$\mathrm{SU(3)}_c\times\mathrm{U(3)}_{L}\times\mathrm{U(3)}_{R}$
symmetry. Due to the non-vanishing quark masses, the symmetry is like
in QCD broken down to $\mathrm{SU(3)}_c \times
\mathrm{U(3)}_{V}$. Since the masses of the quark and the chemical 
potentials will be different, the symmetry
of the Lagrangian density is further reduced to
$\mathrm{SU(3)}_c\times\mathrm{U(1)}_B\times\mathrm{U(1)}_I
\times\mathrm{U(1)}_Y $, where $B$, $I$, $Y$ stand respectively for
baryon, isospin (the $z$-component) and hypercharge number.  An additional
$\mathrm{U(1)}_{A}$ symmetry breaking term is necessary to explain the
large mass of the $\eta$ particle. As was shown by
\citet{tHooft1976} this is due to the non-trivial topological structure
of the vacuum, the instantons. The effect of these instantons can be
modeled by the following effective six-point interaction 
\begin{equation}
  \mathcal{L}_{\mathrm{inst}} = K \left [ \mathrm{det}\, \bar \psi (1
 - \gamma_5 ) \psi + \mathrm{det}\, \bar \psi ( 1 + \gamma_5) \psi \right] \;,
\label{eq:njlthooftint}
\end{equation}
where the determinant acts only in flavor space. In the two-flavor
case one can argue by applying a Fierz transformation on this
interaction, that the scalar diquark interaction is
more attractive than the pseudoscalar diquark interaction
\citep{Rapp1998}. Therefore condensation in the scalar diquark channel
is favored over pseudoscalar diquark condensation. Likewise this is
also expected to hold in the three-flavor case. Therefore from now on
the pseudoscalar diquark interaction will be neglected.

Unfortunately the six-point instanton-induced interaction gives a very
complicated contribution to the effective potential when allowing for
pseudoscalar quark-antiquark condensates in the mean field
approximation or beyond
\citep{Osipov2005}.  Therefore in this chapter this instanton-induced interaction is
left out but it would be interesting to take this term into account in
a further study.

The results that will be presented in this chapter are obtained with the
following choice of parameters which were also used by
\citet{Buballa2005a}
\begin{equation}
\begin{array}{ccc}
 m_{0u} = 5.5\;\mathrm{MeV}\;, & m_{0d} = 5.5\;\mathrm{MeV}\;, & m_{0s} =
112\;\mathrm{MeV} \;,
 \\
 G = 2.319 / \Lambda^2\;,& K = 0 \;, & \Lambda = 602.3\;\mathrm{MeV} \;.
\end{array}
\end{equation}
These are the precise values used in the calculations, but clearly
not all digits are significant implying that a small change
of parameters will not have a big influence on the results.
This choice of parameters gives rise to constituent quark masses
$M_u=M_d=368$ MeV and $M_s=550$ MeV. In that case the pion and kaon
have a mass of respectively 138 MeV and 450 MeV.

%=====================================================================
\section{Effective potential}
To obtain a phase diagram one has to investigate the behavior of order
parameters. A good order parameter vanishes in one phase and is
non-vanishing in another and related to symmetry breaking and
restoration. In massless QCD good order parameters indicating chiral
symmetry breaking are the quark-antiquark condensates $\langle \bar u
u\rangle, \langle \bar d d \rangle$ and $\langle \bar s s
\rangle$. Since quarks in reality have a mass, chiral symmetry is
already explicitly broken in the QCD and NJL Lagrangian. Even if one
uses massless quarks, different quark chemical potentials also break
the chiral symmetry. Therefore the quark-antiquark condensates will be
nonzero in every phase.  So the quark-antiquark condensates are no
good order parameters in a strict sense, but since their values in the
broken phase are much larger than in the approximately restored phase,
one can use these condensates to distinguish different phases. So the
quark-antiquark condensates are ``approximate order parameters''.  The
quark-antiquark condensates are not the only possible order
parameters, others follow naturally from the discussion below in which
it is explained how to calculate the condensates.

As is discussed in Chapter 3, the quantity from which one can derive
thermodynamical quantities is the effective potential. One also can
find the values of certain condensates from this effective potential
and hence determine the phase diagram. To calculate the effective potential of
the NJL model it appears to be useful to introduce 18 auxiliary real scalar fields
$\alpha_a$ and $\beta_a$ and 9 complex scalar fields $\Delta_{AB}$ in
the following way
\begin{equation}
 \mathcal{L} \rightarrow \mathcal{L} - \frac{\alpha_a^2 + \beta_a^2}{4G} -
 \frac{\Delta_{AB}^* \Delta_{AB}}{3G} \;.
\end{equation}
By shifting these auxiliary fields as follows
\begin{align}
  &\alpha_a \rightarrow \alpha_a + 2 G \bar \psi \lambda_a \psi \;,
  &\Delta_{AB} \rightarrow \Delta_{AB} - \frac{3}{2} G \psi^T t_A
 \lambda_B C \gamma_5 \psi \;, \\
  &\beta_a \rightarrow \beta_a + 2 G \bar \psi \lambda_a i \gamma_5 
 \psi \;, 
&
  \Delta^*_{AB} \rightarrow \Delta^*_{AB} + \frac{3}{2}  G \bar \psi t_A
 \lambda_B C \gamma_5 \bar \psi^T \;, 
\end{align}
the Lagrangian density Eqs.~(\ref{eq:lagrnjlfermkin}), (\ref{eq:lagrnjlqbarq}) and
(\ref{eq:lagrnjldiq}) becomes quadratic in the fermion fields
\begin{equation}
\begin{split}
  \mathcal{L}& = \bar \psi \left(i \gamma^\mu \partial_\mu - M_0 -
  \alpha_a \lambda_a - i \gamma_5 \beta_a \lambda_a
  + \mu \gamma_0 \right)\psi - \frac{\alpha_a^2 + \beta_a^2}{4 G}\\ 
  & \quad - \tfrac{1}{2} \Delta_{AB} \bar \psi t_A
 \lambda_B C \gamma_5 \bar \psi^T 
  + \tfrac{1}{2} \Delta^*_{AB} 
 \psi^T t_A \lambda_B C \gamma_5 \psi
  - \frac{\Delta_{AB} \Delta_{AB}^*}{3 G} \;.
\end{split}
\label{eq:njlferm}
\end{equation}
which allows integration over the quark fields. These quark fields $\psi$
have 3 (color) $\times\, 3$ (flavor) $\times\, 4$ (Dirac) $= 36$ components.
As is explained in appendix \ref{nambugorkov} the integration is most easily done
by introducing a two-component Nambu-Gorkov field $\Psi^T =
\left( \psi^T\!,\;\bar \psi \right) / \sqrt{2}$ which has 72 components in
this case. After integration over the fermion fields one obtains an
effective action which depends on the auxiliary fields. To obtain the
effective potential to leading order in $1/N_c$, these auxiliary
fields are replaced by their vacuum expectation values. The leading
order $1/N_c$ approximation is equivalent to the so-called mean field
approximation. From now on $\alpha_a$, $\beta_a$ and $\Delta_{AB}$
stand for the vacuum expectation values of the auxiliary fields. The
quantum fluctuations around the vacuum expectation values of the
auxiliary fields are not taken into account in the mean field
approximation. The vacuum expectation values of the auxiliary fields
are directly related to the quark condensates
\begin{equation}
\begin{split}
  & \alpha_a  = - 2 G \left < \bar \psi \lambda_a \psi 
\right>\,,\;\;\;
  \beta_a  = - 2 G \left < \bar \psi \lambda_a i
  \gamma_5 \psi
  \right> \,, \\
  &\Delta_{AB} = \frac{3}{2}G  \left< \psi^T t_A \lambda_B C \gamma_5
\psi  \right>\,.
\end{split}
\end{equation}
The condensate $\Delta_{AB}$ is complex. To calculate its vacuum
expectation value one should in principle average over all its
phases. Because of the $\mathrm{U}(1)_B$ symmetry of the action, this
averaging will let the vacuum expectation vanish. However, in a real
physical situation the system will just pick one direction giving rise
to a nonzero vacuum expectation value of $\Delta_{AB}$. The phase will
however be undetermined unless a small asymmetry is brought in. This
is situation is similar to the magnetization below the phase
transition temperature in the Ising model. In that case averaging over
all possibilities will also give rise to a zero net
magnetization. However the real system should choose to have either a
finite positive or negative magnetization.  It turns out that
$\Delta_{AB}$ is a good order parameter, giving rise to a BCS gap in
the quasi-particle excitation spectrum indicating color
superconductivity.

In this chapter it is assumed that all condensates are space-time
independent. When this restriction is dropped it is possible to find
the crystalline Larkin-Ovchinnikov-Fulde-Ferrell (LOFF) phase
\citep{Alford2001b}. This is a phase in which quarks with different
momentum magnitudes pair. One can find this phase between the phase
in which chiral symmetry is broken and the color superconducting phases.

After going to imaginary time, the thermal effective potential
$\mathcal{V}$ to leading order in $1/N_c$ (which is equivalent to the
mean-field approximation) reads
\begin{equation}
  \mathcal{V} = \frac{\alpha_a^2 + \beta_a^2}{4G}
+ \frac{\left \vert \Delta_{AB}\right \vert^2 }{3 G} 
-\frac{T}{2} \sum_{p_0 = (2n + 1)\pi T} \int \frac{\mathrm{d}^3 p}{
\left( 2\pi \right)^3} \log \mathrm{det} K \;,
\label{eq:njleffpotential}
\end{equation}
where $K$ is a $72 \times 72$ matrix
\begin{equation}
K = 
\left(
\begin{array}{cc}
\unity_c \otimes \mathcal{D}_1 
&
\Delta_{AB}\, t_A \otimes \lambda_B \otimes \gamma_5 
\\
-\Delta^*_{AB}\,  t_A \otimes \lambda_B \otimes \gamma_5 
&
\unity_c \otimes \mathcal{D}_2
\end{array} \right)\;,
\end{equation}
and 
\begin{eqnarray}
  \!\!\!\! \mathcal{D}_1 \!\!\!\!&=& \!\!\!\!\unity_f \otimes (i \gamma_0 p_0\!+\!
\gamma_i p_i) 
-  \mu \otimes \gamma_0 
  - ( M_0\!+\!\alpha_a \lambda_a ) \otimes \unity_d 
- \beta_a \lambda_a \otimes i \gamma_5 
\,,
\\
  \!\!\!\! \mathcal{D}_2 \!\!\!\! &=& \!\!\!\! \unity_f \otimes (i \gamma_0 p_0\!+\!\gamma_i
p_i) 
 +  \mu \otimes \gamma_0
 - ( M_0\!+\!\alpha_a \lambda_a^T)\otimes \unity_d  
 - \beta_a \lambda_a^T \otimes i \gamma_5
\,.
\end{eqnarray}
The matrix $\unity$ is the identity matrix in color ($c$), flavor ($f$),
or Dirac ($d$) space.

The values of the condensates and the phase diagram are determined by
minimizing the effective potential ${\cal V}$ with respect to the
condensates. To make the minimization procedure easier, one can take
advantage of the fact that certain condensates must vanish. Firstly,
application of QCD inequalities derived by \citet{Weingarten1983}
shows that in QCD at zero temperature, global $\mathrm{SU}(3)_V$
symmetry breaking cannot be driven by condensates of the type $\left <
\bar u i \gamma_5 u \right>$ \citep{Vafa1984a, Son2001}. 
Therefore, $\beta_0$, $\beta_3$ and $\beta_8$ are
zero. Outside the phase in which diquarks condense, it was checked
numerically that this is correct at non-zero temperatures and chemical
potentials as well. At zero chemical potential and temperatures a
theorem by \citet{Vafa1984b} states that all parity violating
condensates should vanish in QCD, this is also in agreement with the
calculations.  Moreover, although perturbative one-gluon exchange
cannot distinguish between $\beta_k$ and $\alpha_k$ condensation with
$k \in
\{1,2,4,5,6,7\}$, pseudoscalar condensation is favored due to the
instanton interaction \citep{Son2001}. One can therefore set all
$\alpha_k$'s again with $k \in \{1,2,4,5,6,7\}$ to zero. Numerically
it was found that this is correct, even in the absence of instanton
interactions.

One can further simplify the minimization procedure by exploiting the
symmetries of the NJL model. The free energy (which is the energy in
the minimum of the effective potential) is invariant under the same
transformations as the Lagrangian density, so applying a
$\mathrm{U}(1)$-flavor transformation to all condensates leaves the
free energy invariant. The $\beta$ condensates transform under a
$\mathrm{U}(1)$-flavor transformation which is parametrized by an
angle $\alpha$ as follows
\begin{eqnarray}
  \beta_{1,4,6} \rightarrow \cos \alpha_{1,4,6} \, \beta_{1,4,6} +
 \sin \alpha_{1,4,6} \,\beta_{2,5,7} 
  \;, \\
  \beta_{2,5,7}
 \rightarrow -\sin \alpha_{1,4,6} \, \beta_{1,4,6} + \cos
 \alpha_{1,4,6} \, \beta_{2,5,7} \;.
\end{eqnarray}
Therefore, using the $\mathrm{U}(1)$-flavor transformations one can
choose the pseudoscalars to condense in the $\beta_2$, $\beta_5$, and
$\beta_7$ channels, and set $\beta_1$, $\beta_4$ and $\beta_6$ to
zero. The phase in which $\beta_2$, $\beta_5$ and/or $\beta_7$ is
non-vanishing is called the $\pi^+/\pi^-$, $K^0 / \bar K^0$ and/or
$K^+/K^-$ condensed phase, respectively. The condensates $\beta_2$,
$\beta_5$ and $\beta_7$ are good order parameters since they break the
$\mathrm{U}(1)_u \times \mathrm{U(1)}_d \times \mathrm{U}(1)_s$
symmetry which is present in the Lagrangian density into
$\mathrm{U(1)_B} \times \mathrm{U(1)}$ if only one pseudoscalar
condensate arises or to $\mathrm{U(1)_B}$ if there are more. Moreover,
these condensates break parity symmetry. To summarize for all
quark-antiquark condensates, only the $\alpha_0, \alpha_3, \alpha_8$
and $\beta_2, \beta_5, \beta_7$ condensates will be considered below.

Because of the global $\mathrm{SU}(3)_c$ symmetry, one can also rotate
away several diquark condensates. Without loss of generality one can
minimize with respect to $\Delta_{22}$, $\Delta_{25}$, $\Delta_{55}$,
$\Delta_{27}$, $ \Delta_{57}$ and $\Delta_{77}$. In principle, all six
diquark condensates can have a phase. It is always possible to remove
two of them by using the two diagonal $\mathrm{SU}(3)_c$
transformations. As long as there is no pseudoscalar condensation, one
can use the $\mathrm{U}(1)$-flavor symmetries to rotate away three
other phases. As a result either $\Delta_{25}, \Delta_{55},
\Delta_{27}$, or $\Delta_{57}$ has a phase \citep{Buballa2005a}.
However, this reduction is not completely possible if pseudoscalar
condensation occurs. By choosing the pseudoscalars to condense in the
$\beta_2$, $\beta_5$ and $\beta_7$ channels, one breaks the
$\mathrm{U}(1)$-flavor symmetry. Hence if pseudoscalar condensation
arises in one channel, one can in general rotate away one phase less
in the diquark sector. If it occurs in more channels, two phases less
can be rotated away.  However, numerically it is found that allowing
for such a complex phase leads to diquark condensation only in the
$\Delta_{22}$, $\Delta_{55}$ and the $\Delta_{77}$ channels. The
$\Delta_{25}$, $\Delta_{27}$, and $\Delta_{57}$ diquark condensates do
not arise or can be rotated away. Moreover it is found that
pseudoscalar condensation in the quark-antiquark channel does not
coexist with color superconductivity, such that one can always take
the diquark condensates to be real.

The different possible color-superconducting phases are
named as follows \citep{Ruster2005}
\begin{eqnarray}
 \Delta_{22}\neq0,\,\;\Delta_{55}\neq0,\,\Delta_{77} \neq 0 & & \mathrm{CFL} \;, 
  \nonumber \\
 \Delta_{77} =0,\;\; \Delta_{22}\neq0,\,\Delta_{55} \neq 0 & & \mathrm{uSC} \;,
  \nonumber \\
 \Delta_{55} = 0,\;\;\Delta_{22}\neq0,\,\Delta_{77} \neq 0 & & \mathrm{dSC} \;,
  \nonumber \\
 \Delta_{22} = 0,\;\;\Delta_{55}\neq0,\,\Delta_{77} \neq 0 & & \mathrm{sSC} \;,
  \nonumber \\
 \Delta_{22} \neq 0, \;\; \Delta_{55}=0,\, \Delta_{77} = 0 & & \mathrm{2SC} \;,
  \nonumber \\
 \Delta_{55} \neq 0, \;\; \Delta_{22}=0,\, \Delta_{77} = 0 & & \mathrm{2SCus} \;,
  \nonumber \\
 \Delta_{77} \neq 0, \;\; \Delta_{22}=0,\, \Delta_{55} = 0 & & \mathrm{2SCds} \;.
\end{eqnarray}
The abbreviation CFL stands for color-flavor locked phase. If there is
exact $\mathrm{SU}(3)_\mathrm{V}$ flavor symmetry, the three diquark
condensates in this phase have equal size and the vacuum is invariant
under a combined rotation in color and flavor space \citep{Alford1999,
Alford1999q}. In the uSC, dSC or sSC phase the up, down or strange
quark always takes part in the diquark condensate, respectively. In
the 2SC phase an up and a down quark form a diquark condensate, in the
2SCus and the 2SCds phase this condensate is formed by an
up and and strange quark and a down and a strange quark, respectively.

To calculate the effective potential one needs to evaluate a
determinant of a $72 \times 72$ matrix. Only in special cases such as
when all masses and chemical potentials are equal and in absence of pseudoscalar
condensation one can perform over the sum over Matsubara frequencies
analytically and hence simplify the effective potential somewhat
further. But in the more general cases which will be discussed in this
chapter this either is not possible or gives rise to very complicated
equations. One is therefore restricted to do numerical calculations.

To calculate the determinant of the effective potential in an
efficient way, one can multiply the matrix $K$ with
$\mathrm{diag}(\unity_c \otimes \unity_f \otimes
\gamma_0, \unity_c \otimes \unity_f \otimes \gamma_0 )$ which leaves
the determinant invariant.  In this way, one obtains a new matrix $K'$
with $i p_0$'s on the diagonal.  By determining the eigenvalues
$\lambda_i$ of the matrix $K'$ with $p_0 = 0$, one can reconstruct the
determinant of $K$ for all values of $p_0$ which is namely
$\prod_{i=1}^{72}(\lambda_i + ip_0)$. After summing over Matsubara
frequencies, one finds using Eq.~(\ref{eq:sumintfermpres})
\begin{equation}
 T \!\!\!\!\!\! \sum_{p_0 = (2 n+1) \pi T} \!\!\!\!\!\! \log
  \mathrm{det} K = \sum_{i=1}^{72} \left[\frac{\lambda_i}{2} + T
  \log\left(1 + e^{-\lambda_i / T} \right) \right] \;.
\end{equation}
All that remains in order to determine the effective potential is to
integrate over three-momentum $p$ up to an ultraviolet cutoff $\Lambda$. 

The speed of the calculation of the effective potential depends
heavily on how fast one can compute the eigenvalues. There are several
ways to speed up the calculation. Firstly, the determinant of $K$ does
not depend on the direction of $\vec p$. Therefore, one can choose 
$\vec p$
to lie in the $z$-direction.  Together with the choice of the
non-vanishing condensates mentioned above, this implies that
$K'(p_0=0)$ becomes a real symmetric matrix, which simplifies the
calculation of the eigenvalues. Secondly, one can interchange rows and
columns of $K'$ without changing its determinant. By doing so, one can
bring $K'$ in a block-diagonal form. One can then determine the
eigenvalues of the blocks separately which is significantly faster
since the time needed to compute eigenvalues numerically scales
cubically with the dimension of the matrix. In the most general case
with diquark condensation, one can always reduce the problem to two
$36 \times 36$ matrices. Moreover, if there is no diquark
condensation, but only pseudoscalar condensation 
\citep{Barducci2005}, the problem can be further reduced to computing the eigenvalues of
two $6 \times 6$ matrices

The eigenvalues were obtained using LAPACK routines \citep{Lapack1999}.
After numerical integration over three-momentum $p$ up to the cutoff,
the condensates were determined by minimizing the effective potential
using MINUIT \citep{James1975}. To be certain that the minimization
procedure did not end up in a local minimum the continuity of the
minimized effective potential as a function of chemical potentials
and/or temperature was always checked.

%=====================================================================
\section{Phase diagrams}
\begin{figure}[htb]
\begin{center}
\begin{tabular}{ccc}
\scalebox{0.4}{\includegraphics{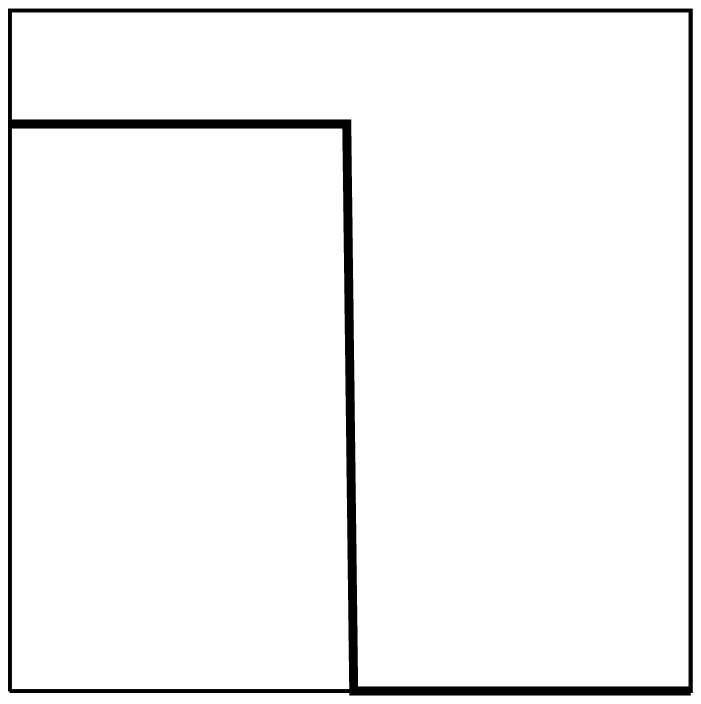}}&
\scalebox{0.4}{\includegraphics{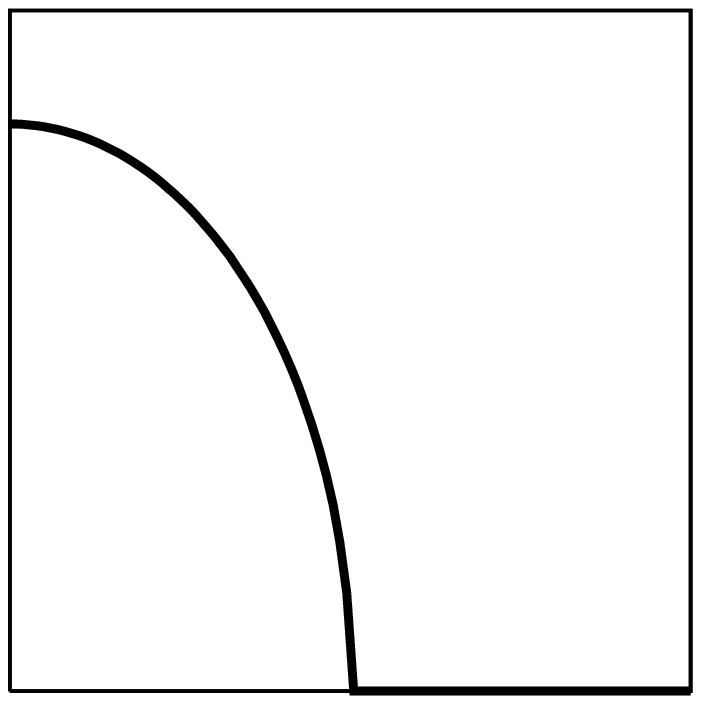}}&
\scalebox{0.4}{\includegraphics{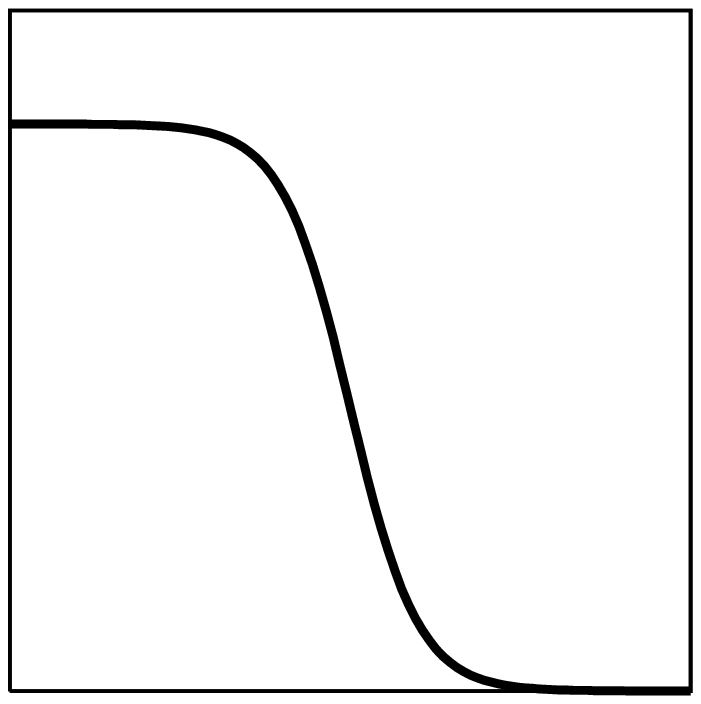}}\\
a & b & c
\end{tabular}
\caption{Typical behavior of condensates near a transition. Fig.~(a)
shows a condensate undergoing a first order transition, Fig.~(b) 
a second order transition and Fig.~(c) a cross-over.
}
\label{fig:trans}
\end{center}
\end{figure}

In this section results for the phase diagrams of the NJL model with
$u$, $d$, and $s$ quarks are presented. The phase diagrams are plotted
as a function of the chemical potentials and temperature.  To
determine the locations of the phase boundaries, the behavior of the
condensates was examined. The different possibilities are displayed in
Fig.~\ref{fig:trans}. If a condensate jumps discontinuously the
transition is first-order (a), and this is indicated by a solid line
in all phase diagrams.  If its derivative has a discontinuity, the
transition is second order (b), and this is indicated by a dotted
line.  If a condensate changes rapidly in a narrow range without
vanishing, in other words no derivative has a discontinuity, there is
a smooth cross-over (c), and this is indicated in the phase diagrams
by a dashed-dotted line at the point were the condensate varies
maximally.

\begin{figure}[htb]
\begin{center}
\scalebox{0.9}{\includegraphics{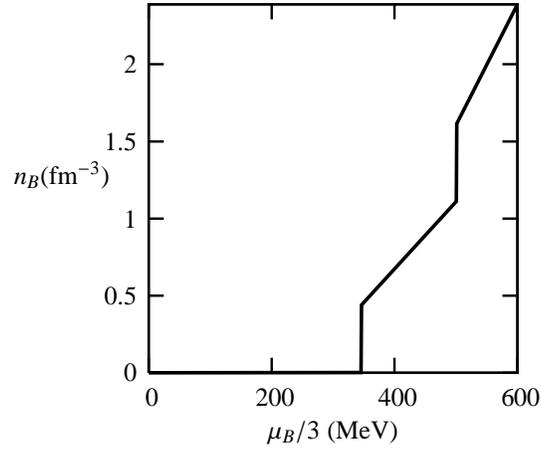}}
\caption{Baryon number density as a function of baryon chemical
potential for $T=0$ and $\mu_u = \mu_d = \mu_s$. The vertical pieces
indicate the presence of a mixed phase. Using Fig.~\ref{fig:mubT} one
can conclude that upon increasing $n_B$, quark matter is subsequently
in a phase with broken chiral symmetry, in a mixed phase of broken
chiral symmetry and 2SC, in a 2SC phase, in a mixed phase of 2SC and
CFL, and finally in a CFL phase.  For comparison, the nuclear matter
density is about $0.17\;\mathrm{fm}^{-3}$.}
\label{fig:numberdens}
\end{center}
\end{figure}
\begin{figure}[htb]
\begin{center}
\scalebox{0.9}{\includegraphics{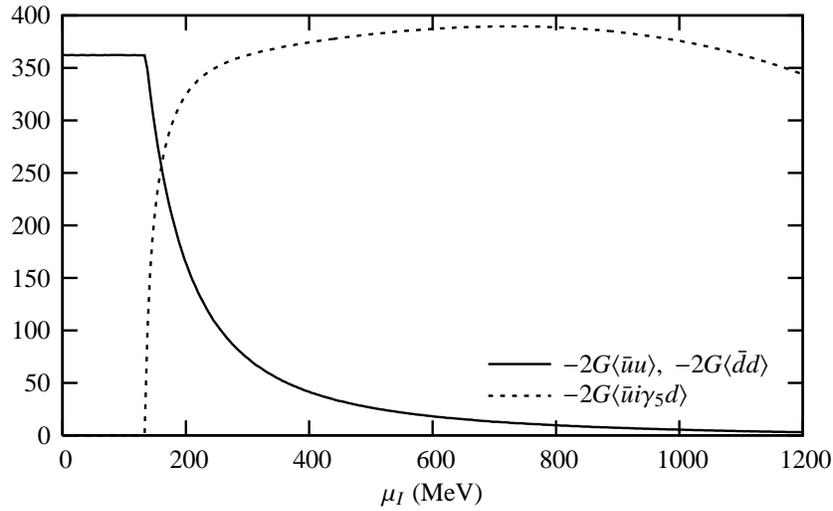}}
\caption{The chiral condensates (solid line) and
pion condensate (dotted line) as a function of $\mu_I = \mu_u - \mu_d$ with
$\mu_u = -\mu_d$.}
\label{fig:cond}
\end{center}
\end{figure}
\clearpage

To calculate the phase diagrams, the effective potential was minimized
on a grid of chemical potentials and or temperatures.  The numerical
uncertainties in the presented phase diagrams are in the order of the
distance between the points in the grid which was about 3 MeV, this
coincides with the thickness of the lines in the phase diagrams. 

One should keep in mind that the relation between chemical potential
and number density is not linear. In Fig.~\ref{fig:numberdens} the
baryon number density as a function of the baryon chemical potential
is displayed.  In that figure it can be seen that at a first-order
phase boundary the number density increases discontinuously. At these
particular densities, quark matter for example can be in a mixed state
of normal and superconducting matter (see e.g.\
\citet{Bedaque2002a} and \citet{Lawley2005}).

A typical outcome of the minimization procedure is displayed in
Fig.~\ref{fig:cond}, in which the chiral and pion condensates are
plotted as a function of $\mu_I$ for $\mu_B = 0$. It can be seen from
the figure that pion condensation arises via a second order transition
for $\mu_I > m_\pi$. It can also be seen that the chiral condensates
$\bar u u$ and $\bar d d$ are small but non-vanishing in the phase
with pion condensation.

\subsection*{Phase diagram with $\mu_B$ vs. $T$}
\begin{figure}[htb]
\begin{center}
\includegraphics{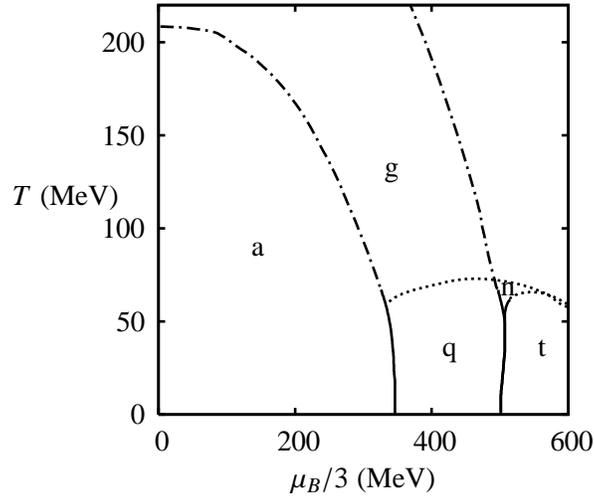}
\caption{Phase diagram for $\mu_u = \mu_d = \mu_s$ as a function of
$\mu_B$ and $T$. First and second-order transitions are indicated by
solid and dotted lines, respectively. A cross-over is indicated by a
dashed line. The letters denote the different phases, where a:
$\bar u u$ + $\bar d d$ + $\bar s s$, g: $\bar s s$, n: 2SC, q: 2SC +
$\bar s s$ and t: CFL.  }
\label{fig:mubT}
\end{center}
\end{figure}

In Fig.~\ref{fig:mubT} the phase diagram of the NJL model as a
function of baryon chemical potential $\mu_B = \mu_u + \mu_d + \mu_s$
and temperature is displayed for $\mu_u = \mu_d = \mu_s$. This phase
diagram is similar to the qualitative QCD phase diagram displayed in
Fig.~\ref{fig:qcdphase}. At low temperatures and baryon chemical
potential the chiral symmetry is broken in the phases denoted by (a)
and (g). By increasing the chemical potential at low temperatures one
enters the color superconducting phases, first the 2SC phase (n/q) and
thereafter again via a first order transition the CFL phase (t).
Three critical endpoints can be seen in this diagram.

\begin{figure}[bht]
\begin{center} \hspace{-1.5cm}
\includegraphics{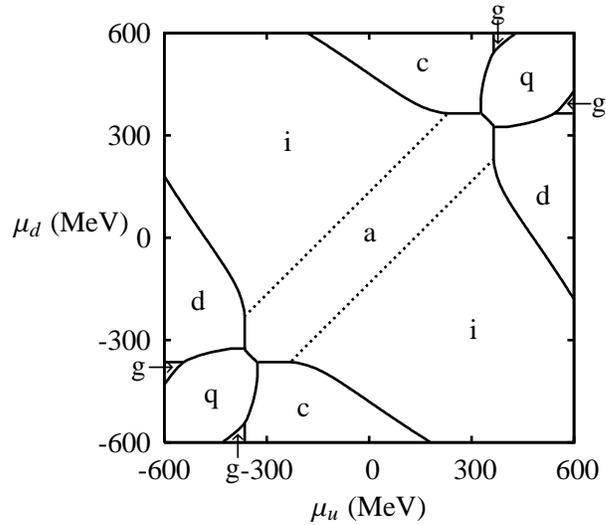}
\caption{Phase diagram for $\mu_s = 0$ and $T=0$ as a function of
$\mu_u$ and $\mu_d$. The
letters denote the different phases, where 
a: $\bar u u$ + $\bar d d$ + $\bar s s$, 
c: $\bar u u$ + $\bar s s$, 
d: $\bar d d$ + $\bar s s$,
g: $\bar s s$, 
i: $\pi^+ / \pi^-$ + $\bar s s$ and
q: 2SC + $\bar s s$.}
\label{fig:muumudt0}
\end{center}
\end{figure}
\clearpage

\subsection*{Phase diagram with $\mu_s=0$ and $T=0$}

In Fig.~\ref{fig:muumudt0} the phase diagram of the NJL model for
$\mu_s=0$ and $T=0$ as a function of $\mu_u$ and $\mu_d$ is
displayed. Outside the 2SC phase (q), the results agree qualitatively
with the two-flavor calculations by \citet{Barducci2004} (see their
Fig.\ 1), where color-superconducting phases were not taken into
account.  Moreover, \citet{Barducci2004} use different parameters, in
addition to a form factor.
One can clearly see that the phase diagram is symmetric under
reflection in the origin. This is because the free energy is invariant
under the transformation $(\mu_u, \mu_d, \mu_s) \rightarrow (-\mu_u,
-\mu_d, -\mu_s)$, that stems from the symmetry between particles and
antiparticles. Fig.~\ref{fig:muumudt0} is also symmetric under
interchange of $u$ and $d$, because of the choice of equal up and down
quark masses. This gives rise to the symmetry of the phase boundaries
with respect to the diagonals.

In general, horizontal and vertical lines in the phase diagrams arise
if the pairing of one type of quark is not changed after a
transition. In this case, the location of the phase boundary is
determined by the properties of other quarks. Therefore, changing the
chemical potential of the unchanged quark species cannot have a big
influence on the location of the phase boundary. This results in the
horizontal and vertical lines. For $T=0$, one always finds these lines
near the values of the constituent quark masses, i.e.\ $\mu_u \approx
M_u$, $\mu_d \approx M_d$ and $\mu_s \approx M_s$ (see for example
\citet{Buballa2005a}). The diagonal lines arise
because at $T=0$ pion condensation can occur if $\vert \mu_u - \mu_d
\vert > m_\pi = 138\;\mathrm{MeV}$ \citep{Son2001}.

The diagram shows that if the chemical potentials are different, the
transition to the 2SC phase (q) remains first order as
was concluded by \citet{Bedaque2002a}.  Moreover, one can see
from Fig.~\ref{fig:muumudt0} that if $\mu_u \neq \mu_d$ it is
possible to go through two first-order transitions before entering the
2SC phase (q) (similar to the situation discussed by
\citet{Toublan2003} without color superconductivity). To have such a
scenario at zero temperature, a minimum difference between $\mu_u$ and
$\mu_d$ is required. In the present case this is about 35 MeV.  Pion
condensation (i) and the 2SC phase (q) are in this diagram separated
by two phase transitions in contrast to the estimated $(\mu_B,\mu_I)$
phase diagram of \citet{He2005} which correctness is therefore
questionable.

\clearpage
\subsection*{Phase diagram with $\mu_d = 0$ and $T=0$}
\begin{figure}[htb]
\begin{center} \hspace{-1.5cm}
\includegraphics{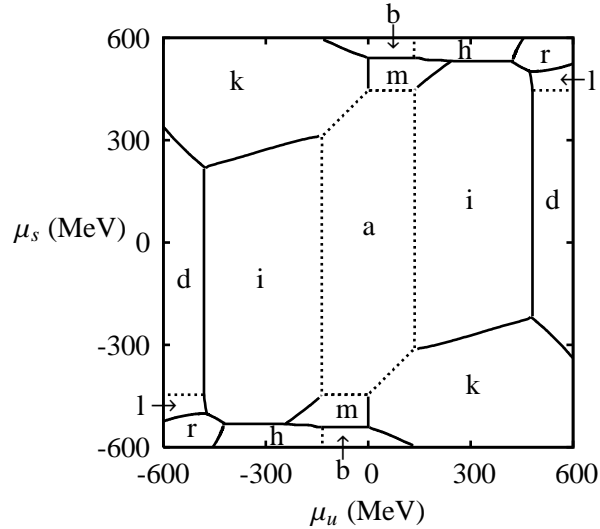}
\caption{Phase diagram for $\mu_d = 0$ and $T = 0$ as a function of
$\mu_u$ and $\mu_s$. First and second-order transitions are indicated
by solid and dotted lines, respectively. The letters denote the
different phases, where a: $\bar u u$ + $\bar d d$ + $\bar s s$, 
b: $\bar u u$ + $\bar d d$, 
d: $\bar d d$ + $\bar s s$, 
h: $\pi^+ / \pi^-$, i: $\pi^+ / \pi^-$ + $\bar s s$, 
k: $K^+/ K^-$ + $\bar d d$,
l: $K^0 / \bar K^0$, 
m: $K^0/\bar K^0$ + $\bar u u$
and r: 2SCus + $\bar d d$.}
\label{fig:muumust0} 
\end{center}
\end{figure}

In Fig.~\ref{fig:muumust0} the phase diagram for $\mu_d = 0$ and
$T=0$ as a function of $\mu_u$ and $\mu_s$ is displayed. Since the up
and down quark masses are much smaller than the strange quark mass,
this diagram is very different from Fig.~\ref{fig:muumudt0}. Besides
the possibility of pion condensation in (h) and (i) also phases in
which the charged kaon (k) and the neutral kaon condense (l)/(m)
arise. The lines separating the charged kaon phase (k) from the
chirally broken phase (a) are diagonal because at $T=0$ kaon
condensation can occur if $\vert \mu_s - \mu_{u, d} \vert > m_K =
450\;\mathrm{MeV}$
\citep{Kogut2001} (the chosen parameter set gives rise to a somewhat low kaon
mass, but this is not relevant for the qualitative features of the phase 
diagram). 
The 2SCus phase appears in (r). This phase is surrounded by
phases in which the pions (h)/(i) and the neutral kaons (l)
condense. One passes a first-order transition when going
from the pion and neutral kaon condensed to the 2SCus phase.

\clearpage
\subsection*{Phase diagrams with $\mu_u \approx \mu_d$}

\begin{figure}[htb]
\begin{center} \hspace{-1.5cm}
\includegraphics{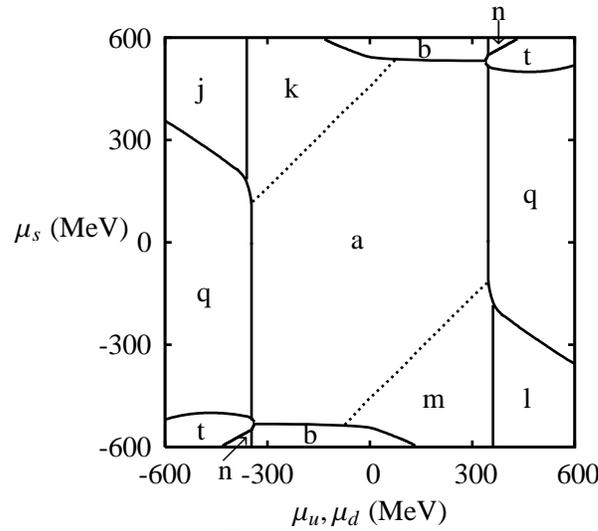}
\caption{Phase diagram for $T = 0$ as a function of $\mu_u = \mu_d +
\epsilon $ and $\mu_s$.  The letters denote
the different phases, where 
a: $\bar u u$ + $\bar d d$ + $\bar s s$,
b: $\bar u u$ + $\bar d d$, 
j: $K^+ / K^-$,  
k: $K^+/ K^-$ + $\bar d d$,
l: $K^0 / \bar K^0$, 
m: $K^0/\bar K^0$ + $\bar u u$,
n: 2SC, 
q: 2SC + $\bar s s$ and 
t: CFL.
}
\label{fig:muudmust0} 
\end{center}
\end{figure}

In Fig.~\ref{fig:muudmust0} the phase diagram for $T=0$ as a function
of the up and down quark chemical potential and the strange quark
chemical potential is displayed. In this phase diagram $\mu_u = \mu_d +
\epsilon$ where $\epsilon$ is a very small positive
number. This $\epsilon$ is necessary because when $\epsilon = 0$ one
is just at a first-order phase boundary between the phase in which the
charged kaons condense (k) and the one in which the neutral kaons
condense (m), as can be seen from Fig.~\ref{fig:muumust0}. This
nonzero value of $\epsilon$ gives rise to a small asymmetry in the
phase diagram. If $\epsilon$ is chosen to be negative, the phases in which
the neutral (l)/(m) and the charged (j)/(k) kaon condenses are
interchanged.

\begin{figure}[htb]
\begin{center} \hspace{-1.5cm}
\includegraphics{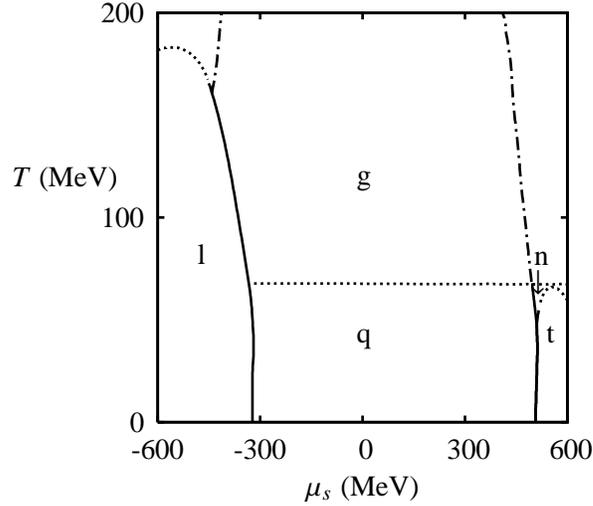}
\caption{Phase diagram as a
function of $\mu_s$ and $T$, for fixed 
$\mu_u = \mu_d = 550\;\mathrm{MeV}$. 
Here a cross-overs is indicated by
a dashed-dotted line.
The letters denote the different phases,
where 
g: $\bar s s$,
l: $K^0 / \bar K^0$, 
n: 2SC,
q: 2SC + $\bar s s$ and 
t: CFL. 
}
\label{fig:muud550must}
\end{center}
\end{figure}

Apart from the additional 2SC (n)/(q) and CFL (t) phases, the results
presented here agree qualitatively with the three-flavor calculations
by \citet{Barducci2005} (see their Fig.\ 7).
\citet{Barducci2005} have used different quark masses
and a different coupling constant, and in addition employed a form
factor to mimic asymptotic freedom. Therefore, one may conclude that
the use of such a form factor does not affect the phase diagram
qualitatively. The phase diagram Fig.\
\ref{fig:muudmust0} cannot simply be obtained by a superposition of
phase diagrams obtained from a calculation with pseudoscalar
condensation, but without superconductivity (such as done by
\citet{Barducci2005}), and
one with superconductivity, but without pseudoscalar condensation
(such as done by \citet{Gastineau2002}). 
Despite the fact that the two types of phases do
not coexist, there is nevertheless competition between them. 
Figure \ref{fig:muudmust0} shows that the $K^0 / \bar K^0$ (l)/(m)
and the $K^+ / K^-$ (j)/(k) phase are separated from the 2SC phase (q) by a
first-order transition. This remains the case at finite temperature as
is illustrated in Fig.~\ref{fig:muud550must}. This figure displays
the phase diagram as a function of $\mu_s$ and temperature, for fixed
$\mu_u = \mu_d = 550\;\mathrm{MeV}$.

Returning to the discussion of Fig.\ \ref{fig:muudmust0}; the line
$\mu_u = \mu_d = \mu_s$ goes through the phase (a) in which chiral
symmetry is spontaneously broken. At some point it enters via a
first-order transition the 2SC + $\bar s s$ phase (q), and finally
goes into the CFL phase (t), again via a first-order transition. If
there is a difference between $\mu_u = \mu_d$ and $\mu_s$, one can see
in Fig.~\ref{fig:muudmust0} that as the densities increase, quark
matter can go directly from a phase with chiral symmetry breaking (a)
to a CFL phase (t) without passing the 2SC phase (q) first. This can
also occur in compact stars as was shown by
\citet{Alford2002}. 

It is also interesting to note that the phases (l)/(j) of kaon
condensation can also occur outside the region of spontaneous chiral
symmetry breaking. Assuming the phase transition towards chiral
symmetry restoration coincides with the deconfinement transition (as
appears to be the case in lattice studies at small baryon chemical
potential and in some models), this would imply that condensation of a
state with quantum numbers of the kaon may persist in the deconfined
phase. This was first observed by \citet{Son2001}, based on a
perturbative calculation at very high isospin chemical potential.

\clearpage
\subsection*{Phase diagrams with $\mu_u = \mu_s$}
\begin{figure}[t]
\begin{center} \hspace{-1.5cm}
\includegraphics{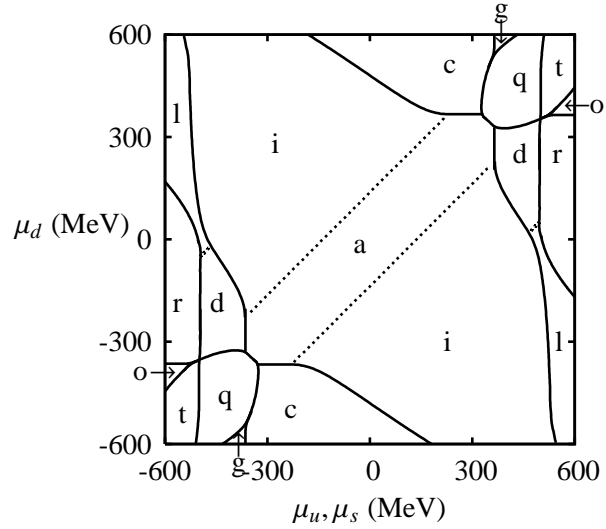}
\caption{Phase diagram for $T = 0$ as a function of $\mu_u=\mu_s$ and
$\mu_d$. The letters denote
the different phases, where 
a: $\bar u u$ + $\bar d d$ + $\bar s s$,
c: $\bar u u$ + $\bar s s$, 
d: $\bar d d$ + $\bar s s$,
g: $\bar s s$, 
i: $\pi^+ / \pi^-$ + $\bar s s$ , 
l: $K^0 / \bar K^0$, 
o: 2SCus, 
q: 2SC + $\bar s s$,
r: 2SCus + $\bar d d$ and
t: CFL.}
\label{fig:muusmudt0} 
\end{center}
\end{figure}

In Fig.\ \ref{fig:muusmudt0} the phase diagram at zero
temperature as a function of $\mu_u = \mu_s$ and $\mu_d$ is shown. This diagram
is similar to Fig.\ \ref{fig:muumudt0} for small strange quark
chemical potentials (below the kaon mass). 
At larger strange quark chemical potentials the
diagrams differ, exhibiting kaon condensation (l) and
diquark condensation involving strange quarks (r)/(o)/(t).

\begin{figure}[t]
\begin{center} \hspace{-1.5cm}
\includegraphics{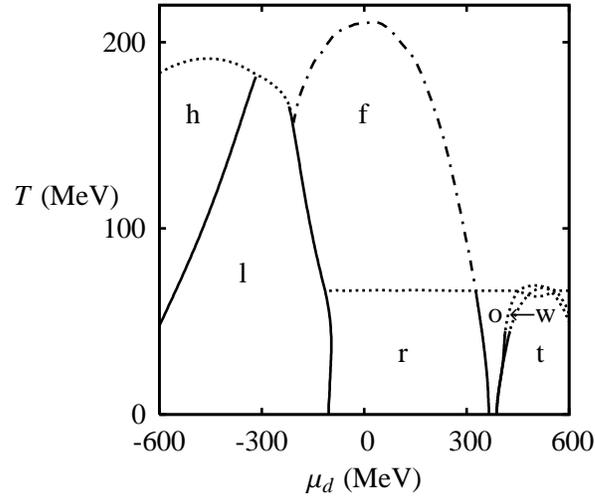}
\caption{Phase diagram as a function of $\mu_d$ and $T$, for fixed
$\mu_u = \mu_s = 550\;\mathrm{MeV}$. 
The letters
denote the different phases, where f: $\bar d d$, h: $\pi^+ / \pi^-$,
l: $K^0 / \bar K^0$, o: 2SCus, r: 2SCus + $\bar d d$, t: CFL and w:
sSC. The upper phase is phase with chiral symmetry in which all condensates
vanish. The lower right corner of this figure is enlarged in Fig.\
\ref{fig:muus550mudtr}.}
\label{fig:mus550mudt}
\end{center}
\end{figure}

\begin{figure}[bht]
\begin{center} 
\hspace{-1.5cm}
\includegraphics{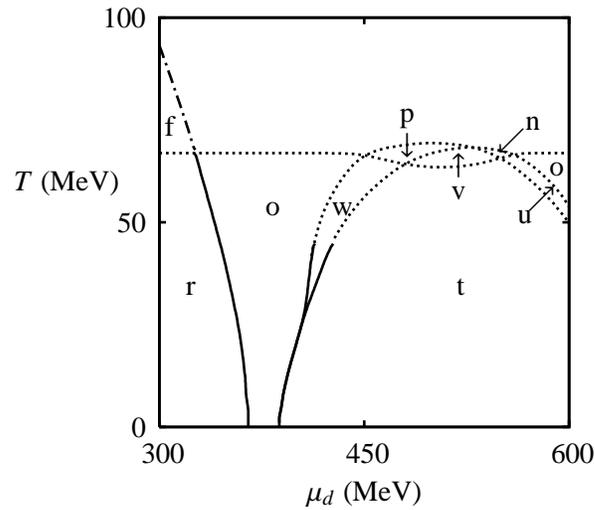}
\caption{Same as Fig.\ \ref{fig:mus550mudt}. The phases that occur are 
f: $\bar d d $, 
n: 2SC,
o: 2SCus, 
p: 2SCds,
r: 2SCus + $\bar d d$, 
t: CFL,
u: uSC,
v: dSC and
w: sSC.}
\label{fig:muus550mudtr} 
\end{center}
\end{figure}

In Fig.\ \ref{fig:mus550mudt} the phase diagram as a function of
$\mu_d$ and $T$, for fixed $\mu_u = \mu_s = 550\;\mathrm{MeV}$ is
displayed. In this figure one can find five critical points. Also, one
can see in this figure that the $K^0 / \bar K^0$ phase (l) is separated
from the 2SCus phase (r) by a first-order transition for all temperatures.
Furthermore, it is interesting that at finite temperature there is a
first-order transition from the phase in which the neutral kaons
condense (l) to the pion condensed phase (h). In Fig.\
\ref{fig:muus550mudtr} the lower-right corner of
Fig.\ \ref{fig:mus550mudt} is enlarged for clarity. In this figure one can find
all possible superconducting phases, including the more exotic uSC (u), dSC (v) and 
sSC (w) phases. For $\mu_u = \mu_d = \mu_s$ one goes from the
CFL phase (t) via the 2SC phase (n) to the chirally restored
phase when raising the temperature. However, small differences 
between $\mu_u = \mu_s$ and $\mu_d$ can cause one to go through completely 
different phases.

\clearpage
%=====================================================================
\section{Summary and Conclusions}
In this chapter the phase diagram of the three-flavor NJL model
including pseudoscalar quark-antiquark condensation and color
superconductivity was studied as a function of the different quark
chemical potentials and temperature. The NJL model has a rich and
interesting phase structure. The pseudoscalar condensed and color
superconducting phases are competing and are separated by a
first-order phase transition. As was discussed, this need not be the
case for other (less conventional) choices of the parameters of the
model.

Furthermore, it was concluded that at zero temperature and zero strange
quark chemical potential, a minimum asymmetry of about 35 MeV
between the up and the down quark chemical potentials is required in
order to have two first-order transitions, when going from the phase
with spontaneous chiral symmetry breaking to the 2SC phase.

The results provide a qualitative check and extension of several
earlier calculations that appeared in the literature. The calculations
in this chapter simultaneously allowed for nine different nonzero
condensates, $\bar u u$, $\bar d d$, $\bar s s$, $\pi^+/\pi^-$,
$K^+/K^-$, $K^0/\bar K^0$, $\Delta_{22}$, $\Delta_{55}$ and
$\Delta_{77}$. The new aspects of the phase diagrams are often located
in regions, where the quark chemical potentials are large and very
different in magnitude for the different flavors.  Although such
situations are not necessarily realized in compact stars or in
heavy-ion collisions. The study of such unusual situations is
nevertheless interesting for a fundamental understanding of the
theory. Moreover a comparison with future lattice data may provide
interesting information.  This is especially relevant for pseudoscalar
condensation in the phase where chiral symmetry is restored and also
for the complicated superconductivity phase structure close to the
cutoff of the model.

This work can be extended in several ways. For example, one can take
into account the instanton-induced interaction
\citep{tHooft1976}, Eq.~(\ref{eq:njlthooftint}) in
the mean field approximation or using the method of
\citet{Osipov2005}.  If one has pseudoscalar condensation, this is
much more difficult than in the normal case though.  Another very useful
extension would be the inclusion of the neutrality
conditions~\citep{Alford2002, Steiner2002}, in which case the phase
structure changes and for instance gapless phases will occur. It would
also be interesting to see how the results depend on the strength of
the diquark coupling and on the quark masses. Furthermore, one
could add the LOFF phase
\citep{Alford2001b}. In this crystalline phase, quarks of different
momentum magnitudes can pair. One could also include vector interactions. In this
case spin-1 diquark condensation (see for example
\citet{Pisarski2000, Buballa2003}) and an induced Lorentz-symmetry
breaking (ISB) phase~\citep{Langfeld1999} are among the
possibilities. It would also be worthwhile to take pseudoscalar diquark
condensation \citep{Buballa2005b} into account. Finally,
one could try to go beyond the mean-field approximation as was done by
\citet{Hufner1994}.

\begin{subappendices}

\section{The Nambu-Gorkov formalism}\label{nambugorkov}
In case a Lagrangian density contains terms which are
proportional to $\psi^T \Gamma \psi$, the integration over
the quark fields is non-trivial. In that case it
useful to apply the Nambu-Gorkov formalism which is 
discussed for the Lagrangian density used in this
chapter, Eq.\ (\ref{eq:njlferm}).

Taking the fermionic part of the Lagrangian Eq.\ (\ref{eq:njlferm}) and 
transposing half of the quark-antiquark term gives
\begin{equation}
\begin{split}
 \mathcal{L}_\mathrm{f} & = 
  \tfrac{1}{2} \bar \psi 
  \left( 
 i \gamma^\mu \partial_\mu - M + \mu \gamma_0
  \right)
  \psi +  \tfrac{1}{2} \psi^T  
  \left( 
  i C \gamma^\mu C \partial_\mu + M^T - \mu \gamma_0
  \right)
\bar \psi^T \\
& \quad 
 + \tfrac{1}{2} \Delta^*_{AB} \psi^T t_A \lambda_B C \gamma_5 \psi 
 - \tfrac{1}{2} \Delta_{AB} \bar \psi  t_A \lambda_B C \gamma_5 \bar \psi^T \;,
  \label{eq:lagrnambugorkov}
\end{split}
\end{equation}
where $M = M_0 - \alpha_a \lambda_a - i \gamma_5 \beta_a \lambda_a$.
By introducing a two-component Nambu--Gorkov field
\begin{equation}
\Psi = \frac{1}{\sqrt{2}} 
  \left( \begin{array}{c} \psi \\ \bar \psi^T \end{array} \right)
  \;,
\end{equation}
one can rewrite the Lagrangian in Eq.\ (\ref{eq:lagrnambugorkov}) 
as $\mathcal{L}_\mathrm{f} = \Psi^T S \Psi$ with
\begin{equation}
  S = \left(
  \begin{array}{cc}
\Delta^*_{AB}  t_A \lambda_B C \gamma_5
&
 i C \gamma^\mu C \partial_\mu + M^T - \mu \gamma_0
\\
 i \gamma^\mu \partial_\mu - M + \mu \gamma_0
&
-\Delta_{AB}  t_A \lambda_B C \gamma_5
\end{array} \right) \;.
\end{equation}
In the path integral the measure $\mathcal{D} \bar \psi \mathcal{D}
\psi$ will be changed into $\mathcal{D} \Psi$.
Integration over the Nambu-Gorkov fields yields that $S_{\mathrm{eff}}
= \frac{1}{2} \log \mathrm{Det}\, \left[ \left(S - S^T \right) / 2
\right] = \frac{1}{2} \log \mathrm{Det}\, S$ since $S$ is antisymmetric in this
case.  The determinant is invariant under interchange of an even
number of rows. Using that one can put the part containing the
kinetic term on the diagonal. The effective action is simplified
further if one multiplies from the left with $\mathrm{diag}\,(1, C)$ and
from the right with $\mathrm{diag}\,(1, C)$ where $C$ is the charge conjugation
matrix, resulting in
\begin{equation}
S_{\mathrm{eff}} = \log \mathrm{Det}\, \left(
  \begin{array}{cc}
 i \gamma^\mu \partial_\mu - M + \mu \gamma_0
&
  \Delta_{AB}  t_A \lambda_B   \gamma_5
\\
-\Delta^*_{AB}  t_A \lambda_B   \gamma_5
&
 i  \gamma^\mu   \partial_\mu - M^T - \mu \gamma_0
\end{array} \right) \;.
\end{equation}
This expression was evaluated further in Eq.~(\ref{eq:njleffpotential}),
using a basis for a complete set of functions which satisfy the 
anti-periodicity conditions for fermions in imaginary time like
in Eq.~(\ref{eq:presfermtomomentum}).

\end{subappendices}
 %njl
\chapter*{Summary}
\addcontentsline{toc}{chapter}{Summary}
\fancyhf{}
\fancyhead[LE,RO]{\thepage} 
\fancyhead[RE,LO]{{\it Summary}}

In this thesis, entitled ``Thermodynamics of QCD-inspired theories''
the behavior of (quark) matter under extreme high temperatures and
densities is investigated. Such extreme conditions can be reached for
example in the early universe shortly after the big bang, during a
heavy-ion collision and inside the core of a neutron star.  They can
be so extreme that normal nuclear matter does longer exist, such that
matter is in a different phase than usual, like for example the
so-called quark-gluon plasma (at high temperatures and/or densities)
or the color-superconducting phase (at high densities).

A general introduction to this thesis is given in Chapter 1. Firstly,
it will be discussed what could happen to matter under such extreme
conditions.  Then, the relevant aspects of quantum chromodynamics
(QCD) (the theory which describes the interactions between the quarks)
will be explained shortly. Thereafter, an overview of the results and
problems of previous calculations of the pressure and the phase diagram
of quark matter are given. The relevance of these kind of calculations
for the understanding of a compact star, heavy ion collisions, and the
state of the universe shortly after the big bang, will be examined as
well. At the end of the chapter the term ``QCD-inspired
theories'' will be explained.

In Chapter 2 it is discussed how one can perform calculations at
finite temperature and densities in a field theory like QCD. For that,
results from statistical physics will be applied. Firstly, a short
review of classical statistical physics is given, which thereafter
will be generalized to a quantum mechanical approach. Using
path-integrals it will be discussed how this approach can be used in
order to investigate quantum field theories at finite temperature and
densities. In calculations at finite temperature and densities one
often has to perform summations over an infinite number of terms. A
method to perform these sums exactly will be discussed. Also it will
be explained in detail how these summations can be computed
numerically for more difficult cases.

The third chapter treats the so-called effective potential and the
$1/N$ expansion. The effective potential is an important quantity,
since the pressure and the phase diagrams can be obtained by
minimizing this potential. The $1/N$ expansion is a method which can
be used to investigate the non-perturbative (in the coupling constant)
behavior of some theories. It will be explained how this expansion can
be obtained by the introduction of auxiliary fields. At the end of
this chapter an important result of this thesis, which is the
occurrence of temperature-dependent divergences in the effective
potential, will be discussed. Such divergences cannot be renormalized
properly. These divergences become independent of the temperature at
the minimum of the effective potential. That is just at the point at
which one has to evaluate the effective potential in order to obtain
the physical quantities like the pressure and the phase diagram. At
the minimum, these divergences are no problem because there they can
be renormalized in a systematic way.

The subject of Chapter 4 is the thermodynamics of the nonlinear sigma
model in two dimensions. This model has some aspects with QCD in
common such as asymptotic freedom. Using the $1/N$ expansion it is
possible to investigate this model in the domain where it is
non-perturbative in the coupling constant. By studying this model one
can learn something about QCD at finite temperature, for example how
the behavior of an asymptotically free theory changes when going from
low to high temperatures. In this chapter, the effective potential
will be calculated to next-to-leading order (NLO) in $1/N$. It turns out
that this effective potential contains temperature-dependent
divergences which can only be renormalized properly at the minimum. By
minimizing the effective potential, the pressure as a function of
temperature is obtained. It is found that $1/N$ is a good
expansion method. Furthermore it turns out that the pressure divided
by the pressure in the limit of infinite temperature is almost
independent of $N$. Similar behavior is found in lattice calculations
of SU($N$) Yang-Mills theory.

Chapter 5 discusses the thermodynamics of the $\mathbb{C}P^{N-1}$
model in two dimensions. This model is an extension of the nonlinear
sigma model in which also electromagnetism is taken into account.
This model shares also some features with QCD, it contains for example
like QCD so-called topological non-trivial vacua. The influence of
these vacua on the pressure is investigated using the same methods as
in Chapter 4. It is found that the non-trivial vacua give a large
contribution in the region in which the pressure as a function of the
temperature raises quickly. This result is a possible indication that
topological non-trivial vacua are important for the thermodynamics of
QCD in the neighborhood of the phase transition.

In Chapter 6 the thermodynamics of the linear and nonlinear sigma
model in four dimensions is investigated. These models are low-energy
effective theories of QCD. The pressure of these models is calculated
to NLO in $1/N$. The most important result of these calculations is a
prediction for the pressure of two-flavor QCD at low temperatures. The
NLO correction is compared to predictions of this correction in the
literature. These predictions turn out to be wrong. Finally a
relatively low upper bound on the mass of the sigma meson is found.
The mass of the sigma meson has not yet been determined very
accurately experimentally. If this upper bound is a result of the
approximations or occurs really in QCD should be investigated further.

The phase diagram of the NJL model is examined in Chapter 7. The NJL
model is a low-energy effective theory of QCD as well. The phase
diagrams are calculated by minimizing the effective potential. The
calculations have been carried out for situations in which the
densities of the up, down and the strange quark are different, such as
in a compact star or in a heavy-ion collision. Next to the phase in
which one can find ordinary matter, different color-superconducting
phases and phases in which the pseudoscalar mesons condense are
found. The results of this chapter are an extension of earlier
calculations in which either only the color-superconducting phases or
the phases in which the pseudoscalar mesons condense were taken into
account. The main result is that these previous calculations are
extended, such that one gets a more complete picture of the phase
diagram.  Furthermore the competition between the
color-superconducting phase and the phase in which pseudoscalar mesons
condense is studied.  It is found that they are separated by a
first-order phase transition. A possible extension of this work would
be the inclusion of color and electric neutrality constraints, such
that obtain the phase diagram of a compact star.

 %summary and outlook
\chapter*{Samenvatting}
\addcontentsline{toc}{chapter}{Samenvatting}
\fancyhf{}
\fancyhead[LE,RO]{\thepage} 
\fancyhead[RE,LO]{{\it Samenvatting}}

Het standaardmodel is {\it de} natuurkundige theorie die de deeltjes
waaruit materie is opgebouwd en hun onderlinge wisselwerkingen (met
uitzondering van de zwaarte\-kracht) beschrijft. Dit model heeft tot
nu toe, op het aantonen dat {\it neutrino's} een hele kleine massa
hebben na, alle experimentele testen glansrijk doorstaan.  Hoewel de
wisselwerkingen tussen de deeltjes onderling dus goed begrepen zijn,
is het gedrag in situaties waarbij een heleboel deeltjes met elkaar
wisselwerken vaak beperkt be\-kend. Een voorbeeld van zo'n situatie is
materie onder extreem hoge temperaturen en dichtheden, zoals tijdens
de oerknal en in hele compacte sterren. In dit proefschrift zijn
enkele berekeningen uitgevoerd die tot een beter begrip van materie
onder zulke extreme omstandigheden kunnen leiden.

\section*{Van atomen naar quarks}
De materie om ons heen is opgebouwd uit atomen. Een atoom bestaat uit
een positief geladen {\it kern} waar {\it elektronen}, die negatief
geladen zijn, zich omheen bevinden. De elektronen blijven in de buurt
van de kern omdat positief en negatief geladen \mbox{deeltjes} elkaar via
de {\it elektromagne\-tische} wisselwerking aantrekken. De
elektromagnetische kracht tussen twee ladingen wordt veroorzaakt door
de uitwisseling van {\it fotonen} (hetzelfde soort deeltjes waaruit
licht bestaat). Deze kracht wordt beschreven door {\it
quantum\-elektrodynamica} (QED), de quantumtheorie van het
elektromagnetisme. QED is het best geteste onderdeel van het
standaardmodel.

De kern bevat bijna alle massa van het atoom en is ook weer opgebouwd
uit deel\-tjes: positief geladen {\it protonen} en neutrale {\it
neutronen}. De kern van het waterstofatoom bestaat uit \'e\'en proton,
terwijl bijvoorbeeld het stabiele goudatoom 79 protonen en 118
neutronen bevat.

Omdat protonen elektrisch geladen zijn, zorgt de elektromagnetische
wisselwer\-king voor een afstotende kracht tussen de
protonen. Desondanks zijn er veel stabiele atoomkernen waarin de
protonen en neutronen toch bij elkaar blijven. Daarom moet er nog een
aantrekkende kracht tussen de protonen en de neutronen bestaan. Deze
kracht wordt de {\it sterke wisselwerking} genoemd. Het blijkt dat de
aantrekkende kracht tussen protonen en neutronen een gevolg is van de
wisselwerkingen tussen de {\it quarks}, dat zijn de deeltjes waaruit
protonen en neutronen zijn opgebouwd.

Een proton is opgebouwd uit drie quarks, twee zogenaamde {\it up
quarks} en een {\it down quark}. Een neutron bestaat uit een up
quark en twee down quarks. Naast de elektrische lading hebben de
quarks nog een zogenaamde {\it kleurlading}. Quarks oefenen een
kracht op elkaar uit door uitwisseling van {\it gluonen}. Deze kracht
is op hele kleine afstand veel sterker dan de elektromagnetische
kracht, waardoor een proton en ook sommige atoomkernen stabiel zijn.

Naast het up en het down quark zijn er nog vier andere zwaardere
soorten quarks bekend, te weten het {\it strange}, {\it charm}, {\it
bottom} en {\it top} quark. Met al deze quarks kunnen allerlei
(instabiele) deeltjes gevormd worden. Maar vanwege de onderlinge
interacties kunnen lang niet alle combinaties van quarks
voorkomen. Bijvoorbeeld, een quark als vrij deeltje is
onmogelijk. Onder normale omstandigheden vormen quarks altijd met z'n
drie\"en een deeltje (bijvoorbeeld een proton of een neutron) of met
z'n twee\"en. In het laatste geval vormt het quark tezamen met een
antiquark een deeltje, een zogenaamd meson. Een voorbeeld hiervan is
het positief geladen {\it pion}, dat een gebonden toestand
van een up quark en een anti-down quark is.

De wisselwerkingen tussen de quarks worden beschreven door {\it
quantumchromodynamica} (QCD) (zie formule \ref{eq:QCDLagrangian}).
QCD is net als QED een onderdeel van het standaardmodel. Met behulp
van deze theorie kunnen bepaalde uitkomsten van botsings\-experimenten
tussen bijvoorbeeld twee protonen worden voorspeld.  De resultaten
van deze experimenten stemmen overeen met de voorspellingen, wat dus
betekent dat de fundamentele wisselwerkingen tussen de quarks in
principe goed begrepen zijn.

\section*{Materie onder extreme omstandigheden}
In dit proefschrift is theoretisch onderzocht wat met materie
opgebouwd uit quarks gebeurt onder extreem hoge temperaturen, zoals
tijdens de oerknal en bij hele grote dichtheden, zoals in een compacte
ster. In beide gevallen hebben we natuurlijk te maken met een heleboel
quarks, veel meer dan in een bot\-singsexperiment tussen twee
protonen. Net zoals het ondoenlijk is om te voorspellen hoe een
individueel watermolekuul zich door water beweegt, kunnen we in dat
geval geen uitspraken doen over het gedrag van ieder individueel
quark. Maar het is wel mogelijk om uitspraken te doen over het gedrag
van alle quarks tezamen, net zoals water te karakteriseren is door
collectieve grootheden, als de temperatuur, de dichtheid en de
druk. De relaties tussen deze grootheden kunnen met behulp van {\it
thermodynamica} (warmteleer) worden uitgerekend en zijn ondere andere
van belang voor het begrijpen van het gedrag van compacte sterren en
het ontstaan van het heelal.

Water kan in verschillende toestanden (ook wel {\it fasen} genoemd)
voorkomen, te weten ijs, vloeibaar en damp. Al deze fasen zijn
natuurlijk opgebouwd uit dezelfde watermolekulen, alleen gedraagt de
verzameling watermolekulen zich in iedere fase compleet anders. In ijs
zijn de watermolekulen relatief sterk aan elkaar gebonden. Als de
temperatuur verhoogd wordt, gaan de watermolekulen sneller trillen, op
een gegeven moment trillen ze zo hard dat ze niet meer in een soort
roostervorm bij elkaar kunnen worden gehouden, water gaat dan over
naar de vloeibare fase. Bij de overgang naar waterdamp gebeurt
ongeveer hetzelfde.

Algemeen wordt aangenomen dat iets dergelijks ook met quarks moet
gebeuren. Zoals eerder is aangegeven, vormen quarks met z'n twee\"en of
met z'n drie\"en een deeltje zoals het neutron. Neem nu eens in
gedachten een groot aantal neutronen in een doos. Als de temperatuur laag
is dan blijven de quarks bij elkaar, maar als de temperatuur verhoogd
wordt kan het gebeuren dat de quarks zo hard gaan trillen, dat ze niet
meer bij elkaar kunnen worden gehouden in een neutron. De materie komt
dan in een nieuwe fase die het {\it quark-gluonplasma} wordt
genoemd. In deze fase hoeven de quarks niet langer met z'n twee\"en of
drie\"en bij elkaar te blijven. De temperatuur waarbij de faseovergang
optreedt is ongeveer twee biljoen graden Celsius, dat is maar liefst
honderdduizend keer de temperatuur in het binnenste van de zon.  Vlak
na de oerknal was het heelal waarschijnlijk eventjes een
quark-gluonplasma.  Er zijn sterke aanwijzingen dat het
quark-gluonplasma onlangs voor een hele korte tijd in het la\-boratorium
is gemaakt tijdens een experiment in het Brookhaven National
Laboratory, New York, waar goudkernen met enorme snelheden tegen
elkaar werden geschoten.

De neutronen in de doos kunnen natuurlijk ook op elkaar gedrukt
worden.  Op een gegeven moment wordt de dichtheid zo groot dat alle
neutronen tegen elkaar aanzitten. Als er dan nog harder gedrukt wordt,
dan gaan de neutronen overlappen, zodat ze hun identiteit als
individuele deeltjes verliezen. In dit geval kan er kan een
faseovergang optreden naar een zogenaamde {\it kleur-supergeleidende}
fase.  Dit gebeurt bij enorm grote dichtheden van ongeveer honderd
biljoen kilogram per liter. In theorie zou die faseovergang bereikt
kunnen worden in het binnenste van een {\it neutronenster}. Een
neutronenster is een uitgebrande ster met een enorm hoge dichtheid.
Tot nu zijn er echter nog geen neutronensterren gevonden die zo'n hoge
dichtheid hebben dat ze zich met zekerheid in de kleur-supergeleidende
fase bevinden.

De bovenstaande beweringen over het gedrag van deeltjes onder extreme
omstandigheden kunnen in principe wat preciezer gedaan worden door
berekeningen met QCD uit te voeren. Het blijkt dat de krachten tussen
de quarks minder sterk worden naarmate de energie van de quarks groter
wordt. Dit fenomeen heet {\it asymptotische vrij\-heid}. Als gevolg
daarvan is het mogelijk {\it analytische} berekeningen (berekeningen
met pen en papier) met QCD bij hele hoge energie\"en uit te voeren.
Maar deze energie\"en zijn nog veel hoger dan de energie\"en die de
quarks hebben rond de faseovergangen.  Een manier om QCD bij lagere
energie\"en te bekijken is met behulp van computersimu\-laties ({\it
lattice QCD}). Het probleem van deze computersimulaties is dat ze niet
goed werken bij grote dichtheden.  Om toch iets te voorspellen en te
begrijpen over het gedrag van QCD bij temperaturen en dichtheden rond
de faseovergangen kun je eenvoudiger theorie\"en bekijken die QCD tot
op zekere hoogte benaderen (zoals in hoofdstuk 6 en 7 van dit
proefschrift) of die bepaalde aspecten met QCD gemeen hebben (zoals in
hoofdstuk 4 en 5) (vandaar de uitdrukking {\it QCD-inspired theories}
in de titel).

Een belangrijke uit te rekenen grootheid is de druk (aangeduid met het
symbool $\mathcal{P}$) van materie die uit quarks is opgebouwd, als
functie van de temperatuur ($T$). Het resultaat van de exacte
berekeningen is afgebeeld in figuur \ref{fig:pertqcdpres}. De grijze
banden geven de foutmarges aan. Duidelijk is te zien dat voor lagere
temperaturen (in het bijzonder rond de faseovergang) de resultaten van
de berekeningen onbetrouwbaar zijn (omdat de krachten tussen de quarks
dan te groot worden). Het resultaat van de computerberekeningen is
voor een verschillend aantal quarks afgebeeld in figuur
\ref{fig:fullQCDpres}. De resultaten van deze computerberekeningen
zijn betrouwbaar voor temperaturen in de buurt van de faseovergang
(ongeveer 170 MeV, dat correspondeert met 2 biljoen graden Celsius) en
hoger. Om de druk bij lagere temperaturen te voorspellen kunnen
QCD-achtige theorie\"en gebruikt worden, zoals in hoofdstuk 6 van dit
proefschrift is gedaan.  In figuur \ref{fig:presN4h1} is het resultaat
van de berekening van de druk afgebeeld. De onderbroken lijn in deze
figuur is een voorspelling is van de druk van QCD bij lage
temperatuur.

Een ander interessant voorbeeld is het {\it fasediagram} van QCD.  Het
fasediagram geeft, afhankelijk van bijvoorbeeld de dichtheid en de
temperatuur, aan in welke fase materie zich bevindt. In figuur
\ref{fig:qcdphase} is het fasediagram van QCD {\it geschetst}.  Op de
horizontale as van dit diagram staat $\mu_B$ afgebeeld, een grootheid
die een directe relatie heeft met de dichtheid (hoe groter $\mu_B$,
hoe groter de dichtheid). Op de verticale as staat de temperatuur. Het
quark-gluonplasma en de kleur-supergeleidende fase zijn in dit dia\-gram
terug te vinden.  Het is belangrijk te weten dat dit diagram een
schets is, het is namelijk niet precies bekend waar de lijnen die de
verschillende fasen van elkaar scheiden in het diagram liggen. Dat
komt doordat dit diagram gebaseerd is op berekeningen aan theorie\"en
die QCD benaderen in een bepaald gebied. In het fasediagram afgebeeld
in figuur \ref{fig:qcdphase} zijn de dichtheden van de verschillende
soorten quarks allemaal gelijk. In hoofdstuk 7 van dit proefschrift
zijn, met behulp van een QCD-achtige theorie, fasedia\-grammen
uitgerekend waarbij de dichtheden van de quarks allemaal verschillend
zijn gekozen.

\section*{Overzicht en resultaten van dit proefschrift}
Dit proefschrift getiteld, ``Thermodynamics of QCD-inspired theories''
(Thermodynamica van QCD-achtige theorie\"en), gaat dus
over het gedrag van materie onder extreme omstandigheden.

In hoofdstuk 1 wordt een algemene inleiding gegeven. Eerst wordt
ingegaan op de vraag wat er zou kunnen gebeuren met materie onder
extreme omstandigheden. Daarna worden in het kort de relevante
aspecten van quantumchromodynamica (QCD) uitgelegd. Vervolgens worden
eerdere berekeningen van de druk en het fasediagram en de problemen
die daar bij komen kijken behandeld. Ook wordt in hoofdstuk 1 aandacht
besteed aan situaties waarbij zulke extreme omstandigheden kunnen
voorkomen, te weten in het heelal vlak na de oer\-knal, tijdens
zware-ionenbotsingen en in neutronensterren. Tot slot wordt uitgelegd
wat er bedoeld wordt met QCD-achtige theorie\"en.

Hoofdstuk 2 gaat over hoe met een zogenaamde {\it
quantumveldentheorie} als QCD berekeningen kunnen worden uitgevoerd bij
eindige temperatuur en dichtheid. Daarvoor wordt een statistische
analyse gebruikt. Eerst wordt er ingegaan op hoe dat in zijn werk gaat
voor een klassieke theorie, daarna worden de resultaten
veralgemeniseerd voor een quantummechanische aanpak. Met behulp van
het zogenaamde {\it pad-integraal} forma\-lisme wordt uitgelegd hoe
dit kan worden gebruikt voor een quantumveldentheorie. In berekeningen
bij eindige temperatuur moeten vaak sommaties over een oneindig aantal
termen worden uitgerekend.  Een methode om dat exact te doen wordt
uitgelegd, en er wordt ingegaan op hoe er met oneindigheden, die in
deze berekeningen kunnen optreden, moet worden omgegaan. Ook wordt ruim
aandacht besteed aan hoe deze sommen voor ingewikkelder gevallen met de
computer kunnen worden berekend.

Het derde hoofdstuk behandelt de zogenaamde {\it effectieve
potentiaal} en de $1/N$ benadering. De effectieve potentiaal is een
belangrijke grootheid, aangezien door het minimaliseren van die
potentiaal de druk en de fasediagrammen kunnen worden verkregen. De
$1/N$ benadering is een methode waarbij inzicht kan worden verkregen
in het lage energie gedrag van de theorie\"en die in dit proefschrift
worden behandeld. Er wordt uitgelegd hoe deze benadering kan worden
afgeleid door het introduceren van extra velden. Tot slot wordt
ingegaan op het feit dat in berekeningen van de effectieve potentiaal
er oneindigheden kunnen ontstaan die van de temperatuur
afhangen. Zulke oneindigheden vormen een grote bedreiging voor de
betekenis en bruikbaarheid van de antwoorden. Een belangrijk resultaat
uit dit proefschrift is dat deze oneindigheden
temperatuursonafhankelijk worden op het minimum van de effectieve
potentiaal. Dat is juist het punt waar de druk en het fasediagram
worden bepaald. In het minimum zijn deze oneindigheden geen probleem
omdat ze dan door subtracties en herdefinities van bepaalde constanten
op een systematische manier kunnen worden verwijderd zonder de
betekenis van de antwoorden te be\"invloeden. Deze procedure wordt ook
wel {\it renormalisatie} genoemd.

Het onderwerp van hoofdstuk 4 is de thermodynamica van het
niet-lineaire sigma model in twee dimensies (een ruimte- en een
tijddimensie). Dit model heeft bepaalde aspecten met QCD gemeen, zoals
asymptotische vrijheid. Maar met behulp van de $1/N$ benadering kan
hier het lage energie gedrag wel analytisch worden berekend.  Door dit
model te bestuderen kunnen we iets leren van het gedrag van QCD bij
eindige temperatuur, bijvoorbeeld hoe een asymptotische vrije theorie
zich gedraagt als er van lage naar hoge energie wordt gegaan.  In dit
hoofdstuk is de effectieve potentiaal uitgerekend met behulp van de
$1/N$ benadering tot en met de eerste niet-triviale correctie (ook wel
next-to-leading order (NLO) genoemd). Er worden
temperatuursafhanke\-lijke oneindigheden gevonden, die blijken te kunnen
worden gerenormaliseerd op het minimum. Vervolgens wordt de druk als
functie van de temperatuur uitgerekend. Met dit resultaat kunnen we
ondere andere concluderen dat de $1/N$ benadering een goede
rekenmethode is. Verder lijkt het gedrag van de druk gedeeld door de
druk in the limiet van oneindige temperatuur, onafhankelijk van
$N$. Een soortgelijk resultaat is gevonden in computerberekeningen aan QCD.

Hoofdstuk 5 gaat over de thermodynamica van het $\mathbb{C}P^{N-1}$
model in twee dimensies. Dit model is een uitbreiding van het
niet-lineaire sigma model waarbij ook elektromagnetisme wordt
meegenomen. Ook dit model heeft weer een aantal aspecten met QCD
gemeen, het bevat bijvoorbeeld net als QCD zogenaamde {\it topologisch
niet-triviale} vacua. Het effect van deze vacua op de druk is
onderzocht. Daarbij wordt gebruik gemaakt van dezelfde methoden als in
hoofdstuk 4. Uit de berekeningen volgt dat de topologisch
niet-triviale vacua voor temperaturen waarbij de druk relatief snel
stijgt een grote bijdrage leveren. Dit resultaat is een mogelijke
aanwijzing dat deze niet-triviale vacua rond de faseovergangen in QCD
belangrijk zouden kunnen zijn.

In hoofdstuk 6 wordt de thermodynamica van het lineaire en het
niet-lineaire sigma model in een tijd- en drie ruimtedimensies
onderzocht. Algemene wordt aangenomen dat deze modellen een goede
benadering van QCD bij lage energie zijn. Opnieuw is met behulp van de
$1/N$ benadering de druk uitgerekend tot en met de eerste
niet-triviale correctie. Het belangrijkste resultaat van deze
berekening is een voorspelling van de druk van QCD met twee soorten
quarks bij lage temperaturen. Dit resultaat is vergeleken met eerdere
schattingen uit de literatuur, waarvan geconcludeerd wordt dat ze
foutief zijn.  Tot slot wordt er een vrij lage bovengrens gevonden op
de massa van het zogenaamde sigma meson. Deze massa is experimenteel
nog niet goed vastgesteld. Of deze bovengrens een gevolg is van de
benaderingsmethode of ook echt in QCD voorkomt zou verder moeten
worden onderzocht.

Hoofdstuk 7 gaat tenslotte over de fasediagrammen van het NJL model.
Dit model geldt ook als een theorie die QCD kan beschrijven bij lage
energie\"en. Door het mini\-maliseren van de effectieve potentiaal
worden de fasediagrammen uitgerekend. De berekeningen zijn uitgevoerd
voor situaties waarbij de dichtheden van de up, down en strange quark
verschillend kunnen zijn, zoals bijvoorbeeld in een ster of in een
zware-ionenbotsing. Naast de fase waarin materie zich bij normale
omstandigheden bevindt, zijn verschillende kleur-supergeleidende fasen
en fasen waarin pseudoscalaire mesonen condenseren gevonden. De
resultaten uit dit hoofdstuk zijn een uitbreiding op eerdere
berekeningen uit de literatuur waarin of alleen rekening met de
kleur-supergeleidende fase of met de fase waarin pseudoscalaire
mesonen condenseren wordt gehouden. In dit hoofdstuk worden beide
mogelijkheden meegenomen.  Het belangrijkste resultaat is dat de
eerdere berekeningen van fasediagrammen zijn uitgebreid. Verder is de
competitie tussen de twee typen fasen onderzocht, een belangrijk
resultaat is dat ze gescheiden zijn door een zogenaamde eerste-orde
faseovergang.
\vspace{0.1cm} \\
Dit onderzoek heeft er toe geleid dat het begrip van het gedrag van
materie onder extreme omstandigheden iets is toegenomen. Toch blijven
er nog een heleboel zaken onbegrepen, zoals bijvoorbeeld de precieze
vorm van het QCD fasediagram. Hopelijk zullen verbeterde
(computer)berekeningen en nieuwe experimenten (bijvoorbeeld bij
CERN in Gen\`eve) ons in de toekomst meer leren.

 %samenvatting
\chapter*{Acknowledgments}
\addcontentsline{toc}{chapter}{Acknowledgments}
\fancyhf{}
\fancyhead[LE,RO]{\thepage} 
\fancyhead[RE,LO]{{\it Acknowledgments}}

Hoewel er maar \'e\'en naam op het kaft van het proefschrift staat hebben
veel mensen op een of andere manier bijgedragen aan dit
werk. Allereerst wil ik mijn copromotor, Dani\"el Boer bedanken voor de
uitstekende begeleiding en de prettige samenwerking in de afgelopen
vier jaar. Ik vond het fijn dat je bijna altijd tijd voor me had en ik
heb veel geleerd van alle discussies die ik met je gevoerd heb. Ik
waardeer het zeer dat je me zo vrij hebt gelaten. Verder heb ik aan de
vele idee\"en en opmerkingen van Jens Andersen veel gehad. Ik vond het
leuk om een aantal artikelen met je te schrijven. Ook mijn promotor
Piet Mulders wil ik bedanken voor zijn ondersteuning.

Mijn werkgever, de Vrije Universiteit, wil ik bedanken voor de prettige
werkomge\-ving en de financi\"ele ondersteuning die mij het onder meer
mogelijk gemaakt heeft om conferenties te bezoeken.

I would like to thank all members of the reading committee,
consisting of Jens Andersen, Ben Bakker, Dirk Rischke, Jan Smit and
Chris van Weert for taking the time to read the manuscript. All your
suggestions were very welcome.

Furthermore, I would like to acknowledge Rob Pisarski for some useful
discussions. Thanks for giving me the opportunity to present part of
my work at a seminar at Brookhaven National Laboratory. I would also
like to thank the theory group of BNL for the pleasant stay during
that week. Furthermore I would like to thank \mbox{Krishna} \mbox{Rajagopal} for the
possibility to visit the Center for Theoretical Physics of MIT for
some days, and Edward Shuryak for giving me the opportunity to give a
seminar at Stony Brook University.

I also appreciate it very much that the organizers of the national
theoretical high-energy physics seminar, the organizers of the Strong
and Electroweak matter conference in Helsinki, the organizers of the
43rd Cracow School of \mbox{theoretical} physics, and the organizers
of the extreme QCD workshop in Swansea gave me the chance to give a
talk about my work. All these meetings were very useful to me.

Ubel Warringa wil ik bedanken voor het grondig doorlezen van de
Nederlandse samenvatting en zijn nuttige opmerkingen.

Graag wil ik bij deze alle medewerkers van de afdeling theoretische
natuurkunde van de Vrije Universiteit bedanken voor de leuke sfeer en
de vele discussies tijdens de koffiepauzes en op andere momenten, in
het bijzonder Fetze Pijlman, Calina Ciuhu, Miranda van Iersel, Cedran
Bomhof, Erik Wessels, Hugo Schouten, \mbox{Klaus Scharnhorst}, Paul
Becherer, David Fokkema, Paul Visser, Klaas Allaart, Taco Visser, Henk
Blok, Ben Bakker, Alessandro Bacchetta, Philipp H\"agler, Gui Milhano,
Hartmut Erzgr\"aber, Adri Lodder en Daan Lenstra.

Tot slot wil ik al mijn vrienden bedanken, mijn familie en niet te
vergeten mijn broertjes en mijn ouders.

\vspace{1.0cm}
Amsterdam, november 2005
 %acknowledgements

\newpage

\fancyhf{} 
\fancyhead[LE,RO]{\thepage} 
\fancyhead[RE]{\itshape\bibname}
\fancyhead[LO]{\itshape\bibname}

\bibliographystyle{authordate1}
\bibliography{thesis}

\end{document}